\documentclass[11pt]{article}
\pdfoutput=1 

\usepackage{jheppub} 

\usepackage[T1]{fontenc} 

\usepackage[export]{adjustbox}
\usepackage{color}
\usepackage{xcolor}
\usepackage{autobreak}
\usepackage{multirow}
\usepackage{subfig}
\usepackage{amsmath}
\usepackage[vcentermath]{youngtab}
\usepackage{ytableau}
\usepackage{pifont}
\usepackage{extarrows}
\usepackage{makecell}
\usepackage{ulem} 
\usepackage{bbm}
\usepackage{diagbox}
\usepackage{tikz}
\usepackage[compat=1.1.0]{tikz-feynhand}
\usetikzlibrary{snakes}
\usetikzlibrary{arrows}
\usetikzlibrary{plotmarks}

\newcommand{\Amp}[7] 
{\begin{tikzpicture}[baseline=0.7cm,scale=0.75,transform shape] \begin{feynhand}
\vertex [particle] (i1) at (0,0.8) {$1^{#1}$};
\vertex [particle] (i2) at (1.6,1.6) {$2^{#2}$};
\vertex [particle] (i3) at (1.6,0) {$3^{#3}$};
\vertex (v1) at (0.9,0.8);
\graph{(i1)--[plain,#4,very thick] (v1)};
\graph{(i2)--[plain,#5,very thick] (v1)};
\graph{(i3)--[plain,#6,very thick] (v1)};
#7
\end{feynhand} \end{tikzpicture}}
\newcommand{\one}{\draw plot[mark=x,mark size=2.7,mark options={rotate=0}] coordinates {(0.65,0.8)};}
\newcommand{\oneone}{\draw plot[mark=x,mark size=2.7,mark options={rotate=0}] coordinates {(0.55,0.8)};\draw plot[mark=x,mark size=2.7,mark options={rotate=0}] coordinates {(0.75,0.8)};}
\newcommand{\two}{\draw plot[mark=x,mark size=2.7,mark options={rotate=45}] coordinates {(0.9+0.7*0.3,0.8+0.8*0.3)};}
\newcommand{\twotwo}{\draw plot[mark=x,mark size=2.7,mark options={rotate=45}] coordinates {(0.9+0.7*0.19,0.8+0.8*0.19)};\draw plot[mark=x,mark size=2.7,mark options={rotate=45}] coordinates {(0.9+0.7*0.42,0.8+0.8*0.42)};}
\newcommand{\three}{\draw plot[mark=x,mark size=2.7,mark options={rotate=45}] coordinates {(0.9+0.7*0.3,0.8-0.8*0.3)};}
\newcommand{\threethree}{\draw plot[mark=x,mark size=2.7,mark options={rotate=45}] coordinates {(0.9+0.7*0.19,0.8-0.8*0.19)};\draw plot[mark=x,mark size=2.7,mark options={rotate=45}] coordinates {(0.9+0.7*0.42,0.8-0.8*0.42)};}

\usepackage[utf8]{inputenc}
\usepackage{times} 
\usepackage{mathptmx} 

\newcommand{\beq}{\begin {equation}}  
\newcommand{\eeq}{\end   {equation}} 
\newcommand{\bea}{\begin {eqnarray}} 
\newcommand{\eea}{\end   {eqnarray}}  
\newcommand{\baa}{\begin {array}   } 
\newcommand{\eaa}{\end   {array}   }     
\newcommand{\bit}{\begin {itemize} }
\newcommand{\eit}{\end   {itemize} }
\newcommand{\be }{\begin {equation}} 
\newcommand{\ee }{\end   {equation}}

\newcommand{\eq}[1]{\begin{equation}\begin{split} #1 \end{split}\end{equation}}

\newcommand{\bbmc}{\mathbbm{C}}

\newcommand{\jhyu}[1]{{\color{cyan}{JHY: #1}}}

\newcommand\undermat[2]{
	\makebox[0.5pt][l]{$\smash{\underbrace{\phantom{%
					\begin{matrix}#2\end{matrix}}}_{ \let\scriptstyle\textstyle\text{ $#1$}}}$}#2}
\newcommand\overmat[2]{
	\makebox[-1pt][l]{$\smash{\overbrace{\phantom{%
					\begin{matrix}#2\end{matrix}}}^{ \let\scriptstyle\textstyle\text{ $#1$}}}$}#2}    

\allowdisplaybreaks

\title{Massive Helicity-Chirality Spinor Formalism from Massless Amplitudes with On-shell Mass Insertion}

\author[a,b]{Yu-Han Ni, }
\author[a,b]{Yi-Ning Wang, }
\author[a]{Chao Wu, }
\author[a, b, c]{Jiang-Hao Yu, }

\affiliation[a]{CAS Key Laboratory of Theoretical Physics, Institute of Theoretical Physics, Chinese Academy of Sciences, Beijing 100190, China}
\affiliation[b]{School of Physical Sciences, University of Chinese Academy of Sciences, Beijing 100049, P.\ R.\ China}
\affiliation[c]{School of Fundamental Physics and Mathematical Sciences, Hangzhou Institute for Advanced
Study, UCAS, Hangzhou 310024, China}

\emailAdd{niyuhan@itp.ac.cn}
\emailAdd{wangyining@itp.ac.cn}
\emailAdd{wuch7@itp.ac.cn}
\emailAdd{jhyu@itp.ac.cn}

\abstract{

We introduce a helicity-chirality spinor formalism to describe scattering amplitudes for particles of any masses and spins. The massive spin-spinors introduced by Arkani-hamed-Huang-Huang have been extended to the spin/helicity-transversality spinors, in which a new quantum number transversality, closely related to chirality, is introduced by extending the Poincare symmetry. The massive helicity-chirality amplitudes can be written by the large and small components of massless spinors $\lambda$ and $\eta$ following the $\lambda \sim \sqrt{E}, \eta \sim \mathbf{m}/\sqrt{E}$ expansion order by order, which formulate the power counting rules of a large energy effective theory. Diagrammatically the mass expansion in amplitudes originates from the on-shell mass insertion: the helicity flip and chirality flip, which completely determines the three-point massive amplitudes. From the chirality-helicity unification at the UV, any massive helicity-chirality amplitude can be one-to-one corresponded to massless helicity amplitudes with (without) additional Higgs insertion. This UV-IR correspondence explains the mass enhancement in the weak decay processes $\pi^+ \to \mu^+ \nu$ and $t \to W^+ b$, and isolates the correct UV of the 3-point massive QED $F\bar{F}\gamma$ amplitudes in Arkani-hamed-Huang-Huang formalism. From massless-massive correspondence, the massless on-shell techniques can be utilized to construct higher-point massive amplitudes, taking $F\bar{F}VS$ and $e^-e^+ \to \mu^- \mu^+$ as examples.

}

\begin{document} 
\maketitle
\flushbottom


\section{Introduction}

Recent years the scattering amplitudes of massless particles have achieved tremendous progress on understanding gauge theories and gravity both theoretically and practically. In the on-shell program, the spinor-helicity variables~\cite{Parke:1986gb,Bern:1996je,Dixon:1996wi,Elvang:2013cua,Cheung:2017pzi,Travaglini:2022uwo,Badger:2023eqz} are introduced to remove huge gauge redundancy and trivialise the kinematical on-shell constraints. The massless three-particle amplitudes are entirely fixed by the consistence condition from little group (LG) scaling~\cite{Benincasa:2007xk,Witten:2003nn}, while the four-particle amplitudes can be determined from consistent factorization due to locality and unitarity. The higher-point amplitudes can be built by either gluing lower-point together in the bootstrap approach or using the recursive relations such as the BCFW construction~\cite{Britto:2004ap,Britto:2005fq}.

The familiar spinor-helicity formalism for massless particles has been extended to describe massive particles~\cite{Kleiss:1985yh,Hagiwara:1985yu,Kleiss:1988xr,Dittmaier:1998nn,Schwinn:2005pi,Schwinn:2006ca,Badger:2005zh,Badger:2005jv,Conde:2016vxs,Conde:2016izb,Basile:2024ydc}, such as the top quark, W and Z bosons in the standard model (SM). Arkani-Hamed, Huang and Huang (AHH) developed the spin-spinor formalism~\cite{Arkani-Hamed:2017jhn} for the massive on-shell amplitudes, which manifests the covariance of the $SU(2)$ massive LG. Following the logic of the massless three-point (3-pt) amplitudes, the Lorentz structures of massive 3-pt amplitudes have been enumerated, although some redundancies need to be removed from over-complete 3-pt amplitudes, by applying equation of motion (EOM) relation (see for example Ref.~\cite{Durieux:2020gip}), etc. Thus different from the massless case, the LG covariance of the Poincar\'e symmetry is not enough to fully determine the 3-pt amplitudes. Once the 3-pt massive amplitudes are obtained, the higher-point amplitudes can be constructed from the basic 3-pt amplitudes, expressed in the LG covariant way by the bolded form. The AHH formalism, together with the massless spinor-helicity, has been successfully applied to describe electroweak theory~\cite{Christensen:2018zcq, Franken:2019wqr, Bachu:2019ehv, Ballav:2020ese, Wu:2021nmq, Ballav:2021ahg, Liu:2022alx, Bachu:2023fjn,Ema:2024rss}, massive QCD or QED~\cite{Ochirov:2018uyq, Falkowski:2020aso,Lazopoulos:2021mna, Christensen:2022nja, Ema:2024vww, Campbell:2023fjg}, effective field theories~\cite{Shadmi:2018xan, Aoude:2019tzn, Durieux:2019eor, Ma:2019gtx, Durieux:2019siw, Li:2020gnx, Li:2020xlh, Li:2020tsi, Li:2020zfq, Durieux:2020gip, Li:2021tsq, AccettulliHuber:2021uoa, Dong:2021vxo, Li:2022tec, Balkin:2021dko, DeAngelis:2022qco, Dong:2022mcv, Ren:2022tvi, Liu:2023jbq, Goldberg:2024eot}, supersymmetry and new physics~\cite{Herderschee:2019ofc, Engelbrecht:2022aao, KNBalasubramanian:2022sae, Johansson:2023ymb,Falkowski:2020fsu,Alves:2021rjc,Bertuzzo:2023slg},  
gravitational wave physics~\cite{Guevara:2018wpp, Chung:2018kqs, Guevara:2019fsj, Bern:2019nnu, Maybee:2019jus, Arkani-Hamed:2019ymq}, etc.



The on-shell construction for massive amplitudes is usually performed by either the gluing procedure or the recursive relation, which correctly characterizes the pole structures. However, challenges have been encountered when a massless internal photon is involved in gluing 3-pt amplitudes with massive fermions. The on-shell constructive calculation~\cite{Christensen:2022nja,Christensen:2024bdt,Christensen:2024nre} on the simplest QED process $e^+e^- \to \mu^+\mu^-$ does not agree with the standard textbook result, indicating the existence of non-trivial ambiguous term. The ambiguity of terms also appears when the electroweak boson scattering amplitudes, such as $W^+W^- \to W^+W^-$, are constructed in the on-shell way. Additional condition has to be imposed to avoid such ambiguities, such as the tree-level unitarity condition~\cite{Liu:2022alx} to reproduce the $W^+W^- \to W^+W^-$ result, the old-fashioned perturbation theory and the all-line transverse recursion relation~\cite{Lai:2023upa, Ema:2024vww} to get the textbook result for $e^+e^- \to \mu^+\mu^-$ , etc. All of these indicate a lack of systematical understanding of the ultraviolet (UV, or large $z$) behaviors of the amplitudes in the AHH formalism, referred as the AHH amplitudes below.

In the AHH formalism, the massive spin-spinors $\lambda^I_\alpha$ and $\tilde{\lambda}^I_{\dot{\alpha}}$ can be decomposed as the direct product of the helicity-spinors $\lambda_\alpha$ and $\eta_\alpha$, and the spin vector basis $\zeta^{\pm I}$. This provides a connection to the massless amplitudes at the UV. For a massive amplitude at the infrared (IR), from the massless decomposition, it is possible to find several massless amplitudes at the UV through momentum deformation, which is called the IR unification. However, there is no way to identify these massless UV amplitudes and isolate the unwanted UV ones.
For example, the 3-pt $F\bar{F}\gamma$ amplitudes in the $e^+e^- \to \mu^+\mu^-$ process contain other massless UVs than the massless QED ones. The task would be how to isolate these unwanted UV amplitudes and thus it is necessary to build an UV-IR correspondence, instead of the IR unification.

In this work, we propose to extend the helicity-spinors $\lambda_\alpha$ and $\eta_\alpha$ to {\it the helicity-transversality spinors}, in which a new quantum number transversality is introduced to be carried by the helicity-spinors $\lambda_\alpha$ and $\eta_\alpha$. This is based on the spin and spinor space decomposition of {\it the spin-transversality (ST) spinor}, introduced by the authors in Ref.~\cite{Ni:2024yrr}. The ST spinor $\lambda_{A}^I = (\lambda^I_\alpha, \tilde{\lambda}^I_{\dot{\alpha}})$ unifies the AHH spin-spinors $\lambda^I_\alpha$ and $\tilde{\lambda}^I_{\dot{\alpha}}$ with different transversality, by the extended Poincar\'e symmetry, the $SO(5,1)$ symmetry. Unlike the AHH formalism only enumerating various possible 3-pt amplitudes, this $SO(5,1)$ symmetry completely determines the Lorentz structures of the 3-pt massive amplitudes. For example, in the AHH formalism, there are four independent Lorentz structures in $F\bar{F}V$ amplitudes, $\left\{\langle\mathbf{13}\rangle [\mathbf{23}], [\mathbf{13}]\langle\mathbf{23}\rangle, \langle\mathbf{13}\rangle \langle\mathbf{23}\rangle, [\mathbf{13}][\mathbf{23}] \right\}$, enumerated by the Poincare symmetry. While the above four Lorentz structures can actually be determined by the $SO(5, 1)$ symmetry, in another way to say, these Lorentz structures can be related by the $SO(5, 1)$  generators. Therefore, the helicity-transversality spinors would entirely fix the 3-pt massive amplitudes.

The transversality is closely related to the chirality. Thus the 3-pt massive amplitudes based on the ST spinors have been extended to describe the chirality flip effects, such as the $\mu$ mass enhancement in the charged pion decay, muon magnetic moment~\cite{Arkani-Hamed:2021xlp,Stockinger:2022ata}, flavor physics~\cite{Hall:1985dx,Gabbiani:1996hi,Misiak:1997ei}, etc. After decomposing the ST spinor into the helicity-transversality spinors, the helicity flip effect, which also parametrizes the mass insertion,  can be taken into account in the massless amplitudes. Therefore, both chirality flip and helicity flip effects can be described by the helicity-transversality spinors, denoted as {\it the on-shell mass insertion}. The on-shell mass insertion provides a systematic expansion on different orders $\eta \sim \frac{\textbf{m}}{\sqrt{E}}$, more precisely, $\lambda\cdot \eta \sim \mathbf{m}$ for massive amplitudes. Therefore, based on the helicity-transversality spinor formalism, any massive amplitude can be decomposed into categories with different helicity and chirality, referred as the massive helicity-chirality (MHC) amplitudes.

These MHC amplitudes help to build the UV-IR correspondence between massive amplitudes and its UV massless ones. We note, in the high energy limit, the chirality and helicity should be unified such that the left-handed (right-handed) chirality corresponds to $-$ ($+$) helicity. Therefore, each $n$-piont MHC amplitude should have a one-to-one correspondence to either $n$-piont massless helicity amplitudes or higher-point massless helicity amplitudes with additional Higgs scalars. When these additional scalars take vacuum expectation values (VEVs), the massless amplitude gives rise to the massive one via momentum deformation. In terms of helicity-transversality spinors, the AHH massive amplitudes can be decomposed into massive amplitudes with different transversality and helicity, which has a one-to-one correspondence to the UV massless amplitudes. To get rid of unwanted UV, one only needs to remove the corresponding 3-pt IR amplitude. For example, removing the unwanted UV in the 3-pt $F\bar{F}\gamma$ amplitudes would give rise to the correct results in the $e^+e^- \to \mu^+\mu^-$ process.

Compared to the AHH amplitudes, the MHC amplitudes contain two layers of mass expansions via chirality flip and helicity flip. Both flips characterize the on-shell mass insertion, or the mass expansion of $\eta$. Therefore, the AHH amplitudes can be treated as a series expansion of the MHC amplitudes order by order with the power counting rules of the mass expansion $\eta$, and with the building blocks of the helicity-transversality spinors. This provides us with an effective field theory (EFT) of massive amplitudes, which we call large energy effective theory (LEET) for any massive amplitudes~\footnote{This large energy effective theory is different from the EFT defined in Ref.~\cite{Dugan:1990de}.}, because it contains necessary ingredients of any EFTs: $\lambda$ and $\eta$ as building blocks, and their scaling behavior $\lambda \sim \sqrt{E}$ and $\eta\sim \frac{\mathbf{m}} {\sqrt{E}}$ indicating a mass expansion power counting rules $\eta\sim \frac{\textbf{m}}{\sqrt{E}}$, and the extended Poincar\'e symmetry. 
Different from other EFTs, the small component $\eta$ is not integrated out, and thus summing over all order of MHC amplitudes would reproduce the full AHH massive amplitudes. From the top-down view, the AHH amplitudes can be expanded into MHC amplitudes order by order. Inversely, from the bottom-up view, we can start from LEET amplitudes to construct the MHC amplitudes, and then by folding the layer of the helicity info and absorbing the chirality info by removing the mass dependence, it recovers the AHH amplitudes. A comparison of massive amplitudes in the above formalisms is as follows
\begin{equation}
\resizebox{.9\linewidth}{!}{
\begin{tabular}{c|cccc}
Formalism & AHH & \makecell{spinor-helicity\\AHH} & \makecell{spin-transversality\\spinor} & \makecell{helicity-transversality\\spinor}  \\
\hline
Building block & $\lambda^I_\alpha$,$\tilde{\lambda}^I_{\dot{\alpha}}$ & $\lambda_{\alpha},\eta_{\alpha},\tilde{\eta}_{\dot{\alpha}},\tilde{\lambda}_{\dot{\alpha}}$ & $\lambda^I_A=(\lambda^I_\alpha \quad \lambda^I_{\dot{\alpha}})$ & $\begin{pmatrix}
-\lambda_{\alpha}  &  \eta_{\alpha}  \\
 \tilde{\eta}_{\dot{\alpha}}  & \tilde{\lambda}_{\dot{\alpha}}
\end{pmatrix}$ \\
\hline
\makecell{Momentum\\conservation} & $\sum_i \mathbf{p}_{i\alpha\dot{\alpha}}=0$ & $\sum_i (p+\eta)_{i\alpha\dot{\alpha}}=0$ & $\sum_i p_{iAB}=0$ & \makecell{$\sum_i p_{i\alpha\dot{\alpha}}=0$\\$\sum_i\eta_{i\alpha\dot{\alpha}}=0$} \\
\hline
Little group & $SU(2)$ & $U(1)_w$ & $U(2)$ & $U(1)_w\times U(1)_z$ \\
\hline
\makecell{Little group\\transformation} & $\lambda^I_\alpha \rightarrow W_{J}^{I} \lambda^{J}$ & \makecell[l]{$\lambda\rightarrow w^{-1}\lambda$\\$\eta\rightarrow w \eta$\\$\tilde{\lambda}\rightarrow w \tilde{\lambda}$\\$\tilde{\eta}\rightarrow w^{-1} \tilde{\eta}$} & \makecell[l]{$\lambda_{\alpha}^{I}\rightarrow e^{-\frac{i}{2}\phi}W^I_J\lambda_{\alpha}^{J}$\\$\tilde{\lambda}_{\alpha}^{I}\rightarrow e^{\frac{i}{2}\phi}W^I_J\tilde{\lambda}_{\alpha}^{J}$} & \makecell[l]{$\lambda\rightarrow w^{-1}z^{-1}\lambda$\\$\eta\rightarrow w z^{-1}\eta$\\$\tilde{\lambda}\rightarrow w z \tilde{\lambda}$\\$\tilde{\eta}\rightarrow w^{-1} z\tilde{\eta}$} \\
\hline
3-point amp & Lorentz structure & HE limit & \makecell{determined by\\ $SO(5,1)$ symmetry} & \makecell{HE structure \\ with $U(1)_z$}\\
\hline
Power counting & No & No & No & Yes \\
\hline
\makecell{UV-IR\\correspondence} & not one-to-one & not one-to-one  & \makecell{one-to-one \\for all-massive amp} & one-to-one \\
\end{tabular} }
\end{equation} 
where amp is an abbreviation for amplitude, and HE denotes high energy limit.

The MHC amplitudes provide a bridge to connect both the massless amplitudes and the AHH massive amplitudes via the UV-IR correspondence. It can be incorporated into the EFT framework, and knowing a few LEET amplitudes could determine the AHH massive amplitudes by symmetry. Therefore, the first few terms of the MHC or LEET amplitudes are enough to obtain full massive amplitudes. We have applied this formalism to several electroweak processes.

The paper is structured as follows. In section~\ref{sec:extendPoincare}, after a brief review over the massless and massive spinors, we present the helicity-transversality spinors for all masses and spins, and then define the on-shell mass insertion via the helicity and chirality flips. In section~\ref{sec:3pt1} we consider the generic 3-pt contact amplitude and utilize the highest weight construction to determine the 3-massive MHC amplitudes. In section~\ref{sec:3pt2}, we also apply the highest weight construction to the 1-massless-2-massive amplitudes and build the one-to-one UV-IR correspondence to eliminate the unwanted UV structure for $F\bar{F}\gamma$ amplitudes. Then various 4-pt MHC amplitudes are constructed using the on-shell constructive method and constructive rules for internal particles are shown in section~\ref{sec:facAmp}. Finally we summarize in section~\ref{sec:sum}. 

\section{Massive Spinor and Massless Decomposition under $U(2)$ Little Group}\label{sec:extendPoincare}

In this section, the spacetime Poincar\'e symmetry is extended to the $ISO(3, 1) \times ISO(2)$ by introducing the internal $U(1)$ symmetry for the complex mass spurion $m$ and $\tilde{m}$. Following Wigner's construction ~\cite{Wigner:1939cj,Bargmann:1948ck}, the one-particle state contains a new quantum number, {\it transversality}. We will discuss the large and small component helicity spinors with transversality for any masses and spins using the spinor-helicity notation~\cite{Cheung:2017pzi}.

\subsection{Review of massless and massive spinors}

In the spinor-helicity formalism, a massless momentum vector $p_\mu$ can be mapped to a bi-spinor matrix  $p_{\alpha\dot{\alpha}}$, which is rank one due to $\textrm{det}\,p_{\alpha\dot{\alpha}} = 0$ and thus can be decomposed as the direct product of two spinors $\lambda, \tilde{\lambda}$,
\bea
p_{\alpha\dot{\alpha}} \equiv p_{\mu} \sigma^{\mu}_{\alpha\dot{\alpha}} = \lambda_{\alpha} \tilde{\lambda}_{\dot{\alpha}},
\eea
where the spinors $\lambda, \tilde{\lambda}$ transform respectively as the $(\frac12, 0)$ and $(0, \frac12)$ representations of the $SL(2,\bbmc)$ group.
For real momenta $p_{\alpha \dot{\alpha}}$ is Hermitian, implying the reality condition, $\tilde{\lambda}_{\dot{\alpha}}= \pm\left(\lambda^{*}\right)_{\dot{\alpha}}$.
While for complex momenta, the Lorentz group is extended to the complexified one $SO(3,1,\mathbbm{C}) \simeq SL(2, \bbmc) \otimes \widetilde{SL}(2, \bbmc)$, and thus $\lambda$ and $\tilde{\lambda}$ are  two independent complex spinors which transform as 
\bea
\lambda_{\alpha} \rightarrow(\Lambda \lambda)_{\alpha}, \quad \tilde{\lambda}_{\dot{\alpha}} \rightarrow(\widetilde{\Lambda} \tilde{\lambda})_{\dot{\alpha}},
\eea
where $\Lambda$, and   $\widetilde{\Lambda}$ are two arbitrary and independent $SL(2, \bbmc)$ matrices.

The spinors $\lambda, \tilde{\lambda}$ also transform under the LG, since one can always rescale 
\bea 
\lambda_{\alpha} \rightarrow w^{-1}\, \lambda_{\alpha}, \quad  \tilde{\lambda}_{\dot{\alpha}} \rightarrow  w \,\tilde{\lambda}_{\dot{\alpha}},
\eea
while keeping $p_{\alpha \dot{\alpha}}$ invariant. More explicitly, one first chooses a standard reference massless momentum $k_{\alpha \dot{\alpha}}=\lambda_{\alpha}^{(k)} \tilde{\lambda}_{\dot{\alpha}}^{(k)}$ and then performs Lorentz transformation from $k$ to momentum $p_{\alpha\dot{\alpha}}  = \lambda_{\alpha} \tilde{\lambda}_{\dot{\alpha}}$ by $\lambda_{\alpha}^{(p)} = \mathcal{L}(p ; k)_{\alpha}^{\beta} \lambda_{\beta}^{(k)}, \, \tilde{\lambda}_{\dot{\alpha}}^{(p)} = \tilde{\mathcal{L}}(p ; k)_{\dot{\alpha}}^{\dot{\beta}} \tilde{\lambda}_{\dot{\beta}}^{(k)}$. Thus the spinor $\lambda$ transforms under the LG for any Lorentz transformation $\Lambda$ as
\bea
\lambda_{\alpha}^{(p)} \to \lambda_{\alpha}^{(\Lambda p)}=w^{-1}(\Lambda, p, k) \Lambda_{\alpha}^{\beta} \lambda_{\beta}^{(p)},
\eea
and similarly $\tilde{\lambda}$. For complex momenta, $w$ is a complex scalar function, and thus the LG is $GL(1, \mathbbm{C})$, while  for real momenta the spinors satisfy a reality condition telling $w^{-1}= \pm(w)^{*}$ and so $w=e^{i \theta}$ is a phase representing the $U(1)$ LG.

For massive particle, its momentum $\mathbf{p}_{\alpha\dot{\alpha}}$ is rank two instead of one due to $\textrm{det}\,\mathbf{p}_{\alpha\dot{\alpha}} = m^2 \neq 0$. Thus it can be written as the sum of two rank-one bi-spinors, and there are usually two ways of decomposing the momenta
\bea
\mathbf{p}_{\alpha\dot{\alpha}} \equiv
\begin{cases}
\lambda_{\alpha}^I \tilde{\lambda}_{\dot{\alpha} I}, \quad I = 1, 2 \\
 p_{\alpha\dot{\alpha}} +\eta_{\alpha\dot{\alpha}} = \lambda_{\alpha} \tilde{\lambda}_{\dot{\alpha}} + \eta_{\alpha} \tilde{\eta}_{\dot{\alpha}}.
\end{cases}
\eea
Here two different kinds of spinors are introduced: 
\bit 
\item {\it the spin-spinors} $\lambda_{\alpha}^I, \tilde{\lambda}_{\dot{\alpha} I}$, introduced by AHH~\cite{Arkani-Hamed:2017jhn}, transforms under the $SU(2)$ ($GL(2, \bbmc)$ for complex momenta) LG, with the index $I = 1, 2$ made manifest; 
\item {\it the helicity-spinors} $\lambda_{\alpha}, \eta_{\alpha},\tilde{\lambda}_{\dot{\alpha}}, \tilde{\eta}_{\dot{\alpha}}$, introduced in Ref.~\cite{Schwinn:2005pi,Schwinn:2006ca,Craig:2011ws}, each transform under $U(1)$ ($GL(1, \bbmc)$ for complex momenta) LG, with the index naturally hidden inside the complex nature of the spinor. 
\eit
These two descriptions are equivalent and can be converted to each other via the induced $GL(1, \bbmc)$ representation of the   $GL(2, \bbmc)$ group discussed in the next subsection.

In the spin-spinor notation, in addition to the Lorentz indices, the LG indices $I$ are made explicit by transforming 
\bea
\lambda^{I} \rightarrow W_{J}^{I} \lambda^{J}, \quad \tilde{\lambda}_{I} \rightarrow\left(W^{-1}\right)_{I}^{J} \tilde{\lambda}_{J}, 
\label{eq:massiveLGTrans}
\eea
while keeping $\mathbf{p}_{\alpha\dot{\alpha}}$ invariant. 
Similar to the metric in the Lorentz space, $\epsilon^{I J}$ and $\epsilon_{I J}$ are introduced in the LG space to raise and lower indices so that one can write $\mathbf{p}_{\alpha \dot{\alpha}}=\lambda_{\alpha}^{I} \tilde{\lambda}_{\dot{\alpha}}^{J} \epsilon_{I J}$. Consider how the mass parameter transforms under the LG. The on-shell mass condition tells
\bea
\mathbf{p}^2 = \mathbf{m}^2 \rightarrow \operatorname{det} \lambda \times \operatorname{det} \tilde{\lambda}= \frac{1}{2}\lambda^{I\alpha} \lambda^J_{\alpha} \tilde{\lambda}_{J\dot{\alpha}} \tilde{\lambda}_I^{\dot{\alpha}} = m\tilde{m}, 
\eea
where $m$ and $\tilde{m}$ are spurions for complex momenta, and they are the mass parameter $m = \tilde{m} = \mathbf{m}$ when the complex momenta are real. For complex momenta, taken $\textrm{det}\lambda = m$ and $\textrm{det}\tilde{\lambda}=\tilde{m}$, the LG belongs to the $GL(2, \mathbbm{C})$ group if $\tilde{m}$ and $m$ are independent, and $SL(2, \bbmc)$ if $|\tilde{m}| = |m| = \mathbf{m}$, while for real momenta the LG reduces to $SU(2)$ with $\tilde{m} = m = \mathbf{m}$. 

In this subsection, following AHH formalism, the LG is taken to be $SL(2, \bbmc)$. In this case, the massive LG has two invariants
\bea
\begin{cases}
m = \frac{1}{2}\lambda_{\alpha I}\lambda^{\alpha I} = \lambda^{\alpha}\eta_{\alpha} , \\
\tilde{m} =  \frac{1}{2} \tilde{\lambda}_{\dot{\alpha}I}\tilde{\lambda}^{\dot{\alpha}I} = \tilde{\lambda}_{\dot{\alpha}}\tilde{\eta}^{\dot{\alpha}}. 
\end{cases}
\label{eq:mass-spurion-def}
\eea
The on-shell spinors automatically satisfy the  EOM using the definition $\mathbf{p}_{\alpha\dot{\alpha}}$ and eq.~\eqref{eq:mass-spurion-def}, which relate $\tilde{\lambda}^{I}$ and $\lambda^{I}$ as 
\bea
\mathbf{p}_{\alpha \dot{\alpha}} \tilde{\lambda}^{\dot{\alpha} I} = \tilde{m} \lambda_{\alpha}^{I}, \quad \mathbf{p}_{\alpha \dot{\alpha}} \lambda^{\alpha I} = -m \tilde{\lambda}_{\dot{\alpha}}^{I}.
\eea
The spin-$s$ particle state in the $(j, 2s - j)$ representation can be written as the direct product of  $\lambda^{I}$ and $\tilde{\lambda}^{I}$
\bea (j, 2s - j) \textrm{ representation}: \quad \lambda^{(I_1}\dots\lambda^{I_{I_j}} 
\tilde{\lambda}^{I_{j+1}}\dots\tilde{\lambda}^{I_{2s})}, 
\eea
where $(I_1\dots I_{2s})$ represents the totally symmetric spin indices.

In the helicity-spinor notation, the LG $SL(2, \mathbbm{C})$ (or its subgroup $SU(2)$) is realized via the induced $U(1)$ subgroup representations of the $SU(2)$ LG for complex (real) massive momenta. To relate the helicity-spinor with the spin-spinor, one introduces two 2-dimensional vectors $\zeta^{+}$ and $\zeta^{-}$ to span the 2-dimensional spin group space. Thus the spin-spinor can be expressed as the direct product of the $SU(2)$ spin group tensor and the $SL(2, \mathbbm{C})$ Lorentz group spinor 
\bea
\begin{cases}
\lambda_{\alpha}^{I} &= -\lambda_{\alpha} \zeta^{-I} + \eta_{\alpha} \zeta^{+I}, \\
\tilde{\lambda}_{\dot{\alpha}}^I &= \tilde{\lambda}_{\dot{\alpha}} \zeta^{+I} + \tilde{\eta}_{\dot{\alpha}} \zeta^{-I}, 
\end{cases}
\label{eq:massless-decompose-su2-ahh}
\eea
where $\zeta^{\pm I}$ is chosen to be the eigenstates of the spin operator along the direction of the spatial momentum, with the normalization $\epsilon_{IJ} \zeta^{+I} \zeta^{-J} =1$.

\subsection{Massive particle with extended Little group}
\label{sec:littlegroup}

In this subsection, we propose that the massive spinors $\lambda_{\alpha}^{I}$ can be extended to the one with additional complex phase, leading to the $U(2)$ LG transformation instead of the $SU(2)$ one. This extension is motivated by promoting the mass spurions $m$ and $\tilde{m}$ to be complex, even for real momenta.

Recall the two mass spurions $m$ and $\tilde{m}$ are defined as 
\begin{equation}
m \equiv \frac{1}{2}\lambda_{\alpha I}\lambda^{\alpha I},\quad \tilde{m} \equiv \frac{1}{2} \tilde{\lambda}_{\dot{\alpha}I}\tilde{\lambda}^{\dot{\alpha}I}. 
\end{equation}
For the real momentum, since $\mathbf{p}_{\alpha\dot{\alpha}}$ is Hermitian, the spinor variables satisfy the complex conjugation relation $\tilde{\lambda}_{\dot{\alpha}I} = \lambda_{\alpha}^{I*}$~\footnote{When the momenta become complex, this conjugation relation breaks down and the two mass spurions have no relation. }. 
In the AHH formalism, the massive spinors carry the Lorentz and LG indices and the mass spurions become real for real momenta satisfying the following condition $m = \tilde{m}= \mathbf{m}$. 
Suppose the mass spurions become complex while keeping $|\tilde{m}| = |m| = \mathbf{m}$, thus the condition becomes $\tilde{m} = m^*$, indicating that $m$ and $\tilde{m}$ carry opposite phase. We introduce an internal $U(1)$ symmetry that acts as
\bea
\label{eq:D-act-on-m}
m\rightarrow e^{-i\phi}m\;, \quad \;\tilde{m}\rightarrow e^{i\phi}\tilde{m}. 
\eea
Similar to $\mathbf{p}$ as a generator, we can think of both $m$ and $\tilde{m}$ as generators. The role of generators $m$ and $\tilde{m}$ is simply to multiply themselves. From the adjoint acting eq.~\eqref{eq:D-act-on-m} we find $e^{i\phi D_-}me^{-i\phi D_-}=me^{-i\phi}$ and its dual. Then the $U(1)$ generator $D_-$, $m$ and $\tilde{m}$ compose the Lie algebra of the $ISO(2)$ group:
\begin{equation}
    [D_-,m]=-m\;,\quad [D_-,\tilde{m}]=+\tilde{m}\;,\quad [m,\tilde{m}]=0.
\end{equation}
In this case, similar to the supercharge $Q$ and $\bar{Q}$ carrying the $U(1)$ R-charge in supersymmetry, we can treat the $m$ and $\tilde{m}$ as variables in an internal $ISO(2)$ space. However, unlike supercharge, the $m, \tilde{m}$ commute.


The spacetime Poincar\'e symmetry should be extended to accommodate this new $U(1)$ symmetry by considering the commutator among the Poincar\'e generators and new $m, \tilde{m}, D_-$ generators. $[m,\tilde{m}] = 0$ tells us that these new generators commute with the ones in the Poincar\'e algebra. Following the Coleman-Mandula theorem~\cite{Coleman:1967ad}, the spacetime Poincaré group can be extended to $ISO(2) \times ISO(3, 1)$. The commutator relations are shown in appendix~\ref{app:6D}, which also shows that the four-momentum $\mathbf{p}^\mu$ can be extended with two additional internal components with $(\mathbf{p}^\mu, m, \tilde{m})$.

Let us consider the single-particle state under this extended group. According to Wigner's construction, the single-particle state can be constructed by the induced representation via the LG. 
For the Poincar\'e group, the massive single-particle state is described by momentum $\mathbf{p}$, mass $\mathbf{m}$, spin $s$ and its component $\sigma$, denoted as $|\mathbf{p}, \mathbf{m}^2, s, \sigma\rangle$. 
In this new group, an additional $U(1)$ phase would introduce a new quantum number $t\in \frac{1}{2}\mathbbm{Z}$, where the half-integers are contained because we want to use them to describe the fermions, to label this $U(1)$ representation. Since neither $m$ nor $\tilde{m}$, but the combination $m \tilde{m}$ are invariant under this $U(1)$ symmetry, we would consider the LG transformation for the group $SO(2) \times ISO(3, 1)$~\footnote{In this case, the physical state is described by the $U(2)$ LG keeping $\mathbf{p}_\mu$ invariant. There is an equivalent description on physical state. Taking the maximal symmetry $ISO(2) \times ISO(3, 1)$, the physical state should be described by $|\mathbf{p}, m, \tilde{m}, s, s_z \rangle$ which belongs to the $SU(2)$ LG while keeping $p^{M}=(p^{\mu},m,\tilde{m})$ invariant. In this way, the transversality transformation is under the $SO(2) \times SO(3, 1)$ spinor space, not the LG space. }. The single-particle state is lifted from $|\mathbf{p}, \mathbf{m}^2, s, \sigma\rangle$ to $|\mathbf{p}, m\tilde{m}, t, s, \sigma\rangle$ such that
\begin{equation}
    D_{-}|\mathbf{p}, m\tilde{m}, t, s, \sigma\rangle = \ t \ |\mathbf{p}, m\tilde{m}, t, s, \sigma\rangle. 
\end{equation}
Here the new quantum number $t$ is denoted as {\it the transversality}, and the Casimir $\mathbf{p}^2$ eigenvalue $\mathbf{m}^2$ is replaced by $m\tilde{m}$. 
The LG transformation of the states is $U(2)$ instead of $SU(2)$. For a Lorentz transformation $\Lambda$ and $SO(2)$ rotation (parameterized by $\phi$), we have: 
\beq
U(\Lambda,\phi)|\mathbf{p}, m\tilde{m}, t, s, \sigma\rangle = \sum_{\sigma'}|\Lambda \mathbf{p}, m\tilde{m}, t, s, \sigma'\rangle W_{\sigma' \sigma}(\Lambda, \mathbf{p}) e^{-it\phi}. 
\label{eq:states-little-group-trans}
\eeq
Here $W_{\sigma' \sigma}(\Lambda, \mathbf{p})$ is the LG corresponding to $\Lambda$ and $\mathbf{p}$.

The LG transformation can also be applied to the spinor-helicity $\lambda_{\alpha}^{I}$ with an additional $U(1)$ phase, which can be absorbed due to the complex nature of spinors. Similar to the transformation for the state in eq.~\eqref{eq:states-little-group-trans}, the massive spinors have the following $U(2)$ LG transformation
\begin{equation}
    \lambda_{\alpha}^{I}\rightarrow e^{-\frac{i}{2}\phi}W^I_J\lambda_{\alpha}^{J}\;,\quad\tilde{\lambda}_{\dot{\alpha}I}\rightarrow e^{\frac{i}{2}\phi}(W^{-1})_I^J\tilde{\lambda}_{\dot{\alpha}J}. 
\end{equation}
Thus the massive spinors carry not only the Lorentz indices $(j_1, j_2)$ and spin indices~\footnote{In the 4-dimensional quantum field theory Language, the fields carry $SO(3, 1)$ representation $(j_1, j_2)$ while particle carries $SU(2)$ LG index. Thus the wave functions are needed to relate the spacetime group and the corresponding LG. For spin-$1/2$ particle, the spinor $\lambda_{\alpha}^I$ is chosen to be the wave function of the fermion. }, but also the transversality index $t$. In the following, let us denote the representation of the massive spinors as $[t, j_1, j_2]$. We take $\lambda^I$ and $\tilde{\lambda}^I$ as the following $[t, j_1, j_2]$ irreducible representation
\bea
{\textrm{spin-}}1/2: && \quad \tilde{\lambda}^I \ \sim \ [1/2, 0, 1/2], \ \ \lambda^I \ \sim \  [-1/2, 1/2, 0]. \label{eq:primaryhalf-transversality}
\eea  
Note that the $\lambda^I$ and $\tilde{\lambda}^I$ are spin-$\frac12$ wave-functions for fermion with left-handed $L$ and right-handed $R$ chirality, or anti-fermion with opposite chirality. 
Thus transversality is closely related to chirality: the $L$ ($R$) chirality corresponds to $t = -\frac12$ ($+\frac12$) for fermion, verse vice for anti-fermion, but with difference: although $\lambda^I$ and $\tilde{\lambda}^I$ carry both transversality and chirality, $m$ and $\tilde{m}$ carry transversality but no chirality. 

Let us construct the higher-spin spinors with transversality. Note that different from the spin-spinors in the Poincar\'e group, although the mass spurions are both Lorentz and spin singlets, they carry transversality, but not chirality. So the massive spinors with transversality should contain mass spurions as an ingredient, but massive spinors with chirality do not need to contain mass spurions. Let us first consider the general spin-$s$ spinors with chirality. 
From the basic spinors, the spin-$s$ {\it spin-chirality spinors} can be written as
\beq
\textrm{spin}-s: \quad \lambda^{(I_1}_{\alpha_1}\dots\lambda^{I_{2s-2t}}_{\alpha_{2s-2t}}\tilde{\lambda}^{I_{2s-2t+1}}_{\dot{\alpha}_{1}} \tilde{\lambda}^{I_{2s+2t})}_{\dot{\alpha}_{2s+2t}}. \label{eq:wavefunc}
\eeq
Listing all possible $\lambda$ and $\tilde{\lambda}$s, we obtain a total of $2s+1$ ($t=-s,\dots, s$) spinors~\footnote{In general the single-particle state shouldn't give such $t=-s,\dots, s$ constraint and the value of $t$ can be any half-integer. The constraint in fact comes from the 4D reduction of the SO(5,1).}. Further, each spin-chirality spinor with fixed $t$ has $2s + 1$ spin components and all these are related by the $SU(2)$ Lie algebra generators. 
Unlike the AHH spin-spinors which only take one of the above $2s + 1$ spinors by the EOM conversion, the spin-chirality spinors take all $2s + 1$ ones running over all the transversality. The constraint $t = -s,\dots, s$ suggests that the $U(1)$ is not enough, and there should be a larger group that relates different transversality.


In the following let us consider the generators of the $ISO(2) \times ISO(3, 1)$ group, and try to extend this group further. 
In the spinor-helicity formalism, any amplitudes can be expressed using the $\lambda_{\alpha I},\tilde{\lambda}_{\dot{\alpha} I}$. The generators of the $ISO(2) \times ISO(3, 1)$ group can be induced by the infinitesimal transformation of the spinors for any function $M(\lambda_{\alpha I},\tilde{\lambda}_{\dot{\alpha} I})$: 
 \bea
 M(\lambda_{\alpha I},\tilde{\lambda}_{\dot{\alpha} I}) &\rightarrow& M(\lambda_{\alpha}^I - i\omega^{IJ}\lambda_{\alpha J},\tilde{\lambda}_{\dot{\alpha} I} - i\omega^{IJ}\tilde{\lambda}_{\dot{\alpha}J }) = M(\lambda_{\alpha I},\tilde{\lambda}_{\dot{\alpha} I}) - \frac{1}{2}\omega^{IJ} J_{IJ} M(\lambda_{\alpha I}, \tilde{\lambda}_{\dot{\alpha} I}), \\
 M(\lambda_{\alpha I}, \tilde{\lambda}_{\dot{\alpha} I}) &\rightarrow& M(\lambda_{\alpha}^I - \frac{i}{2}\phi\lambda_{\alpha}^I, \tilde{\lambda}_{\dot{\alpha} I} - \frac{i}{2}\phi\tilde{\lambda}_{\dot{\alpha}I}) = M(\lambda_{\alpha I}, \tilde{\lambda}_{\dot{\alpha} I}) - \phi D_-M(\lambda_{\alpha I}, \tilde{\lambda}_{\dot{\alpha} I}). 
 \eea
 Here we choose the parameter $\omega^{IJ}$ to be symmetric so that $\omega^{IJ}\epsilon_{IJ} = 0$ and then there would be no phase term in the transformation. To obtain the $U(1)$ part of the $U(2)$ LG, we simply replace $\omega^{IJ} J_{IJ}$ by $\phi D_{-}$, and then find
\bea
    J_{IJ} &\equiv& \frac{i}{2}\left(\lambda_{\alpha I}\frac{\partial}{\partial \lambda^{J}_{\alpha}} + \lambda_{\alpha J}\frac{\partial}{\partial \lambda^{I}_{\alpha}} + \tilde{\lambda}_{\dot{\alpha} I}\frac{\partial}{\partial \tilde{\lambda}^{J}_{\dot{\alpha}}} + \tilde{\lambda}_{\dot{\alpha} J}\frac{\partial}{\partial \tilde{\lambda}^{I}_{\dot{\alpha}}}\right), \\
D_- &\equiv&\frac{1}{2}\left(-\lambda_{\alpha}^{I}\frac{\partial}{\partial \lambda_{\alpha}^{I} }+\tilde{\lambda}_{\dot{\alpha}}^{I}\frac{\partial}{\partial \tilde{\lambda}_{\dot{\alpha}}^{I}}\right), 
\eea
where the $SU(2)$ part can be extracted by redefining $J$ as the more familiar forms
\begin{equation}
    J_i \equiv \frac{i}{2}J_{IJ}\sigma_{iK}^{J}\epsilon^{KI}, 
\end{equation}
with $\sigma_{iK}^{J}$ the Pauli matrices.
These generators give the spinor information e.g. we could use $J^2$ to find the amplitudes that correspond to the definite particle spin with different spin components. Moreover, we could use the different form of the $SU(2)$ generators to acquire the components information by defining
\begin{equation}
    J_{\pm}\equiv J_{1}\pm i J_{2}. 
\end{equation}

Similar to $J_\pm$, let us try to relate the spin-chirality spinors of different transversality. The well-known CP transformation chirality of particle between the left-handness and right-handness. So CP could relate the spinors with transversality $t$ and $-t$. 
Put it more generally, we need a transformation to relate amplitudes with different transversality, and thus construct the following generators:
\begin{equation}
    T^{+}_{\alpha \dot{\alpha}}\equiv \tilde{\lambda}_{\dot{\alpha}}^{I}\frac{\partial}{\partial \lambda^{\alpha I}}, \quad 
    T^{-}_{\alpha \dot{\alpha}}\equiv \lambda_{\alpha}^{I}\frac{\partial}{\partial \tilde{\lambda}^{\dot{\alpha}I}}. 
\end{equation}
Here $T^{\pm}$ simply replace $\lambda_{\alpha}^I$($\tilde{\lambda}_{\dot{\alpha}}^I$) by $\tilde{\lambda}_{\dot{\alpha}}^I$($\lambda_{\alpha}^I$) so they do change one unit of transversality. The action of $T^{\pm}_{\alpha,\dot{\alpha}}$ to the basic spinors takes the form
\beq
{T^+}^{\alpha}_{\dot{\alpha}}\circ \lambda_{\beta}^I = \delta_{\beta}^{\alpha}\tilde{\lambda}_{\dot{\alpha}}^I\;, \;{T^-}^{\alpha}_{\dot{\alpha}}\circ \lambda_{\beta}^I=0, 
\eeq
\beq
{T^-}^{\alpha}_{\dot{\alpha}}\circ \tilde{\lambda}_{\dot{\beta}}^I = \delta_{\dot{\beta}}^{\dot{\alpha}}\lambda_{\alpha}^I\;, \;{T^+}^{\alpha}_{\dot{\alpha}}\circ \tilde{\lambda}_{\dot{\beta}}^I = 0. 
\eeq
Therefore, the $T^{\pm}$ generator relates the spinors with different transversality. This indicates that the $T^{\pm}_{\mu}$ generator and our extended spacetime group form a larger group. Together with the Lorentz algebra
\bea
[L_{\alpha\beta}, L_{\gamma\delta}] &=& i (\epsilon_{\gamma\beta} L_{\beta\delta} + \epsilon_{\delta\beta} L_{\alpha\gamma} + \epsilon_{\gamma\alpha} L_{\beta\delta} + \epsilon_{\delta\alpha} L_{\beta \gamma}), 
\eea
the commutator relations take the form
\bea
[(T^+)_{\alpha\dot{\alpha}}, (T^-)_{\beta\dot{\beta}} ] &=& -\frac{i}{2}( \epsilon_{\alpha\beta} \tilde{L}_{\dot{\alpha}\dot{\beta}} +\tilde{\epsilon}_{\dot{\alpha}\dot{\beta}} L_{\alpha\beta} ) -\frac{1}{2}\epsilon_{\alpha\beta}\tilde{\epsilon}_{\dot{\alpha}\dot{\beta}} D_-, \\
{[D_-, {(T^{\pm})}_{\alpha\dot{\alpha}} ]} &=& \pm{(T^{\pm})}_{\alpha\dot{\alpha}}, \\
{[L_{\beta\gamma}, (T^{\pm})_{\alpha\dot{\alpha}}}] &=& -i\epsilon_{\beta\alpha}(T^{\pm})_{\gamma\dot{\alpha}}-i\epsilon_{\gamma\alpha}(T^{\pm})_{\beta\dot{\alpha}}. 
\eea
The above algebra is in the closed form. Thus the generators $\{T^{\pm}, D_{-},L\}$ can be organized in the form of the regular $SO(5, 1)$ algebra with one kind of generators $ \{L_{MN} \} $ for $M, N=0, 1, 2, 3, 4, 5$: 
\bea    [L_{QR}, L_{MN}] = i\left(\eta_{QM}L_{NR} - \eta_{QN}L_{MR} - \eta_{RM}L_{NQ} + \eta_{RN}L_{MQ}\right),  
\eea
where $\eta_{MN}$ is the metric in the 6-dimensional (6D) space time which is chosen to be $diag(1, -1, -1,\allowbreak -1, -1, -1)$. 
Note this group is isomorphism to the 4-dimensional (4D) conformal group. For the conformal group, the primary and descendent representation can be used, although now the ladder generators are not the $(P^\mu, K^\mu)$ generators, but $(T^{+\mu}, T^{-\mu})$.

When extending $SO(5, 1)$ to $ISO(5, 1)$, more generators are considered such as $\{T^{\pm}, D_{-}, L, P, m,\allowbreak  \tilde{m}\}$, then the commutator relations
\bea
{[P_{\alpha\dot{\alpha}}, {T^+}^{\beta}_{\dot{\beta}}]} &=& -\tilde{m} \delta_{\alpha}^{\beta} \tilde{\epsilon}_{\dot{\alpha}\dot{\beta}}, \\
{[P_{\alpha\dot{\alpha}}, {T^-}^{\dot{\beta}}_{\beta}]} &=& -m \tilde{\delta}^{\dot{\beta}}_{\dot{\alpha}} \epsilon_{\alpha\beta}, \\
{[m, {T^+}^{\alpha}_{\dot{\alpha}}]} &=& - P^{\alpha}_{\dot{\alpha}}, \\
{[\tilde{m}, {T^-}^{\dot{\alpha}}_{\alpha}]} &=& P^{\dot{\alpha}}_{\alpha},\\
{[m, D_{-}]} &=&  m, \\
{[\tilde{m}, D_{-}]} &=& -\tilde{m}, 
\eea
give a $ISO(5, 1)$ Lie algebra. If we define $m = p_4 + ip_5$, $\tilde{m} = p_4 - ip_5$, the $ISO(5,1)$ symmetry in 4D spacetime can be viewed as the Poincar\'e symmetry in 6D spacetime. In this case, the massive particle description above can be viewed as a massless particle in extended 6D spacetime. This indicates the 4D amplitudes do contain some 6D structure. However, different from the 6D description~\footnote{In 6D spinor-helicity formalism, the LG of the 6D Poincar\'e symmetry should be $SU(2) \times SU(\overline{2})$, in which only the first $SU(2)$ is taken to be the spin group in 4D spacetime. If we introduce momenta $p_{AB}$, since it can be decomposed as direct product of fundamental and anti-fundamental representations, we should also have inequivalent $\overline{\lambda}^{\overline{I}}_B$ transforming under another $SU(\overline{2})$ LG space, which constitutes $SU(2) \times SU(\overline{2})$ LG for $ISO(5, 1)$. However, since the massive scattering amplitudes do not contain transformation in another $SU(\overline{2})$, the $ISO(5, 1)$ symmetry should be broken. }, the LG transformation of the massive scattering amplitudes is still $U(2)$, which explicitly break the $ISO(5, 1)$ symmetry. In the following, we will see the spin-chirality spinors satisfy the $SO(5, 1)$ symmetry, but not all the spinors satisfy this maximal symmetry.

The $SO(5, 1)$ symmetry would relate different components of the spin-chirality spinor $\lambda^I_A = \{\lambda^I_\alpha, \lambda^I_{\dot{\alpha}} \}$. For general spin-$s$ particle, the spin-chiraility spinor takes the form $\lambda^{(I_1}_{A_1}\cdots \lambda^{I_{2s})}_{A_{2s}}$, and the components are related by the $T^\pm$ generator
\begin{eqnarray}
\begin{array}{|c|c|c|c|c|c|}
\hline
t & -s & -s+1 & -s+2 & ... & s \\
\hline
\textrm{spin-chirality spinor}& \lambda^{(I_1}...\lambda^{I_{2s})} & \lambda^{(I_1}...\lambda^{I_{2s-1}} \tilde{\lambda}^{I_{2s})} & \lambda^{(I_1}...\lambda^{I_{2s-2}} \tilde{\lambda}^{I_{2s-1}} \tilde{\lambda}^{I_{2s})} & ... & \tilde{\lambda}^{(I_1} ... \tilde{\lambda}^{I_{2s})} \\
\hline
\end{array}
\end{eqnarray}
Note that in this spinor construction, we have removed all the mass spurions. Adding mass spurions into the above spinors would change the transversality. In general, we could add mass spurions back and then the spinors with the form 
\bea
\lambda^{(I_1}_{\alpha_1}...\lambda^{I_{2j_1}}_{\alpha_{2j_1}}\tilde{\lambda}^{I_{2s-2t+1}}_{\dot{\alpha}_{1}}\tilde{\lambda}^{I_{2j_2})}_{\dot{\alpha}_{2j_2}}m^a \tilde{m}^b,
\eea
which take the transversality $t = j_2 + b -j_1 - a$. In the aspect of the representation theory, they also belong to an irreducible representation of the $SO(5, 1)$ group.



In the following let us construct all possible irreducible representations of the $SO(5, 1)$ group. 
The commutator $[D_-,T^\pm_{\mu}]$ tell us that $T^+$ ($T^-$) will increase (lower) the eigenvalue of $D_-$. For a finite dimension $SO(5, 1)$ representation, after finite-time action of $T^+$ the eigenvalue of $D_-$ will reach its maximum $\Delta$. The subspace of the representation with transversality $\Delta$ could be decomposed by the $SO(3, 1)$ subgroup because $\Delta$ is invariant under the action of $SO(3, 1)$. We could find a $SO(3, 1)$ representation $(J_1, J_2)$ with transversality $\Delta$. Then using $T^-$ action we could find the total representation. Thus we say that three quantum numbers $(\Delta, J_1, J_2)$ label an irreducible representation of the $SO(5, 1)$ group. We also say that the maximal $D_-$-valued subspace of the representation primary and the rest part descendant which are obtained by the action of $T^-$ on the primary subspace.

\begin{figure}[htbp]
    \centering
    \includegraphics[scale=0.65]{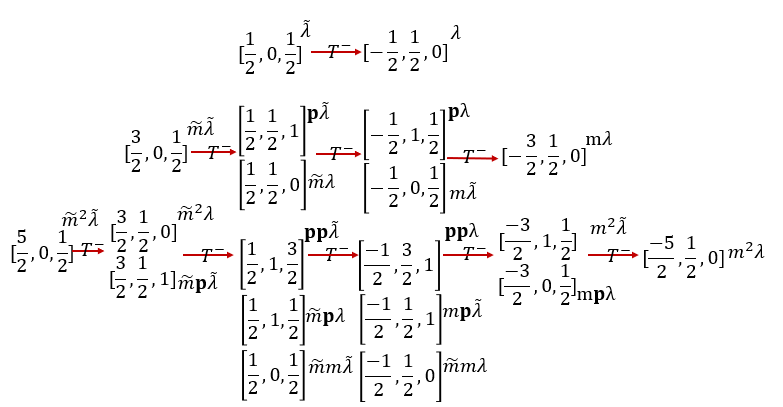}
    \caption{The $(\Delta, 0, \frac{1}{2})$ representation of the $SO(5, 1)$ symmetry with $\Delta = \frac{1}{2},\frac{3}{2},\frac{5}{2}$, and its reduction to the $[t, j_1, j_2]$ representations of the $SO(2) \times SO(3, 1)$ symmetry. The spinors are also shown above each representation, in which the symmetric spinor indices are hidden. }
    \label{fig:spin-half-chain}
\end{figure}

\begin{figure}[htbp]
    \centering
    \includegraphics[scale=0.65]{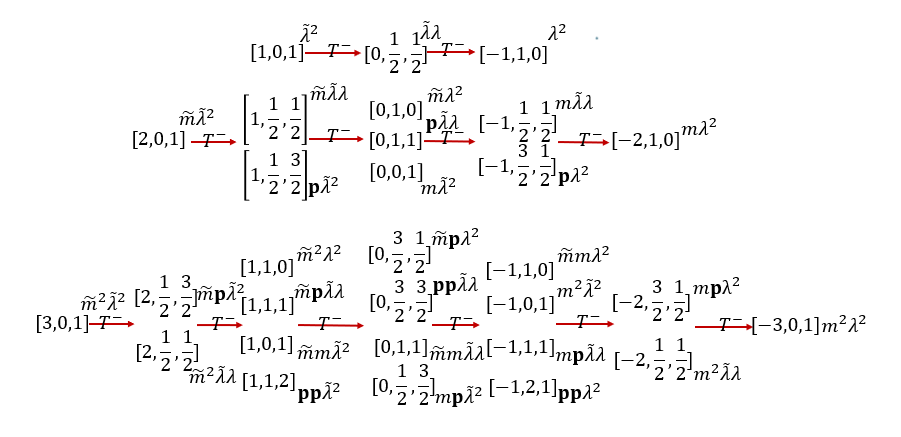}
    \caption{Same as figure~\ref{fig:spin-half-chain} but $(\Delta, 0, 1)$ with $\Delta = 1, 2, 3$. }
    \label{fig:spin-one-chain}
\end{figure}

To construct the most general spinors, let us first identify the highest weight representation, which can only contain $\tilde{\lambda}$ and $\tilde{m}$, and thus the form should be 
\bea
\tilde{\lambda}^{(I_{2s-2t+1}}_{\dot{\alpha}_{1}}\dots\tilde{\lambda}^{I_{2s})}_{\dot{\alpha}_{2s}}\tilde{m}^b, 
\eea 
in which the three quantum number read as $(s+b, 0, s)$. 
For all the descendant representations, we only care about those $J_1 = 0$ representations. For the particle states, the 6D momentum $p^{M} = (\mathbf{p}^\mu, m, \tilde{m})$ now is massless and its LG structure is $SO(4)$. In order to relate to the 4D massive particle states, the $SO(3)$ subgroup of the $SO(4)$ is the LG, so the single-particle states carry quantum number spin-$s$. Note that $b$ comes from the number of 6D momentum and it corresponds to the direct product of the 6D $p^{M}$ and the spin-chirality spinor. For the spin-chirality spinor, it takes $\Delta = J_2 = s, J_1 = 0$. Here we show the more general $SO(5, 1)$ representations, contain the product of the spin-chirality spinor and 6D momentum. Take $s = \frac{1}{2}, \Delta = \frac{1}{2}, \frac{3}{2}, \frac{5}{2}$ we obtain the spinor in figure~\ref{fig:spin-half-chain} and similarly $s = 1, \Delta=1, 2, 3$ in figure~\ref{fig:spin-one-chain} by action of the $T^-$ generators. Note that we can list infinite representations with increasing  $\Delta$ for such $SO(5, 1)$ group, but not all of them are utilized in the massive amplitudes. 

\begin{figure}[htb]
    \centering
    \includegraphics[scale=0.65]{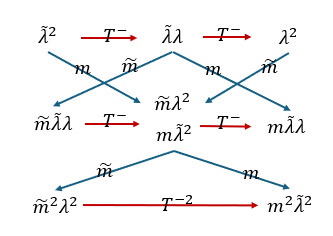}
    \caption{The ST spinor for spin-1 particle. Same as figure~\ref{fig:spin-half-chain} but $(\Delta,0,1)$ with $\Delta=1,2,3$ and the selection condition in eq.~\eqref{eq:so51condition}. }
    \label{fig:spin-one-wavefunction}
\end{figure}

Although the massive spin-$s$ spinors with
different transversality can be related by the $SO(5, 1)$ symmetry,  the massive spinors containing the 6D momentum would contain redundant components under the U(2) LG. 
This is because the 6D momentum contains trivial 4D Lorentz invariant $m$ and $\tilde{m}$. The following procedure can be performed: 
\bit 
\item Since the 4D momentum is $\mathbf{p}^2=m\tilde{m}$ so all the 4D momentum should be thought as an additional structure. Thus we should only keep the spinors with the form $\lambda^{(I_1}_{\alpha_1}...\lambda^{I_{2j_1}}_{\alpha_{2j_1}}\tilde{\lambda}^{I_{2s-2t+1}}_{\dot{\alpha}_{1}}\tilde{\lambda}^{I_{2j_2})}_{\dot{\alpha}_{2j_2}}m^a \tilde{m}^b$. 
\item Since the mass terms appear in the form of $m\tilde{\lambda}^I = -\mathbf{p}\lambda^I$ and $\tilde{m}\lambda^I = \mathbf{p}\tilde{\lambda}^I$, there should be no $m$($\tilde{m}$) if there is no right(left)-handed spinor $\tilde{\lambda}^I$ ($\lambda^I$) in the amplitude. The $m$ and $\tilde{m}$ should not appear altogether because of $\mathbf{p}^2=m\tilde{m}$. 
\item There exist finite different mass structures but infinite momentum structures. The momentum structure can be dropped in 4D representations. 
\eit 
All of these constrain the quantum numbers: $0 \leq a \leq 2j_2, 0 \leq b \leq 2j_1, ab = 0$. For the mass structure, note that $b\leq 2j_1$ and $j_1 \leq s $ thus the highest power of $\tilde{m}$ is $3s$. Finally, the transversality $t = j_2 - j_1 + b - a$ satisfies: 
\beq
-(j_1 + j_2) = j_2 - j_1 - 2j_2 \leq t \leq j_2 - j_1 + 2j_1 = j_1 + j_2. 
\eeq
Since $s = j_1 + j_2$ we find the constraint is nothing but $-s \leq t \leq s$ as we have mentioned before. Note that these representations $(t, j_1, j_2)$ of $SO(2) \times SO(3, 1)$ we selected satisfy condition $t = -s,\dots, s, j_1 + j_2 = s, \Delta - s = |t - (j_2 - j_1)|$. The $\Delta - s$ in the last condition counts the number of 6D momentum and equals the number of $m$ or $\tilde{m}$. With all of these, we find that the spinors of $s = 1, \Delta = 1, 2, 3$ should take the form as shown in figure~\ref{fig:spin-one-wavefunction}.

There is an equivalent but simpler construction on all the massive spinors with all possible transversality. Note for each $(j_1, j_2)$, the spin-chirality spinors only describe one kind of transversality, and there should have $2s + 1$ spinors running over all the transversality. So the spin-chirality spinor should be extended to {\it the ST spinor} with {\it the selection condition}: 
\bea
s = j_1 + j_2, \  |t| \leq s, \  s \leq \Delta \leq 3s, \  |t - j_2 + j_1| = \Delta - s. 
\label{eq:so51condition}
\eea
For spin-$1/2$ particle, we contract $\lambda^{I}$ and $\tilde{\lambda}^{I}$ with $\mathbf{p}_{\alpha \dot{\alpha}}$, and utilize the EOM
$\mathbf{p}_{\alpha \dot{\alpha}} \tilde{\lambda}^{\dot{\alpha} I} = \tilde{m} \lambda_{\alpha}^{I}, \ \mathbf{p}_{\alpha \dot{\alpha}} \lambda^{\alpha I} = -m \tilde{\lambda}_{\dot{\alpha}}^{I}$, and we obtain other transversality
\bea
{\textrm{spin-}}1/2: && \quad \tilde{m}\lambda^I \ \sim \ [1/2, 1/2, 0], \ \ m \tilde{\lambda}^I \ \sim \ [-1/2, 0, 1/2]. \label{eq:descendanthalf-transversality}
\eea
This provides us a systematic way of constructing the ST spinors: first start from primary representation with the highest $t$, and apply $T^-$ to obtain all primary representations, then contract indices with $\mathbf{p}$ to obtain descendant representations, and finally select the ones satisfying the condition in eq.~\eqref{eq:so51condition}. For spin-$1$, the complete ST spinors are
\bea
\begin{tabular}{c|c|c|c}
\hline
& $t = 1$ & $t = 0$ & $t = -1$ \\
\hline
$\Delta = 1$ & $\tilde{\lambda}^{(I_{1}}_{\dot{\alpha}_{1}}\tilde{\lambda}^{I_{2})}_{\dot{\alpha}_{2}}$ & $\lambda^{(I_{1}}_{\alpha_{1}}\tilde{\lambda}^{I_{2})}_{\dot{\alpha}_{1}}$ & $\lambda^{(I_{1}}_{\alpha_{1}}\lambda^{I_{2})}_{\alpha_{2}}$ \\
$\Delta = 2$ & $\lambda^{(I_{1}}_{\alpha_{1}}\tilde{\lambda}^{I_{2})}_{\dot{\alpha}_{1}}\tilde{m}$  & $\tilde{\lambda}^{(I_{1}}_{\dot{\alpha}_{1}}\tilde{\lambda}^{I_{2})}_{\dot{\alpha}_{2}}m,\lambda^{(I_{1}}_{\alpha_{1}}\lambda^{I_{2})}_{\alpha_{2}}\tilde{m}$ & $\lambda^{(I_{1}}_{\alpha_{1}}\tilde{\lambda}^{I_{2})}_{\dot{\alpha}_{1}}m$ \\
$\Delta = 3$ & $\lambda^{(I_{1}}_{\alpha_{1}}\lambda^{I_{2})}_{\alpha_{2}}\tilde{m}^2$  &  & $\tilde{\lambda}^{(I_{1}}_{\dot{\alpha}_{1}}\tilde{\lambda}^{I_{2})}_{\dot{\alpha}_{2}}m^2$ \\
\hline
\end{tabular}
\eea

\begin{figure}[htbp]
    \centering
    \includegraphics[scale=0.65]{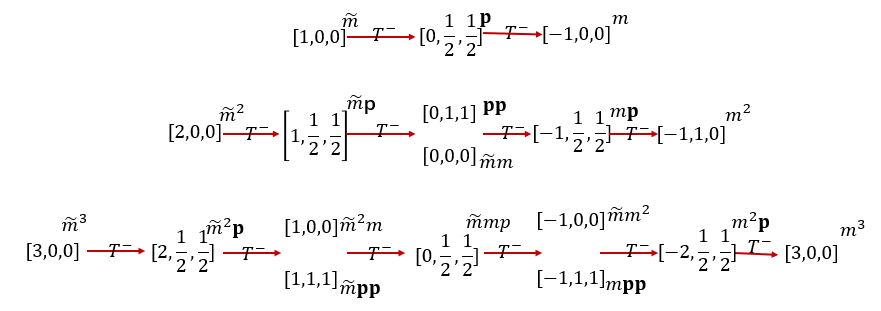}
    \caption{Same as figure~\ref{fig:spin-half-chain} but the $(\Delta,0,0)$ representations for spin-$0$ particle with $\Delta = 1, 2, 3$. }
    \label{fig:spin-0-chain}
\end{figure}
\begin{figure}[htb]
    \centering
    \includegraphics[scale=0.65]{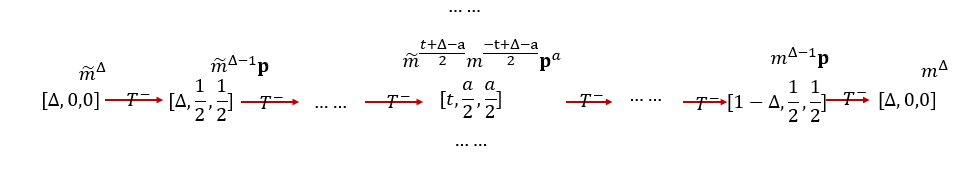}
    \caption{Same as figure~\ref{fig:spin-half-chain} but the $(\Delta, 0, 0)$ representations for spin-$0$ particle with general $\Delta$. }
    \label{fig:spin-0-chain-Delta}
\end{figure}

Other than ST spinors, the momentum structure can also be organized in the $SO(5, 1)$ representation. Usually, there is an infinite chain because the power of 4D momentum is not limited. Just notice that for the spin-0 spinors, the $SO(5, 1)$ structure reads as $(\Delta, 0, 0)$ for $\Delta = 1, 2, 3$ shown in figure~\ref{fig:spin-0-chain}. For general $\Delta$, it should be decomposed as
\bea
  (\Delta, 0, 0) &=& \mathop{\bigoplus}\limits_{-\Delta\leq t\leq\Delta,0\leq k\leq[\frac{\Delta-|t|}{2}]}[t, \frac{\Delta-|t|-2k}{2}, \frac{\Delta-|t|-2k}{2}], 
\eea
where $[\dots]$ is the least integer function. 
Note that for any $\Delta$ we always have $t = 0$ components and the dropped momentum structure is contained in the column which reads as $(m\tilde{m})^l \mathbf{p}^k$. The general momentum structure is shown in figure~\ref{fig:spin-0-chain-Delta}. 

\subsection{Massive helicity-transversality spinors in induced representations}
\label{sec:induce}

We have constructed the massive ST spinors based on the residue symmetry of the $SO(5,1)$ group. Now we would like to investigate their connection with the helicity spinors carrying the transversality, called the helicity-transversality spinors, which shows a clear power counting rule on mass expansion. 

We introduce the induced representation of the LG $U(2)$ ($GL(2, \mathbbm{C})$) for real (complex) momenta. Similar to the eq.~\eqref{eq:massless-decompose-su2-ahh}, the spin-spinor carrying transversality can be decomposed as the direct product of the $U(2)$ spin group tensor and two massless $SL(2,\mathbbm{C})$ Lorentz group spinors
\bea
\begin{cases}
\lambda_{\alpha}^{I} &= -\lambda_{\alpha} \zeta^{-I} z +\eta_{\alpha} \zeta^{+I} z, \\
\tilde{\lambda}_{\dot{\alpha}}^I &= \tilde{\lambda}_{\dot{\alpha}} \zeta^{+I} z^{-1} +\tilde{\eta}_{\dot{\alpha}} \zeta^{-I} z^{-1},
\end{cases}
\label{eq:massless-decompose-u2}
\eea
where $z$ represents the $U(1)$ transformation while $\zeta^{+}$ and $\zeta^{-}$ span the $SU(2)$ spin group space. In the Lorentz space, the massive spinor is decomposed into large-energy massless spinor $\lambda$ and small-energy massless spinor $\eta$, with the following energy scaling behavior 
\begin{eqnarray} 
\lambda, \tilde{\lambda} \sim \sqrt{2E}, \quad \eta, \tilde{\eta} \sim \frac{\mathbf{m}}{\sqrt{2E}}.
\label{eq:LambdaEtaEnergy}
\end{eqnarray}
which can be obtained by decomposing $\mathbf{p}=p+\eta$ and $\mathbf{m}^2=2p\cdot\eta$, with eq.~\eqref{eq:helicity-spinor-rep}. 
This energy scaling would serve as the power counting rules for any massive states and scattering amplitudes.

In the induced representation, the $SU(2)$ $W$ matrix in eq.~\eqref{eq:massiveLGTrans} reduces to the $U(1)_w$ transformation
\bea
(W^{-1})^I_J=\zeta^{-I} \zeta^+_J e^{-i\theta} +\zeta^{+I} \zeta^-_J e^{i\theta},
\eea
while the $z$ scaling $U(1)_z$ is kept. 
Thus the induced $U(1)_w \times U(1)_z$ LG scaling for the above eq.~\eqref{eq:massless-decompose-u2} takes
\bea
&z (W^{-1})^I_J  \lambda_\alpha^J \rightarrow  z e^{i\theta}  \eta_{\alpha} {\zeta^+}^{I}- z e^{-i\theta}\lambda_{\alpha} {\zeta^-}^{I} ,\\
&z^{-1}(W^{-1})^I_J \tilde{\lambda}_{\dot{\alpha}}^J \rightarrow z^{-1} e^{-i\theta}\tilde{\eta}_{\dot{\alpha}} {\zeta^-}^I+z^{-1} e^{i\theta}\tilde{\lambda}_{\dot{\alpha}} {\zeta^+}^I.
\eea

For later convenience, we would consider the massless spinors with complex momenta. The induced group $U(1)_w \times U(1)_z$ would be lifted as $GL(1, \mathbbm{C})_w \times GL(1, \mathbbm{C})_z$. 
Under the LG $GL(1, \mathbbm{C})_w$ and $GL(1, \mathbbm{C})_z$ they transform
\begin{eqnarray}
\begin{cases}
w z \circ \tilde{\lambda}_{\dot{\alpha}} \zeta_p^{+} = w_p z^{-1} \tilde{\lambda}_{\dot{\alpha}} \zeta^{+}, & 
w z \circ \lambda_{\alpha} \zeta_p^{-} = w_p^{-1} z  \lambda_{\alpha} \zeta^{-},  \\
w z \circ \eta_{\alpha} \zeta_{\eta}^{+} = w_{\eta} z \eta_{\alpha} \zeta^{+}, & 
w z \circ \tilde{\eta}_{\dot{\alpha}} \zeta_{\eta}^{-} = w_{\eta}^{-1} z^{-1}  \tilde{\eta}_{\dot{\alpha}} \zeta^{-}. 
\end{cases} 
\label{eq:GL1-p-eta-transform}
\end{eqnarray}
Note that $\zeta^\pm$ belongs to the reducible representation in the $U(1)_w$ LG, which can reduce to one-dimensional irreducible phases and thus can be absorbed.
Removing the $\zeta^\pm$ structure, the LG transformation takes the form
\bea
(wz,wz^{-1}) \circ \begin{pmatrix}
\tilde{\lambda}   & \eta    \\
\lambda   & \tilde{\eta}  
\end{pmatrix}
&=& \begin{pmatrix}
w z  \tilde{\lambda}  &  w  z^{-1} \eta   \\
w^{-1} z^{-1}  \lambda   &  w^{-1} z\tilde{\eta} 
\end{pmatrix}.
\label{eq:GL1-wz-transform}
\eea

Decomposing the above transformation, the eigenvalues and eigenstates of the $GL(1,\mathbbm{C})_{z} \times GL(1,\mathbbm{C})_{w}$ group are
\bea
w (h=+\frac12): \,\, \begin{pmatrix}
\tilde{\lambda} \\ \eta
\end{pmatrix}, \,\,
w^{-1}(h=-\frac12):\,\, \begin{pmatrix}
\tilde{\eta}  \\ \lambda 
\end{pmatrix};\quad 
z(t=+\frac12): \,\, \begin{pmatrix}
\tilde{\lambda} \\ \tilde{\eta} 
\end{pmatrix}, \,\,
z^{-1}(t=-\frac12):\,\, \begin{pmatrix}
 \eta\\ \lambda
\end{pmatrix},
\label{eq:wz-eigenstates}
\eea
where two quantum numbers $h$ and $t$ are defined for the $GL(1,\mathbbm{C})_{z}$ and $GL(1,\mathbbm{C})_{w}$ groups,  respectively. The physical consequence is described as follows
\bit 
\item The $GL(1,\mathbbm{C})_{w}$ group is referred to as {\it the helicity group}, with the quantum number $h$, since eigenstates are aligned with helicities of $\lambda$ and $\eta$;
\item The $GL(1,\mathbbm{C})_{z}$ group is referred to as {\it the transversality group}, with the quantum number $t$, because the fundamental and anti-fundamental irreducible representations of $GL(1,\mathbbm{C})_{z}$ correspond to square and angle brackets respectively.
\eit

Under the $GL(1,\mathbbm{C})_{z} \times GL(1,\mathbbm{C})_{w}$ transformation, $m$ and $\tilde{m}$ are no longer invariant, but instead covariant
\begin{eqnarray}
\begin{cases}
m = \lambda^\alpha \eta_\alpha \rightarrow z^{-2} m, \\
\tilde{m} = \tilde{\lambda}^\alpha \tilde{\eta}_\alpha \rightarrow z^2 \tilde{m},
\end{cases}
\end{eqnarray}
and the only invariant becomes $\mathbf{p}^2 = \tilde{m} m = \mathbf{m}^2$, where $\mathbf{m}$ is just a mass parameter, and $\tilde{m} $ and $m$ are spurion masses. Given the spurion masses, the EOM would relate spinors for the same helicity but different transversality, as follows
\bea \label{eq:EOM1}
\begin{cases}
  p_{\alpha \dot{\alpha}} \tilde{\eta}^{\dot{\alpha}}=-\tilde{m} \lambda_{\alpha}\\
  \eta_{\alpha \dot{\alpha}} \tilde{\lambda}^{\dot{\alpha}}=\tilde{m} \eta_{\alpha}\\
\end{cases},\quad 
\begin{cases}
  p_{\alpha \dot{\alpha}} \eta^{\alpha}=-m \tilde{\lambda}_{\dot{\alpha}}\\
  \eta_{\alpha \dot{\alpha}} \lambda^{\alpha}=m \tilde{\eta}_{\dot{\alpha}}\\
\end{cases}.
\eea
This equation gives the on-shell version of the Dirac equation. For majorana fermion, the corresponding EOM is given in appendix~\ref{app:particle}, which also discusses the EOM of spin-1 particle.

Using the decomposed helicity spinors, the generators for the $U(2)$ LG can be re-expressed as
\begin{eqnarray}
J^{+} &=& -\eta_{\alpha} \frac{\partial}{\partial \lambda_{\alpha}} +\tilde{\lambda}_{\dot{\alpha}} \frac{\partial}{\partial \tilde{\eta}_{\dot{\alpha}}}, \\
J^{-} &=& -\lambda_{\alpha} \frac{\partial}{\partial\eta_{\alpha}} +\tilde{\eta}_{\dot{\alpha}} \frac{\partial}{\partial\tilde{\lambda}_{\dot{\alpha}}}, \\
J^{3} &=& \frac{1}{2} \left( \Big( \eta_{\alpha} \frac{\partial}{\partial\eta_{\alpha}} +\tilde{\lambda}_{\dot{\alpha}} \frac{\partial}{\partial \tilde{\lambda}_{\dot{\alpha}}} \Big) -\Big( \lambda_{\alpha} \frac{\partial}{\partial\lambda_{\alpha}} +\tilde{\eta}_{\dot{\alpha}} \frac{\partial}{\partial \tilde{\eta}_{\dot{\alpha}}} \Big) \right),  \\
D_- &=& \frac{i}{2} \left( \Big( \tilde{\lambda}_{\dot{\alpha}} \frac{\partial}{\partial \tilde{\lambda}_{\dot{\alpha}}} + \tilde{\eta}_{\dot{\alpha}} \frac{\partial}{\partial \tilde{\eta}_{\dot{\alpha}}} \Big) - \Big( \lambda_{\alpha} \frac{\partial}{\partial \lambda_{\alpha}} + \eta_{\alpha} \frac{\partial}{\partial \eta_{\alpha}} \Big) \right).  \end{eqnarray}
Acting on the helicity spinors obtain the eigenstates for helicity and transversality
\bea
J^{3} \circ \tilde{\eta}_{\dot{\alpha}} &=& -\frac{1}{2} \tilde{\eta}_{\dot{\alpha}}, \
J^{3} \circ \lambda_{\alpha} = -\frac{1}{2} \lambda_{\alpha}, \\
J^{3} \circ \tilde{\lambda}_{\dot{\alpha}} &=& \frac{1}{2} \tilde{\lambda}_{\dot{\alpha}}, \
J^{3} \circ \eta_{\alpha} = \frac{1}{2} \eta_{\alpha}, 
\eea
\bea
D_- \circ \tilde{\eta}_{\dot{\alpha}} &=& \frac{i}{2} \tilde{\eta}_{\dot{\alpha}}, \
D_- \circ \lambda_{\alpha} = -\frac{i}{2} \lambda_{\alpha}, \\
D_- \circ \tilde{\lambda}_{\dot{\alpha}} &=& \frac{i}{2} \tilde{\lambda}_{\dot{\alpha}}, \
D_- \circ \eta_{\alpha} = -\frac{i}{2} \eta_{\alpha}. 
\eea
Similarly the ladder operators change the helicity state as
\bea
J^{+} \circ \tilde{\eta}_{\dot{\alpha}} &=& \tilde{\lambda}_{\dot{\alpha}}, \
J^{+} \circ (-\lambda_{\alpha}) = \eta_{\alpha}, \label{eq:OperatorJ1}\\
J^{-} \circ \tilde{\lambda}_{\dot{\alpha}} &=& \tilde{\eta}_{\dot{\alpha}}, \
J^{-} \circ \eta_{\alpha} = -\lambda_{\alpha}. 
\label{eq:OperatorJ2}
\eea
The $T^\pm$ generators take the form
\bea
{T^+}^{\alpha}_{\dot{\alpha}} &=& \tilde{\lambda}_{\dot{\alpha}} \frac{\partial}{\partial \eta_{\alpha}} -\tilde{\eta}_{\dot{\alpha}} \frac{\partial}{\partial \lambda_{\alpha}}, \label{eq:generatorT1} \\
{T^-}^{\dot{\alpha}}_{\alpha} &=& \eta_{\alpha} \frac{\partial}{\partial \tilde{\lambda}_{\dot{\alpha}}} -\lambda_{\alpha} \frac{\partial}{\partial \tilde{\eta}_{\dot{\alpha}}}. \label{eq:generatorT2} 
\eea
The ladder operators for the transversality change  takes
\bea
{T^{+}}^{\alpha}_{\dot{\alpha}} \circ (-\lambda_{\alpha}) &=& \tilde{\eta}_{\dot{\alpha}}, \ 
{T^{+}}^{\alpha}_{\dot{\alpha}} \circ \eta_{\alpha} = \tilde{\lambda}_{\dot{\alpha}}, \\
{T^{-}}_{\alpha}^{\dot{\alpha}} \circ \tilde{\lambda}_{\dot{\alpha}} &=& \eta_{\alpha}, \
{T^{-}}_{\alpha}^{\dot{\alpha}} \circ \tilde{\eta}_{\dot{\alpha}} = -\lambda_{\alpha}.
\eea

From the above generators and ladder operators, we can construct the helicity-transversality spinors for general spin-$s$ particle. Obviously one can obtain the helicity-transversality spinors by expanding the ST spinors using eq.~\eqref{eq:massless-decompose-u2}. Here we present two alternative ways of constructing the helicity-transversality spinors with $\Delta = 0$, or saying the helicity-chirality spinors.

The first alternative way is starting from the highest weight spinor with $h=t=s$ taking the form $\tilde{\lambda}^{2s}$. Then acting on the ladder operators $J^\pm$ and $T^\pm$, we obtain other transversality and helicity components. According to the $w, z$ scaling properties defined in eq.~\eqref{eq:wz-eigenstates}, we tabulate all the eigenstates of the $(h, t)$ quantum number ranging from $-s$ to $s$:
\begin{eqnarray} \label{eq:SpinorTable}
\begin{array}{|c|c|c|c|c|}
\hline
& t=s & t=s-1 & t=s-2 & ... \\
\hline
h=s & \tilde{\lambda}^{2s} & \eta \tilde{\lambda}^{2s-1} & \eta^{2} \tilde{\lambda}^{2s-2} & ... \\
\hline
h=s-1 & \tilde{\lambda}^{2s-1} \tilde{\eta} & -\lambda \tilde{\lambda}^{2s-1}, \eta \tilde{\eta} \tilde{\lambda}^{2s-2} & -\lambda \eta \tilde{\lambda}^{2s-2}, \eta^2 \tilde{\eta} \tilde{\lambda}^{2s-3} & ... \\
\hline
h=s-2 & \tilde{\lambda}^{2s-2} \tilde{\eta}^{2} & -\lambda \tilde{\lambda}^{2s-2} \tilde{\eta}, \eta \tilde{\lambda}^{2s-3} \tilde{\eta}^{2} & \lambda^2 \tilde{\lambda}^{2s-2}, -\lambda \eta \tilde{\lambda}^{2s-3} \tilde{\eta}, \eta^2 \tilde{\lambda}^{2s-4} \tilde{\eta}^2 & ... \\
\hline
\vdots & \vdots & \vdots  & \vdots & \ddots \\
\hline
\end{array}
\end{eqnarray}
Here each component can be labeled by the quantum number $[s, h, t]$.

Applying to the spin-$\frac12$ fermion, the table takes the form
\bea
\begin{array}{|c|c|c|}
\hline
& t = -1/2  & t = +1/2 \\
\hline
h = -1/2 & \lambda_{\alpha} & \tilde{\eta}_{\dot{\alpha}} \\
\hline
h = +1/2 & \eta_{\alpha}   & \tilde{\lambda}_{\dot{\alpha}} \\
\hline
\end{array}\label{eq:spin-half-states-tab}
\eea
 we obtain the massive fermion helicity-chirality states
\bea
\psi^{s = \frac12}_{\alpha} \equiv \left(
\begin{array}{cc}
-\lambda_{\alpha}  &  \eta_{\alpha}  \\
 \tilde{\eta}_{\dot{\alpha}}  & \tilde{\lambda}_{\dot{\alpha}}
\end{array}
\right).
\label{eq:spin-half-states}
\eea
For spin-1 particle, all the components take the form
\bea
\begin{array}{|c|c|c|c|}
\hline
& t = -1 & t = 0 & t = +1 \\
\hline
h = -1 & \lambda_{\alpha_1}\lambda_{\alpha_2} & \lambda_{\alpha_1}\tilde{\eta}_{\dot{\alpha}_2} & \tilde{\eta}_{\alpha_1} \tilde{\eta}_{\alpha_2} \\
\hline
h = 0 & \lambda_{\alpha_1}\eta_{\alpha_2},\ \eta_{\alpha_1}\lambda_{\alpha_2}  & \lambda_{\alpha_1}\tilde{\lambda}_{\dot{\alpha}_2}, \ \eta_{\alpha_1}\tilde{\eta}_{\dot{\alpha}_2} & \tilde{\lambda}_{\alpha_1}\tilde{\eta}_{\dot{\alpha}_2} ,\ \tilde{\eta}_{\alpha_1}\tilde{\lambda}_{\dot{\alpha}_2} \\
\hline
h = +1 & \eta_{\alpha_1}\eta_{\alpha_2}  & \eta_{\alpha_1}\tilde{\lambda}_{\dot{\alpha}_2} & \tilde{\lambda}_{\alpha_1} \tilde{\lambda}_{\alpha_2} \\
\hline
\end{array}\label{eq:spin-1-states-redund}
\eea
Notice that this table contains redundant components, such that for the helicity 0 state, the spin-0 state $\eta_{\alpha_1}\tilde{\eta}_{\dot{\alpha}_2}+\lambda_{\alpha_1}\tilde{\lambda}_{\dot{\alpha}_2}$ should be removed. 

To remove this kind of states, we impose the condition realized by the Pauli-Lubanski operator $W^2$ in Ref.~\cite{Jiang:2020rwz, Li:2022abx,Li:2023pfw} that only the spin-$s$ eigenstate remains. 
In this case, the $W^2$ acting on the $(t=0,h=0)$ states gives
\bea
W^2
\begin{pmatrix}
\lambda_{\alpha_1}\tilde{\lambda}_{\dot{\alpha}_2} \\ \eta_{\alpha_1}\tilde{\eta}_{\dot{\alpha}_2}
\end{pmatrix} 
= -\mathbf{p}^2 s(s+1)
\begin{pmatrix}
\lambda_{\alpha_1}\tilde{\lambda}_{\dot{\alpha}_2} \\ \eta_{\alpha_1}\tilde{\eta}_{\dot{\alpha}_2}
\end{pmatrix}. 
\eea
After diagonalizing the matrix, the eigensystem takes the form
\bea \label{eq:h0t0state}
\begin{cases}
s = 1 : & \eta_{\alpha_1}\tilde{\eta}_{\dot{\alpha}_2}-\lambda_{\alpha_1}\tilde{\lambda}_{\dot{\alpha}_2},  \\
s = 0 : &\eta_{\alpha_1}\tilde{\eta}_{\dot{\alpha}_2} +\lambda_{\alpha_1}\tilde{\lambda}_{\dot{\alpha}_2}. 
\end{cases}
\eea
The diagonalized matrix $\mathrm{diag}[2,0]$ gives the eigenvalue of $s(s+1)$, so that the state corresponding to the first component $2$ is the required spin-1 state.
After removing such redundancies, we obtain
\bea
\textrm{spin-}1: \quad \left(
\begin{array}{ccc}
\lambda_{\alpha_1}\lambda_{\alpha_2}   &  -\lambda_{\alpha_1}\tilde{\eta}_{\dot{\alpha}_2} &   \tilde{\eta}_{\alpha_1} \tilde{\eta}_{\alpha_2} \\
-\lambda_{\alpha_1}\eta_{\alpha_2} -\eta_{\alpha_1}\lambda_{\alpha_2}  & \eta_{\alpha_1}\tilde{\eta}_{\dot{\alpha}_2}-\lambda_{\alpha_1}\tilde{\lambda}_{\dot{\alpha}_2}  & -\tilde{\lambda}_{\alpha_1}\tilde{\eta}_{\dot{\alpha}_2}-\tilde{\eta}_{\alpha_1}\tilde{\lambda}_{\dot{\alpha}_2} \\
\eta_{\alpha_1}\eta_{\alpha_2} & \eta_{\alpha_1}\tilde{\lambda}_{\dot{\alpha}_2} & \tilde{\lambda}_{\alpha_1} \tilde{\lambda}_{\alpha_2}
\end{array}
\right). 
\label{eq:spin-1-states}
\eea

There is a third way of constructing the spinors based on the Young diagrams. The helicity wave function can be obtained by filling the Young diagram with $\pm$ signs, where each sign represents a helicity of $\pm \frac{1}{2}$. The $(t=0,s=1)$ representation, for example, is composed of three states
\bea
h=1 && \arraycolsep=0pt\def\arraystretch{1}\begin{array}{cc}
{\color{orange} \young(+)} & \young(+)
\end{array} \sim \tilde{\lambda} \eta \zeta^{+(I_1} \zeta^{+I_2)}, \nonumber \\
h=0 && \arraycolsep=0pt\def\arraystretch{1}\begin{array}{cc}
{\color{orange} \young(+)} & \young(-)
\end{array} \sim (-\tilde{\lambda} \lambda +\tilde{\eta} \eta) \zeta^{+(I_1} \zeta^{-I_2)}, \\
h=-1 && \arraycolsep=0pt\def\arraystretch{1}\begin{array}{cc}
{\color{orange} \young(-)} & \young(-)
\end{array} \sim -\tilde{\eta} \lambda \zeta^{-(I_1} \zeta^{-I_2)}, \nonumber
\eea
according with the middle column in eq.~\eqref{eq:spin-1-states}. More details of the Young diagram formulation are given in appendix~\ref{app:Young}.

\subsection{Mass insertion the on-shell way: chirality and helicity flips}
\label{sec:MassInsert}

In above, we provide several ways of constructing the spin-$s$ massive spinors with transversality. We note in terms of the helicity-transversality spinors, the large and small components have clear power counting rules, and thus they can be related to the massless helicity spinors. We would like to construct the spin-$s$ helicity-transversality spinors diagrammatically based on the power counting rules. 


Let us first discuss the power counting rules for massive spin-$s$ helicity-transversality spinor. 
Using the energy scaling behavior of the $\lambda_{\alpha}, \eta_{\alpha}$ helicity spinors
\begin{eqnarray} \label{eq:LambdaEtaEnergy2}
\lambda, \tilde{\lambda} \sim \sqrt{2E}; \quad \eta, \tilde{\eta} \sim \frac{\mathbf{m}}{\sqrt{2E}},
\end{eqnarray}
we can list the energy powers of the spin-$s$ state, e.g. the spin-1 state with transversality $t = 0$ reads
\begin{eqnarray}\label{eq:scaleofpv}
\begin{cases}
(h = +1): & \epsilon_+ = \frac{\eta \tilde{\lambda}}{\mathbf{m}} \sim 1, \\
(h = 0): & \epsilon_{0} -\epsilon_{L} = -\frac{\lambda \tilde{\lambda}}{\mathbf{m}} \sim \frac{2E}{\mathbf{m}}, \\
(h = 0'): & \epsilon_L = \frac{\eta \tilde{\eta}}{\mathbf{m}} \sim \frac{\mathbf{m}}{2E} ,\\
(h = -1): & \epsilon_- = -\frac{\lambda \tilde{\eta}}{\mathbf{m}} \sim 1.
\end{cases} 
\end{eqnarray}
Similarly, any component of a massive state can be decomposed into large energy components and small energy components.

Based on the power counting rules, we observe that the ladder operation is equivalent to the helicity flip and transversality flip between large component $\lambda$ and small component $\eta$:
\bea
\textrm{helicity flip}&:& 
\quad \lambda \sim \sqrt{2E} \leftrightarrow \eta \sim \frac{\mathbf{m}}{\sqrt{2E}},\\
\textrm{transversality flip}&:& 
\quad \lambda \sim \sqrt{2E} \leftrightarrow \tilde{\eta} \sim \frac{\mathbf{m}}{\sqrt{2E}}.
\eea
We note that the transversality flip is quite similar to the chirality flip defined in textbooks. Here we present the connection between transversality and chirality: the left-handed chirality corresponds to the transversality $t= -\frac12$, while right-handed chirality corresponds to $t= +\frac12$. However, the chirality for fermion and anti-particle have different notation, thus we have 
\begin{equation}
    \begin{tabular}{c|c|c|c}
    \hline
     & spinors & chirality & transversality  \\
    \hline
    \multirow{2}{*}{particle}  &  $\lambda,\eta$ & L & $-\frac{1}{2}$   \\
    & $\tilde{\lambda},\tilde{\eta}$  & R & $+\frac{1}{2}$ \\
    \hline
    \multirow{2}{*}{anti-particle} & $\tilde{\lambda},\tilde{\eta}$ & L & $+\frac{1}{2}$  \\
    & $\lambda,\eta$ & R & $-\frac{1}{2}$  \\
    \hline
    \end{tabular}
\end{equation}

In the following, we will present the helicity-transversality spinors using the notion of the helicity and transversality flips diagrammatically based on the action of the $T^\pm$, $J^\pm$ and $m (\tilde{m})$ generators in the $ISO(5,1)$ symmetry. Note that the generators $m$ and $\tilde{m}$ change transversality. Given the EOM, the product between internal $p$ and external $\tilde{\lambda}$ is equivalent to the product of internal $L$ fermion and mass insertion, this transversality flip is related to the chirality flip. In the following let us look at how the helicity and chirality flip is related to the on-shell mass insertion diagrammatically.

We begin with the large energy components, which are located at the diagonal elements in eq.~\eqref{eq:SpinorTable}. For spin-$\frac{1}{2}$ particle, the large energy components are $\lambda$ and $\tilde{\lambda}$. We can use red and cyan lines to represent the left-handed and right-handed fermions. Diagrammatically, for particle $i$, we have
\begin{equation} \begin{aligned} \label{eq:fermion}
\text{Fermion}: 
\begin{tikzpicture}[baseline=0.7cm] \begin{feynhand}
\vertex [particle] (i1) at (1.5,0.8) {$i^{-\frac{1}{2}}$};
\vertex [dot] (v1) at (0,0.8) {};
\graph{(v1)--[plain,red,very thick] (i1)};
\end{feynhand} \end{tikzpicture}&=|i\rangle,&
\begin{tikzpicture}[baseline=0.7cm] \begin{feynhand}
\vertex [particle] (i1) at (1.5,0.8) {$i^{+\frac{1}{2}}$};
\vertex [dot] (v1) at (0,0.8) {};
\graph{(v1)--[plain,cyan,very thick] (i1)};
\end{feynhand} \end{tikzpicture}&=|i],\\
\text{Anti-fermion}: 
\begin{tikzpicture}[baseline=0.7cm] \begin{feynhand}
\vertex [particle] (i1) at (1.5,0.8) {$i^{-\frac{1}{2}}$};
\vertex [dot] (v1) at (0,0.8) {};
\graph{(v1)--[plain,cyan,very thick] (i1)};
\end{feynhand} \end{tikzpicture}&=|i\rangle,&
\begin{tikzpicture}[baseline=0.7cm] \begin{feynhand}
\vertex [particle] (i1) at (1.5,0.8) {$i^{+\frac{1}{2}}$};
\vertex [dot] (v1) at (0,0.8) {};
\graph{(v1)--[plain,red,very thick] (i1)};
\end{feynhand} \end{tikzpicture}&=|i],
\end{aligned} \end{equation}
where $|i\rangle=\lambda_i$ and $|i]=\tilde{\lambda}_i$. They can be related by the commutator of helicity and transversality operator operator as
\begin{equation} \label{eq:TJ}
\lambda\ 
\mathrel{\vcenter{\hbox{\begin{tikzpicture}
    \node[minimum width=1cm,minimum height=1ex,anchor=south,align=center] (a){\text{\footnotesize\vphantom{hg}$[T^-,J^-]$}\\[0.5ex] \footnotesize\vphantom{hg}$[T^+,J^+]$};
  \draw[<-] ([yshift=0.35ex]a.west) -- ([yshift=0.35ex]a.east);
  \draw[->] ([yshift=-0.35ex]a.west) -- ([yshift=-0.35ex]a.east);
\end{tikzpicture}}}}\ 
\tilde{\lambda}.
\end{equation}

Then we use the flips to generate the small energy components.
Since all transversality states are generated by the commutator $[T^\pm,J^\pm]$, we do not need the transversality flip. Given the energy scaling behavior, both helicity and chirality flip would change the order of scaling
\bea
\textrm{helicity flip}&:& 
\quad \lambda \sim \sqrt{2E} \leftrightarrow \eta \sim \frac{\mathbf{m}}{\sqrt{2E}},\\
\textrm{chiraliy flip}&:& 
\quad \lambda \sim \sqrt{2E} \leftrightarrow \tilde{m}\lambda \sim \mathbf{m}\sqrt{2E},
\eea
which indicates the \textit{mass insertion}. In the mass expansion power counting, we denote the expansion parameter as $\varepsilon_{\eta}\sim \frac{\mathbf{m}}{E}$. When we flip the states from the left-hand side to the right-hand side of the above equations, the power counting will get an additional $\varepsilon_{\eta}$. Diagrammatically, let us lay out how they indicate mass insertion, 
\begin{equation}
\begin{tabular}{c|ccc}
 & state & power counting & diagram \\
\hline
helicity flip & $\lambda\rightarrow\eta,\tilde{\lambda}\rightarrow\tilde{\eta}$ & $\varepsilon_\eta$ & 
\begin{tikzpicture}[baseline=0.7cm] \begin{feynhand}
\vertex [particle] (i1) at (1,0.8) {};
\vertex [particle] (v1) at (0,0.8) {};
\graph{(v1)--[plain,cyan,very thick](i1)};
\draw plot[mark=x,mark size=3.5] coordinates {(0.5,0.8)};
\end{feynhand} 
\end{tikzpicture}
\begin{tikzpicture}[baseline=0.7cm] \begin{feynhand}
\setlength{\feynhanddotsize}{0.8mm}
\vertex [particle] (i1) at (1,0.8) {};
\vertex [particle] (v1) at (0,0.8) {};
\graph{(v1)--[plain,red,very thick](i1)};
\draw plot[mark=x,mark size=3.5] coordinates {(0.5,0.8)};
\end{feynhand} 
\end{tikzpicture}\\
\hline
chirality flip & $\lambda\rightarrow m\lambda,\tilde{\lambda}\rightarrow \tilde{m}\tilde{\lambda}$ & $\varepsilon_\eta$ & 
\begin{tikzpicture}[baseline=0.7cm] \begin{feynhand}
\setlength{\feynhanddotsize}{0.8mm}
\vertex [particle] (i1) at (1,0.8) {};
\vertex [particle] (v1) at (0,0.8) {};
\vertex (v2) at (0.5,0.8);
\graph{(v1)--[plain,cyan,very thick](v2)--[plain,red,very thick](i1)};
\draw[very thick] plot[mark=x,mark size=3.5] coordinates {(0.5,0.8)};
\end{feynhand} 
\end{tikzpicture}
\begin{tikzpicture}[baseline=0.7cm] \begin{feynhand}
\vertex [particle] (i1) at (1,0.8) {};
\vertex [particle] (v1) at (0,0.8) {};
\vertex (v2) at (0.5,0.8);
\graph{(v1)--[plain,red,very thick](v2)--[plain,cyan,very thick](i1)};
\draw[very thick] plot[mark=x,mark size=3.5] coordinates {(0.5,0.8)};
\end{feynhand} 
\end{tikzpicture}\\
\end{tabular}
\end{equation}
where thin and thick crosses denote helicity and chirality flips. Here we use red and cyan lines to represent different chirality, so the color is changed by a thick cross but not a thin cross.



For convenience, we focus on the flips for fermion, because the flips for anti-fermion can be easily derived by the exchange of colors. The helicity flip keeps the chirality, so the color does not change. We have
\begin{equation} \begin{aligned} \label{eq:flipF1}
\begin{tikzpicture}[baseline=0.7cm] \begin{feynhand}
\vertex [particle] (i1) at (1.5,0.8) {$i^{+\frac{1}{2}}$};
\vertex [dot] (v1) at (0,0.8) {};
\graph{(v1)--[plain,red,very thick] (i1)};
\draw plot[mark=x,mark size=3.5] coordinates {(0.6,0.8)};
\end{feynhand} \end{tikzpicture}
&=J^+\circ|i\rangle=-|\eta_i\rangle,&
\begin{tikzpicture}[baseline=0.7cm] \begin{feynhand}
\vertex [particle] (i1) at (1.5,0.8) {$i^{-\frac{1}{2}}$};
\vertex [dot] (v1) at (0,0.8) {};
\graph{(v1)--[plain,cyan,very thick] (i1)};
\draw plot[mark=x,mark size=3.5] coordinates {(0.6,0.8)};
\end{feynhand} \end{tikzpicture}
&=J^-\circ|i]=|\eta_i],& \\
\end{aligned} \end{equation}
where $|\eta_i\rangle=\eta_i$ and $|\eta_i]=\tilde{\eta}_i$, and $\times$ denotes the heilicity mass insertion. In the left side of the two diagrams, the red and cyan lines represent the states with $h=-\frac{1}{2}$ and $+\frac{1}{2}$, as shown in eq.~\eqref{eq:fermion}.  Therefore, the helicity mass insertion just changes the helicity. 

The chirality flip is simply multiplying the states by $m$ or $\tilde{m}$. In this case, the color should be changed as 
\begin{equation} \begin{aligned}
\begin{tikzpicture}[baseline=0.7cm] \begin{feynhand}
\vertex [particle] (i1) at (1.5,0.8) {$i^{-\frac{1}{2}}$};
\vertex [dot] (v1) at (0,0.8) {};
\vertex (v2) at (0.6,0.8);
\graph{(v1)--[plain,red,very thick] (v2)--[plain,cyan,insertion={[style=black]0},very thick] (i1)};
\end{feynhand} \end{tikzpicture}
&=\tilde{m}_i |i\rangle,&
\begin{tikzpicture}[baseline=0.7cm] \begin{feynhand}
\vertex [particle] (i1) at (1.5,0.8) {$i^{+\frac{1}{2}}$};
\vertex [dot] (v1) at (0,0.8) {};
\vertex (v2) at (0.6,0.8);
\graph{(v1)--[plain,cyan,very thick] (v2)--[plain,red,insertion={[style=black]0},very thick] (i1)};
\end{feynhand} \end{tikzpicture}
&=m_i |i],\\
\begin{tikzpicture}[baseline=-0.1cm] \begin{feynhand}
\vertex [particle] (i1) at (1.5,0) {$i^{+\frac{1}{2}}$};
\vertex [dot] (v1) at (0,0) {};
\vertex (v2) at (0.8,0);
\graph{(v1)--[plain,red,very thick] (v2)--[plain,cyan,insertion={[style=black]0},very thick] (i1)};
\draw plot[mark=x,mark size=3.5] coordinates {(0.4,0)};
\end{feynhand} \end{tikzpicture}
&=-\tilde{m}_i |\eta_i\rangle ,&
\begin{tikzpicture}[baseline=-0.1cm] \begin{feynhand}
\vertex [particle] (i1) at (1.5,0) {$i^{-\frac{1}{2}}$};
\vertex [dot] (v1) at (0,0) {};
\vertex (v2) at (0.8,0);
\graph{(v1)--[plain,cyan,very thick] (v2)--[plain,red,insertion={[style=black]0},very thick] (i1)};
\draw plot[mark=x,mark size=3.5] coordinates {(0.4,0)};
\end{feynhand} \end{tikzpicture}
&=m_i |\eta_i], \\
\end{aligned} \end{equation}
where the bold cross denotes the chirality mass insertion. Since the commutator gives $[m,J^{-}]=[\tilde{m},J^{+}]=0$, the chirality flip and helicity flip are interchangeable. Therefore, we have
\beq
\begin{tikzpicture}[baseline=-0.1cm] \begin{feynhand}
\vertex [particle] (i1) at (1.5,0) {$i^{+\frac{1}{2}}$};
\vertex [dot] (v1) at (0,0) {};
\vertex (v2) at (0.8,0);
\graph{(v1)--[plain,cyan,very thick] (v2)--[plain,red,insertion={[style=black]0},very thick] (i1)};
\draw plot[mark=x,mark size=3.5] coordinates {(0.4,0)};
\end{feynhand} \end{tikzpicture}=
\begin{tikzpicture}[baseline=-0.1cm] \begin{feynhand}
\vertex [particle] (i1) at (1.5,0) {$i^{+\frac{1}{2}}$};
\vertex [dot] (v1) at (0,0) {};
\vertex (v2) at (0.4,0);
\graph{(v1)--[plain,cyan,very thick] (v2)--[plain,red,insertion={[style=black]0},very thick] (i1)};
\draw plot[mark=x,mark size=3.5] coordinates {(0.8,0)};
\end{feynhand} \end{tikzpicture}.
\eeq
Notice that the repeated transversality flips $m^2$ or $\tilde{m}^2$ gives the components with transversality $\pm\frac{3}{2}$, which are not involved in the definition of fermion. Therefore, each spinor has at most one transversality flip. In summary, we draw diagrams for the complete ST spinors for spin-$\frac12$ particle 
\bea \label{eq:Fdiagram}
\includegraphics[scale=0.48,valign=c]{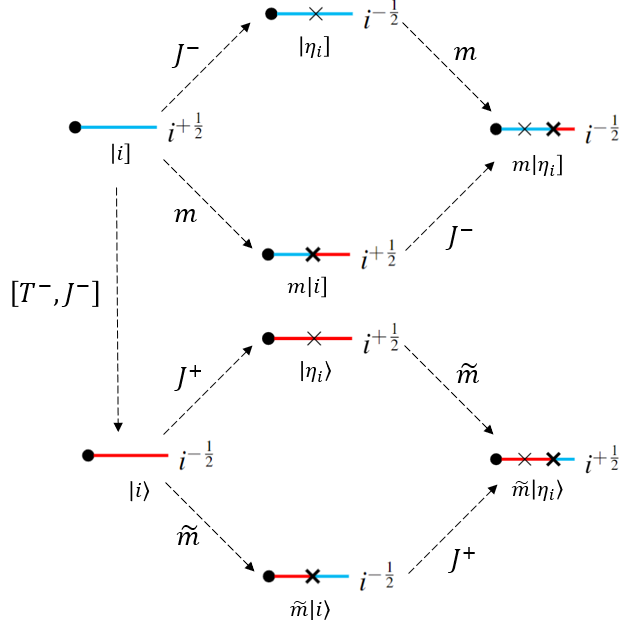}.
\eea

To illustrate the chirality of $s=1$ particle, we need three colors to represent the single-particle state. Similarly, we still use cyan and red lines to represent $+1$ and $-1$ chirality, while the brown line stands for $0$ chirality. We have
\begin{equation} \begin{aligned}
\text{Vector}: 
\begin{tikzpicture}[baseline=0.7cm] \begin{feynhand}
\vertex [particle] (i1) at (1.5,0.8) {$i^{-1}$};
\vertex [dot] (v1) at (0,0.8) {};
\graph{(v1)--[plain,red,very thick] (i1)};
\end{feynhand} \end{tikzpicture}&=|i\rangle^2,&
\begin{tikzpicture}[baseline=0.7cm] \begin{feynhand}
\vertex [particle] (i1) at (1.5,0.8) {$i^{0}$};
\vertex [dot] (v1) at (0,0.8) {};
\graph{(v1)--[plain,brown,very thick] (i1)};
\end{feynhand} \end{tikzpicture}&=|i]|i\rangle,&
\begin{tikzpicture}[baseline=0.7cm] \begin{feynhand}
\vertex [particle] (i1) at (1.5,0.8) {$i^{+1}$};
\vertex [dot] (v1) at (0,0.8) {};
\graph{(v1)--[plain,cyan,very thick] (i1)};
\end{feynhand} \end{tikzpicture}&=|i]^2,\\
\text{Anti-vector}: 
\begin{tikzpicture}[baseline=0.7cm] \begin{feynhand}
\vertex [particle] (i1) at (1.5,0.8) {$i^{-1}$};
\vertex [dot] (v1) at (0,0.8) {};
\graph{(v1)--[plain,cyan,very thick] (i1)};
\end{feynhand} \end{tikzpicture}&=|i\rangle^2,&
\begin{tikzpicture}[baseline=0.7cm] \begin{feynhand}
\vertex [particle] (i1) at (1.5,0.8) {$i^{0}$};
\vertex [dot] (v1) at (0,0.8) {};
\graph{(v1)--[plain,brown,very thick] (i1)};
\end{feynhand} \end{tikzpicture}&=|i]|i\rangle,&
\begin{tikzpicture}[baseline=0.7cm] \begin{feynhand}
\vertex [particle] (i1) at (1.5,0.8) {$i^{+1}$};
\vertex [dot] (v1) at (0,0.8) {};
\graph{(v1)--[plain,red,very thick] (i1)};
\end{feynhand} \end{tikzpicture}&=|i]^2.\\
\end{aligned} \end{equation}

Again, we focus on the vector boson, but not the anti-vector boson. The action of helicity flip is similar to eq.~\eqref{eq:flipF1}. The new thing is that we can perform the helicity flip twice for vector boson. For the zero chirality component, it generates not only the component with two helicity mass insertion, but also itself. Diagrammatically, it gives
\begin{equation} \begin{aligned}
J^+\circ J^-\circ|i]|i\rangle
\rightarrow
\begin{cases}
\begin{tikzpicture}[baseline=0.7cm] \begin{feynhand}
\vertex [particle] (i1) at (1.5,0.8) {$i^{0}$};
\vertex [dot] (v1) at (0,0.8) {};
\graph{(i1)--[plain,brown,very thick] (v1)};
\draw plot[mark=x,mark size=3.5] coordinates {(0.4,0.8)};
\draw plot[mark=x,mark size=3.5] coordinates {(0.85,0.8)};
\end{feynhand} 
\end{tikzpicture}=-|\eta_i]|\eta_i\rangle,\\
\begin{tikzpicture}[baseline=0.7cm] \begin{feynhand}
\vertex [particle] (i1) at (1.5,0.8) {$i^0$};
\vertex [dot] (v1) at (0,0.8) {};
\graph{(i1)--[plain,brown,very thick] (v1)};
\end{feynhand} \end{tikzpicture}=|i]|i\rangle.\\
\end{cases}
\end{aligned} \end{equation}
The brown line between the two crosses represents the state with $h=-1$. In fact, it is difficult to read out helicity information for spin-1 particles. In appendix~\ref{app:3pt}, we introduce double-line formalism to represent spin-1 particles, which provides more information in the diagrams and can solve this problem.

Then we consider the chirality flip for vector boson. There are four kinds of chirality flips for the large components:
\begin{equation} \begin{aligned}
\begin{tikzpicture}[baseline=-0.1cm] \begin{feynhand}
\vertex [particle] (i1) at (1.5,0) {$i^{-1}$};
\vertex [dot] (v1) at (0,0) {};
\vertex (v2) at (0.6,0);
\graph{(v1)--[plain,red,very thick] (v2)--[plain,brown,insertion={[style=black]0},very thick] (i1)};
\end{feynhand} \end{tikzpicture}
&=\tilde{m}_i |i\rangle^2,&
\begin{tikzpicture}[baseline=-0.1cm] \begin{feynhand}
\vertex [particle] (i1) at (1.5,0) {$i^{+1}$};
\vertex [dot] (v1) at (0,0) {};
\vertex (v2) at (0.6,0);
\graph{(v1)--[plain,cyan,very thick] (v2)--[plain,brown,insertion={[style=black]0},very thick] (i1)};
\end{feynhand} \end{tikzpicture}
&=m_i |i]^2,\\
\begin{tikzpicture}[baseline=-0.1cm] \begin{feynhand}
\vertex [particle] (i1) at (1.5,0) {$i^{0}$};
\vertex [dot] (v1) at (0,0) {};
\vertex (v2) at (0.6,0);
\graph{(v1)--[plain,brown,very thick] (v2)--[plain,cyan,insertion={[style=black]0},very thick] (i1)};
\end{feynhand} \end{tikzpicture}
&=\tilde{m}_i |i]|i\rangle,&
\begin{tikzpicture}[baseline=-0.1cm] \begin{feynhand}
\vertex [particle] (i1) at (1.5,0) {$i^{0}$};
\vertex [dot] (v1) at (0,0) {};
\vertex (v2) at (0.6,0);
\graph{(v1)--[plain,brown,very thick] (v2)--[plain,red,insertion={[style=black]0},very thick] (i1)};
\end{feynhand} \end{tikzpicture}
&=m_i |i]|i\rangle. \\
\end{aligned} \end{equation}
In the last line, we flip $t=0$ state to give $t=\pm 1$ states. The chirality flip can act on the different spinors of the single-particle states. Therefore, the massive vector can have twice the chirality flip as
\begin{equation} \begin{aligned}
\begin{tikzpicture}[baseline=-0.1cm] \begin{feynhand}
\vertex [particle] (i1) at (2.1,0) {$i^{-1}$};
\vertex [dot] (v1) at (0,0) {};
\vertex (v2) at (0.6,0);
\vertex (v3) at (1.2,0);
\graph{(v1)--[plain,red,very thick] (v2)--[plain,brown,insertion={[style=black]0},very thick] (v3)--[plain,cyan,insertion={[style=black]0},very thick] (i1)};
\end{feynhand} \end{tikzpicture}
&=\tilde{m}_i^2 |i\rangle^2, \\
\begin{tikzpicture}[baseline=-0.1cm] \begin{feynhand}
\vertex [particle] (i1) at (2.1,0) {$i^{+1}$};
\vertex [dot] (v1) at (0,0) {};
\vertex (v2) at (0.6,0);
\vertex (v3) at (1.2,0);
\graph{(v1)--[plain,cyan,very thick] (v2)--[plain,brown,insertion={[style=black]0},very thick] (v3)--[plain,red,insertion={[style=black]0},very thick] (i1)};
\end{feynhand} \end{tikzpicture}
&=m_i^2 |i]^2.\\
\end{aligned} \end{equation}
The twice chirality flip does not change the $t=0$ state, so there are only two diagrams. 

Finally, we obtain the helicity-transversality spinors for spin-$1$ particle. Starting with $|i]^2$, we have 
\bea \label{eq:Vdiagram}
\includegraphics[scale=0.7,valign=c]{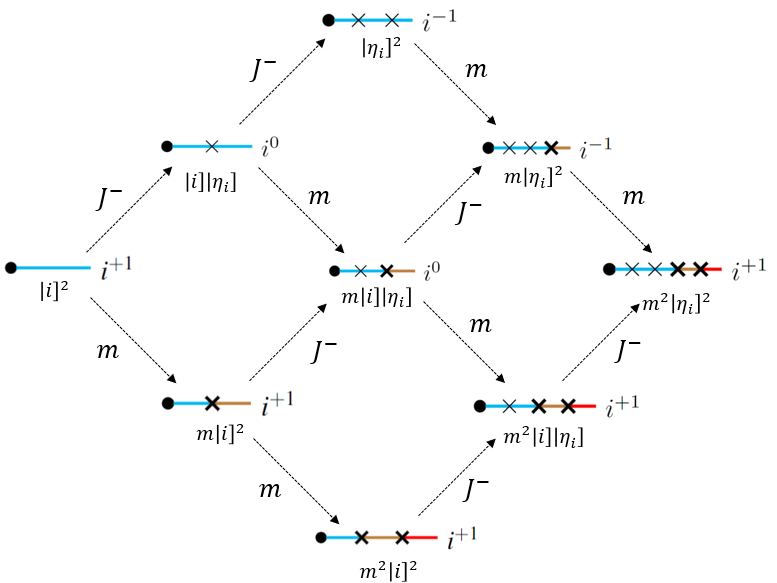}.
\eea
The other helicity-transversality spinors generated from $|i]|i\rangle$ and $|i\rangle^2$ can be given in a similar way.

For the higher-spin particle, the line with different colors becomes a tedious way to represent the chirality. Therefore, we ignore the chirality information to illustrate the two kinds of flips for the states with arbitrary spins. Start from the highest weight component
\begin{equation} \begin{aligned} \label{eq:external}
\begin{tikzpicture}[baseline=0.7cm] \begin{feynhand}
\vertex [particle] (i1) at (1.5,0.8) {$i^h$};
\vertex [dot] (v1) at (0,0.8) {};
\graph{(i1)--[plain,very thick] (v1)};
\end{feynhand} \end{tikzpicture}
=|i]^{s+h}|i\rangle^{s-h}.
\end{aligned} \end{equation}
Note there is no $\eta$ in this component. From the helicity flip, the ladder operators tell the next component read
\begin{equation} \begin{aligned} \label{eq:MassInsertion}
\begin{tikzpicture}[baseline=0.7cm] \begin{feynhand}
\vertex [particle] (i1) at (1.5,0.8) {$i^{h-1}$};
\vertex [dot] (v1) at (0,0.8) {};
\graph{(i1)--[plain,very thick] (v1)};
\draw plot[mark=x,mark size=3.5] coordinates {(0.55,0.8)};
\end{feynhand} \end{tikzpicture}
&=J^-\circ|i]^{s+h}|i\rangle^{s-h}=|i]^{s+h-1}|i\rangle^{s-h}|\eta_i],\\
\begin{tikzpicture}[baseline=0.7cm] \begin{feynhand}
\vertex [particle] (i1) at (1.5,0.8) {$i^{h+1}$};
\vertex [dot] (v1) at (0,0.8) {};
\graph{(i1)--[plain,very thick] (v1)};
\draw plot[mark=x,mark size=3.5] coordinates {(0.55,0.8)};
\end{feynhand} \end{tikzpicture}
& =J^+\circ|i]^{s+h}|i\rangle^{s-h}=-|i]^{s+h}|i\rangle^{s-h-1}|\eta_i\rangle.
\end{aligned} \end{equation}
Performing the helicity flip repeatedly, we will find the other subleading components. For a spin-$s$ particle with helicity $h$, we need at most $2s$ flips to get all components. There are $s+h$ flips from $J^-$ and $s-h$ flips from $J^+$.

The chirality flip gives the next component with more $m$ or $\tilde{m}$. Diagrammatically, it gives
\begin{equation} \begin{aligned}
\begin{tikzpicture}[baseline=0.7cm] \begin{feynhand}
\vertex [particle] (i1) at (1.5,0.8) {$i^{h}$};
\vertex [dot] (v1) at (0,0.8) {};
\graph{(i1)--[plain,very thick] (v1)};
\draw[very thick] plot[mark=x,mark size=3.5] coordinates {(0.65,0.8)};
\end{feynhand} \end{tikzpicture}
=m_i|i]^{s+h}|i\rangle^{s-h}
\textrm{ or }\tilde{m}_i|i]^{s+h}|i\rangle^{s-h}.
\end{aligned} \end{equation}
Although we can perform chirality flips repeatedly to give the expression with arbitrary $t\in \frac{1}{2}\mathbbm{Z}$, only the state with $|t|\le s$ has the physical meaning. Performing the chirality flip repeatedly, we can find the subleading components with other transversality. In contrast to the helicity flip, $m$ and $\tilde{m}$ will not act on the single-particle state simultaneously.

\section{Large Energy Effective Theory for Massive 3-pt Amplitudes}
\label{sec:3pt1}

In the above section, the single-particle state of the massive spin-$s$ particle is defined to be transformed under the $U(2)$ LG. Scattering amplitude for $n$ particles should be LG covariant for each external particle under the $U(2)$ LG, so it can be written in terms of the $U(2)$ helicity-transversality spinors $\lambda_{\alpha}, \eta_{\alpha}$.   
The $n$-point scattering amplitude can thus be decomposed into
\begin{equation} \begin{aligned} \label{eq:MStoMH}
\mathcal{M}
&=\sum_{\mathcal{H}} \prod_{i=1}^n \left((\zeta^+_i)^{s_i+h_i}(\zeta^-_i)^{s_i-h_i}\right) \mathcal{M}^{\mathcal{H}}(\lambda,\tilde{\lambda},\eta,\tilde{\eta}),
\end{aligned} \end{equation}
where $\mathcal{H}=(h_1,\cdots,h_n)$ denotes the helicity category, and $\mathcal{M}^{\mathcal{H}}$ represents the scattering amplitude for external particles characterized by the quantum numbers $(s_i, h_i, t_i)$, with $h_i\in\{s_i,s_i-1,\cdots,-s_i\}$. Then the energy scaling behavior from the dimension analysis $\mathcal{M}_n \sim E^{4-n}$ indicates a further expansion of the amplitude
\begin{eqnarray}
\mathcal{M}^{\mathcal{H}} (\lambda,\tilde{\lambda},\eta,\tilde{\eta}) &\simeq& 
\left. \mathcal{M}^{\mathcal{H}}\right|_{\eta,\tilde{\eta}=0}
+\left.\frac{\partial\mathcal{M}^{\mathcal{H}}}{\partial\eta}\right|_{\eta,\tilde{\eta}=0}\eta+\left.\frac{\partial\mathcal{M}^{\mathcal{H}}}{\partial\tilde{\eta}}\right|_{\eta,\tilde{\eta}=0}\tilde{\eta}
+\cdots \\
&\sim& \sum_k E^{4-n} \left(\frac{\mathbf{m}}{E} \right)^k,
\end{eqnarray}
where the amplitude is expanded order by order using the power counting $\varepsilon_\eta = \frac{\mathbf{m}}{E}$ of external particles with large components $\lambda_{\alpha}, \tilde{\lambda}_{\dot{\alpha}}$ and small components $\eta_{\alpha},\tilde{\eta}_{\dot{\alpha}}$. 
This defines an effective field theory, denoted as LEFT, with three ingredients
\bit
\item building blocks: the large and small components of helicity-transversality spinors $\lambda_{\alpha}, \tilde{\lambda}_{\dot{\alpha}}, \eta_{\alpha},\allowbreak \tilde{\eta}_{\dot{\alpha}}$, and internal momenta $p_{\alpha\dot{\alpha}}$ and $\eta_{\alpha\dot{\alpha}}$, with the scaling behavior $\lambda_{\alpha}, \tilde{\lambda}_{\dot{\alpha}} \sim \sqrt{E}$ and $\eta_{\alpha}, \tilde{\eta}_{\dot{\alpha}}\sim \frac{\mathbf{m}}{\sqrt{E}}$;
\item symmetry: LG covariant $U(2)$ and other internal symmetries;
\item power counting: mass expansion $\eta_{\alpha},\tilde{\eta}_{\dot{\alpha}}\sim \varepsilon_\eta$ and $m,\tilde{m}\sim \varepsilon_\eta$. 
\eit
Note that the large component is not integrated out here, and therefore this large energy effective theory can describe the amplitudes in the high energy limit.


Furthermore, the scattering amplitudes should satisfy the momentum conservation law order by order in the effective theory. To avoid the breakdown of power counting rules, the momentum conservation should take the form
\bea
 \begin{cases}
\sum_i p_{i,\alpha\dot{\alpha}}=0, \\
\sum_i \eta_{i,\alpha\dot{\alpha}}=0,
\end{cases}
\label{eq:momenta-conserv}
\eea
for both real and complex momenta. This is crucial for a correct description of the massive amplitudes.
With this momentum conservation, we can organize the massive amplitudes using the large energy effective theory.

The on-shell mass insertion can be applied to the massive amplitudes in the large energy effective theory. The numbers of the helicity flip and chirality flip in an amplitude diagram clearly count the orders and powers of the large energy effective theory amplitudes. In the following we first construct the 3-pt massive amplitudes using the chirality and helicity flips.


\subsection{Generic 3-pt amplitudes}
\label{sec:generic3pt}

In the on-shell amplitude framework, the basic building block is the 3-pt scattering amplitudes, determined by LG covariance and locality. Similar to the single-particle states, we first consider how the helicity scaling determines the 3-pt contact amplitudes. Under the helicity transformation, 3-pt contact terms depend solely on helicity category and locality, independent of spin or energy scaling. There are five kinds of contact terms classified by mass: all-massless, 2-massless-1-massive, 1-massless-2-massive with unequal masses, 1-massless-2-massive with equal mass and all-massive.

First of all, all-massless amplitudes take the form
\begin{eqnarray}
\mathcal{A}^{\mathcal{H}}_{3} = \begin{cases}
\langle12\rangle^{h_3-h_1-h_2} \langle23\rangle^{h_1-h_2-h_3} \langle31\rangle^{h_2-h_1-h_3} & \text{for } h \equiv h_1 + h_2 + h_3 \leq 0 , \\
{[12]}^{h_1+h_2-h_3} {[23]}^{h_2+h_3-h_1} {[31]}^{h_3+h_1-h_2} & \text{for } h \geq 0 , 
\end{cases}\label{eq:3ml}
\end{eqnarray}
which follows from its helicity category and locality constraints. Momentum conservation and EOM imply that either $|1\rangle \propto |2\rangle \propto |3\rangle$ or $|1] \propto |2] \propto |3]$, and the LG scaling determines the exponent of each bracket. Locality necessitates that any poles are spurious or absent, thus determining the choice of bracket based on the helicity summation $h$. The energy scaling of the amplitude is $\mathcal{A}^{\mathcal{H}}_3 \sim E^{|h|}$, otherwise the energy scaling would be $\mathcal{A}^{\mathcal{H}}_3 \sim E^{-|h|}$, leading to a real pole in contact amplitudes.

The 2-massless-1-massive amplitudes take the same form as eq.~\eqref{eq:3ml} due to the momentum conservation given by eq.~\eqref{eq:momenta-conserv}. Momentum conservation for a 2-massless-1-massive amplitudes indicates $\eta = 0$, meaning that the spurion mass should have $\begin{cases}
m_i = \langle i\eta_i\rangle \rightarrow 0 \\
\Tilde{m}_i = [\eta_i i] \rightarrow \infty
\end{cases}$ or $\begin{cases}
m_i \rightarrow \infty \\
\Tilde{m}_i \rightarrow 0
\end{cases}$. Thus, the spurion mass should not appear in the kinematic structure and the amplitude is composed of $\lambda$ or $\tilde{\lambda}$. Along with locality, it is in the same form as eq.~\eqref{eq:3ml}.

The 1-massless-2-massive amplitudes $[\mathcal{M}^{\mathcal{H}}_3]$ take the form
\begin{eqnarray}
\mathcal{M}^{\mathcal{H}}_3 = \begin{cases}
\langle12\rangle^{h_3-h_1-h_2} \langle23\rangle^{h_1-h_2-h_3} \langle31\rangle^{h_2-h_1-h_3} f(m_1,m_2,\Tilde{m}_1,\Tilde{m}_2) & \text{for } h \leq 0, \\
{[12]}^{h_1+h_2-h_3} {[23]}^{h_2+h_3-h_1} {[31]}^{h_3+h_1-h_2} f(m_1,m_2,\Tilde{m}_1,\Tilde{m}_2) & \text{for } h \geq 0, 
\end{cases}
\end{eqnarray}
where $f(m_1,m_2,\tilde{m}_1,\tilde{m}_2)$ is an undetermined function of spurion masses. Here let us assume $h\leq0$ without loss of generality for derivation. Momentum conservation implies the following relations
\begin{eqnarray} \label{eq:kinematic1}
|1]\propto|2]\propto|3],\quad
|\eta_1]=|\eta_2],\quad
|\eta_1\rangle=-|\eta_2\rangle. 
\end{eqnarray} 
Since $\eta_1=-\eta_2$, we note that the following structures share the same scaling behavior $w_3^{-1} z_1 w_1$ under the $GL(1, \mathbbm{C})_w \times GL(1, \mathbbm{C})_z$ LG: 
\bea
\langle3\eta_1\rangle \sim \frac{\langle23\rangle}{\langle12\rangle \Tilde{m}_1} \sim \langle3\eta_2\rangle. 
\eea
This tells us that all the kinematic structures containing $\eta$ should be proportional to spurion masses, because
\begin{eqnarray}
\langle3\eta_1\rangle &=& \frac{\langle3\eta_1\rangle (\mathbf{m}_1^2 -\mathbf{m}_2^2)}{2(p_1+p_2)\cdot\eta_1} = -\frac{\langle23\rangle}{\langle12\rangle} \frac{{\mathbf{m}}_1^2 -{\mathbf{m}}_2^2}{\mathbf{m}_1^2}m_1, \quad 
{[3{\eta}_1]} = -\frac{\langle12\rangle}{\langle23\rangle} {\Tilde{m}}_1, \\
\langle2{\eta}_1\rangle &=& \frac{\langle23\rangle}{\langle13\rangle} \frac{{\mathbf{m}}_2^2}{{\Tilde{m}}_1} = -{m}_2, \quad
{[2{\eta}_1]} = \frac{\langle13\rangle}{\langle23\rangle} {\Tilde{m}}_1 = {\Tilde{m}}_2. 
\end{eqnarray}
Therefore, independent non-vanishing brackets should only contain $\lambda$ and spurion masses.

Finally, all massive 3-pt amplitudes have more possible expressions, because there are four kinds of 3-particle kinematics that satisfy the momentum conservation in eq.~\eqref{eq:momenta-conserv}:
\beq \label{eq:all-massive-kinematics}
\left\{\begin{aligned}
&|1]\propto|2]\propto|3]\\
&|\eta_1\rangle\propto|\eta_2\rangle\propto|\eta_3\rangle
\end{aligned}\right.,\;
\left\{\begin{aligned}
&|1]\propto|2]\propto|3]\\
&|\eta_1]\propto|\eta_2]\propto|\eta_3]
\end{aligned}\right.,\;
\left\{\begin{aligned}
&|1\rangle\propto|2\rangle\propto|3\rangle\\
&|\eta_1\rangle\propto|\eta_2\rangle\propto|\eta_3\rangle
\end{aligned}\right.,\;
\left\{\begin{aligned}
&|1\rangle\propto|2\rangle\propto|3\rangle\\
&|\eta_1]\propto|\eta_2]\propto|\eta_3]
\end{aligned}\right..
\eeq
We choose the first and the last kinematics, so that all massive 3-pt amplitudes take the following generic form
\begin{eqnarray}
[\mathcal{M}_3^{\mathcal{H}}]_{k,l} = \begin{cases}
\langle12\rangle^{a_1} \langle23\rangle^{a_2} \langle31\rangle^{a_3} [\eta_1\eta_2]^{b_1} [\eta_2\eta_3]^{b_2} [\eta_3\eta_1]^{b_3} f(m,\Tilde{m}), \\
[12]^{a_1} [23]^{a_2} [31]^{a_3} \langle\eta_1\eta_2\rangle^{b_1} \langle\eta_2\eta_3\rangle^{b_2} \langle\eta_3\eta_1\rangle^{b_3} f(m,\Tilde{m}), 
\end{cases} 
\end{eqnarray}
where $b_i=0$ if there is any massless particle, agreed with the previous cases. If we choose the other two kinematics in eq.~\eqref{eq:all-massive-kinematics}, the form would become more complicated than the above expressions. For convenience, we do not discuss them.

In the AHH formalism, instead of the helicity scaling, the $SU(2)$ covariance transformation is utilized to determine the 3-pt massive amplitude. Here 
for completeness, we review how the massive 3-pt amplitude is constructed in the AHH formalism. The massive amplitude $M$ can be viewed as a spin-$s_i$ representation for massive particle $i$. Therefore, we can decompose the LG covariant amplitude into $\lambda^I$ and a LG invariant. For 3-massive amplitude, we have
\begin{equation} \begin{aligned}
M^{(I_1\cdots I_{2s_1}),(J_1\cdots J_{2s_2}),(K_1\cdots K_{2s_3})}=&\lambda_{1,\alpha_1}^{(I_1}\cdots\lambda_{1,\alpha_{2s_1}}^{I_{2s_1})}\lambda_{2,\beta_1}^{(J_1}\cdots\lambda_{2,\beta_{2s_2}}^{J_{2s_2})}\lambda_{3,\gamma_1}^{(K_1}\cdots\lambda_{3,\gamma_{2s_3}}^{K_{2s_3})} \\
&\times \mathcal{I}^{(\alpha_1\cdots\alpha_{2s_1}),(\beta_1\cdots\beta_{2s_2}),(\gamma_1\cdots\gamma_{2s_3})}. 
\end{aligned} \end{equation}
In the AHH formalism, $\tilde{\lambda}^I$ can be converted to $\lambda^I$, because of the EOM. The LG invariant $\mathcal{I}$ can be written in terms of symmetric tensor $\mathcal{O}$, antisymmetric tensor $\varepsilon$. We have
\begin{equation}
\mathcal{I}^{(\alpha_1\cdots\alpha_{2s_1}),(\beta_1\cdots\beta_{2s_2}),(\gamma_1\cdots\gamma_{2s_3})}=\sum_{i=0}^{1}\sum_{\sigma_i}g_{\sigma_{i}}(\mathcal{O}^{s_1+s_2+s_3-i}\varepsilon^i)^{(\alpha_1\cdots\alpha_{2s_1}),(\beta_1\cdots\beta_{2s_2}),(\gamma_1\cdots\gamma_{2s_3})}_{\sigma_i}, 
\end{equation}
where $\sigma_i$ labels all permutations of spinor indices. 
Take $F\bar{F}V$ amplitude as an example, it has the form
\begin{equation}
\mathcal{I}^{\alpha,\beta,(\gamma_1\gamma_2)}=g_1\mathcal{O}^{\alpha(\gamma_1}\mathcal{O}^{\gamma_2)\beta}
+g_2\mathcal{O}^{\alpha\beta}\mathcal{O}^{(\gamma_1\gamma_2)}
+g_3 \mathcal{O}^{\alpha(\gamma_1}\varepsilon^{\gamma_2)\beta}
+g_4 \mathcal{O}^{\beta(\gamma_1}\varepsilon^{\gamma_2)\alpha}
+g_5\mathcal{O}^{(\gamma_1\gamma_2)}\varepsilon^{\alpha\beta}, 
\end{equation}
where
\begin{equation}
\mathcal{O}^{\alpha\beta}=\mathbf{p}_{1}^{\dot{\beta}(\alpha} \mathbf{p}_{2\dot{\beta}}^{\beta)}. 
\end{equation}
Contracting with $\lambda^I_{\alpha}\lambda^J_{\beta}\lambda^{(K_1}_{\gamma_1}\lambda^{K_2)}_{\gamma_2}$, the five products of tensors correspond to different Lorentz structures
\begin{equation} \begin{aligned}
\mathcal{O}^{\alpha(\gamma_1}\mathcal{O}^{\gamma_2)\beta}&\rightarrow\langle\mathbf{1}|\mathbf{p}_1\mathbf{p}_2-\mathbf{p}_2\mathbf{p}_1|\mathbf{3}\rangle\langle\mathbf{2}|\mathbf{p}_1\mathbf{p}_2-\mathbf{p}_2\mathbf{p}_1|\mathbf{3}\rangle, \\
\mathcal{O}^{\alpha\beta}\mathcal{O}^{(\gamma_1\gamma_2)}&\rightarrow\langle\mathbf{1}|\mathbf{p}_1\mathbf{p}_2-\mathbf{p}_2\mathbf{p}_1|\mathbf{2}\rangle\langle\mathbf{3}|\mathbf{p}_1\mathbf{p}_2-\mathbf{p}_2\mathbf{p}_1|\mathbf{3}\rangle, \\
\mathcal{O}^{\alpha(\gamma_1}\varepsilon^{\gamma_2)\beta}&\rightarrow\langle\mathbf{1}|\mathbf{p}_1\mathbf{p}_2-\mathbf{p}_2\mathbf{p}_1|\mathbf{3}\rangle\langle\mathbf{2}\mathbf{3}\rangle, \\
\mathcal{O}^{\beta(\gamma_1}\varepsilon^{\gamma_2)\alpha}&\rightarrow\langle\mathbf{2}|\mathbf{p}_1\mathbf{p}_2-\mathbf{p}_2\mathbf{p}_1|\mathbf{3}\rangle\langle\mathbf{1}\mathbf{3}\rangle, \\
\mathcal{O}^{(\gamma_1\gamma_2)}\varepsilon^{\beta\alpha}&\rightarrow\langle\mathbf{3}|\mathbf{p}_1\mathbf{p}_2-\mathbf{p}_2\mathbf{p}_1|\mathbf{3}\rangle\langle\mathbf{1}\mathbf{2}\rangle. \\
\end{aligned} \end{equation}
However, there is an EOM redundancy among these contractions. Using EOM and momentum conservation, these Lorentz structures can be reduced to a compact form, in which momentum $\mathbf{p}$ is converted to mass $\mathbf{m}$. For example, we have
\begin{equation} \begin{aligned}
\langle\mathbf{1}|\mathbf{p}_1\mathbf{p}_2-\mathbf{p}_2\mathbf{p}_1|\mathbf{3}\rangle\langle\mathbf{2}\mathbf{3}\rangle=
2\mathbf{m}_1 \mathbf{m}_3 [\mathbf{13}]\langle\mathbf{23}\rangle+(\mathbf{m}_2^2-\mathbf{m}_1^2-\mathbf{m}_3^2)\langle\mathbf{13}\rangle\langle\mathbf{23}\rangle. 
\end{aligned} \end{equation}
The mass $\mathbf{m}_i$ can be absorbed into the coefficients, so we just consider the brackets. Similarly for other terms, we find only four independent Lorentz structures: $\langle\mathbf{13}\rangle [\mathbf{23}]$, $[\mathbf{13}]\langle\mathbf{23}\rangle$, $\langle\mathbf{13}\rangle \langle\mathbf{23}\rangle$ and $[\mathbf{13}][\mathbf{23}]$. Thus, reorganizing the AHH $F\bar{F}V$ amplitude we obtain
\bea
M(\mathbf{1}^{\frac{1}{2}}, \mathbf{2}^{\frac{1}{2}}, \mathbf{3}^{1}) = g_1^\prime\langle\mathbf{13}\rangle [\mathbf{23}] + g_2^\prime[\mathbf{13}]\langle\mathbf{23}\rangle + g_3^\prime\langle\mathbf{13}\rangle \langle\mathbf{23}\rangle + g_4^\prime[\mathbf{13}][\mathbf{23}]. 
\eea
In the following we will show the above four Lorentz structures, enumerated by the Poincare symmetry, can actually be determined by the $SO(5, 1)$ symmetry. In another way to say, these Lorentz structures can be related by the $SO(5, 1)$  generators. Furthermore, the $SO(5, 1)$ symmetry would also give the mass dependent behaviors of the these Lorentz structures.

\subsection{Highest-weight construction for 3-massive amplitudes}
\label{sec:highest-weight}

The above $GL(1,\mathbbm{C})$ LG covariance can only determine the general form of a 3-pt amplitude. To determine the kinematic structure completely, the $GL(2,\mathbbm{C})$ covariance should be utilized by investigating the helicity and transversality flip, with the on-shell mass insertion in the 3-pt amplitudes.

Any 3-massive amplitudes can be decomposed into the massless ones with all the possible combinations of the helicity and chirality quantum numbers. 
Similar to the ladder operators $T^{\pm}$ and $J^{\pm}$ acting on the single-particle states, we start from the amplitude carrying the spin, helicity and transversality $(s, h, t)$ with the minimal mass insertion, defined as {\it highest weight} amplitude, and then act the ladder operators to obtain other different helicity amplitudes.

The mass insertion construction for 3-massive amplitudes is performed based on the following procedure: 


1. {\it highest weight representation}: Basically, an amplitude can be written as the contraction of the wave functions of external particles and the internal structure: 
\begin{equation}
\text{amplitude}=\text{wave function} \times \text{internal structure}. 
\end{equation}
The internal structure of a 3-pt amplitude does not include poles, so it always gives the local expression. When all the wave functions of the external particles are of the highest weight $t=h$, we call the amplitude the highest weight representation. Thus, the highest weight representation is characterized by $(s_i,h_i=t_i)$ with $-s_i\le h_i\le +s_i$. 

If the 3-pt amplitude satisfies the following condition
\bea \label{eq:LocalCondition}
\Delta s_i\equiv s_1+s_2+s_3-2s_i\ge 0,\quad i=1, 2, 3, 
\eea
its internal structure will be trivial, i.e. a combination of Levi-Civita tensor $\epsilon_{\alpha\beta}$ and $\epsilon_{\dot{\alpha}\dot{\beta}}$.\footnote{For the theory with spin $\le 1$, there are six kinds of 3-pt amplitudes, i.e. $SSS$, $VSS$, $VVS$, $VVV$, $F\bar{F}S$ and $F\bar{F}V$. Only the $VVS$ amplitude does not satisfy the locality condition. The SM does not have such an interaction, so each 3-pt massive amplitude in the SM can be expressed as a contraction of single-particle states.}Similar to the all-massless amplitudes determined from $U(1)$ LG scaling,  we can use $U(1)_w\times U(1)_t$ LG transformation to determine the highest weight representations. Here we focus on the representation with the maximal helicity $h_i=+s_i$ and use $\tilde{\lambda}$ to construct amplitudes. Consider the following ansatz 
\begin{equation} 
[12]^{y_{12}}[23]^{y_{23}}[31]^{y_{31}}. 
\end{equation}
To determine the unknown $y_{ij}$ exponential variables, we only need to consider the $U(1)_w$ LG covariance for massive particles. In the $U(1)_w$ LG of a helicity-$s$ particle, the scaling behavior is $w^{2s}$. So we have
\begin{equation}  \begin{aligned}
2s_1=y_{12}+y_{31},\\ 
2s_2=y_{12}+y_{23},\\
2s_3=y_{23}+y_{31}. 
\end{aligned}  \end{equation} 
Solving this equation, we find
\begin{equation} \begin{aligned}
y_{12}=s_1+s_2-s_3=\Delta s_3,\\ 
y_{23}=s_2+s_3-s_1=\Delta s_1,\\
y_{31}=s_3+s_1-s_2=\Delta s_2. 
\end{aligned} \end{equation} 
Therefore, the highest weight amplitude with maximal helicity is determined by the LG covariance
\begin{equation} \label{eq:HighAmp2}
\textrm{highest weight amp:} \quad [12]^{\Delta s_3}[23]^{\Delta s_1}[31]^{\Delta s_2}.
\end{equation}
For example, the 3-massive $F\bar{F}S$ amplitude with maximal $h$ is
\begin{equation} \tilde{\lambda}_1^{\dot{\alpha}}\tilde{\lambda}_2^{\dot{\beta}}\times \epsilon_{\dot{\alpha}\dot{\beta}}=[12].
\end{equation}

Remember that, as shown in eq.~\eqref{eq:TJ}, we have already used the commutator $[T_i^-,J_i^-]$ to find other highest weight components of single-particle state. At the amplitude level, we can perform
\begin{equation}
[T_i^-,J_i^-]^{\alpha}_{\dot{\alpha}} [T_j^-,J_j^-]^{\dot{\alpha}}_{\alpha} 
\end{equation}
on the amplitudes with maximal transversality to find other highest weight amplitudes. This operation is symmetric in $i\leftrightarrow j$, and it is equivalent to the replacement $[ij]\rightarrow \langle ij\rangle$, which changes transversality and helicity simultaneously. For 3-pt amplitudes, there are three replacements with $i,j=1,2,3$. We act these replacements on eq.~\eqref{eq:HighAmp2} and find the highest weight representations with lower helicity
\begin{equation} \begin{aligned}
\begin{tabular}{c|c}
\hline
helicity $=$ transversality & highest weight representation \\
\hline
$(s_1-1,s_2-1,s_3)$ & $\Delta s_3 \langle12\rangle [12]^{\Delta s_3-1}[23]^{\Delta s_1}[31]^{\Delta s_2}$ \\
$(s_1,s_2-1,s_3-1)$ & $\Delta s_1 \langle23\rangle [12]^{\Delta s_3}[23]^{\Delta s_1-1}[31]^{\Delta s_2}$ \\
$(s_1-1,s_2,s_3-1)$ & $\Delta s_2 \langle31\rangle [12]^{\Delta s_3}[23]^{\Delta s_1}[31]^{\Delta s_2-1}$ \\
\hline
\end{tabular}
\end{aligned} \end{equation}
Performing this repeatedly, we can find all the highest weight representations. For the amplitude with general spins $(s_1,s_2,s_3)$, the number of highest weight amplitudes is 
\begin{equation}
    n_{HW}=\prod_{i=1}^3 (\Delta s_i+1).
\end{equation}

For the 3-massive $F\bar{F}S$ amplitude, $n_{HW}=2$, so one replacement gives all highest weight amplitudes:
\begin{equation} \begin{aligned}
\begin{tikzpicture}[baseline=0.7cm] \begin{feynhand}
\vertex [particle] (i1) at (-0.2,0.8) {$1^{+\frac{1}{2}}$};
\vertex [particle] (i2) at (1.6,1.6) {$2^{+\frac{1}{2}}$};
\vertex [particle] (i3) at (1.6,0) {$3^{0}$};
\vertex (v1) at (0.9,0.8);
\graph{(i1)--[plain,cyan,very thick] (v1)};
\graph{(i2)--[plain,red,very thick] (v1)};
\graph{(i3)--[plain,brown,very thick] (v1)};
\end{feynhand} \end{tikzpicture}=[12] \quad \xrightarrow{[T_1^-,J_1^-][T_2^-,J_2^-]} \quad 
\begin{tikzpicture}[baseline=0.7cm] \begin{feynhand}
\vertex [particle] (i1) at (-0.2,0.8) {$1^{-\frac{1}{2}}$};
\vertex [particle] (i2) at (1.6,1.6) {$2^{-\frac{1}{2}}$};
\vertex [particle] (i3) at (1.6,0) {$3^{0}$};
\vertex (v1) at (0.9,0.8);
\graph{(i1)--[plain,red,very thick] (v1)};
\graph{(i2)--[plain,cyan,very thick] (v1)};
\graph{(i3)--[plain,brown,very thick] (v1)};
\end{feynhand} \end{tikzpicture}=\langle12\rangle,
\end{aligned} \end{equation}
while $[T_1^-,J_1^-][T_3^-,J_3^-]\langle12\rangle=[T_2^-,J_2^-][T_3^-,J_3^-]\langle12\rangle=0$. Similarly, for the $F\bar{F}V$ amplitude, there are four highest weight representations:
\begin{equation} \begin{aligned}
\begin{tikzpicture}[baseline=0.7cm] \begin{feynhand}
\vertex [particle] (i1) at (-0.2,0.8) {$1^{+\frac{1}{2}}$};
\vertex [particle] (i2) at (1.6,1.6) {$2^{+\frac{1}{2}}$};
\vertex [particle] (i3) at (1.6,0) {$3^{+1}$};
\vertex (v1) at (0.9,0.8);
\graph{(i1)--[plain,cyan,very thick] (v1)};
\graph{(i2)--[plain,red,very thick] (v1)};
\graph{(i3)--[plain,cyan,very thick] (v1)};
\end{feynhand} \end{tikzpicture}&=[23][31],&
\begin{tikzpicture}[baseline=0.7cm] \begin{feynhand}
\vertex [particle] (i1) at (-0.2,0.8) {$1^{-\frac{1}{2}}$};
\vertex [particle] (i2) at (1.6,1.6) {$2^{+\frac{1}{2}}$};
\vertex [particle] (i3) at (1.6,0) {$3^{0}$};
\vertex (v1) at (0.9,0.8);
\graph{(i1)--[plain,red,very thick] (v1)};
\graph{(i2)--[plain,red,very thick] (v1)};
\graph{(i3)--[plain,brown,very thick] (v1)};
\end{feynhand} \end{tikzpicture}&={[23]\langle 31\rangle},\\
\begin{tikzpicture}[baseline=0.7cm] \begin{feynhand}
\vertex [particle] (i1) at (-0.2,0.8) {$1^{+\frac{1}{2}}$};
\vertex [particle] (i2) at (1.6,1.6) {$2^{-\frac{1}{2}}$};
\vertex [particle] (i3) at (1.6,0) {$3^{0}$};
\vertex (v1) at (0.9,0.8);
\graph{(i1)--[plain,cyan,very thick] (v1)};
\graph{(i2)--[plain,cyan,very thick] (v1)};
\graph{(i3)--[plain,brown,very thick] (v1)};
\end{feynhand} \end{tikzpicture}&={\langle 23\rangle[31]},&
\begin{tikzpicture}[baseline=0.7cm] \begin{feynhand}
\vertex [particle] (i1) at (-0.2,0.8) {$1^{-\frac{1}{2}}$};
\vertex [particle] (i2) at (1.6,1.6) {$2^{-\frac{1}{2}}$};
\vertex [particle] (i3) at (1.6,0) {$3^{-1}$};
\vertex (v1) at (0.9,0.8);
\graph{(i1)--[plain,red,very thick] (v1)};
\graph{(i2)--[plain,cyan,very thick] (v1)};
\graph{(i3)--[plain,red,very thick] (v1)};
\end{feynhand} \end{tikzpicture}&=
\langle23\rangle\langle31\rangle.\\
\end{aligned} \end{equation}
In this step, we do not consider the 3-particle kinematics $|1\rangle \propto |2\rangle \propto |3\rangle$ or $|1] \propto |2] \propto |3]$, so $[23]\langle 31\rangle$ and $\langle 23\rangle[31]$ are not zero. In this form, we can focus on the application of mass insertion. 

If the condition eq.~\eqref{eq:LocalCondition} is not satisfied, we need to consider the non-trivial internal structure, i.e. momentum structure. For 3-pt amplitudes, it comes from the spin-$0$ single-particle states. Suppose that $\Delta s_3=s_1+s_2-s_3<0$. There is a new constraint for particle 3, $-s_1-s_2\le h_3 \le s_1+s_2$. In this case, the maximal helicity is $s_1+s_2$, so we use $\tilde{\lambda}_3^{2s_3-\Delta s}\lambda_3^{\Delta s}$ to give the highest weight representation
\bea \label{eq:SpinFactor}
\tilde{\lambda}_1^{2s_1} \tilde{\lambda}_2^{2s_2} \tilde{\lambda}_3^{2s_3+\Delta s_3}\lambda_3^{-\Delta s_3}\times \epsilon^{2s_1+2s_2} (p_1+\eta_1)^{-\Delta s_3}=
[23]^{2s_2}[31]^{2s_1} [3|p_1+\eta_1|3\rangle^{-\Delta s_3},
\eea
where $p_1+\eta_1$ is the momentum structure from particle 1.\footnote{We can choose the internal structure to be $p_2+\eta_2$, but the momentum conservation shows it gives the same amplitude as the choice of particle 1.} Then we can also perform $[T_i^-,J_i^-]\allowbreak \times[T_j^-,J_j^-]$ to derive other highest weight representations, in which $[3|p_1+\eta_1|3\rangle$ is unchanged. In this case, the number of highest weight representations is $n_{HW}=2s_1\times 2s_2$ . As an example, the amplitude with $(s_1,s_2,s_3)=(0,\frac{1}{2},\frac{3}{2})$ has two highest weight representations:
\begin{equation} \begin{aligned}
\begin{tikzpicture}[baseline=0.7cm] \begin{feynhand}
\setlength{\feynhandblobsize}{5mm}
\vertex [particle] (i1) at (0,0.8) {$1^{0}$};
\vertex [particle] (i2) at (1.6,1.6) {$2^{+\frac{1}{2}}$};
\vertex [particle] (i3) at (1.6,0) {$3^{+\frac{1}{2}}$};
\vertex [ringblob] (v1) at (0.9,0.8) {$p_1$};
\graph{(i1)--[plain,very thick] (v1)};
\graph{(i2)--[plain,very thick] (v1)};
\graph{(i3)--[plain,very thick] (v1)};
\end{feynhand} \end{tikzpicture}&=[23][3|p_1+\eta_1|3\rangle,&\\
\begin{tikzpicture}[baseline=0.7cm] \begin{feynhand}
\setlength{\feynhandblobsize}{5mm}
\vertex [particle] (i1) at (0,0.8) {$1^{0}$};
\vertex [particle] (i2) at (1.6,1.6) {$2^{-\frac{1}{2}}$};
\vertex [particle] (i3) at (1.6,0) {$3^{-\frac{1}{2}}$};
\vertex [ringblob] (v1) at (0.9,0.8) {$p_1$};
\graph{(i1)--[plain,very thick] (v1)};
\graph{(i2)--[plain,very thick] (v1)};
\graph{(i3)--[plain,very thick] (v1)};
\end{feynhand} \end{tikzpicture}&=\langle23\rangle[3|p_1+\eta_1|3\rangle.&
\end{aligned} \end{equation}

2. {\it helicity mass insertion}: After obtaining the highest weight helicity with leading contributions, we apply the ladder operator for helicity, namely $J^\pm$. As shown in eqs.~\eqref{eq:OperatorJ1} and \eqref{eq:OperatorJ2}, the operation of $J^\pm$ can be viewed as the replacements $|\lambda]\rightarrow|\eta]$ and $|\lambda\rangle\rightarrow-|\eta\rangle$, which would change helicity but not transversality. Thus the replacements on the highest weight helicity amplitudes would give rise to different helicity amplitudes. Note that we should sum over all possible kinematics structures with replaced spinors. Acting the replacement $|i]\rightarrow|\eta_i]$ with $i=1,2,3$ on eq.~\eqref{eq:HighAmp2}, we get
\begin{equation} \begin{aligned}
\begin{tabular}{c|c}
\hline
helicity & amplitude \\
\hline
$(s_1-1,s_2,s_3)$ & $\Delta s_3 [\eta_1 2] [12]^{\Delta s_3-1}[23]^{\Delta s_1}[31]^{\Delta s_2}+\Delta s_2 [3\eta_1] [12]^{\Delta s_3}[23]^{\Delta s_1}[31]^{\Delta s_2-1}$ \\
$(s_1,s_2-1,s_3)$ & $\Delta s_3 [1\eta_2] [12]^{\Delta s_3-1}[23]^{\Delta s_1}[31]^{\Delta s_2}+\Delta s_1 [\eta_23][12]^{\Delta s_3}[23]^{\Delta s_1-1}[31]^{\Delta s_2}$ \\
$(s_1,s_2,s_3-1)$ & $\Delta s_1 [2\eta_3][12]^{\Delta s_3}[23]^{\Delta s_1-1}[31]^{\Delta s_2}+\Delta s_2 [\eta_31][12]^{\Delta s_3}[23]^{\Delta s_1}[31]^{\Delta s_2-1}$ \\
\hline
\end{tabular}
\end{aligned} \end{equation}
They still have the maximal transversality $t_i=s_i$. For the highest weight representation with lower helicity, we should also consider the replacement $|i\rangle\rightarrow-|\eta_i\rangle$. Performing these replacements repeatedly, a highest weight repsentation will give $\prod_i(2s_i+1)-1$ structures with helicity mass insertions. In total, the 3-pt amplitude will have $n_{HW}\prod_i(2s_i+1)$ helicity structures, where $n_{HW}$ denotes the number of highest weight representations.

\begin{figure}[htbp]
\centering
\includegraphics[width=0.7\linewidth,valign=c]{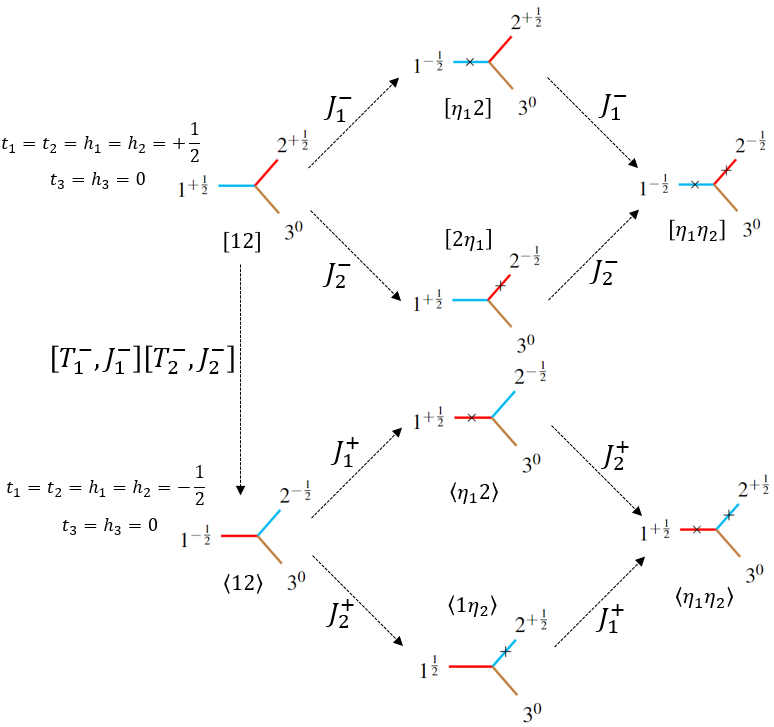}
\caption{The helicity structures for $F\bar{F}S$ amplitudes.}
\label{fig:helicity-FFS}
\end{figure}

For $F\bar{F}S$ amplitude, we can apply the replacement $|\lambda]\rightarrow|\eta]$ to the highest weight representation $[12]$ and get
\begin{equation}
\begin{tikzpicture}[baseline=0.7cm] \begin{feynhand}
\vertex [particle] (i1) at (-0.2,0.8) {$1^{-\frac{1}{2}}$};
\vertex [particle] (i2) at (1.6,1.6) {$2^{+\frac{1}{2}}$};
\vertex [particle] (i3) at (1.6,0) {$3^{0}$};
\vertex (v1) at (0.9,0.8);
\graph{(i1)--[plain,cyan,very thick] (v1)};
\graph{(i2)--[plain,red,very thick] (v1)};
\graph{(i3)--[plain,brown,very thick] (v1)};
\draw plot[mark=x,mark size=2.7] coordinates {(0.55,0.8)};
\end{feynhand} \end{tikzpicture}=[\eta_1 2],\quad
\begin{tikzpicture}[baseline=0.7cm] \begin{feynhand}
\vertex [particle] (i1) at (-0.2,0.8) {$1^{+\frac{1}{2}}$};
\vertex [particle] (i2) at (1.6,1.6) {$2^{-\frac{1}{2}}$};
\vertex [particle] (i3) at (1.6,0) {$3^{0}$};
\vertex (v1) at (0.9,0.8);
\graph{(i1)--[plain,cyan,very thick] (v1)};
\graph{(i2)--[plain,red,very thick] (v1)};
\graph{(i3)--[plain,brown,very thick] (v1)};
\draw plot[mark=x,mark size=2.7,mark options={rotate=45}] coordinates {(0.9+0.7*0.33,0.8+0.8*0.33)};
\end{feynhand} \end{tikzpicture}=[1\eta_2].
\end{equation}
Applying these replacements repeatedly for $[12]$ and $\langle12\rangle$, we get eight helicity structures. The result is shown in figure~\ref{fig:helicity-FFS}, in which the number and location of helicity mass insertion are changed by the replacement. For $F\bar{F}V$ amplitude, the helicity structures are given in appendix~\ref{app:3pt}. 



Moreover, if the highest-weight amplitude has the form in eq.~\eqref{eq:SpinFactor}, $J^\pm$ would also give the amplitudes in all helicity categories. For the universal structure $[3|p_1+\eta_1|3\rangle$, we only need to act $J^\pm$ on $|3]$ and $|3\rangle$, because $p+\eta$ is a $SU(2)_W$-singlet.

3. {\it Chirality mass insertion}:  Now we choose one helicity category and perform the chirality flip to the amplitudes in step 2. The chirality flip will insert spurion mass $m$ or $\tilde{m}$ into the amplitudes and change the chirality. Performing the chirality flips repeatedly, each term in the last step gives $\prod_i(2s_i+1)-1$ structures with chirality mass insertions.

For $F\bar{F}S$ amplitude, we choose the helicity category $(h_1,h_2,h_3)=(-\frac{1}{2},+\frac{1}{2},0)$. In this helicity category, there are two helicity structures $[\eta_1 2]$ and $-\langle 1 \eta_2\rangle$, which correspond to the highest weight representation $(+\frac12,+\frac12,0)$ and $(-\frac12,-\frac12,0)$. As shown in subsection~\ref{sec:MassInsert}, each massive fermion can have at most one transversality flip. For the highest weight representation $(+\frac12,+\frac12,0)$, we have
\begin{equation} \begin{aligned}
\begin{tikzpicture}[baseline=0.7cm] \begin{feynhand}
\vertex [particle] (i1) at (-0.2,0.8) {$1^{-\frac{1}{2}}$};
\vertex [particle] (i2) at (1.6,1.6) {$2^{+\frac{1}{2}}$};
\vertex [particle] (i3) at (1.6,0) {$3^{0}$};
\vertex (v1) at (0.9,0.8);
\graph{(i1)--[plain,cyan,very thick] (v1)};
\graph{(i2)--[plain,red,very thick] (v1)};
\graph{(i3)--[plain,brown,very thick] (v1)};
\draw plot[mark=x,mark size=2.7,mark options={rotate=0}] coordinates {(0.6,0.8)};
\end{feynhand} \end{tikzpicture}&=[\eta_1 2],&
\begin{tikzpicture}[baseline=0.7cm] \begin{feynhand}
\vertex [particle] (i1) at (-0.2,0.8) {$1^{-\frac{1}{2}}$};
\vertex [particle] (i2) at (1.6,1.6) {$2^{+\frac{1}{2}}$};
\vertex [particle] (i3) at (1.6,0) {$3^{0}$};
\vertex (v1) at (0.9,0.8);
\vertex (v2) at (0.9+0.7*0.33,0.8+0.8*0.33);
\graph{(i1)--[plain,cyan,very thick] (v1)};
\graph{(i2)--[plain,cyan,very thick](v2)--[plain,red,very thick] (v1)};
\graph{(i3)--[plain,brown,very thick] (v1)};
\draw plot[mark=x,mark size=2.7,mark options={rotate=0}] coordinates {(0.6,0.8)};
\draw[very thick] plot[mark=x,mark size=2.7,mark options={rotate=45}] coordinates {(0.9+0.7*0.35,0.8+0.8*0.35)};
\end{feynhand} \end{tikzpicture}&=m_2[\eta_1 2],&\\
\begin{tikzpicture}[baseline=0.7cm] \begin{feynhand}
\vertex [particle] (i1) at (-0.2,0.8) {$1^{-\frac{1}{2}}$};
\vertex [particle] (i2) at (1.6,1.6) {$2^{+\frac{1}{2}}$};
\vertex [particle] (i3) at (1.6,0) {$3^{0}$};
\vertex (v1) at (0.9,0.8);
\vertex (v2) at (0.5,0.8);
\graph{(i1)--[plain,red,very thick](v2)--[plain,cyan,very thick] (v1)};
\graph{(i2)--[plain,red,very thick] (v1)};
\graph{(i3)--[plain,brown,very thick] (v1)};
\draw plot[mark=x,mark size=2.7,mark options={rotate=0}] coordinates {(0.75,0.8)};
\draw[very thick] plot[mark=x,mark size=2.7] coordinates {(v2)};
\end{feynhand} \end{tikzpicture}&=m_1[\eta_1 2],&
\begin{tikzpicture}[baseline=0.7cm] \begin{feynhand}
\vertex [particle] (i1) at (-0.2,0.8) {$1^{+\frac{1}{2}}$};
\vertex [particle] (i2) at (1.6,1.6) {$2^{-\frac{1}{2}}$};
\vertex [particle] (i3) at (1.6,0) {$3^{0}$};
\vertex (v1) at (0.9,0.8);
\vertex (v2) at (0.5,0.8);
\vertex (v3) at (0.9+0.7*0.33,0.8+0.8*0.33);
\graph{(i1)--[plain,red,very thick](v2)--[plain,cyan,very thick] (v1)};
\graph{(i2)--[plain,cyan,very thick](v3)--[plain,red,very thick] (v1)};
\graph{(i3)--[plain,brown,very thick] (v1)};
\draw plot[mark=x,mark size=2.7,mark options={rotate=0}] coordinates {(0.75,0.8)};
\draw[very thick] plot[mark=x,mark size=2.7] coordinates {(v2)};
\draw[very thick] plot[mark=x,mark size=2.7,mark options={rotate=45}] coordinates {(v3)};
\end{feynhand} \end{tikzpicture}&=m_1 m_2[\eta_1 2].&
\end{aligned} \end{equation}
Similarly, we can start with the highest-weight representation $(+\frac{1}{2},+\frac{1}{2},0)$ and perform chirality flips. Finally, we derive the $F\bar{F}S$ amplitudes in this helicity category as
\begin{equation} \begin{aligned} \label{eq:resultFFS}
\mathcal{M}^{(-\frac{1}{2},+\frac{1}{2},0)}
=&\varepsilon_\eta(-c_1\langle 1\eta_2\rangle+c_2[\eta_1 2])
+\varepsilon_\eta^2\left(-(c_3 \tilde{m}_1+c_4 \tilde{m}_2)\langle 1\eta_2\rangle+(c_5 m_1+c_6 m_2)[\eta_1 2]\right)\\
&+\varepsilon_\eta^3\left(-c_7 \tilde{m}_1 \tilde{m}_2\langle 1\eta_2\rangle+c_8 m_1 m_2[\eta_1 2]\right),\\
\end{aligned} \end{equation}
where the coefficient $c_i$ is a function of $\mathbf{m}$ but not $m$ and $\tilde{m}$, and thus does not carry transversality. Here we impose the rescaling $\lambda,\tilde{\lambda}\rightarrow\varepsilon_{\eta}\lambda,\varepsilon_{\eta}\tilde{\lambda}$ to give the manifest power counting. Since this amplitude define a power expansion on $\eta$, it belongs to the EFT amplitude. To achieve this expansion, we impose the momentum conservation given by eq.~\eqref{eq:momenta-conserv}, so some structures may vanish, e.g. $\langle12\rangle[12]=0$ for $F\bar{F}V$ amplitude.

For other helicity categories, the amplitudes with all the possible transversality can be also derived by the chirality flips. The procedure can be illustrated by
\begin{equation}
\includegraphics[width=0.9\linewidth,valign=c]{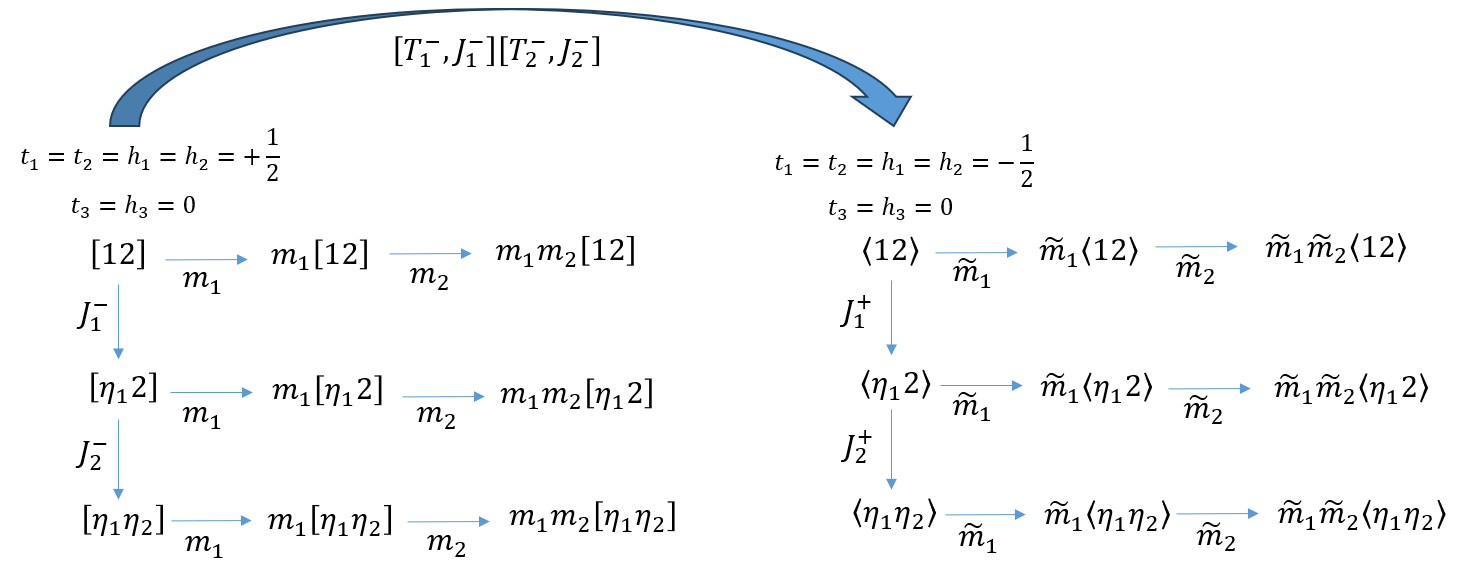}
\end{equation}
Note that the helicity structures are related by $J^\pm$, so they share the same coefficients. For example, in the helicity category $(-\frac{1}{2},-\frac{1}{2},0)$, the $F\bar{F}S$ amplitude should be
\begin{equation} \begin{aligned}  \label{eq:resultFFS2}
\mathcal{M}^{(-\frac{1}{2},-\frac{1}{2},0)}
=&c_1\langle 12\rangle+\varepsilon_\eta(c_3 \tilde{m}_1+c_4 \tilde{m}_2)\langle 12\rangle+\varepsilon_\eta^2(c_2[\eta_1 \eta_2]+c_7 \tilde{m}_1 \tilde{m}_2\langle 12\rangle)\\
&+\varepsilon_\eta^3(c_5 m_1+c_6 m_2)[\eta_1 \eta_2]+\varepsilon_\eta^4 c_8 m_1 m_2[\eta_1 \eta_2],
\end{aligned} \end{equation}
where coefficients $c_i$ are the same as the ones in eq.~\eqref{eq:resultFFS}.

Due to the $SU(2)$ LG covariance, the chirality amplitude with one helicity is enough to determine the bolded form.
Recovering the $SU(2)$ LG covariance from either eq.~\eqref{eq:resultFFS} or eq.~\eqref{eq:resultFFS2}, we obtain the bolded ST form
\begin{equation} \begin{aligned}
\mathcal{M}(\mathbf{1}^{\frac12},\mathbf{2}^{\frac12},\mathbf{1}^0)
=&(c_1+c_3 \tilde{m}_1+c_5 m_1+c_7 \tilde{m}_1 \tilde{m}_2)\langle \mathbf{12}\rangle+(c_2+c_4 \tilde{m}_1+c_6 m_1+c_8 \tilde{m}_1 \tilde{m}_2)[\mathbf{12}].
\end{aligned} \end{equation}

In summary, to obtain the 3-massive amplitudes, it is necessary to determine the chirality amplitude for a given helicity. This can be obtained by starting from the highest weight amplitude, and then acting $J^\pm$ on the amplitude to reach the amplitude with the targeting helicity, finally acting $m$ or $\tilde{m}$ to obtain the chirality amplitude with all the chirality mass insertion.


\subsection{Chirality-helicity unification and UV-IR correspondence}

We have discussed how to construct amplitudes with arbitrary helicity and transversality, but the coefficients of these amplitudes are still not yet determined. Let us consider the UV origins of these coefficients.

It is well-known that the helicity is equal to chirality for massless amplitudes. Therefore, any UV massless amplitudes $\mathcal{A}$ should match to the IR structures in which transversality is equal to helicity. Since the massless scalar boson does not carry chirality information, the UV amplitudes can have extra scalar bosons. This is the so-called "chirality-helicity unification". More specifically, for 3-pt massive amplitude, the UV-IR correspondence tells
\begin{equation} \label{eq:AtoM}
\mathcal{A}(1^{h_1}, 2^{h_2}, 3^{h_3};\overbrace{4^{0},\cdots,n^0}^{\textrm{extra scalar}})\rightarrow
\mathcal{M}(\mathbf{1}^{h_1=t_1}, \mathbf{2}^{h_2=t_2}, \mathbf{3}^{h_3=t_3}).
\end{equation}
Given a massive amplitude, to build the correspondence, it is necessary to determine the UV massless amplitudes, which indicates how many extra scalars should be determined. The procedure can be laid out as follows 
\bit 
\item For any 3-massive amplitudes, we exhaust all the possible helicity combinations, such as $2 \times 2 \times 1= 4$ for the $FFS$ amplitudes. 
\item For each helicity category $(h_1, h_2, h_3)$, it always has corresponding $n$-point UV amplitude with helicity category $(h_1, h_2, h_3; 0, \cdots, 0)$, in which external particles contain additional $n-3$ scalars. For renormalizable UV theory, the total helicity $h = \sum_i h_i$ should satisfy the following constraint $|h| \le |n-4|$~\footnote{Later we will show there would be an additional constraint for EFT contact UV amplitudes. }~\cite{Azatov:2016sqh}. Thus it provides correspondence to the UV amplitudes with additional scalars.  Note there is a maximal number of extra scalars for a 3-massive amplitude. For example, there would be maximal two extra scalars for $F\bar{F}V$ amplitudes. 
\item Given the UV massless amplitudes with fixed numbers of extra scalars, we can match to IR massive amplitude via the $\eta$ deformation, as well as the on-shell Higgsing~\cite{Balkin:2021dko,Bresciani:2023jsu}. 
\eit

\begin{figure}[htbp]
\centering
\includegraphics[width=1\linewidth,valign=c]{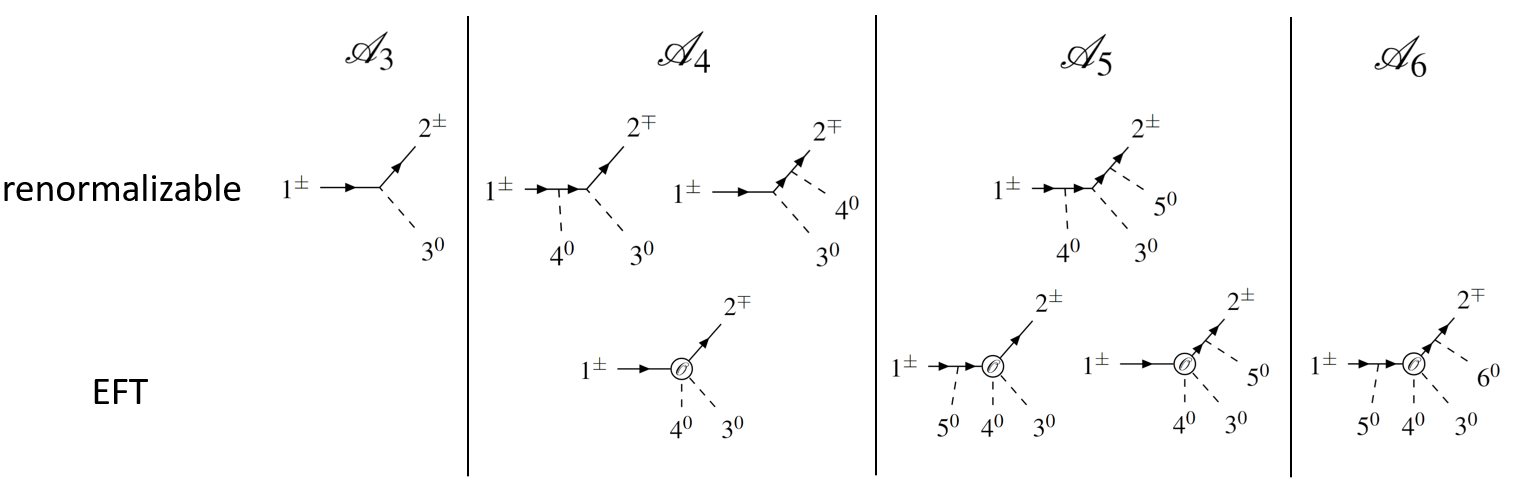}
\caption{All possible UV diagrams in renormalizable theory and EFT with dimension-6 operator $\mathcal{O}=\psi\bar{\psi}\varphi^2 D$. }
\label{fig:UVdiagram1}
\end{figure}

Consider the UV origins of the IR $F\bar{F}S$ amplitudes. First we list all the possible helicity combinations, and determine the coefficients from the helicity-chirality unification $h_i = t_i$. Then, from total helicity, we obtain the $n$-pt UV massless amplitudes. This connection is shown as follows
\begin{equation}
\begin{tabular}{c|c|c|c|c}
\hline
$(h_1,h_2,h_3)$ & coefficient with $h_i=t_i$ & total helicity & renormalizable & EFT \\
\hline
$(-\frac{1}{2},-\frac{1}{2},0)$ &
$c_1,c_8$  & $1$ & $\mathcal{A}_3,\mathcal{A}_5$ & $\mathcal{A}_5$ \\
$(-\frac{1}{2},+\frac{1}{2},0)$ &
$c_4,c_5$  & $0$ & $\mathcal{A}_4$ & $\mathcal{A}_4,\mathcal{A}_6$ \\
$(+\frac{1}{2},-\frac{1}{2},0)$ &
$c_3,c_6$  & $0$ & $\mathcal{A}_4$ & $\mathcal{A}_4,\mathcal{A}_6$ \\
$(+\frac{1}{2},+\frac{1}{2},0)$ &
$c_2,c_7$  & $-1$ & $\mathcal{A}_3,\mathcal{A}_5$ & $\mathcal{A}_5$ \\
\hline
\end{tabular}
\end{equation} 
Here we draw all the UV diagrams in figure~\ref{fig:UVdiagram1} and classify the UV theories. 

Let us analyse the UV diagrams by the numbers of external scalars. First consider the 3-pt UV amplitude, which has a direct correspondence to the 
amplitude with the highest weight $\langle12\rangle$. Thus the coefficient $c_1$ is equal to the Yukawa coupling $y$. Similarly for $[12]$ in another helicity category, we obtain $c_2=y$.

\begin{figure}[htbp]
\centering
\includegraphics[width=0.95\linewidth,valign=c]{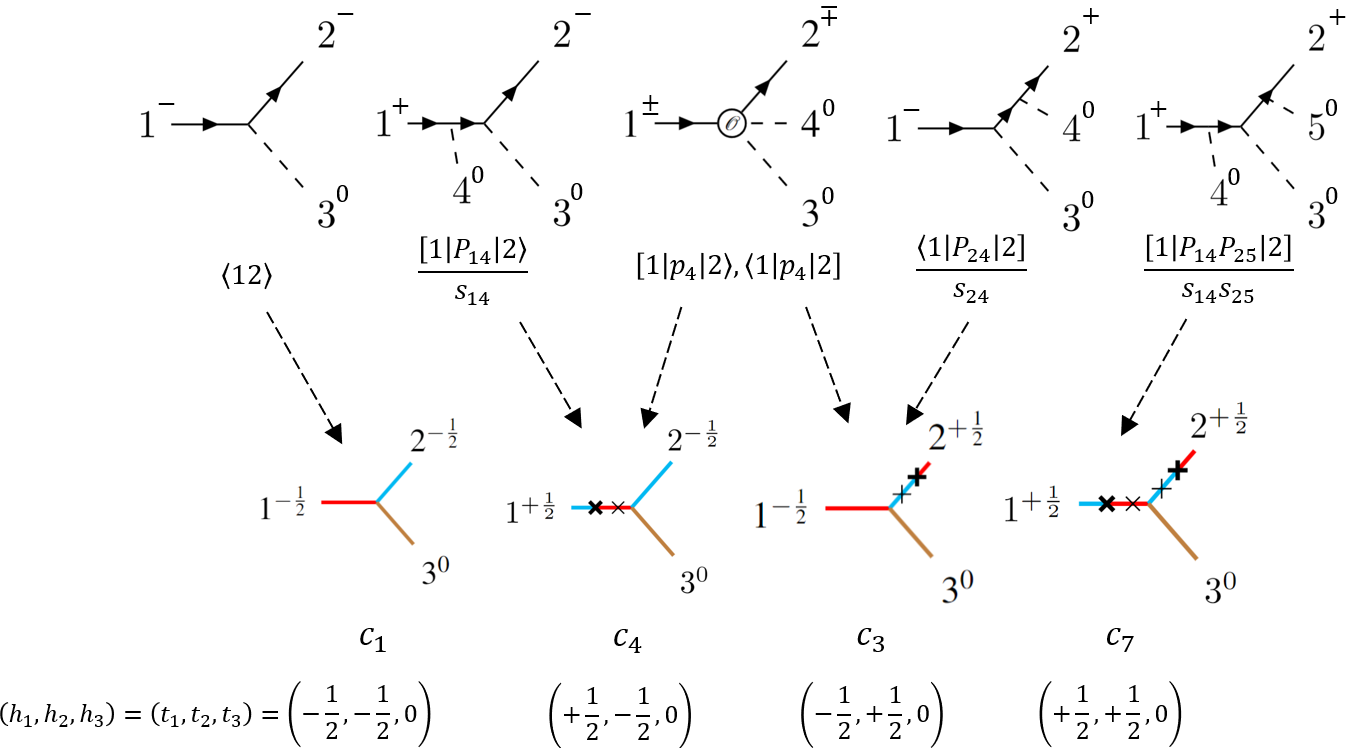}
\caption{The UV massless and IR massive amplitude correspondence for $F\bar{F}S$ amplitudes.}
\label{fig:UV-FFS}
\end{figure}

Then we consider the terms with mass insertions, which correspond to higher point UV amplitudes with extra scalars. Note that the helicity and chirality mass insertions should act on the same particle, e.g. $\tilde{m}_2\langle 1\eta_2\rangle$. This term $c_4$ corresponds to the 4-pt UV amplitude with an extra Higgs boson. Taking the on-shell limit of the Higgs momentum $p_4 \to \eta_2$, the UV amplitude reduces to
\begin{equation} \label{eq:limit4pt}
\lim_{p_4\rightarrow \eta_2} v \mathcal{A}_4=
\begin{tikzpicture}[baseline=0.7cm] \begin{feynhand}
\vertex [particle] (i1) at (-0.2,0.8) {$1^{-\frac{1}{2}}$};
\vertex [particle] (i2) at (1.6,1.6) {$2^{+\frac{1}{2}}$};
\vertex [particle] (i3) at (1.6,0) {$3^{0}$};
\vertex (v1) at (0.9,0.8);
\vertex (v2) at (0.9+0.7*0.4,0.8+0.8*0.4);
\graph{(i1)--[plain,red,very thick] (v1)};
\graph{(i2)--[plain,red,very thick](v2)--[plain,cyan,very thick] (v1)};
\graph{(i3)--[plain,brown,very thick] (v1)}; 
\draw plot[mark=x, mark size=2.7,mark options={rotate=45}] coordinates {(0.9+0.7*0.18,0.8+0.8*0.18)};
\draw[very thick] plot[mark=x,mark size=2.7,mark options={rotate=45}] coordinates {(v2)}; 
\end{feynhand} \end{tikzpicture},
\end{equation}
where $v$ denotes the VEV of the extra Higgs boson. From the 4-pt UV diagrams, $c_4$ may have three possible UV origins: 
\begin{itemize}
\item \textit{On-shell Higgsing: } The mass insertion is from the Higgs boson, which gives masses for fermions in the SM. Taking the on-shell limit in eq.~\eqref{eq:limit4pt}, the following 4-pt massless amplitude contributes to $F\bar{F}S$ amplitude, 
\begin{equation} \label{eq:Higgsing}
\begin{tikzpicture}[baseline=0.7cm] \begin{feynhand}
\setlength{\feynhandarrowsize}{3.5pt}
\vertex [particle] (i1) at (0,0.8) {$1$}; 
\vertex [particle] (i2) at (1.6,1.6) {$2$}; 
\vertex [particle] (i3) at (1.6,0) {$3$}; 
\vertex [particle] (i4) at (1.6,0.8) {$v$}; 
\vertex (v1) at (0.9,0.8); 
\vertex (v2) at (0.9+0.7*0.33,0.8+0.8*0.33); 
\graph{(i1)--[fer](v1)--[fer](v2)--[fer](i2)};
\graph{(i3)--[sca] (v1)}; 
\graph{(i4)--[sca] (v2)}; 
\end{feynhand} \end{tikzpicture}
=\lim_{p_4\rightarrow\eta_2} v y^2\frac{\langle 1|P_{24}|2]}{s_{24}}
=\frac{y}{\mathbf{m}_2}\tilde{m}_2\langle 1\eta_2\rangle, 
\end{equation}
where the pole becomes fermion mass $s_{24}=(p_2+p_4)^2\rightarrow \mathbf{m}_2^2=y^2 v^2$. Note the Yukawa interaction also contributes to $c_1=y$, so $c_1$ and $c_4$ are related in this case. 

\item \textit{Heavy fermion mixing:} If the external fermion is mixed with another heavy fermion, there would be no relation between $c_1$ and $c_4$. Taking the Yukawa coupling as $y^\prime$, the 4-pt amplitude reduces to
\begin{equation} \begin{aligned}
\begin{tikzpicture}[baseline=0.7cm] \begin{feynhand}
\setlength{\feynhandarrowsize}{3.5pt}
\vertex [particle] (i1) at (0,0.8) {$1$};
\vertex [particle] (i2) at (1.6,1.6) {$2$};
\vertex [particle] (i3) at (1.6,0) {$3$};
\vertex [particle] (i4) at (1.6,0.8) {$v$};
\vertex (v1) at (0.9,0.8);
\vertex (v2) at (0.9+0.7*0.33,0.8+0.8*0.33);
{\setlength{\feynhandlinesize}{1.3pt}
\propagator [fer] (v1) to (v2);}
\graph{(i1)--[fer](v1)};
\graph{(v2)--[fer](i2)};
\graph{(i3)--[sca] (v1)};
\graph{(i4)--[sca] (v2)};
\end{feynhand} \end{tikzpicture}
=\lim_{p_4\rightarrow\eta_2} v y^{\prime 2}\frac{\langle 1|P_{24}|2]}{s_{24}-\mathbf{m}_{\Psi}^2}
= \frac{vy^{\prime2}\tilde{m}_2\langle 1\eta_2\rangle}{\mathbf{m}^2_2-\mathbf{m}^2_{\Psi}},
\end{aligned} \end{equation}
where the thick line denotes possible heavy fermion $\Psi$.  

\item \textit{EFT:} In the standard model effective field theory, consider the dimension-6 operator $\mathcal{O}= \psi\bar{\psi}\varphi^2 D$~\cite{Grzadkowski:2010es}, in which two fermions are both left-handed. The 4-pt amplitude with operator insertion reduces to 
\begin{equation} 
\begin{tikzpicture}[baseline=0.7cm] \begin{feynhand}
\setlength{\feynhandblobsize}{4mm}
\setlength{\feynhandarrowsize}{3.5pt}
\vertex [particle] (i1) at (0,0.8) {$1$};
\vertex [particle] (i2) at (1.6,1.6) {$2$};
\vertex [particle] (i3) at (1.6,0) {$3$};
\vertex [particle] (i4) at (1.6,0.8) {$v$};
\vertex [ringblob] (v1) at (0.9,0.8) {\scriptsize $\mathcal{O}$};
\graph{(i1)--[fer] (v1)--[fer](i2)};
\graph{(i3)--[sca] (v1)};
\graph{(i4)--[sca] (v1)};
\end{feynhand} \end{tikzpicture}=\lim_{p_4\rightarrow\eta_2}v\frac{C_{\mathcal{O}}}{\Lambda^2}\langle 1|p_4|2]=\frac{vC_{\mathcal{O}}}{\Lambda^2}\tilde{m}_2\langle 1\eta_2\rangle,
\end{equation} 
where $C_{\mathcal{O}}$ is the Wilson coefficients, and $\Lambda$ is the cutoff, such as $\mathbf{m}_W$ in the pion decay case. 
\end{itemize}
Here we note that in the heavy fermion mixing case, the UV theory would contain massive particles. This can be viewed as the UV realization of the EFT operators. In the following, we will not discuss such massive particles in the UV. Instead, we would like to consider the EFT operators with massless particles. Finally let us summarize the UV-IR correspondence with all the possible massless UVs and IR amplitudes in figure~\ref{fig:UV-FFS}.

For the $F\bar{F}V$ amplitude, we also use the condition $h_i=t_i$ to select the terms corresponding to UV amplitudes. For example, in helicity category $(-\frac{1}{2},+\frac{1}{2},+1)$, the $F\bar{F}V$ amplitude is
\begin{equation} \begin{aligned} \label{eq:FFV1}
&\mathcal{M}^{\mathcal{H}=(-\frac{1}{2},+\frac{1}{2},+1)}\\
=&-(c_1+c_2 \tilde{m}_1+c_3 \tilde{m}_2+c_4 \tilde{m}_3+c_5 \tilde{m}_1 \tilde{m}_2+c_6 \tilde{m}_1 \tilde{m}_3+c_7 \tilde{m}_2 \tilde{m}_3+c_8 \tilde{m}_3^2+c_9 \tilde{m}_1 \tilde{m}_2 \tilde{m}_3\\
&+c_{10} \tilde{m}_1 \tilde{m}_3^2+c_{11} \tilde{m}_2 \tilde{m}_3^2+c_{12} \tilde{m}_1 \tilde{m}_2 \tilde{m}_3^2)\langle\eta_2 \eta_3\rangle\langle\eta_3 1\rangle-(c_{13}+c_{14} \tilde{m}_1+c_{15} m_2+c_{16} \tilde{m}_3\\
&+c_{17} m_3+c_{18} \tilde{m}_1 m_2+c_{19} \tilde{m}_1 \tilde{m}_3+c_{20} \tilde{m}_1 m_3+c_{21} m_2 \tilde{m}_3+c_{22} m_2 m_3+c_{23} \tilde{m}_1 m_2 \tilde{m}_3\\
&+c_{24} \tilde{m}_1 m_2 m_3)[23]\langle\eta_3 1\rangle+(c_{25}+c_{26} m_1+c_{27} \tilde{m}_2+c_{28} \tilde{m}_3+c_{29} m_3+c_{30} m_1 \tilde{m}_2\\
&+c_{31} m_1 \tilde{m}_3+c_{32} m_1 m_3+c_{33} \tilde{m}_2 \tilde{m}_3+c_{34} \tilde{m}_2 m_3+c_{35} m_1 \tilde{m}_2 \tilde{m}_3+c_{36} m_1 \tilde{m}_2 m_3)\langle\eta_2 \eta_3\rangle[3 \eta_1]\\
&+(c_{37}+c_{38} m_1+c_{39} m_2+c_{40} \tilde{m}_3+c_{41} m_1 m_2+c_{42} m_1 \tilde{m}_3+c_{43} m_2 \tilde{m}_3+c_{44} \tilde{m}_3^2+c_{45} m_1 m_2 \tilde{m}_3\\
&+c_{46} m_1 \tilde{m}_3^2+c_{47} m_2 \tilde{m}_3^2+c_{48} m_1 m_2 \tilde{m}_3^2)[23][3\eta_1].\\
\end{aligned} \end{equation}
Diagrammatically, the terms with only helicity mass insertion are given in eq.~\eqref{eq:FFV-hflip}. Here we list some typical diagrams with chirality mass insertions. For $c_{15}$ and $c_{16}$ structures, we have
\begin{equation} \begin{aligned}
\begin{tikzpicture}[baseline=0.55cm,scale=0.8,transform shape] \begin{feynhand}
\vertex [particle] (i1) at (-0.2,0.8) {$1^{-\frac{1}{2}}$};
\vertex [particle] (i2) at (1.6,1.6) {$2^{+\frac{1}{2}}$};
\vertex [particle] (i3) at (1.6,0) {$3^{+1}$}; 
\vertex (v1) at (0.9,0.8);
\vertex (v2) at (0.9+0.7*0.35,0.8+0.8*0.35);
\graph{(i1)--[plain,red,very thick] (v1)}; 
\graph{(i2)--[plain,cyan,very thick](v2)--[plain,red,very thick] (v1)};
\graph{(i3)--[plain,brown,very thick] (v1)}; 
\draw[very thick] plot[mark=x,mark size=2.7,mark options={rotate=45}] coordinates {(v2)}; 
\draw plot[mark=x,mark size=2.7,mark options={rotate=45}] coordinates {(0.9+0.7*0.3,0.8-0.8*0.3)}; 
\end{feynhand} \end{tikzpicture}=-m_2[23]\langle\eta_3 1\rangle,\quad
\begin{tikzpicture}[baseline=0.55cm,scale=0.8,transform shape] \begin{feynhand}
\vertex [particle] (i1) at (-0.2,0.8) {$1^{-\frac{1}{2}}$};
\vertex [particle] (i2) at (1.6,1.6) {$2^{+\frac{1}{2}}$};
\vertex [particle] (i3) at (1.6,0) {$3^{+1}$}; 
\vertex (v1) at (0.9,0.8);
\vertex (v2) at (0.9+0.7*0.41,0.8-0.8*0.41); 
\graph{(i1)--[plain,red,very thick] (v1)}; 
\graph{(i2)--[plain,red,very thick] (v1)}; 
\graph{(i3)--[plain,cyan,very thick](v2)--[plain,brown,very thick] (v1)}; 
\draw[very thick] plot[mark=x,mark size=2.7,mark options={rotate=45}] coordinates {(v2)}; 
\draw plot[mark=x,mark size=2.7,mark options={rotate=45}] coordinates {(0.9+0.7*0.18,0.8-0.8*0.18)}; 
\end{feynhand} \end{tikzpicture}=-\tilde{m}_3[23]\langle\eta_3 1\rangle. 
\end{aligned} \end{equation}

In eq.~\eqref{eq:FFV1}, there are four terms satisfying $h_i=t_i$: $c_{11}$, $c_{16}$, $c_{35}$, $c_{38}$. For $c_{16}$, the corresponding UV amplitude is a massless $F\bar{F}V$ amplitude
\begin{equation} \begin{aligned}
\frac{[23]^2}{[12]}=\frac{[23][2|3|\eta_3\rangle}{[12]\langle3\eta_3\rangle}=\frac{\tilde{m}_3}{\mathbf{m}_3^2}[23]\langle\eta_3 1\rangle. 
\end{aligned} \end{equation}
The other terms correspond to higher-point UV amplitudes. 

In addition to massless $F\bar{F}V$ amplitude, the massless $F\bar{F}S$ amplitude also determines the coefficients of massive $F\bar{F}V$ structure. In helicity category $(+\frac{1}{2},+\frac{1}{2},0)$, $c_{14}$ and $c_{27}$ correspond to the 3-pt UV amplitude
\begin{equation}
[12]=b_1\frac{\tilde{m}_1}{\mathbf{m}_1^2}[23]\langle3\eta_1\rangle+b_2\frac{\tilde{m}_2}{\mathbf{m}_2^2}\langle\eta_2 3\rangle[31], 
\end{equation}
where $b_1+b_2=1$ are determined by matching from massless UV. This matching gives the on-shell version of the Higgs mechanism. 

\begin{figure}[htbp]
\centering
\includegraphics[width=0.95\linewidth,valign=c]{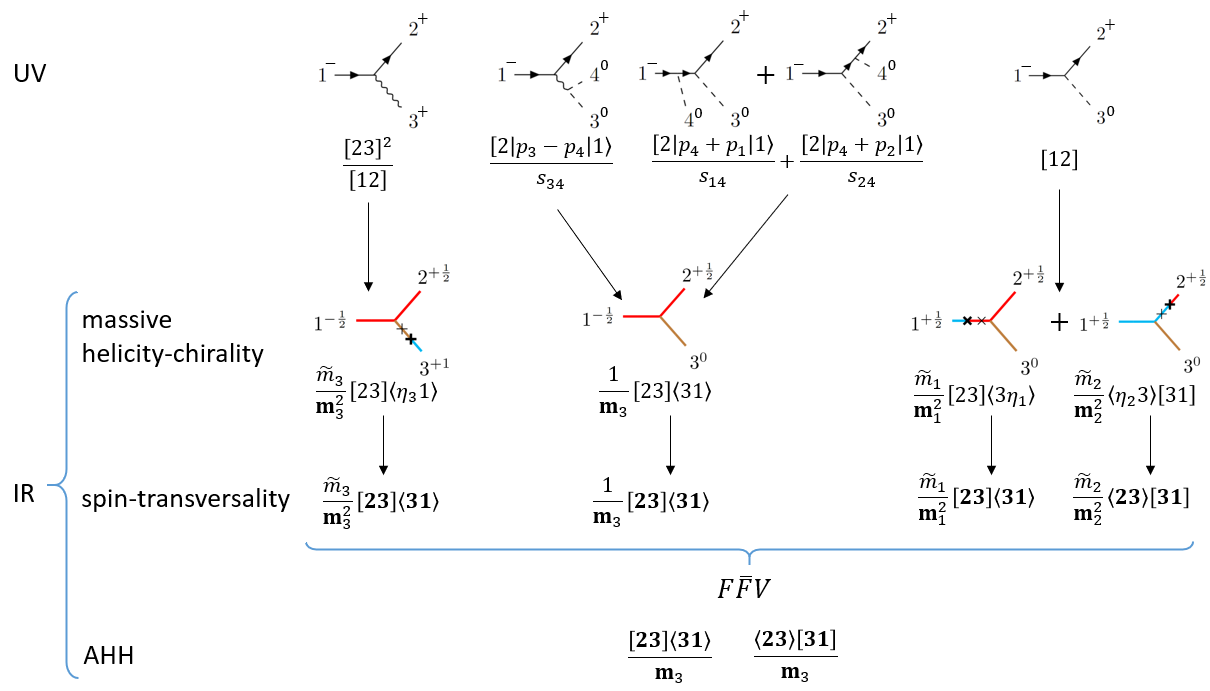}
\caption{The UV-IR correspondence for $F\bar{F}V$ amplitudes. There exist three types of IR massive amplitudes: massive helicity-chirality, spin-transversality, and AHH. These amplitudes are respectively expressed in terms of helicity-transversality spinors, spin-transversality spinors, and spin-spinors. 
}
\label{fig:UV-FFV}
\end{figure}

Then we consider the helicity category $(-\frac{1}{2},+\frac{1}{2},0)$, which includes the highest weight represetation $[23]\langle31\rangle$. The corresponding UV amplitude is massless $F\bar{F}VS$ amplitude. Note that $[23]\langle31\rangle$ should vanish due to momentum conservation eq.~\eqref{eq:momenta-conserv}. In the on-shell limit, the 4-point UV amplitude does not reduce to $[23]\langle31\rangle$, but the one with two chirality mass insertions, i.e. $[2\eta_3]\langle\eta_3 1\rangle$. This UV amplitude has two contributions: the diagram including gauge boson and fermions, which are given in figure~\ref{fig:UV-FFV}. They can both reduce to the IR structure. For the gauge boson exchange diagram, we have
\begin{equation} \label{eq:ReduceFFVS}
\lim_{p_4\rightarrow \eta_3} vg^2\frac{[2|p_3-p_4|1\rangle}{s_{34}}=\frac{g}{\mathbf{m}_3}[2\eta_3]\langle\eta_3 1\rangle, 
\end{equation}
where $g$ is gauge coupling and $\mathbf{m}_3=gv$ is the mass of vector boson. The fermion exchange diagrams have two channels $s_{14}$ and $s_{24}$. Suppose that the internal fermions are different in the two channels and the Yukawa couplings are $y_1$ and $y_2$ separately. The 4-pt UV amplitude reduces to
\begin{equation}
\sum_{i=1}^2\lim_{p_4\rightarrow \eta_i} v y_i^2\frac{[2|p_4+p_i|i\rangle}{s_{i4}}
=\frac{y_1}{\mathbf{m}_1}[2|\eta_1|1\rangle+\frac{y_2}{\mathbf{m}_2}[2|\eta_2|1\rangle
=-\frac{1}{v}[2\eta_3]\langle\eta_3 1\rangle, 
\end{equation}
where $\mathbf{m}_i=y_i v$ is the fermion mass. This formula should be viewed as the on-shell version of the Goldstone equivalence theorem, because there is no gauge boson in the UV amplitude. This result should match eq.~\eqref{eq:ReduceFFVS}, so we have a constriant, $\frac{g}{\mathbf{m}_V}=\frac{y}{\mathbf{m}_F}$.

Until now, we have given the correspondence between the UV and IR amplitudes, which are expressed in terms of the helicity-transversality spinors. Actually, we can enlarge this correspondence to involve IR structure with ST spinors and spin-spinors. Thus, the complete UV-IR correspondence has three stages: 
\begin{itemize}
\item \textit{UV to MHC amplitude:} 
In this stage, massless amplitudes match to the MHC amplitudes with $h_i=t_i$, as shown in eq.~\eqref{eq:AtoM}, via the on-shell Higgsing. The helicity information from UV becomes the transversality information at the IR. The coefficients of the MHC amplitudes are determined by the UV couplings. 

\item \textit{MHC amplitude to ST amplitude: }  
Note that the MHC amplitudes related by $J^{\pm}$ have the same coefficients. Therefore, they can be unified into one ST amplitude. The amplitudes with different helicity categories are viewed as the components of the ST amplitude. For example, the $F\bar{F}V$ amplitude have the following unification: 
\begin{equation}
\begin{tabular}{c|c|c|c|c}
\hline
\multirow{2}{*}{ST spinor} & \multicolumn{4}{c}{helicity} \\
\cline{2-5}
& $(-\frac{1}{2},-\frac{1}{2},0)$ & $(-\frac{1}{2},+\frac{1}{2},-1)$ & $(-\frac{1}{2},+\frac{1}{2},+1)$ & $(+\frac{1}{2},+\frac{1}{2},0)$ \\
\hline
$[\mathbf{23}]\langle\mathbf{31}\rangle$ & $[\eta_23]\langle31\rangle$ & $[2\eta_3]\langle31\rangle$ & $-[23]\langle\eta_3 1\rangle$ & $-[23]\langle3\eta_1\rangle$ \\
\hline
\end{tabular}
\end{equation}
where only the representative helicity categories with their leading terms are listed. 

\item \textit{ST amplitude to AHH amplitude: } 
Consider $SU(2)$ LG, which is a subgroup of the $U(2)$ LG. In the representation of $SU(2)$, the ST spinors reduce to the spin-spinors in the AHH amplitude. In this reduction, the spurion mass $m$ and $\tilde{m}$ are fixed at the physical mass $\mathbf{m}$. For the $F\bar{F}V$ case, the ST amplitudes with one chirality mass insertion have the following correspondence:
\begin{equation}
\begin{tabular}{c|c|c|c|c}
\hline
\multirow{2}{*}{spin-spinor} & \multicolumn{4}{c}{transversality} \\
\cline{2-5}
& $(-\frac{1}{2},+\frac{1}{2},0)$ & $(+\frac{1}{2},+\frac{1}{2},0)$ & $(-\frac{1}{2},-\frac{1}{2},0)$ & $(-\frac{1}{2},+\frac{1}{2},-1)$ \\
\hline
$[\mathbf{23}]\langle\mathbf{31}\rangle$ & $\tilde{m}_1[\mathbf{23}]\langle\mathbf{31}\rangle$ & $m_2[\mathbf{23}]\langle\mathbf{31}\rangle$ & $m_3[\mathbf{23}]\langle\mathbf{31}\rangle$ & $\tilde{m}_3[\mathbf{23}]\langle\mathbf{31}\rangle$ \\
\hline
\end{tabular}
\end{equation}
When we replace $m$ and $\tilde{m}$ by $\mathbf{m}$, there would be IR unification. For example, for the $F\bar{F}S$ amplitude, as shown in eq.~\eqref{eq:Higgsing}, the coefficients give $c_4\rightarrow y=c_1$, indicating the same UV origin for these two terms. Thus we should take them as one contribution at the IR.

\end{itemize}


\subsection{Mass enhancement in top decays}

%
%





Let us consider the top decay, which can be described by the amplitude $\mathcal{M}(\bar{t},b,W^+)$. This is a 3-massive amplitude with $(s_1,s_2,s_3)=(\frac{1}{2},\frac{1}{2},1)$. In the SM, $W^+$ boson couples to the left-handed top and bottom quarks, so the transversality is $(t_1,t_2,t_3)=(+\frac{1}{2},-\frac{1}{2},0)$. 

Let us focus on the leading contributions for all the helicity categories. The leading amplitudes are these without chirality flip, and thus all the terms with chirality mass insertion are neglected. 
From Appendix~\ref{app:3pt} we can read out the leading results for each helicity. We expand these expressions with the following power counting  $\lambda \sim \sqrt{E}, \eta \sim \mathbf{m}/\sqrt{E}$ and obtain the following scaling behaviors
\begin{equation} \begin{aligned} \label{eq:tbW}
\mathcal{M}^{(+\frac{1}{2},-\frac{1}{2},-1)}
&=\langle23\rangle[\eta_3 1]\sim \mathbf{m}_W E,\qquad
\mathcal{M}^{(+\frac{1}{2},+\frac{1}{2},-1)}
=-\langle\eta_2 3\rangle[\eta_3 1]\sim \mathbf{m}_b \mathbf{m}_W,\\
\mathcal{M}^{(+\frac{1}{2},-\frac{1}{2},+1)}
&=-\langle2\eta_3\rangle[3 1]\sim \mathbf{m}_W E,\qquad
\mathcal{M}^{(+\frac{1}{2},+\frac{1}{2},+1)}
=\langle\eta_2 \eta_3\rangle[3 1]\sim \mathbf{m}_b \mathbf{m}_W,\\
\mathcal{M}^{(-\frac{1}{2},-\frac{1}{2},-1)}
&=\langle23\rangle[\eta_3 \eta_1]\sim \mathbf{m}_W E,\qquad
\mathcal{M}^{(-\frac{1}{2},+\frac{1}{2},-1)}
=-\langle\eta_2 3\rangle[\eta_3 \eta_1]\sim \mathbf{m}_b \mathbf{m}_W, \\
\mathcal{M}^{(-\frac{1}{2},-\frac{1}{2},+1)}
&=-\langle2\eta_3\rangle[3 \eta_1]\sim \mathbf{m}_W E,\qquad
\mathcal{M}^{(-\frac{1}{2},+\frac{1}{2},+1)}
=\langle\eta_2 \eta_3\rangle[3 \eta_1]\sim \mathbf{m}_b \mathbf{m}_W, \\
\mathcal{M}^{(+\frac{1}{2},-\frac{1}{2},0)}
&=\frac{1}{\sqrt{2}}(\langle23\rangle[31]-\langle2\eta_3\rangle[\eta_31])\sim E^2+\mathcal{O}(\mathbf{m}_t^2), \\
\mathcal{M}^{(+\frac{1}{2},+\frac{1}{2},0)}
&=\frac{1}{\sqrt{2}}(-\langle\eta_2 3\rangle[31]+\langle\eta_2\eta_3\rangle[\eta_31])\sim \mathbf{m}_b E+\mathcal{O}(\mathbf{m}_b \mathbf{m}_t^2 E^{-1}), \\
\mathcal{M}^{(-\frac{1}{2},-\frac{1}{2},0)}
&=\frac{1}{\sqrt{2}}(\langle23\rangle[3\eta_1]-\langle2\eta_3\rangle[\eta_3\eta_1])\sim \mathbf{m}_t E+\mathcal{O}(\mathbf{m}_t\mathbf{m}^2_W E^{-1}), \\
\mathcal{M}^{(-\frac{1}{2},+\frac{1}{2},0)}
&=\frac{1}{\sqrt{2}}(-\langle\eta_2 3\rangle[3\eta_1]+\langle\eta_2\eta_3\rangle[\eta_3\eta_1])\sim \mathbf{m}_b \mathbf{m}_t+\mathcal{O}(\mathbf{m}_b \mathbf{m}_t\mathbf{m}^2_W E^{-2}). \\
\end{aligned} \end{equation}
where the $\sqrt{2}$ comes from the normalization of single-particle state. There is a hierarchy among the three massive particles, i.e. $\mathbf{m}_t$, $\mathbf{m}_W\gg \mathbf{m}_b$, so the amplitudes with $\eta_2$ and $\mathbf{m}_b$ are not the leading contributions. 

The 3-pt amplitudes can be obtained explicitly by assigning momenta in the center of mass frame, or here the top quark rest frame. In this frame, $E=\mathbf{m}_t$, so $\tilde{\eta}_1$ give the same order contribution as $\tilde{\lambda}_1$. The particle state with $(s,h)$ should match to the one with $(s,s_z)$. We pick $h=+\frac{1}{2}$ state of top quark along the $z$ axis
\begin{equation} \begin{aligned}
\tilde{\lambda}_1=\sqrt{\mathbf{m}_t}
\begin{pmatrix}
    1 \\ 0 
\end{pmatrix},\quad
\tilde{\eta}_1=\sqrt{\mathbf{m}_t}
\begin{pmatrix}
    0 \\ 1 
\end{pmatrix}.
\end{aligned} \end{equation}
This frame is also the center-of-mass frame of the bottom quark and $W$ boson, so we have $\theta=\theta_2=\pi-\theta_3$ and $\varphi=\varphi_2=\pi+\varphi_3$. According to eq.~\eqref{eq:helicity-spinor-rep}, we have
\begin{equation} \begin{aligned}
\lambda_2&=\sqrt{E_b+P}
\begin{pmatrix}
    -e^{-i\varphi}\sin{\frac{\theta}{2}} \\ \cos{\frac{\theta}{2}} 
\end{pmatrix}, &
\eta_2&=\frac{\mathbf{m}_b}{\sqrt{E_b+P}}
\begin{pmatrix}
    \cos{\frac{\theta}{2}} \\ e^{i\varphi}\sin{\frac{\theta}{2}}
\end{pmatrix}, \\
\lambda_3&=\sqrt{E_W+P}
\begin{pmatrix}
    e^{-i\varphi}\cos{\frac{\theta}{2}} \\ \sin{\frac{\theta}{2}} 
\end{pmatrix}, &
\eta_3&=\frac{\mathbf{m}_W}{\sqrt{E_W+P}}
\begin{pmatrix}
    \sin{\frac{\theta}{2}} \\ -e^{i\varphi}\cos{\frac{\theta}{2}}
\end{pmatrix},\\
\tilde{\lambda}_3&=\sqrt{E_W+P}
\begin{pmatrix}
    \sin{\frac{\theta}{2}} \\ -e^{i\varphi}\cos{\frac{\theta}{2}} 
\end{pmatrix}, &
\tilde{\eta}_3&=\frac{\mathbf{m}_W}{\sqrt{E_W+P}}
\begin{pmatrix}
    e^{-i\varphi}\cos{\frac{\theta}{2}} \\ \sin{\frac{\theta}{2}}
\end{pmatrix}, 
\end{aligned} \end{equation}
where $\lambda_2\propto \eta_3$ and $\eta_2\propto \lambda_3$. Substituting into eq.~\eqref{eq:tbW}, we find
\begin{equation} \begin{aligned} \label{eq:tbW2}
\mathcal{M}^{(+\frac{1}{2},-\frac{1}{2},-1)}
&=\mathbf{m}_W X_1 e^{-i\varphi}\sin{\frac{\theta}{2}},\qquad
\mathcal{M}^{(-\frac{1}{2},-\frac{1}{2},-1)}
=\mathbf{m}_W X_1 \cos{\frac{\theta}{2}}, \\
\mathcal{M}^{(-\frac{1}{2},+\frac{1}{2},+1)}
&=\mathbf{m}_W X_2 (-e^{i\varphi}\sin{\frac{\theta}{2}}),\qquad
\mathcal{M}^{(+\frac{1}{2},+\frac{1}{2},+1)}
=\mathbf{m}_W X_2 e^{2i\varphi}\cos{\frac{\theta}{2}}, \\
\mathcal{M}^{(-\frac{1}{2},-\frac{1}{2},0)}
&=\frac{E_W+P}{\sqrt{2}}X_1(-e^{-i\varphi}\sin{\frac{\theta}{2}}),\quad
\mathcal{M}^{(+\frac{1}{2},-\frac{1}{2},0)}
=\frac{E_W+P}{\sqrt{2}}X_1\cos{\frac{\theta}{2}}, \\
\mathcal{M}^{(+\frac{1}{2},+\frac{1}{2},0)}
&=-\frac{E_W-P}{\sqrt{2}}X_2 e^{i\varphi}\sin{\frac{\theta}{2}},\quad
\mathcal{M}^{(-\frac{1}{2},+\frac{1}{2},0)}
=-\frac{E_W-P}{\sqrt{2}}X_2(-\cos{\frac{\theta}{2}}), \\
\mathcal{M}^{(+\frac{1}{2},-\frac{1}{2},+1)}
&=\mathcal{M}^{(-\frac{1}{2},-\frac{1}{2},+1)}
=\mathcal{M}^{(-\frac{1}{2},+\frac{1}{2},-1)}
=\mathcal{M}^{(+\frac{1}{2},+\frac{1}{2},-1)}
=0, \\
\end{aligned} \end{equation}
where $X_1=\sqrt{(E_b+P)\mathbf{m}_t}$ and $X_2=\sqrt{(E_b-P)\mathbf{m}_t}$. Note that $P-E_b=\mathcal{O}(\mathbf{m}_b^2)$, which implies $X_2$ is a subleading quantity. On the other hand, $E_W+P=\mathbf{m}_t-E_b+P=\mathbf{m}_t+\mathcal{O}(\mathbf{m}_b^2)$ is a leading quantity. Consequently, only the helicity categories in the first and third lines of eq.~\eqref{eq:tbW2} provide leading contributions, and they exhibit mass enhancement from either $\mathbf{m}_t$ or $\mathbf{m}_W$. According to $s_{3,z}$, the leading contributions give the ratio
\begin{equation} \begin{aligned}
\left(\sum_{h_1}\left|\mathcal{M}^{(h_1,-\frac{1}{2},-1)}\right|^2\right):
\left(\sum_{h_1}\left|\mathcal{M}^{(h_1,-\frac{1}{2},0)}\right|^2\right)
\approx 1:\frac{\mathbf{m}_t^2}{2\mathbf{m}_W^2}. 
\end{aligned} \end{equation}

As shown in eq.~\eqref{eq:tbW2}, some helicity categories give zero results at the rest frame of the top quark. We can check this result by calculating $s_{3,z}+s_{2,z}$ in the $\theta=0$ frame. There are three typical helicity categories, and we find
\begin{equation} \begin{aligned}
\mathcal{M}^{(+\frac{1}{2},-\frac{1}{2},-1)}:-\frac{1}{2}+1=-\frac{1}{2}, \\
\mathcal{M}^{(+\frac{1}{2},-\frac{1}{2},+1)}:-\frac{1}{2}-1=-\frac{3}{2}, \\
\mathcal{M}^{(-\frac{1}{2},-\frac{1}{2},0)}:-\frac{1}{2}-0=-\frac{1}{2}. \\
\end{aligned} \end{equation}
According to the angular momentum conservation, $|s_{2,z}+s_{3,z}|$ should not be larger than the spin of the top quark. Therefore, the helicity category $(+\frac{1}{2},-\frac{1}{2},+1)$ will not contribute to the rest frame of the top quark. The two non-vanishing contributions can be illustrated by
\begin{equation}
\begin{tikzpicture} \begin{feynhand}
\vertex [particle] (i1) at (1.1,0) {};
\vertex [particle] (i2) at (1.9,0) {};
\graph{(i1)--[bos] (i2)};
\path 
(0.9,-0.3) node(A0) [rectangle] {\footnotesize$\Leftarrow$}
(0.3,-0.3) node(A1) [rectangle] {\footnotesize$\Rightarrow$}
(1.5,-0.3) node(A2) [rectangle] {\footnotesize$\Leftarrow$}
(0.9,0.3) node(B0) [rectangle] {\footnotesize$t$}
(-0.2,0) node(B1) [rectangle] {\footnotesize$b$}
(2.1,0) node(B2) [rectangle] {\footnotesize$W^+$};
\fill (0.9,0) circle (0.1cm);
\draw[->] (0.6,0) -- (0,0);
\end{feynhand} \end{tikzpicture}\qquad
\begin{tikzpicture} \begin{feynhand}
\vertex [particle] (i1) at (0.7,0) {};
\vertex [particle] (i2) at (-0.1,0) {};
\graph{(i1)--[bos] (i2)};
\path 
(0.9,-0.3) node(A0) [rectangle] {\footnotesize$\Leftarrow$}
(0.3,-0.3) node(A1) [rectangle] {\footnotesize$0$}
(1.5,-0.3) node(A2) [rectangle] {\footnotesize$\Leftarrow$}
(0.9,0.3) node(B0) [rectangle] {\footnotesize$t$}
(-0.3,0) node(B1) [rectangle] {\footnotesize$W^+$}
(1.95,0) node(B2) [rectangle] {\footnotesize$b$};
\fill (0.9,0) circle (0.1cm);
\draw[->] (1.2,0) -- (1.8,0);
\end{feynhand} \end{tikzpicture}
\end{equation}
where the arrow $\Rightarrow$ represents the direction of spin.  


It is interesting to think about whether the massless limit of $W$ boson for $\mathcal{M}(\bar{t},b,W^+)$ is a 1-massless 2-massive amplitude. For the zero helicity $W$ boson, it is the case due to the Goldstone equivalence theorem. We have 
\begin{equation}
\lim_{\eta_3,\tilde{\eta}_3\rightarrow0} \mathcal{M}^{(+\frac{1}{2},+\frac{1}{2},0)}=\langle \eta_2 3\rangle[31]-\langle \eta_2 \eta_3\rangle[\eta_3 1]=
m_2 [12]-\tilde{m}_1\langle\eta_1 \eta_2\rangle, 
\end{equation}
where $m_2[12]$ and $\tilde{m}_1\langle\eta_1 \eta_2\rangle$ are 1-massless 2-massive amplitude with unequal masses. 

However, the negative helicity $W$ boson gives a different answer~\footnote{Considering new physics beyond SM, there can have interaction with $\pm1$-chirality vector boson. This amplitude can have a massless limit as
\begin{equation}
\lim_{\eta_3,\tilde{\eta}_3\rightarrow0} \mathcal{M}^{(+\frac{1}{2},-\frac{1}{2},-1)}_{\textrm{BSM}}=\langle23\rangle\langle3\eta_1\rangle. 
\end{equation}}
\begin{equation}
\lim_{\eta_3,\tilde{\eta}_3\rightarrow0} \mathcal{M}^{(+\frac{1}{2},-\frac{1}{2},-1)}=\lim_{\eta_3,\tilde{\eta}_3\rightarrow0}\langle23\rangle[\eta_3 1]=0. 
\end{equation}
For massless particles, transversality is equal to chirality. Since the massless limit does not change the chirality, the $0$-chirality $W$ boson cannot match to the massless particle. One may wonder if we flip the $0$-chirality $W$ boson at IR can give a non-vanishing result, e.g. $m_3\langle23\rangle[\eta_3 1]$. Suppose that the high-energy behavior of the coefficient is $c\sim 1/\mathbf{m}_3^2$, the massless limit does give a non-zero amplitude. However, this expression has a spurious pole, so it is not legal for the 1-massless 2-massive amplitude with unequal masses. If we consider the massless limit for all three particles, the massive amplitude will reduce to the all-massless amplitudes, in which the spurious pole could exist.


%
%


\section{One-massless-two-massive Amplitudes}
\label{sec:3pt2}

In this section, we focus on one-massless 3-pt amplitudes. When two massive particles have unequal mass, the construction is similar to the all-massive amplitudes. If they have equal mass, the 3-pt amplitudes would have the gauge invariance, and thus the $\mathbf{x}$-factor is introduced to indicate such gauge invariance behavior. In Ref.~\cite{Ni:2024yrr}, we have shown that the ST spinor formalism is not enough to describe the $\mathbf{x}$-factor. Thus the helicity-transversality spinor in this work is necessarily introduced, which would separate $\mathbf{x}$ into $x$ and $\bar{x}$ and help to build the one-to-one massless-massive correspondence. This would isolate the massive amplitudes corresponding to the unwanted UV structure, and give rise to the correct 3-pt $F\bar{F}\gamma$ amplitudes. 


\subsection{3-pt amplitudes with unequal masses and charged pion decay}

Suppose that particle 3 is a massless particle with $h_3>0$ and particles 1 and 2 have unequal masses. We apply the similar procedure in subsection~\ref{sec:highest-weight} to construct the 3-pt amplitudes. The single-particle state of particle 3 is $|3]^{h_3}$, and there is no mass insertion for it. In this case, we find the amplitude with $s_1+s_2<h_3$ do not have a highest weight representation, because there is no $|3\rangle$ to help us construct amplitudes. For example, the 3-pt amplitude with $(s_1,s_2,h_3)=(0,0,+1)$ does not exist when the scalar masses are not equal. 

For the amplitude with $s_1+s_2>h_3$, we can use the highest weight construction to give the 3-pt ampliutdes. We begin with the representation with maximal transversality as
\begin{equation} \label{eq:HighAmp}
[12]^{s_1+s_2-h_3}[23]^{s_2-s_1+h_3}[31]^{s_1-s_2+h_3}. 
\end{equation}
Notice that this expression is valid when we have $s_2+h_3 \ge s_1$ or $s_1+h_3\ge s_2$, otherwise we would consider the structures including $[1|p_2+\eta_2|1\rangle$ or $[1|p_1+\eta_1|2\rangle$. Then we perform $[T_1^-,J_1^-] [T_2^-,J_2^-]$, which is equivalent to the replacement $[12]\rightarrow\langle12\rangle$, to obtain other highest weight representations.   If $h_3<0$, we should begin with the presentation with minimal transversality and use $[T_1^+,J_1^+] [T_2^+,J_2^+]$ to obtain other representations. 

For example, consider the $S\bar{F}F$ amplitude with $(s_1,s_2,h_3)=(0,\frac{1}{2},-\frac{1}{2})$. The only highest weight representation is
\begin{equation} \begin{aligned}
\begin{tikzpicture}[baseline=0.7cm] \begin{feynhand}
\vertex [particle] (i1) at (0,0.8) {$1^{0}$};
\vertex [particle] (i2) at (1.6,1.6) {$2^{-\frac{1}{2}}$};
\vertex [particle] (i3) at (1.6,0) {$3^{-\frac{1}{2}}$};
\vertex (v1) at (0.9,0.8);
\graph{(i1)--[plain,brown,very thick] (v1)};
\graph{(i2)--[plain,cyan,very thick] (v1)};
\graph{(i3)--[plain,red] (v1)};
\end{feynhand} \end{tikzpicture}=\langle 23\rangle,
\end{aligned} \end{equation}
where the thin line represents the massless particle. Given the highest weight representation, we can perform the helicity mass insertion to obtain other helicity amplitudes. We have
\begin{equation} \label{eq:UneuqalBasis}
\begin{tabular}{c|c}
\diagbox{Helicity}{Transversality} & $(0,-\frac{1}{2},-\frac{1}{2})$ \\
\hline
$(0,-\frac{1}{2},-\frac{1}{2})$ & 
\begin{tikzpicture}[baseline=0.7cm] \begin{feynhand}
\vertex [particle] (i1) at (0,0.8) {$1^{0}$};
\vertex [particle] (i2) at (1.6,1.6) {$2^{-\frac{1}{2}}$};
\vertex [particle] (i3) at (1.6,0) {$3^{-\frac{1}{2}}$};
\vertex (v1) at (0.9,0.8);
\graph{(i1)--[plain,brown,very thick] (v1)};
\graph{(i2)--[plain,cyan,very thick] (v1)};
\graph{(i3)--[plain,red] (v1)};
\end{feynhand} \end{tikzpicture}$=\langle23\rangle$ \\
\hline
$(0,+\frac{1}{2},-\frac{1}{2})$ & 
\begin{tikzpicture}[baseline=0.7cm] \begin{feynhand}
\vertex [particle] (i1) at (0,0.8) {$1^{0}$};
\vertex [particle] (i2) at (1.6,1.6) {$2^{+\frac{1}{2}}$};
\vertex [particle] (i3) at (1.6,0) {$3^{-\frac{1}{2}}$};
\vertex (v1) at (0.9,0.8);
\graph{(i1)--[plain,brown,very thick] (v1)};
\graph{(i2)--[plain,cyan,very thick] (v1)};
\graph{(i3)--[plain,red] (v1)};
\draw plot[mark=x,mark size=2.7,mark options={rotate=45}] coordinates {(0.9+0.7*0.33,0.8+0.8*0.33)};
\end{feynhand} \end{tikzpicture}$=-\langle\eta_2 3\rangle$ \\
\end{tabular}
\end{equation}
Then we perform the chirality mass insertions for $SF\bar{F}$ amplitude,
\begin{equation} \begin{aligned} \label{eq:FFSunequal}
\begin{tikzpicture}[baseline=0.7cm] \begin{feynhand}
\vertex [particle] (i1) at (0,0.8) {$1^{0}$};
\vertex [particle] (i2) at (1.6,1.6) {$2^{-\frac{1}{2}}$};
\vertex [particle] (i3) at (1.6,0) {$3^{-\frac{1}{2}}$};
\vertex (v1) at (0.9,0.8);
\graph{(i1)--[plain,brown,very thick] (v1)};
\graph{(i2)--[plain,cyan,very thick] (v1)};
\graph{(i3)--[plain,red] (v1)};
\end{feynhand} \end{tikzpicture}&=\langle23\rangle,&
\begin{tikzpicture}[baseline=0.7cm] \begin{feynhand}
\vertex [particle] (i1) at (0,0.8) {$1^{0}$};
\vertex [particle] (i2) at (1.6,1.6) {$2^{-\frac{1}{2}}$};
\vertex [particle] (i3) at (1.6,0) {$3^{-\frac{1}{2}}$};
\vertex (v1) at (0.9,0.8);
\vertex (v2) at (0.9+0.7*0.35,0.8+0.8*0.35);
\graph{(i1)--[plain,brown,very thick] (v1)};
\graph{(i2)--[plain,red,very thick](v2)--[plain,cyan,insertion={[style=black,size=1.8pt]0},very thick] (v1)};
\graph{(i3)--[plain,red] (v1)};
\end{feynhand} \end{tikzpicture}&=m_2\langle23\rangle,&\\
\begin{tikzpicture}[baseline=0.7cm] \begin{feynhand}
\vertex [particle] (i1) at (0,0.8) {$1^{0}$};
\vertex [particle] (i2) at (1.6,1.6) {$2^{+\frac{1}{2}}$};
\vertex [particle] (i3) at (1.6,0) {$3^{-\frac{1}{2}}$};
\vertex (v1) at (0.9,0.8);
\graph{(i1)--[plain,brown,very thick] (v1)};
\graph{(i2)--[plain,cyan,very thick] (v1)};
\graph{(i3)--[plain,red] (v1)};
\draw plot[mark=x,mark size=2.7,mark options={rotate=45}] coordinates {(0.9+0.7*0.33,0.8+0.8*0.33)};
\end{feynhand} \end{tikzpicture}&=-\langle\eta_2 3\rangle,&
\begin{tikzpicture}[baseline=0.7cm] \begin{feynhand}
\vertex [particle] (i1) at (0,0.8) {$1^{0}$};
\vertex [particle] (i2) at (1.6,1.6) {$2^{+\frac{1}{2}}$};
\vertex [particle] (i3) at (1.6,0) {$3^{-\frac{1}{2}}$};
\vertex (v1) at (0.9,0.8);
\vertex (v2) at (0.9+0.7*0.35,0.8+0.8*0.35);
\graph{(i1)--[plain,brown,very thick] (v1)};
\graph{(i2)--[plain,cyan,very thick] (v1)};
\graph{(i3)--[plain,red] (v1)};
\graph{(i2)--[plain,red,very thick](v2)--[plain,cyan,very thick] (v1)};
\draw plot[mark=x,mark size=2.7,mark options={rotate=45}] coordinates {(0.9+0.7*0.15,0.8+0.8*0.15)};
\draw[very thick] plot[mark=x,mark size=2.7,mark options={rotate=45}] coordinates {(v2)};
\end{feynhand} \end{tikzpicture}&=-m_2\langle\eta_2 3\rangle.&\\
\end{aligned} \end{equation}
The final result is
\begin{equation} \begin{aligned}
\mathcal{M}^{(0,-\frac{1}{2},-\frac{1}{2})}
=&c_1\langle 23\rangle+c_2\tilde{m}_2\langle 23\rangle, \\
\mathcal{M}^{(0,+\frac{1}{2},-\frac{1}{2})}
=&-c_1\langle\eta_2 3\rangle-c_2\tilde{m}_2\langle\eta_2 3\rangle. \\
\end{aligned} \end{equation}
They are the massless limit of the all-massive $F\bar{F}S$ amplitude shown in eqs.~\eqref{eq:resultFFS} and \eqref{eq:resultFFS2}, with relabeling the external lines.

As an application, we consider the charged pion decay, i.e. $\pi^+\rightarrow \mu^+\nu_\mu$. In this case, both $\mu^+$ and $\nu_\mu$ are left-handed, which is equivalent to $(t_1,t_2,t_3)=(0,+\frac12,-\frac12)$. It gives a constraint on coefficients: $c_1=0$. Thus the amplitude must have one chirality mass insertion as
\begin{equation} \begin{aligned} \label{eq:piondecay}
\mathcal{M}^{(0,-\frac{1}{2},-\frac{1}{2})}_{\pi^+\rightarrow \mu^+\nu_\mu}
&=c_2\tilde{m}_2\langle23\rangle, \\
\mathcal{M}^{(0,+\frac{1}{2},-\frac{1}{2})}_{\pi^+\rightarrow \mu^+\nu_\mu}
&=-c_2\tilde{m}_2\langle\eta_2 3\rangle, \\
\end{aligned} \end{equation}
where the superscript denotes helicity. This result exhibits mass enhancement from chirality mass insertion $\tilde{m}_2$.

Notice that the charged pion decay is a weak decay, so the amplitudes are related to the one with $W$ boson. Let us consider the W decay with the same decay product, i.e. $W^+\rightarrow \mu^+\nu_\mu$. This is also a 1-massless 2-massive amplitude, but it has a different spin $s_1=1$. In this decay channel, $\mu^+$ and $\nu_\mu$ are also left-handed particles, so the transversality should be $(t_1,t_2,t_3)=(0,+\frac{1}{2},-\frac{1}{2})$ too. The amplitudes with $h_1=+1$ or $0$ can be expressed as
\begin{equation} \begin{aligned} \label{eq:Wdecay}
\mathcal{M}^{(+1,-\frac{1}{2},-\frac{1}{2})}_{W^+\rightarrow \mu^+\nu_\mu}&=-[1\eta_2]\langle 3\eta_1\rangle, &
\mathcal{M}^{(0,-\frac{1}{2},-\frac{1}{2})}_{W^+\rightarrow \mu^+\nu_\mu}&=[1\eta_2]\langle 31\rangle-[\eta_1\eta_2]\langle 3\eta_1\rangle, &\\
\mathcal{M}^{(+1,+\frac{1}{2},-\frac{1}{2})}_{W^+\rightarrow \mu^+\nu_\mu}&=-[12]\langle 3\eta_1\rangle, & 
\mathcal{M}^{(0,+\frac{1}{2},-\frac{1}{2})}_{W^+\rightarrow \mu^+\nu_\mu}&=[12]\langle 31\rangle-[\eta_1 2]\langle 3\eta_1\rangle, &\\
\end{aligned} \end{equation}
where we drop the coefficients. Similarly, we can obtain the amplitudes with $h_1=-1$. This process does not have chirality mass insertion, but it is closely related to charged pion decay in helicity $h_1=0$. For example, $[1 \eta_2]\langle13\rangle=-[2 \eta_2]\langle23\rangle=\tilde{m}_2\langle23\rangle$, which is equal to the first line of eq.~\eqref{eq:piondecay}.

\subsection{3-pt amplitudes with equal mass}

Now we turn to the case with equal mass, and particle 3 is still massless. The momentum conservation shows an additional relation, 
\begin{equation} \label{eq:EqualCondition}
\langle3|\eta_1|3]=(\eta_1+p_3)^2=(\eta_1-p_1-p_2)^2=\mathbf{m}_2^2-\mathbf{m}_1^2=0, 
\end{equation}
so we must have $\lambda_3\propto \eta_1$ or $\tilde{\lambda}_3\propto \tilde{\eta}_1$. Combining this with eq.~\eqref{eq:kinematic1}, there are four possible kinematics. We choose the following two relations to give the 3-particle kinematics
\begin{equation} \label{eq:kinematic2}
\begin{cases}
\lambda_1\propto\lambda_2\propto\lambda_3\\
\tilde{\eta}_1\propto\tilde{\eta}_2\propto\tilde{\lambda}_3
\end{cases}\text{or }
\begin{cases}
\tilde{\lambda}_1\propto\tilde{\lambda}_2\propto\tilde{\lambda}_3\\
\eta_1\propto\eta_2\propto\lambda_3
\end{cases}. 
\end{equation}
The other two choices, e.g. $\lambda_1\propto\lambda_2\propto\lambda_3\propto\eta_1\propto\eta_1$, will lead to $\mathbf{m}_1^2=\mathbf{m}_2^2=0$, so they are illegal. 

The additional relation eq.~\eqref{eq:EqualCondition} also shows the proportionality between $|3]$ and $\eta_1|3\rangle$ for the first kinematics ($|3\rangle$ and $\eta_1|3]$ for the second kinematics). Defining the ratio as the $x$-factor, we have  
\begin{equation} \begin{aligned}
|3]\propto x\eta_1|3\rangle,\quad
|3\rangle\propto \frac{1}{\bar{x}}\eta_1|3]. 
\end{aligned} \end{equation}
Both $x$ and $\bar{x}$ carry $+1$ helicity. To make the $x$-factor dimensionless, we should multiply the right-hand side of the above equation by mass or inverse mass. There are some possible constructions for $x$-factor: 
\begin{equation} \begin{aligned}
x&\in \left\{
\frac{|3]\tilde{m}_1}{\eta_1|3\rangle},
\frac{|3]\tilde{m}_2}{\eta_1|3\rangle},
\frac{|3]m_1}{\eta_1|3\rangle},
\frac{|3]m_2}{\eta_1|3\rangle},
\frac{|3]\mathbf{m}}{\eta_1|3\rangle}
\right\}, \\
\bar{x}&\in \left\{
\frac{\eta_1|3]}{|3\rangle\tilde{m}_1},
\frac{\eta_1|3]}{|3\rangle\tilde{m}_2},
\frac{\eta_1|3]}{|3\rangle m_1},
\frac{\eta_1|3]}{|3\rangle m_2},
\frac{\eta_1|3]}{|3\rangle\mathbf{m}}
\right\}, 
\end{aligned} \end{equation}
where $\mathbf{m}=\mathbf{m}_1=\mathbf{m}_2$. Each one carries a different transversality of particles 1 and 2. 
 
For our purposes, the $x$-factor with $t_1=t_2=0$ is better than the other, because it will not include chirality flips for massive particles. In this definition, we take the $x$-factor to be
\begin{equation} \begin{aligned} \label{eq:t0x}
x&=\frac{\mathbf{m}[1 3]}{[1|\eta_1|3\rangle}=\frac{\mathbf{m}}{\tilde{m}_1}\frac{[1 3]}{\langle\eta_13\rangle}=\frac{[23][31]}{\mathbf{m}[12]},\\
\bar{x}&=\frac{\langle 1|\eta_1|3]}{\mathbf{m}\langle13\rangle}=\frac{m_1}{\mathbf{m}}\frac{[\eta_1 3]}{\langle13\rangle}=\frac{\mathbf{m}\langle12\rangle}{\langle23\rangle\langle31\rangle}.\\
\end{aligned} \end{equation}
This $x$-factor can be viewed as the contraction of a polarization vector and a momentum, i.e. $\varepsilon\cdot \mathbf{p}$. Although this expression includes spurious poles, it will not be changed under helicity flip for massive particles.

In two 3-particle kinematics, the power counting of $x$-factor and its inverse are different. We have
\begin{equation}
\begin{tabular}{c|cccc}
kinematics & $x$-factor & helicity & power counting & diagram \\
\hline
\multirow{2}{*}{\makecell{$\lambda_1\propto\lambda_2\propto\lambda_3$\\$\tilde{\eta}_1\propto\tilde{\eta}_2\propto\tilde{\lambda}_3$}} & $x$ & $+1$ & $\varepsilon_\eta^{-1}$ & 
\begin{tikzpicture}[baseline=0.7cm] \begin{feynhand}
\vertex [particle] (i1) at (1,0.8) {};
\vertex [particle] (v1) at (0,0.8) {};
\vertex [ringdot,cyan] (v2) at (0.5,0.8) {};
\graph{(v1)--[plain,cyan](v2)--[plain,cyan](i1)};
\end{feynhand} 
\end{tikzpicture}\\
& $x^{-1}$ & $-1$ & $\varepsilon_\eta$ & 
\begin{tikzpicture}[baseline=0.7cm] \begin{feynhand}
\setlength{\feynhanddotsize}{0.8mm}
\vertex [particle] (i1) at (1,0.8) {};
\vertex [particle] (v1) at (0,0.8) {};
\vertex [crossdot,red] (v2) at (0.5,0.8) {};
\graph{(v1)--[plain,red](v2)--[plain,red](i1)};
\end{feynhand} 
\end{tikzpicture}\\
\hline
\multirow{2}{*}{\makecell{$\tilde{\lambda}_1\propto\tilde{\lambda}_2\propto\tilde{\lambda}_3$\\$\eta_1\propto \eta_2\propto\lambda_3$}} & $\bar{x}$ & $+1$ & $\varepsilon_\eta$ & 
\begin{tikzpicture}[baseline=0.7cm] \begin{feynhand}
\setlength{\feynhanddotsize}{0.8mm}
\vertex [particle] (i1) at (1,0.8) {};
\vertex [particle] (v1) at (0,0.8) {};
\vertex [crossdot,cyan] (v2) at (0.5,0.8) {};
\graph{(v1)--[plain,cyan](v2)--[plain,cyan](i1)};
\end{feynhand} 
\end{tikzpicture}\\
& $\bar{x}^{-1}$ & $-1$ & $\varepsilon_\eta^{-1}$ & 
\begin{tikzpicture}[baseline=0.7cm] \begin{feynhand}
\vertex [particle] (i1) at (1,0.8) {};
\vertex [particle] (v1) at (0,0.8) {};
\vertex [ringdot,red] (v2) at (0.5,0.8) {};
\graph{(v1)--[plain,red](v2)--[plain,red](i1)};
\end{feynhand} 
\end{tikzpicture}\\
\end{tabular}
\end{equation}
where $\epsilon_{\eta}\sim \frac{\mathbf{m}}{E}$, the cyan and red line still correspond to positive and negative helicity. Diagrammatically, we use new marks $\otimes$ and $\circ$ to represent the power counting $\varepsilon_{\eta}$ and $\varepsilon^{-1}_{\eta}$. Note that $\otimes$ has the same power counting as mass insertion $\times$, but the $x$-factor with the definition in eq.~\eqref{eq:t0x} does not change the helicity or chirality for mass particles. Therefore, $\otimes$ and $\circ$ should not be identified as a helicity or chirality mass insertion. If we choose another definition for $x$-factor, the representation of massive states should be changed, which will be further discussed in subsection~\ref{sec:QED3pt}. 

In contrast to the case with unequal mass, we can use both $x$-factor and spinors to represent the massless single-particle state. If $h_3>0$, the helicity state of particle 3 can denoted by $|3]^{2h_3-2k}x^{k}$ (or $|3]^{2h_3-2k}\bar{x}^{k}$) with $0\le k\le h_3$, so the representation is not unique. Diagrammatically, we have
\begin{equation} \begin{aligned}
\begin{tikzpicture}[baseline=0.7cm] \begin{feynhand}
\vertex [particle] (i1) at (2.8,0.8) {$3^{h_3>0}$};
\vertex [dot] (v1) at (0,0.8) {};
\vertex [ringdot,color=cyan] (v2) at (1,0.8) {};
\vertex [ringdot,color=cyan] (v3) at (1.5,0.8) {};
\graph{(v1)--[plain,cyan](v2)--[plain,cyan](v3)--[plain,cyan](i1)};
\end{feynhand} 
\path (1.35,1.25) node(A0) [rectangle] {\scriptsize $k$th "$\circ$"};
\draw[snake=brace] (0.75,1.02) -- (1.8,1.02);
\end{tikzpicture}&=|3]^{2h_3-2k}x^{k}, &
\begin{tikzpicture}[baseline=0.7cm] \begin{feynhand}
\setlength{\feynhandblobsize}{1.5mm}
\setlength{\feynhanddotsize}{0.8mm}
\vertex [particle] (i1) at (2.8,0.8) {$3^{h_3<0}$};
\vertex [blob] (v1) at (0,0.8) {};
\vertex [crossdot,color=red] (v2) at (1,0.8) {};
\vertex [crossdot,color=red] (v3) at (1.5,0.8) {};
\graph{(v1)--[plain,red](v2)--[plain,red](v3)--[plain,red](i1)};
\end{feynhand} 
\path (1.35,1.25) node(A0) [rectangle] {\scriptsize $k$th "$\otimes$"};
\draw[snake=brace] (0.75,1.02) -- (1.8,1.02);
\end{tikzpicture}
&=|3\rangle^{-2h_3+2k}x^{-k}, \\
\begin{tikzpicture}[baseline=0.7cm] \begin{feynhand}
\setlength{\feynhandblobsize}{1.5mm}
\setlength{\feynhanddotsize}{0.8mm}
\vertex [particle] (i1) at (2.8,0.8) {$3^{h_3>0}$};
\vertex [blob] (v1) at (0,0.8) {};
\vertex [crossdot,color=cyan] (v2) at (1,0.8) {};
\vertex [crossdot,color=cyan] (v3) at (1.5,0.8) {};
\graph{(v1)--[plain,cyan](v2)--[plain,cyan](v3)--[plain,cyan](i1)};
\end{feynhand} 
\path (1.35,1.25) node(A0) [rectangle] {\scriptsize $k$th "$\otimes$"};
\draw[snake=brace] (0.75,1.02) -- (1.8,1.02);
\end{tikzpicture}&=|3]^{2h_3-2k}\bar{x}^{k}, &
\begin{tikzpicture}[baseline=0.7cm] \begin{feynhand}
\vertex [particle] (i1) at (2.8,0.8) {$3^{h_3<0}$};
\vertex [dot] (v1) at (0,0.8) {};
\vertex [ringdot,color=red] (v2) at (1,0.8) {};
\vertex [ringdot,color=red] (v3) at (1.5,0.8) {};
\graph{(v1)--[plain,red](v2)--[plain,red](v3)--[plain,red](i1)};
\end{feynhand} 
\path (1.35,1.25) node(A0) [rectangle] {\scriptsize $k$th "$\circ$"};
\draw[snake=brace] (0.75,1.02) -- (1.8,1.02);
\end{tikzpicture}&=|3\rangle^{-2h_3+2k}\bar{x}^{-k}, 
\end{aligned} \end{equation}
where negative helicity states are also listed. These different representations of particle 3 do have a physical meaning, when we choose an appropriate basis for 3-pt amplitudes. Let us consider the case with $s_1=s_2$ and $h_3=+1$, and write the amplitude in terms of $\tilde{\lambda}$ and $\tilde{\eta}$ of particles 1 and 2. The state $x^{h_3}$ corresponds to the electric charge of the massive particle, which is a universal structure for all spin. The state $|3]x^{h_3-1}$ corresponds to the magnetic dipole moment, which exists for the spinning particles. The other states would correspond to electric quadrupole moment and more for higher spin particles. 


Now we consider the construction of 3-pt amplitudes in this case. We choose the first kinematics in eq.~\eqref{eq:kinematic2}. To be consistent with the case with unequal mass, we fix the state $x^{h_3}$ for particle 3 (namely $k=h_3$), while particles 1 and 2 can be represented by $\tilde{\lambda}$, $\tilde{\eta}$, $\lambda$ and $\eta$. Therefore, the procedure of highest-weight construction can be applied to it. We choose the highest weight representation for particles 1 and 2 to construct amplitudes. If $s_1\ge s_2$, we can begin with
\begin{equation}
x^{h_3}[12]^{s_1+s_2} [1|p_2+\eta_2|1\rangle^{\frac{s_1-s_2}{2}}, 
\end{equation}
which is similar to the construction in eq.~\eqref{eq:SpinFactor}. In the case with equal mass, the amplitude with $s_1+s_2<h_3$ is constructible, because all spurious poles are absorbed into the $x$-factor. 

Then we perform the replacement $[12]\rightarrow\langle12\rangle$ to give other highest weight amplitudes. For example, the $F\bar{F}\gamma$ amplitude with positive helicity vector boson has two structures
\begin{equation} \begin{aligned} \label{eq:FFVequal}
\begin{tikzpicture}[baseline=0.7cm] \begin{feynhand}
\vertex [particle] (i1) at (0,0.8) {$1^{+\frac{1}{2}}$};
\vertex [particle] (i2) at (1.6,1.6) {$2^{+\frac{1}{2}}$};
\vertex [particle] (i3) at (1.6,0) {$3^{+1}$};
\vertex (v1) at (0.9,0.8);
\vertex [ringdot,cyan] (v2) at (1.075,0.6) {};
\graph{(i1)--[plain,cyan,very thick] (v1)};
\graph{(i2)--[plain,red,very thick] (v1)};
\graph{(i3)--[plain,cyan] (v2)--[plain,cyan] (v1)};
\end{feynhand} \end{tikzpicture}=x[12],\quad
\begin{tikzpicture}[baseline=0.7cm] \begin{feynhand}
\vertex [particle] (i1) at (0,0.8) {$1^{-\frac{1}{2}}$};
\vertex [particle] (i2) at (1.6,1.6) {$2^{-\frac{1}{2}}$};
\vertex [particle] (i3) at (1.6,0) {$3^{+1}$};
\vertex (v1) at (0.9,0.8);
\vertex [ringdot,cyan] (v2) at (1.075,0.6) {};
\graph{(i1)--[plain,red,very thick] (v1)};
\graph{(i2)--[plain,cyan,very thick] (v1)};
\graph{(i3)--[plain,cyan] (v2)--[plain,cyan] (v1)};
\end{feynhand} \end{tikzpicture}=x\langle12\rangle.
\end{aligned} \end{equation}

Using the helicity mass insertion, we derive the amplitudes in other helicity categories as
\begin{equation} \label{eq:equalFFV}
\begin{tabular}{c|cc}
\diagbox[width=2.4cm,innerleftsep=.4cm,innerrightsep=.4cm]{$h_i$}{$t_i$} & $(+\frac{1}{2},+\frac{1}{2},+1)$ & $(-\frac{1}{2},-\frac{1}{2},+1)$ \\
\hline
$(+\frac{1}{2},+\frac{1}{2},+1)$ & 
\begin{tikzpicture}[baseline=0.7cm] \begin{feynhand}
\vertex [particle] (i1) at (0,0.8) {$1^{+\frac{1}{2}}$};
\vertex [particle] (i2) at (1.6,1.6) {$2^{+\frac{1}{2}}$};
\vertex [particle] (i3) at (1.6,0) {$3^{+1}$};
\vertex (v1) at (0.9,0.8);
\vertex [ringdot,cyan] (v2) at (1.075,0.6) {};
\graph{(i1)--[plain,cyan,very thick] (v1)};
\graph{(i2)--[plain,red,very thick] (v1)};
\graph{(i3)--[plain,cyan] (v2)--[plain,cyan] (v1)};
\end{feynhand} \end{tikzpicture}$=x[12]$ & 
\begin{tikzpicture}[baseline=0.7cm] \begin{feynhand}
\vertex [particle] (i1) at (0,0.8) {$1^{+\frac{1}{2}}$};
\vertex [particle] (i2) at (1.6,1.6) {$2^{+\frac{1}{2}}$};
\vertex [particle] (i3) at (1.6,0) {$3^{+1}$};
\vertex (v1) at (0.9,0.8);
\vertex [ringdot,cyan] (v2) at (1.075,0.6) {};
\graph{(i1)--[plain,red,very thick] (v1)};
\graph{(i2)--[plain,cyan,very thick] (v1)};
\graph{(i3)--[plain,cyan] (v2)--[plain,cyan] (v1)};
\draw plot[mark=x,mark size=2.7,mark options={rotate=45}] coordinates {(0.9+0.7*0.33,0.8+0.8*0.33)};
\draw plot[mark=x,mark size=2.7] coordinates {(0.65,0.8)};
\end{feynhand} \end{tikzpicture}$=x\langle\eta_1\eta_2\rangle$ \\
\hline
$(-\frac{1}{2},+\frac{1}{2},+1)$ & 
\begin{tikzpicture}[baseline=0.7cm] \begin{feynhand}
\vertex [particle] (i1) at (0,0.8) {$1^{-\frac{1}{2}}$};
\vertex [particle] (i2) at (1.6,1.6) {$2^{+\frac{1}{2}}$};
\vertex [particle] (i3) at (1.6,0) {$3^{+1}$};
\vertex (v1) at (0.9,0.8);
\vertex [ringdot,cyan] (v2) at (1.075,0.6) {};
\graph{(i1)--[plain,cyan,very thick] (v1)};
\graph{(i2)--[plain,red,very thick] (v1)};
\graph{(i3)--[plain,cyan] (v2)--[plain,cyan] (v1)};
\draw plot[mark=x,mark size=2.7] coordinates {(0.65,0.8)};
\end{feynhand} \end{tikzpicture}$=x[\eta_1 2]$ & 
\begin{tikzpicture}[baseline=0.7cm] \begin{feynhand}
\vertex [particle] (i1) at (0,0.8) {$1^{-\frac{1}{2}}$};
\vertex [particle] (i2) at (1.6,1.6) {$2^{+\frac{1}{2}}$};
\vertex [particle] (i3) at (1.6,0) {$3^{+1}$};
\vertex (v1) at (0.9,0.8);
\vertex [ringdot,cyan] (v2) at (1.075,0.6) {};
\graph{(i1)--[plain,red,very thick] (v1)};
\graph{(i2)--[plain,cyan,very thick] (v1)};
\graph{(i3)--[plain,cyan] (v2)--[plain,cyan] (v1)};
\draw plot[mark=x,mark size=2.7,mark options={rotate=45}] coordinates {(0.9+0.7*0.33,0.8+0.8*0.33)};
\end{feynhand} \end{tikzpicture}$=-x\langle1\eta_2\rangle$ \\
\hline
$(+\frac{1}{2},-\frac{1}{2},+1)$ & 
\begin{tikzpicture}[baseline=0.7cm] \begin{feynhand}
\vertex [particle] (i1) at (0,0.8) {$1^{+\frac{1}{2}}$};
\vertex [particle] (i2) at (1.6,1.6) {$2^{-\frac{1}{2}}$};
\vertex [particle] (i3) at (1.6,0) {$3^{+1}$};
\vertex (v1) at (0.9,0.8);
\vertex [ringdot,cyan] (v2) at (1.075,0.6) {};
\graph{(i1)--[plain,cyan,very thick] (v1)};
\graph{(i2)--[plain,red,very thick] (v1)};
\graph{(i3)--[plain,cyan] (v2)--[plain,cyan] (v1)};
\draw plot[mark=x,mark size=2.7,mark options={rotate=45}] coordinates {(0.9+0.7*0.33,0.8+0.8*0.33)};
\end{feynhand} \end{tikzpicture}$=x[1\eta_2]$ & 
\begin{tikzpicture}[baseline=0.7cm] \begin{feynhand}
\vertex [particle] (i1) at (0,0.8) {$1^{+\frac{1}{2}}$};
\vertex [particle] (i2) at (1.6,1.6) {$2^{-\frac{1}{2}}$};
\vertex [particle] (i3) at (1.6,0) {$3^{+1}$};
\vertex (v1) at (0.9,0.8);
\vertex [ringdot,cyan] (v2) at (1.075,0.6) {};
\graph{(i1)--[plain,red,very thick] (v1)};
\graph{(i2)--[plain,cyan,very thick] (v1)};
\graph{(i3)--[plain,cyan] (v2)--[plain,cyan] (v1)};
\draw plot[mark=x,mark size=2.7] coordinates {(0.65,0.8)};
\end{feynhand} \end{tikzpicture}$=-x\langle\eta_1 2\rangle$ \\
\hline
$(-\frac{1}{2},-\frac{1}{2},+1)$ & 
\begin{tikzpicture}[baseline=0.7cm] \begin{feynhand}
\vertex [particle] (i1) at (0,0.8) {$1^{-\frac{1}{2}}$};
\vertex [particle] (i2) at (1.6,1.6) {$2^{-\frac{1}{2}}$};
\vertex [particle] (i3) at (1.6,0) {$3^{+1}$};
\vertex (v1) at (0.9,0.8);
\vertex [ringdot,cyan] (v2) at (1.075,0.6) {};
\graph{(i1)--[plain,cyan,very thick] (v1)};
\graph{(i2)--[plain,red,very thick] (v1)};
\graph{(i3)--[plain,cyan] (v2)--[plain,cyan] (v1)};
\draw plot[mark=x,mark size=2.7,mark options={rotate=45}] coordinates {(0.9+0.7*0.33,0.8+0.8*0.33)};
\draw plot[mark=x,mark size=2.7] coordinates {(0.65,0.8)};
\end{feynhand} \end{tikzpicture}$=x[\eta_1\eta_2]$ & 
\begin{tikzpicture}[baseline=0.7cm] \begin{feynhand}
\vertex [particle] (i1) at (0,0.8) {$1^{-\frac{1}{2}}$};
\vertex [particle] (i2) at (1.6,1.6) {$2^{-\frac{1}{2}}$};
\vertex [particle] (i3) at (1.6,0) {$3^{+1}$};
\vertex (v1) at (0.9,0.8);
\vertex [ringdot,cyan] (v2) at (1.075,0.6) {};
\graph{(i1)--[plain,red,very thick] (v1)};
\graph{(i2)--[plain,cyan,very thick] (v1)};
\graph{(i3)--[plain,cyan] (v2)--[plain,cyan] (v1)};
\end{feynhand} \end{tikzpicture}$=x\langle12\rangle$ \\
\end{tabular}
\end{equation}


Choosing the helicity category $(+\frac{1}{2},-\frac{1}{2},+1)$. We perform the chirality flip for the highest weight representation $(-\frac{1}{2},-\frac{1}{2},0)$ and get
\begin{equation} \begin{aligned} \label{eq:FFgamma-Insertion2}
\begin{tikzpicture}[baseline=0.7cm] \begin{feynhand}
\vertex [particle] (i1) at (-0.2,0.8) {$1^{+\frac{1}{2}}$};
\vertex [particle] (i2) at (1.6,1.6) {$2^{-\frac{1}{2}}$};
\vertex [particle] (i3) at (1.6,0) {$3^{+1}$};
\vertex (v1) at (0.9,0.8);
\vertex [ringdot,cyan] (v2) at (1.075,0.6) {};
\graph{(i1)--[plain,red,very thick] (v1)};
\graph{(i2)--[plain,cyan,very thick] (v1)};
\graph{(i3)--[plain,cyan] (v2)--[plain,cyan] (v1)};
\draw plot[mark=x,mark size=2.7,mark options={rotate=0}] coordinates {(0.6,0.8)};
\end{feynhand} \end{tikzpicture}&=-x\langle\eta_1 2\rangle,&
\begin{tikzpicture}[baseline=0.7cm] \begin{feynhand}
\vertex [particle] (i1) at (-0.2,0.8) {$1^{+\frac{1}{2}}$};
\vertex [particle] (i2) at (1.6,1.6) {$2^{-\frac{1}{2}}$};
\vertex [particle] (i3) at (1.6,0) {$3^{+1}$};
\vertex (v1) at (0.9,0.8);
\vertex (v2) at (0.9+0.7*0.35,0.8+0.8*0.35);
\vertex [ringdot,cyan] (v3) at (1.075,0.6) {};
\graph{(i1)--[plain,red,very thick] (v1)};
\graph{(i2)--[plain,red,very thick](v2)--[plain,cyan,very thick] (v1)};
\graph{(i3)--[plain,cyan] (v3)--[plain,cyan] (v1)};
\draw plot[mark=x,mark size=2.7,mark options={rotate=0}] coordinates {(0.6,0.8)};
\draw[very thick] plot[mark=x,mark size=2.7,mark options={rotate=45}] coordinates {(0.9+0.7*0.33,0.8+0.8*0.33)};
\end{feynhand} \end{tikzpicture}&=-x\tilde{m}_2\langle\eta_1 2\rangle,&\\
\begin{tikzpicture}[baseline=0.7cm] \begin{feynhand}
\vertex [particle] (i1) at (-0.2,0.8) {$1^{+\frac{1}{2}}$};
\vertex [particle] (i2) at (1.6,1.6) {$2^{-\frac{1}{2}}$};
\vertex [particle] (i3) at (1.6,0) {$3^{+1}$};
\vertex (v1) at (0.9,0.8);
\vertex (v2) at (0.48,0.8);
\vertex [ringdot,cyan] (v3) at (1.075,0.6) {};
\graph{(i1)--[plain,cyan,very thick](v2)--[plain,red,very thick] (v1)};
\graph{(i2)--[plain,cyan,very thick] (v1)};
\graph{(i3)--[plain,cyan] (v3)--[plain,cyan] (v1)};
\draw[very thick] plot[mark=x,mark size=2.7] coordinates {(v2)};
\draw plot[mark=x,mark size=2.7,mark options={rotate=0}] coordinates {(0.72,0.8)};
\end{feynhand} \end{tikzpicture}&=-x\tilde{m}_1\langle\eta_1 2\rangle,&
\begin{tikzpicture}[baseline=0.7cm] \begin{feynhand}
\vertex [particle] (i1) at (-0.2,0.8) {$1^{+\frac{1}{2}}$};
\vertex [particle] (i2) at (1.6,1.6) {$2^{-\frac{1}{2}}$};
\vertex [particle] (i3) at (1.6,0) {$3^{+1}$};
\vertex (v1) at (0.9,0.8);
\vertex (v2) at (0.48,0.8);
\vertex (v3) at (0.9+0.7*0.33,0.8+0.8*0.33);
\vertex [ringdot,cyan] (v4) at (1.075,0.6) {};
\graph{(i1)--[plain,cyan,very thick](v2)--[plain,red,very thick] (v1)};
\graph{(i2)--[plain,red,very thick](v3)--[plain,cyan,very thick] (v1)};
\graph{(i3)--[plain,cyan] (v4)--[plain,cyan] (v1)};
\draw[very thick] plot[mark=x,mark size=2.7] coordinates {(v2)};
\draw[very thick] plot[mark=x,mark size=2.7,mark options={rotate=45}] coordinates {(v3)};
\draw plot[mark=x,mark size=2.7,mark options={rotate=0}] coordinates {(0.72,0.8)};
\end{feynhand} \end{tikzpicture}&=-x\tilde{m}_1\tilde{m}_2\langle\eta_1 2\rangle.&
\end{aligned} \end{equation}
Finally, we derive the $F\bar{F}\gamma$ amplitudes in this helicity category as
\begin{equation} \begin{aligned} \label{eq:equalFFV2}
\mathcal{M}^{(-\frac{1}{2},+\frac{1}{2},+1)}
=&x[1 \eta_2](c_1+c_2 m_1 +c_3 m_2 +c_4 m_1 m_2)\\
&-x\langle\eta_1 2\rangle(c_5+c_6 \tilde{m}_1 +c_7 \tilde{m}_2 +c_8 \tilde{m}_1 \tilde{m}_2).
\end{aligned} \end{equation}

If we choose the second kinematics in eq.~\eqref{eq:kinematic2}, the massless state should be expressed by $\bar{x}$. Diagrammatically, we replace $\circ\rightarrow \otimes$ and get
\begin{equation} \begin{aligned}  \label{eq:FFgamma-Insertion3}
\begin{tikzpicture}[baseline=0.7cm] \begin{feynhand}
\setlength{\feynhanddotsize}{0.8mm}
\vertex [particle] (i1) at (-0.2,0.8) {$1^{+\frac{1}{2}}$};
\vertex [particle] (i2) at (1.6,1.6) {$2^{-\frac{1}{2}}$};
\vertex [particle] (i3) at (1.6,0) {$3^{+1}$};
\vertex (v1) at (0.9,0.8);
\vertex [crossdot,cyan] (v2) at (1.075,0.6) {};
\graph{(i1)--[plain,red,very thick] (v1)};
\graph{(i2)--[plain,cyan,very thick] (v1)};
\graph{(i3)--[plain,cyan] (v2)--[plain,cyan] (v1)};
\draw plot[mark=x,mark size=2.7,mark options={rotate=0}] coordinates {(0.6,0.8)};
\end{feynhand} \end{tikzpicture}&=-\bar{x}\langle\eta_1 2\rangle, &
\begin{tikzpicture}[baseline=0.7cm] \begin{feynhand}
\setlength{\feynhanddotsize}{0.8mm}
\vertex [particle] (i1) at (-0.2,0.8) {$1^{+\frac{1}{2}}$};
\vertex [particle] (i2) at (1.6,1.6) {$2^{-\frac{1}{2}}$};
\vertex [particle] (i3) at (1.6,0) {$3^{+1}$};
\vertex (v1) at (0.9,0.8);
\vertex (v2) at (0.9+0.7*0.35,0.8+0.8*0.35);
\vertex [crossdot,cyan] (v3) at (1.075,0.6) {};
\graph{(i1)--[plain,red,very thick] (v1)};
\graph{(i2)--[plain,red,very thick](v2)--[plain,cyan,very thick] (v1)};
\graph{(i3)--[plain,cyan] (v3)--[plain,cyan] (v1)};
\draw plot[mark=x,mark size=2.7,mark options={rotate=0}] coordinates {(0.6,0.8)};
\draw[very thick] plot[mark=x,mark size=2.7,mark options={rotate=45}] coordinates {(0.9+0.7*0.33,0.8+0.8*0.33)};
\end{feynhand} \end{tikzpicture}&=-\bar{x}\tilde{m}_2\langle\eta_1 2\rangle. &\\
\end{aligned} \end{equation}
Similarly, we can give the $F\bar{F}\gamma$ amplitudes in this helicity category,
\begin{equation} \begin{aligned}
\mathcal{M}^{(-\frac{1}{2},+\frac{1}{2},+1)}
=&\bar{x}[1 \eta_2](c_1^{\prime}+c_2^{\prime} m_1 +c_3^{\prime} m_2 +c_4^{\prime} m_1 m_2)\\
&-\bar{x}\langle\eta_1 2\rangle(c_5^{\prime}+c_6^{\prime} \tilde{m}_1 +c_7^{\prime} \tilde{m}_2 +c_8^{\prime} \tilde{m}_1 \tilde{m}_2). 
\end{aligned} \end{equation}

\subsection{UV origins from UV-IR correspondence}

Now we focus on the UV-IR correspondence for 1-massless 2-massive amplitudes. For the case with unequal masses, the correspondence is similar to the all-massive amplitudes in the last section. When the massive particles have equal mass,  gauge invariance plays the role by the $x$-factors in the amplitude. 


Recall that the massive amplitude with helicity category $(h_1,h_2,h_3)$ corresponds to the $n$-point UV amplitudes with helicity category $(h_1,h_2,h_3;0,\cdots,0)$. We consider the UV massless amplitudes in the renormalizable theory and EFT. In this work, we limit our discussion to the dimension 6 EFT operators~\footnote{In principle, there would be massless amplitudes for higher dimensional EFT operators~~\cite{Shadmi:2018xan, Aoude:2019tzn, Durieux:2019eor, Ma:2019gtx, Durieux:2019siw, Li:2020gnx, Li:2020xlh, Li:2020tsi, Li:2020zfq, Durieux:2020gip, Li:2021tsq, AccettulliHuber:2021uoa, Dong:2021vxo, Li:2022tec, Balkin:2021dko, DeAngelis:2022qco, Dong:2022mcv, Ren:2022tvi}. The total helicity of the UV amplitudes is constrained by the external particle number $n$ for $n$-point amplitudes. Here we neglect these contributions. }. The total helicity $h = \sum_i h_i$ should have the following constraint: 
\begin{equation} \begin{aligned} \label{eq:h-and-n}
&\textrm{renormalizable}:& -|n-4|\le h&\le +|n-4|, & &\textrm{for }n\ge 3, &\\
&\textrm{EFT}:& h_{\mathcal{O}}-(n-n_{\mathcal{O}})\le h&\le n+(n-n_{\mathcal{O}}), & &\textrm{for }n\ge n_{\mathcal{O}}, &
\end{aligned} \end{equation} 
where $n_\mathcal{O}$ and $h_\mathcal{O}$ are the minimal particle number and the associated total helicity of an effective operator $\mathcal{O}$. For example, the effective operator $\psi\bar{\psi}X\varphi$ can correspond to the massless amplitude $[23][31]$, which gives $n_\mathcal{O}=4$ and $h_\mathcal{O}=2$. Therefore, for a given helicity category $(h_1,h_2,h_3)$ at IR we can use this relation to give the corresponding UV amplitudes.

\begin{figure}[htbp]
\centering
\includegraphics[width=1\linewidth,valign=c]{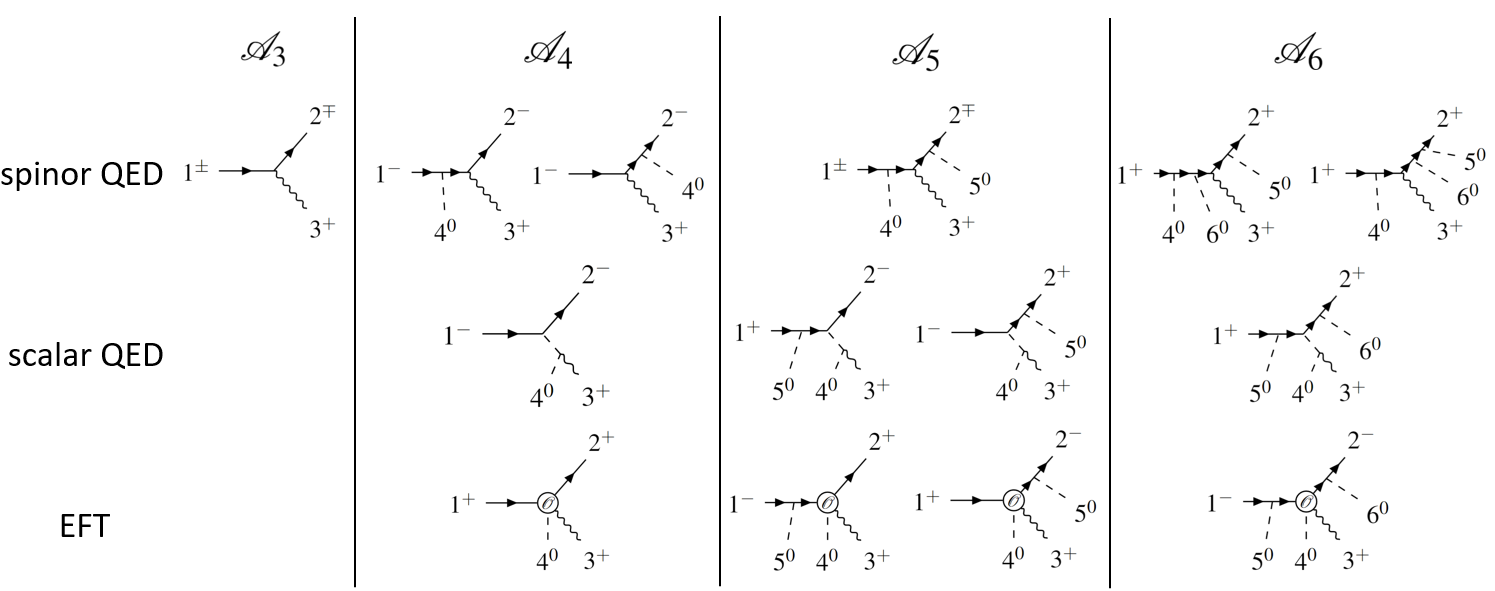}
\caption{All possible UV diagrams in spinor QED, scalar QED and EFT.}
\label{fig:UVdiagram2}
\end{figure}

For $F\bar{F}\gamma$ amplitude, there are in total $2 \times 2  = 4$ helicity amplitudes for positive helicity photon. 
As shown in eq.~\eqref{eq:equalFFV2}, the general $F\bar{F}\gamma$ amplitude has 8 terms in a given kinematics. These terms should be classified in different helicity categories by the relation $h_i=t_i$. Given a helicity category, from the total helicity $h = h_1 + h_2 + h_3$, we can determine the renormalizable and EFT amplitudes by using eq.~\eqref{eq:h-and-n}. Therefore, there is a relation between the coefficients $c_i$ and the massless amplitude $\mathcal{A}_n$. The correspondence among these can be shown as below
\begin{equation}
\begin{tabular}{c|c|c|c|c}
\hline
$(h_1,h_2,h_3)$ & coeff. with $h_i=t_i$ & total helicity & renormalizable & EFT \\
\hline
$(+\frac{1}{2},+\frac{1}{2},+1)$ &
$c_1,c_8,c_1^{\prime},c_8^{\prime}$  & $2$ & $\mathcal{A}_6$ & $\mathcal{A}_4$ \\
$(-\frac{1}{2},+\frac{1}{2},+1)$ &
$c_2,c_7,c_2^\prime,c_7^\prime$  & $1$ & $\mathcal{A}_3,\mathcal{A}_5$ & $\mathcal{A}_5$ \\
$(+\frac{1}{2},-\frac{1}{2},+1)$ &
$c_3,c_6,c_3^{\prime},c_6^{\prime}$  & $1$ & $\mathcal{A}_3,\mathcal{A}_5$ & $\mathcal{A}_5$ \\
$(-\frac{1}{2},-\frac{1}{2},+1)$ &
$c_4,c_5,c_4^{\prime},c_5^{\prime}$  & $0$ & $\mathcal{A}_4$ & $\mathcal{A}_6$ \\
\hline
\end{tabular}
\end{equation} 
where $\mathcal{A}_n$ denotes the massless amplitudes in the helicity category $(h_1,h_2,h_3;0,\cdots,0)$, with $n-3$ external scalars. The massless amplitude for the effective operator is $\psi\bar{\psi}X\varphi$, which is the 4-pt contact amplitude for the anomalous magnetic moment, and other EFT amplitudes $\mathcal{A}_5$ and $\mathcal{A}_6$ would have one or two scalar insertion. From the helicity categories, we can draw all the relevant diagrams for these massless amplitudes, which is shown in figure~\ref{fig:UVdiagram2}. 

\begin{figure}[htbp]
\centering
\includegraphics[width=0.95\linewidth,valign=c]{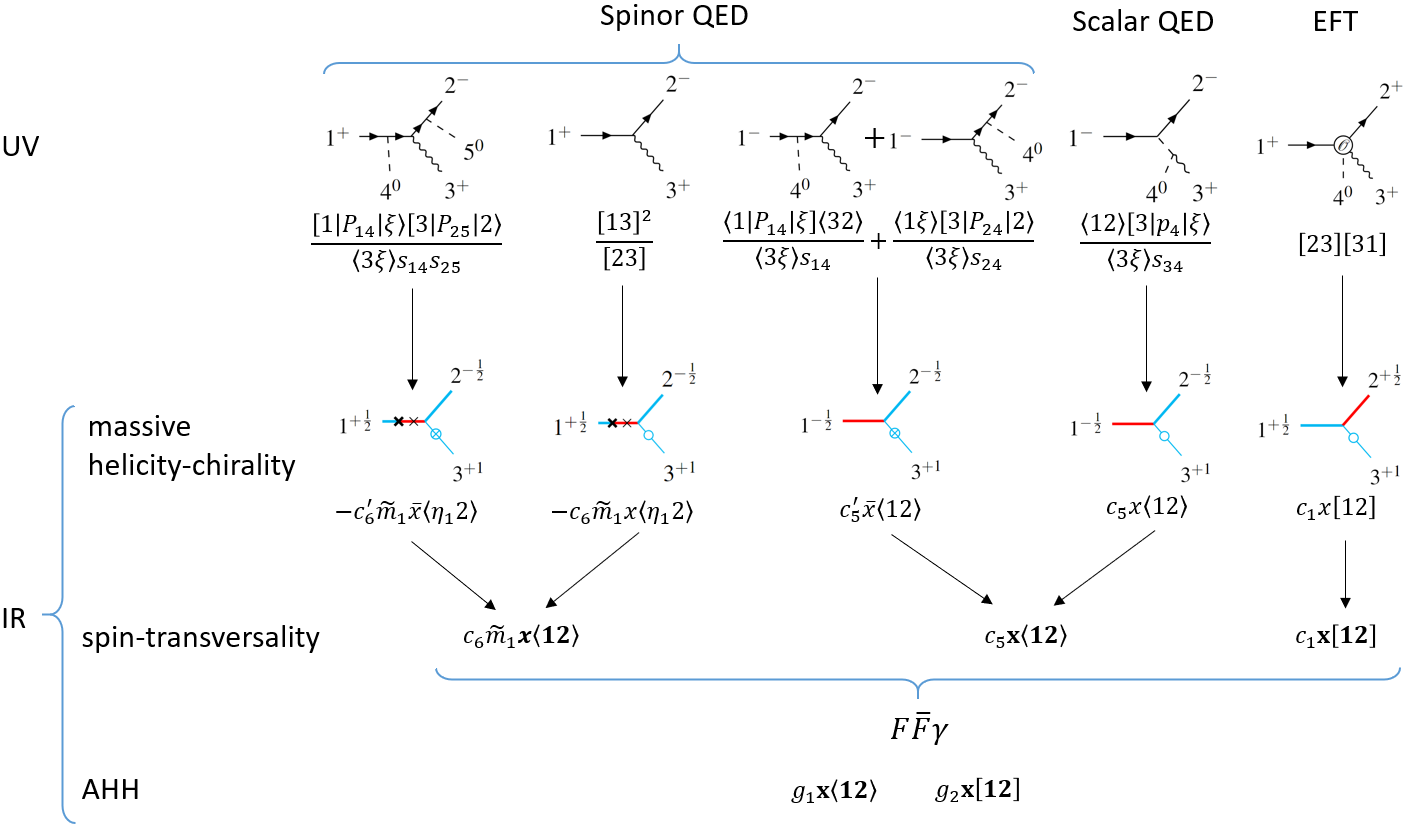}
\caption{The UV massless and IR massive amplitude correspondence for $F\bar{F}\gamma$ amplitudes. }
\label{fig:UV-FFgamma}
\end{figure}

From figure~\ref{fig:UVdiagram2}, based on the type of the main vertices, all the diagrams can be classified into three types of massless UV origins: 
\begin{itemize}

\item \textit{Spinor QED}: Let us calculate these diagrams in each helicity category. First in the helicity category $(+\frac12,-\frac12,+1)$, there are the 3-pt and 5-pt UV amplitudes. Using the on-shell limit for extra Higgs bosons, the corresponding UV amplitudes reduce to: 
\bea
\frac{[13]^2}{[12]} &=& \frac{[13]\langle\xi2\rangle}{\langle\xi 3\rangle}\rightarrow \frac{\tilde{m}}{\mathbf{m}_1}x\langle\eta_1 2\rangle,\\
\lim_{p_4\rightarrow \eta_1,p_5\rightarrow \eta_2}\frac{[1|P_{14}|\xi\rangle[3|P_{25}|2\rangle}{\langle\xi 3\rangle s_{14}s_{25}} &=& \frac{\tilde{m}_1\langle\eta_1\xi\rangle[3|\eta_2|2\rangle}{\langle\xi 3\rangle \mathbf{m}^4}\rightarrow\frac{\tilde{m}_1}{\mathbf{m}^3}\bar{x}\langle \eta_1 2\rangle. \label{eq:5ptQED}
\eea
From the result, we note the 3-pt and 5-pt amplitudes correspond to the $c_6$ and $c_6^\prime$ respectively in eq.~\eqref{eq:equalFFV2}. 
Here we write the UV amplitude in the form with reference spinor $|\xi\rangle$, so each UV amplitude corresponds to a unique topology, as shown in figure~\ref{fig:UV-FFgamma}. In the on-shell limit, the reference spinor is absorbed into $x$ and $\bar{x}$.


Similarly, in helicity category $(-\frac12,+\frac12,+1)$, $c_7$ and $c_7^{\prime}$ also correspond to the 3-pt and 5-pt UV amplitudes. For the other two helicity categories, $c_5^\prime$ and $c_8^\prime$ correspond to 4-pt UV amplitudes, while $c_5$ and $c_8$ do not have corresponding UV.

\item \textit{Scalar QED}: This UV is also a renormalizable theory, but it would contribute to different coefficients than the one obtained by fermion QED. In the helicity category $(-\frac12,-\frac12,+1)$, it is the 4-pt UV $F\bar{F}\gamma S$ amplitude. Here we would consider a different on-shell limit, in which the momentum of extra Higgs is taken to be $p$, not $\eta$. In this new on-shell limit, the UV amplitude reduces to
\begin{equation} \label{eq:deform2}
\lim_{p_4\rightarrow p_1}\frac{\langle12\rangle[3|p_4|\xi\rangle}{\langle3\xi\rangle s_{34}}
=\frac{\langle12\rangle[3|p_1|\xi\rangle}{\langle3\xi\rangle s_{31}}
\rightarrow\frac{1}{\mathbf{m}}x\langle12\rangle, 
\end{equation}
where the pole $s_{31}=p_2^2$ at IR is deformed to the mass $\mathbf{m}^2$. Comparing with the MHC amplitude, we note this diagram contributes to $c_4$. 

Similarly, for the other three helicity categories, the 5-pt and 6-pt UV amplitudes correspond to the coefficients $c_5,c_6,c_7$. 

\item \textit{EFT}: The higher-dimensional contact amplitudes can also contribute to both $c_i$ and $c_i^{\prime}$. In helicity category $(+\frac12,+\frac12,+1)$, the 4-pt diagram corresponds to the UV amplitude $[23][31]$, which contributes to $c_4$. This is nothing but the dimension-6 operator $\psi\bar{\psi}X\varphi$ with minimal particle number. Using the on-shell limit as in eq.~\eqref{eq:5ptQED}, we find $c_5^{\prime}$ corresponds to the 6-pt UV amplitudes in the helicity category $(-\frac12,-\frac12,+1)$.
Then we consider helicity category $(+\frac12,-\frac12,+1)$. In principle, $c_3,c_6,c_3^{\prime},c_6^{\prime}$ correspond to 5-pt UV amplitudes with one scalar insertion, but they are constrained by the 3-particle kinematics. Taking the on-shell limit, the possible 5-pt UV amplitude reduces to
\begin{equation}
\lim_{p_5\rightarrow\eta_2}\frac{[13][3|P_{25}|2\rangle}{s_{25}}=\frac{m_2}{\mathbf{m}^2}[13][3\eta_2]=0. 
\end{equation}
Note that $[3\eta_2]=0$ for the first kinematics in eq.~\eqref{eq:kinematic2}, and $[13]=0$ for the second kinematics. Therefore, there is no such matching for EFT operators. Similarly, there is no corresponding UV for $c_2,c_7,c_2^{\prime},c_7^{\prime}$ indicating zero coefficients for EFT operators.

\end{itemize} 
The above three kinds of UV contributions to the coefficients are summarized in the following table:
\begin{equation}  \label{eq:equalFFV-UV}
\begin{tabular}{c|c|c|c}
\hline
MHC amplitude & spinor QED & scalar QED & EFT \\
\hline
$\mathcal{M}^{(-\frac{1}{2},-\frac{1}{2},+1)}$ &
$\mathcal{A}_4\rightarrow c_5^{\prime}$ & $\mathcal{A}_4\rightarrow c_5$ & $\mathcal{A}_6\rightarrow c_5^{\prime}$ \\
$\mathcal{M}^{(+\frac{1}{2},-\frac{1}{2},+1)}$ &
$\mathcal{A}_3\rightarrow c_6,\mathcal{A}_5\rightarrow c_6^{\prime}$  & $\mathcal{A}_5\rightarrow c_6$ & - \\
$\mathcal{M}^{(-\frac{1}{2},+\frac{1}{2},+1)}$ &
$\mathcal{A}_3\rightarrow c_7,\mathcal{A}_5\rightarrow c_7^{\prime}$ & $\mathcal{A}_5\rightarrow c_7$ & - \\
$\mathcal{M}^{(+\frac{1}{2},+\frac{1}{2},+1)}$ &
$\mathcal{A}_6\rightarrow c_8^{\prime}$ & $\mathcal{A}_6\rightarrow c_8$ & $\mathcal{A}_4\rightarrow c_1$ \\
\hline
\end{tabular}
\end{equation} 
This UV-IR correspondence can be extended to the IR structure with ST spinors and spin-spinors. We list the complete UV-IR correspondence in the three stages as in figure \ref{fig:UV-FFgamma}.

\subsection{Minimal coupling in massive QED}
\label{sec:QED3pt}

In this subsection, we use the UV-IR correspondence to give 3-pt amplitudes in massive QED, which gives the minimal coupling between photon and massive fermion. The UV theory should be the massless spinor QED, so we just need to consider the correspondence in the second column of eq.~\eqref{eq:equalFFV-UV}, i.e. $c_5,c_6,c_7,c_8,c_6^\prime,c_7^\prime$ and their corresponding UV amplitudes.  The non-minimal coupling has a connection to the $c_1,c_2,c_3,c_4$ structures (and corresponding $c_i^{\prime}$ structures)~\footnote{To see the connection, we can use the bolded form. A typical non-minimal coupling for $F\bar{F}\gamma$ amplitude is $\mathbf{x}^2 \langle\mathbf{1}3\rangle\langle3\mathbf{2}\rangle$. We can reduce it to $\mathbf{x}^2 \langle\mathbf{1}3\rangle\langle3\mathbf{2}\rangle= -\mathbf{x} \langle\mathbf{1}|\mathbf{p}_1|3]\langle3\mathbf{2}\rangle
= \mathbf{x} \langle\mathbf{1}|\mathbf{p}_1(\mathbf{p}_1+\mathbf{p}_2)|\mathbf{2}\rangle
=  \mathbf{m}^2 \mathbf{x}\langle\mathbf{12}\rangle-m_1 m_2\mathbf{x}[\mathbf{12}]$. The last term $-m_1 m_2\mathbf{x}[\mathbf{12}]$ corresponds to $c_4$ and $c_4^\prime$. Similarly, we can get the connection to $c_1,c_2,c_3$ by applying chirality mass insertion to $\mathbf{x}^2 \langle\mathbf{1}3\rangle\langle3\mathbf{2}\rangle$.}. In the massive QED theory,  we do not take them in the calculation, so the amplitude would only have minimal coupling.

\begin{figure}[htbp]
\centering
\includegraphics[width=0.95\linewidth,valign=c]{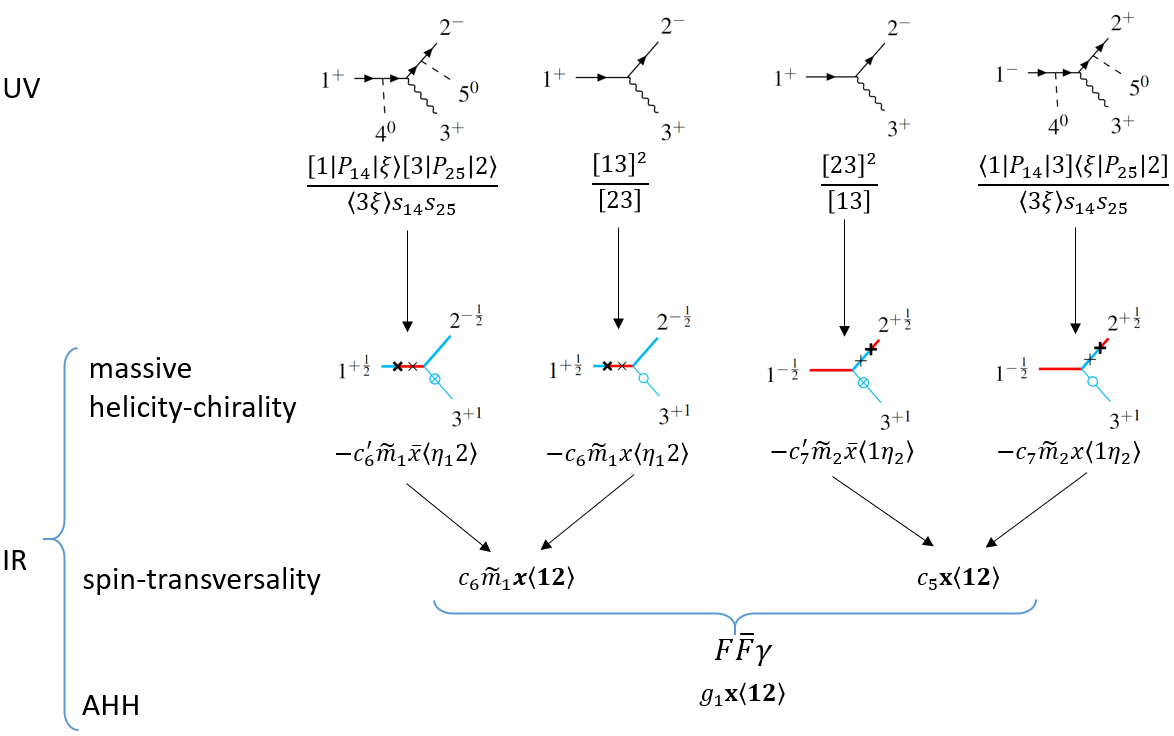}
\caption{The UV-IR correspondence for QED amplitudes. The IR can be massive helicity-chirality, spin-transversality, or AHH amplitude. }
\label{fig:UV-QED}
\end{figure}

Now we should select the needed UV structures. Since the minimal coupling in massless QED corresponds to helicity categories $(\pm\frac{1}{2},\mp\frac{1}{2},+1)$, we do not need to consider $c_5$ and $c_8$~\footnote{In the other two helicity categories there is mass dependence in the amplitude, which is sub-leading and thus are neglected. In principle, the 5-pt amplitudes are also suppressed, but due to the different kinematics choices, it is possible that the 5-pt amplitude's contributions are at the leading order, and thus cannot be neglected. }. Note that the 3-pt structure may vanish in a given kinematics choice~\footnote{For example, in the kinematics $|1]\propto|2]\propto|3]$, $|\eta_1\rangle\propto|\eta_2\rangle\propto|3\rangle$, the 3-pt UV amplitude $\frac{[13]^2}{[12]}$ vanishes but the 5-pt UV amplitude after deformation and Higgsing would survive. In this case, the 5-pt UV amplitude rather than the 3-pt one is needed to do matching in helicity category $(+\frac{1}{2},-\frac{1}{2},+1,0,0)$. In another kinematics, verse vice. }, so both 3-pt and 5-pt UV amplitudes are needed to determine the coefficients
\bea
c_6=c_7=c_6^{\prime}=c_7^{\prime}=\frac{e}{\mathbf{m}},
\label{eq:c67-mdep}
\eea
where $e$ is the coupling constant in QED.
Therefore, the non-vanishing terms in the two kinematics are
\begin{equation} \label{eq:FFgamma2}
\renewcommand{\arraystretch}{1.5}
\begin{tabular}{c|c|c}
helicity & \makecell{$\lambda_1\propto\lambda_2\propto\lambda_3$\\$\tilde{\eta}_1\propto\tilde{\eta}_2\propto\tilde{\lambda}_3$} & \makecell{$\tilde{\lambda}_1\propto\tilde{\lambda}_2\propto\tilde{\lambda}_3$\\$\eta_1\propto \eta_2\propto\lambda_3$} \\
\hline
$(+\frac{1}{2},-\frac{1}{2},+1)$ &
$-c_6 \tilde{m}_1 x\langle \eta_1 2\rangle-c_7 \tilde{m}_2 x\langle \eta_1 2\rangle$  & 
$-c_6^{\prime}\tilde{m}_1 \bar{x}\langle \eta_1 2\rangle-c_7^{\prime}\tilde{m}_2 \bar{x}\langle \eta_1 2\rangle$ \\
$(-\frac{1}{2},+\frac{1}{2},+1)$ &
$-c_6 \tilde{m}_1 x\langle 1\eta_2 \rangle-c_7 \tilde{m}_2 x\langle 1\eta_2\rangle$  & 
$-c_6^{\prime}\tilde{m}_1 \bar{x}\langle 1\eta_2 \rangle-c_7^{\prime}\tilde{m}_2 \bar{x}\langle 1\eta_2\rangle$ \\
\end{tabular}
\end{equation} 
Let us consider the complete UV-IR correspondence for the minimal coupling by involving the ST spinor formalism and AHH formalism. 
\bit 
\item Recover the $U(2)$ LG covariance, these amplitudes have a bolded form as the ST amplitude,
\begin{equation}
\mathcal{M}(\mathbf{1}^{\frac12},\mathbf{2}^{\frac12},3^{+1})=c_6 \tilde{m}_1 \mathbf{x}\langle\mathbf{12}\rangle+c_7\tilde{m}_2 \mathbf{x}\langle\mathbf{12}\rangle. 
\end{equation} 
Notice that $x$ and $\bar{x}$ are converted to one bolded $\mathbf{x}$, so the power counting for such $F\bar{F}\gamma$ is lost in the ST spinor formalism. If we want to analyze this equal-mass amplitude with clear power counting, the MHC amplitude in eq.~\eqref{eq:FFgamma2} is needed. 
\item Furthermore, when reducing to the $SU(2)$ LG, the spurion masses $\tilde{m}_1$ and $\tilde{m}_2$ are fixed to the physical mass $\mathbf{m}$. The ST amplitude is unified to AHH amplitude, 
\bea
M_{AHH}(\mathbf{1}^{\frac{1}{2}}, \mathbf{2}^{\frac{1}{2}}, 3^{+1})&=(c_6 \mathbf{m}+c_7 \mathbf{m})\mathbf{x}\langle\mathbf{12}\rangle \to e\mathbf{x}\langle\mathbf{12}\rangle. 
\eea
Thus $c_6$ and $c_7$ structures reduce to the same AHH amplitude, which is expected because the QED is vectorlike. Since either $c_6$ or $c_7$ MHC structure ($c_6^{\prime}$ and $c_7^{\prime}$ in other kinematics) can give the AHH amplitude, we can choose one of them to get the AHH minimal couplings. 
\eit 
The three-stage correspondence from both 3-pt and 5-pt UV amplitudes is shown in figure~\ref{fig:UV-QED}. 


In the above $x$ definition, although it has the advantage of zero transversality, the MHC amplitudes for the minimal coupling carry the $m$ dependence which is canceled by the coefficients in eq.~\eqref{eq:c67-mdep}. To remove such apparent $m$ dependence, let us consider a different definition of $x$-factor. 
%
Suppose we choose
\bea \label{eq:neq0x}
x=\frac{[23][31]}{m_1[12]},\quad
\bar{x}=\frac{\tilde{m}_2\langle12\rangle}{\langle23\rangle\langle31\rangle},
\eea
where $x$ carries $(t_1,t_2)=(+1,0)$ and $x$ carries $(t_1,t_2)=(0,+1)$. Using these new $x$-factors, the chirality of massive fermion is changed as follows
\begin{equation}
\begin{tabular}{c|c|c}
\hline
amplitude & $x$-factor with $t_1=t_2=0$ & $x$-factor with new definition \\
\hline
$-x\langle\eta_1 2\rangle$ &
\begin{tikzpicture}[baseline=0.7cm] \begin{feynhand}
\vertex [particle] (i1) at (-0.2,0.8) {$1^{+\frac{1}{2}}$};
\vertex [particle] (i2) at (1.6,1.6) {$2^{-\frac{1}{2}}$};
\vertex [particle] (i3) at (1.6,0) {$3^{+1}$};
\vertex (v1) at (0.9,0.8);
\vertex [ringdot,cyan] (v3) at (1.075,0.6) {};
\graph{(i1)--[plain,red,very thick] (v1)};
\graph{(i2)--[plain,cyan,very thick] (v1)};
\graph{(i3)--[plain,cyan] (v3)--[plain,cyan] (v1)};
\draw plot[mark=x,mark size=2.7,mark options={rotate=0}] coordinates {(0.6,0.8)};
\end{feynhand} \end{tikzpicture} & 
\begin{tikzpicture}[baseline=0.7cm] \begin{feynhand}
\vertex [particle] (i1) at (-0.2,0.8) {$1^{+\frac{1}{2}}$};
\vertex [particle] (i2) at (1.6,1.6) {$2^{-\frac{1}{2}}$};
\vertex [particle] (i3) at (1.6,0) {$3^{+1}$};
\vertex (v1) at (0.9,0.8);
\vertex [ringdot,cyan] (v3) at (1.075,0.6) {};
\graph{(i1)--[plain,cyan,very thick] (v1)};
\graph{(i2)--[plain,cyan,very thick] (v1)};
\graph{(i3)--[plain,cyan] (v3)--[plain,cyan] (v1)};
\draw plot[mark=x,mark size=2.7,mark options={rotate=0}] coordinates {(0.6,0.8)};
\end{feynhand} \end{tikzpicture} \\
$-\bar{x}\langle\eta_1 2\rangle$ & 
\begin{tikzpicture}[baseline=0.7cm] \begin{feynhand}
\setlength{\feynhanddotsize}{0.8mm}
\vertex [particle] (i1) at (-0.2,0.8) {$1^{+\frac{1}{2}}$};
\vertex [particle] (i2) at (1.6,1.6) {$2^{-\frac{1}{2}}$};
\vertex [particle] (i3) at (1.6,0) {$3^{+1}$};
\vertex (v1) at (0.9,0.8);
\vertex [crossdot,cyan] (v2) at (1.075,0.6) {};
\graph{(i1)--[plain,red,very thick] (v1)};
\graph{(i2)--[plain,cyan,very thick] (v1)};
\graph{(i3)--[plain,cyan] (v2)--[plain,cyan] (v1)};
\draw plot[mark=x,mark size=2.7,mark options={rotate=0}] coordinates {(0.6,0.8)};
\end{feynhand} \end{tikzpicture} & 
\begin{tikzpicture}[baseline=0.7cm] \begin{feynhand}
\setlength{\feynhanddotsize}{0.8mm}
\vertex [particle] (i1) at (-0.2,0.8) {$1^{+\frac{1}{2}}$};
\vertex [particle] (i2) at (1.6,1.6) {$2^{-\frac{1}{2}}$};
\vertex [particle] (i3) at (1.6,0) {$3^{+1}$};
\vertex (v1) at (0.9,0.8);
\vertex [crossdot,cyan] (v2) at (1.075,0.6) {};
\graph{(i1)--[plain,red,very thick] (v1)};
\graph{(i2)--[plain,red,very thick] (v1)};
\graph{(i3)--[plain,cyan] (v2)--[plain,cyan] (v1)};
\draw plot[mark=x,mark size=2.7,mark options={rotate=0}] coordinates {(0.6,0.8)};
\end{feynhand} \end{tikzpicture} \\
\hline
\end{tabular}
\end{equation}
where red and cyan lines still represent left-handed and right-handed chirality spin $\frac12$ particles. This indicates that when we use this new definition of $x$, the external lines for particles 1 and 2 should not be identified as the single-particle wave function in subsection~\ref{sec:MassInsert}.

Given the chirality, comparing with eqs.~\eqref{eq:FFgamma-Insertion2} and~\eqref{eq:FFgamma-Insertion3}, we find that the diagrams with new $x$ have fewer mass insertion: 
\begin{equation}
\begin{tabular}{c|c|c}
\hline
transversality & $x$-factor with $t_1=t_2=0$ & $x$-factor with new definition \\
\hline
$(+\frac12,-\frac12,+1)$ &
\begin{tikzpicture}[baseline=0.7cm] \begin{feynhand}
\vertex [particle] (i1) at (-0.2,0.8) {$1^{+\frac{1}{2}}$};
\vertex [particle] (i2) at (1.6,1.6) {$2^{-\frac{1}{2}}$};
\vertex [particle] (i3) at (1.6,0) {$3^{+1}$};
\vertex (v1) at (0.9,0.8);
\vertex (v2) at (0.48,0.8);
\vertex [ringdot,cyan] (v3) at (1.075,0.6) {};
\graph{(i1)--[plain,cyan,very thick](v2)--[plain,red,very thick] (v1)};
\graph{(i2)--[plain,cyan,very thick] (v1)};
\graph{(i3)--[plain,cyan] (v3)--[plain,cyan] (v1)};
\draw[very thick] plot[mark=x,mark size=2.7] coordinates {(v2)};
\draw plot[mark=x,mark size=2.7,mark options={rotate=0}] coordinates {(0.72,0.8)};
\end{feynhand} \end{tikzpicture}$=-x\tilde{m}_1\langle\eta_1 2\rangle$ & 
\begin{tikzpicture}[baseline=0.7cm] \begin{feynhand}
\vertex [particle] (i1) at (-0.2,0.8) {$1^{+\frac{1}{2}}$};
\vertex [particle] (i2) at (1.6,1.6) {$2^{-\frac{1}{2}}$};
\vertex [particle] (i3) at (1.6,0) {$3^{+1}$};
\vertex (v1) at (0.9,0.8);
\vertex [ringdot,cyan] (v3) at (1.075,0.6) {};
\graph{(i1)--[plain,cyan,very thick] (v1)};
\graph{(i2)--[plain,cyan,very thick] (v1)};
\graph{(i3)--[plain,cyan] (v3)--[plain,cyan] (v1)};
\draw plot[mark=x,mark size=2.7,mark options={rotate=0}] coordinates {(0.6,0.8)};
\end{feynhand} \end{tikzpicture}$-x\langle\eta_1 2\rangle$ \\
$(-\frac12,+\frac12,+1)$ & 
\begin{tikzpicture}[baseline=0.7cm] \begin{feynhand}
\setlength{\feynhanddotsize}{0.8mm}
\vertex [particle] (i1) at (-0.2,0.8) {$1^{-\frac{1}{2}}$};
\vertex [particle] (i2) at (1.6,1.6) {$2^{+\frac{1}{2}}$};
\vertex [particle] (i3) at (1.6,0) {$3^{+1}$};
\vertex (v1) at (0.9,0.8);
\vertex (v2) at (0.9+0.7*0.35,0.8+0.8*0.35);
\vertex [crossdot,cyan] (v3) at (1.075,0.6) {};
\graph{(i1)--[plain,red,very thick] (v1)};
\graph{(i2)--[plain,red,very thick](v2)--[plain,cyan,very thick] (v1)};
\graph{(i3)--[plain,cyan] (v3)--[plain,cyan] (v1)};
\draw plot[mark=x,mark size=2.7,mark options={rotate=0}] coordinates {(0.6,0.8)};
\draw[very thick] plot[mark=x,mark size=2.7,mark options={rotate=45}] coordinates {(0.9+0.7*0.33,0.8+0.8*0.33)};
\end{feynhand} \end{tikzpicture}$=-\bar{x}\tilde{m}_2\langle\eta_1 2\rangle$ & 
\begin{tikzpicture}[baseline=0.7cm] \begin{feynhand}
\setlength{\feynhanddotsize}{0.8mm}
\vertex [particle] (i1) at (-0.2,0.8) {$1^{+\frac{1}{2}}$};
\vertex [particle] (i2) at (1.6,1.6) {$2^{-\frac{1}{2}}$};
\vertex [particle] (i3) at (1.6,0) {$3^{+1}$};
\vertex (v1) at (0.9,0.8);
\vertex [crossdot,cyan] (v2) at (1.075,0.6) {};
\graph{(i1)--[plain,red,very thick] (v1)};
\graph{(i2)--[plain,red,very thick] (v1)};
\graph{(i3)--[plain,cyan] (v2)--[plain,cyan] (v1)};
\draw plot[mark=x,mark size=2.7,mark options={rotate=0}] coordinates {(0.6,0.8)};
\end{feynhand} \end{tikzpicture}$=-\bar{x}\langle\eta_1 2\rangle$ \\
\hline
\end{tabular}
\end{equation}
The QED MHC amplitude can be reorganized by the replacement $x\rightarrow \frac{m_1}{\mathbf{m}}x$ and $\bar{x}\rightarrow \frac{m_2}{\mathbf{m}}\bar{x}$:
\bea
\mathcal{M}^{(+\frac{1}{2},-\frac{1}{2},+1)}_{\text{QED}} &=& -e x\langle \eta_1 2\rangle-\frac{e}{\mathbf{m}^2} m_1\tilde{m}_2 x\langle \eta_1 2\rangle, \quad -\frac{e}{\mathbf{m}^2}\tilde{m}_1 m_2 \bar{x}\langle \eta_1 2\rangle-e \bar{x}\langle \eta_1 2\rangle, \\
\mathcal{M}^{(-\frac{1}{2},+\frac{1}{2},+1)}_{\text{QED}} &=& -e  x\langle 1\eta_2 \rangle-\frac{e}{\mathbf{m}^2}m_1\tilde{m}_2 x\langle 1\eta_2\rangle, \quad -\frac{e}{\mathbf{m}^2}\tilde{m}_1 m_2 \bar{x}\langle 1\eta_2 \rangle-e \bar{x}\langle 1\eta_2\rangle. 
\eea
Hence the first term without chirality mass insertion $m_1\tilde{m}_2$ and $\tilde{m}_1 m_2$ become a natural choice to describe the minimal couplings. They can be written diagrammatically
\begin{equation} \label{eq:equalFFV1}
\begin{tabular}{c|c|c}
helicity & \makecell{$\lambda_1\propto\lambda_2\propto\lambda_3$\\$\tilde{\eta}_1\propto\tilde{\eta}_2\propto\tilde{\lambda}_3$} & \makecell{$\tilde{\lambda}_1\propto\tilde{\lambda}_2\propto\tilde{\lambda}_3$\\$\eta_1\propto \eta_2\propto\lambda_3$} \\
\hline
$(+\frac{1}{2},-\frac{1}{2},+1)$ & 
\begin{tikzpicture}[baseline=0.7cm] \begin{feynhand}
\vertex [particle] (i1) at (0.15,0.8) {$1^{+\frac{1}{2}}$};
\vertex [particle] (i2) at (0.9+0.7*0.75,0.8+0.8*0.75) {$2^{-\frac{1}{2}}$};
\vertex [particle] (i3) at (0.9+0.7*0.75,0.8-0.8*0.75) {$3^{+1}$};
\vertex (v1) at (0.9,0.8);
\vertex [ringdot,cyan] (v3) at (0.9+0.7*0.2,0.8-0.8*0.2) {};
\graph{(i1)--[plain,cyan,very thick] (v1)};
\graph{(i2)--[plain,cyan,very thick] (v1)};
\graph{(i3)--[plain,cyan] (v3)--[plain,cyan] (v1)};
\draw plot[mark=x,mark size=3.5,mark options={rotate=0}] coordinates {(0.75,0.8)};
\end{feynhand} \end{tikzpicture}$=-x \langle\eta_1 2\rangle$ & 
\begin{tikzpicture}[baseline=0.7cm] \begin{feynhand}
\setlength{\feynhanddotsize}{0.8mm}
\vertex [particle] (i1) at (0.15,0.8) {$1^{+\frac{1}{2}}$};
\vertex [particle] (i2) at (0.9+0.7*0.75,0.8+0.8*0.75) {$2^{-\frac{1}{2}}$};
\vertex [particle] (i3) at (0.9+0.7*0.75,0.8-0.8*0.75) {$3^{+1}$};
\vertex (v1) at (0.9,0.8);
\vertex [crossdot,cyan] (v3) at (0.9+0.7*0.2,0.8-0.8*0.2) {};
\graph{(i1)--[plain,red,very thick] (v1)};
\graph{(i2)--[plain,red,very thick] (v1)};
\graph{(i3)--[plain,cyan] (v3)--[plain,cyan] (v1)};
\draw plot[mark=x,mark size=3.5,mark options={rotate=0}] coordinates {(0.75,0.8)};
\end{feynhand} \end{tikzpicture}$=-\bar{x}\langle \eta_1 2\rangle$ \\
$(-\frac{1}{2},+\frac{1}{2},+1)$ & 
\begin{tikzpicture}[baseline=0.7cm] \begin{feynhand}
\vertex [particle] (i1) at (0.15,0.8) {$1^{-\frac{1}{2}}$};
\vertex [particle] (i2) at (0.9+0.7*0.75,0.8+0.8*0.75) {$2^{+\frac{1}{2}}$};
\vertex [particle] (i3) at (0.9+0.7*0.75,0.8-0.8*0.75) {$3^{+1}$};
\vertex (v1) at (0.9,0.8);
\vertex [ringdot,cyan] (v3) at (0.9+0.7*0.2,0.8-0.8*0.2) {};
\graph{(i1)--[plain,cyan,very thick] (v1)};
\graph{(i2)--[plain,cyan,very thick] (v1)};
\graph{(i3)--[plain,cyan] (v3)--[plain,cyan] (v1)};
\draw plot[mark=x,mark size=3.5,mark options={rotate=45}] coordinates {(0.9+0.7*0.15,0.8+0.8*0.15)};
\end{feynhand} \end{tikzpicture}$=-x \langle1\eta_2 \rangle$ & 
\begin{tikzpicture}[baseline=0.7cm] \begin{feynhand}
\setlength{\feynhanddotsize}{0.8mm}
\vertex [particle] (i1) at (0.15,0.8) {$1^{-\frac{1}{2}}$};
\vertex [particle] (i2) at (0.9+0.7*0.75,0.8+0.8*0.75) {$2^{+\frac{1}{2}}$};
\vertex [particle] (i3) at (0.9+0.7*0.75,0.8-0.8*0.75) {$3^{+1}$};
\vertex (v1) at (0.9,0.8);
\vertex [crossdot,cyan] (v3) at (0.9+0.7*0.2,0.8-0.8*0.2) {};
\graph{(i1)--[plain,red,very thick] (v1)};
\graph{(i2)--[plain,red,very thick] (v1)};
\graph{(i3)--[plain,cyan] (v3)--[plain,cyan] (v1)};
\draw plot[mark=x,mark size=3.5,mark options={rotate=45}] coordinates {(0.9+0.7*0.15,0.8+0.8*0.15)};
\end{feynhand} \end{tikzpicture}$=-\bar{x}\langle 1\eta_2\rangle$ \\ \hline
$(+\frac{1}{2},-\frac{1}{2},-1)$ & 
\begin{tikzpicture}[baseline=0.7cm] \begin{feynhand}
\setlength{\feynhanddotsize}{0.8mm}
\vertex [particle] (i1) at (0.15,0.8) {$1^{+\frac{1}{2}}$};
\vertex [particle] (i2) at (0.9+0.7*0.75,0.8+0.8*0.75) {$2^{-\frac{1}{2}}$};
\vertex [particle] (i3) at (0.9+0.7*0.75,0.8-0.8*0.75) {$3^{-1}$};
\vertex (v1) at (0.9,0.8);
\vertex [crossdot,red] (v3) at (0.9+0.7*0.2,0.8-0.8*0.2) {};
\graph{(i1)--[plain,red,very thick] (v1)};
\graph{(i2)--[plain,red,very thick] (v1)};
\graph{(i3)--[plain,red] (v3)--[plain,red] (v1)};
\draw plot[mark=x,mark size=3.5,mark options={rotate=45}] coordinates {(0.9+0.7*0.15,0.8+0.8*0.15)};
\end{feynhand} \end{tikzpicture}$=x^{-1} [1\eta_2]$ &
\begin{tikzpicture}[baseline=0.7cm] \begin{feynhand}
\vertex [particle] (i1) at (0.15,0.8) {$1^{+\frac{1}{2}}$};
\vertex [particle] (i2) at (0.9+0.7*0.75,0.8+0.8*0.75) {$2^{-\frac{1}{2}}$};
\vertex [particle] (i3) at (0.9+0.7*0.75,0.8-0.8*0.75) {$3^{-1}$};
\vertex (v1) at (0.9,0.8);
\vertex [ringdot,red] (v3) at (0.9+0.7*0.2,0.8-0.8*0.2) {};
\graph{(i1)--[plain,cyan,very thick] (v1)};
\graph{(i2)--[plain,cyan,very thick] (v1)};
\graph{(i3)--[plain,red] (v3)--[plain,red] (v1)};
\draw plot[mark=x,mark size=3.5,mark options={rotate=45}] coordinates {(0.9+0.7*0.15,0.8+0.8*0.15)};
\end{feynhand} \end{tikzpicture}$=\bar{x}^{-1}[1\eta_2]$ \\
$(-\frac{1}{2},+\frac{1}{2},-1)$ & 
\begin{tikzpicture}[baseline=0.7cm] \begin{feynhand}
\setlength{\feynhanddotsize}{0.8mm}
\vertex [particle] (i1) at (0.15,0.8) {$1^{-\frac{1}{2}}$};
\vertex [particle] (i2) at (0.9+0.7*0.75,0.8+0.8*0.75) {$2^{+\frac{1}{2}}$};
\vertex [particle] (i3) at (0.9+0.7*0.75,0.8-0.8*0.75) {$3^{-1}$};
\vertex (v1) at (0.9,0.8);
\vertex [crossdot,red] (v3) at (0.9+0.7*0.2,0.8-0.8*0.2) {};
\graph{(i1)--[plain,red,very thick] (v1)};
\graph{(i2)--[plain,red,very thick] (v1)};
\graph{(i3)--[plain,red] (v3)--[plain,red] (v1)};
\draw plot[mark=x,mark size=3.5,mark options={rotate=0}] coordinates {(0.75,0.8)};
\end{feynhand} \end{tikzpicture}$=x^{-1} [\eta_1 2]$ &
\begin{tikzpicture}[baseline=0.7cm] \begin{feynhand}
\vertex [particle] (i1) at (0.15,0.8) {$1^{-\frac{1}{2}}$};
\vertex [particle] (i2) at (0.9+0.7*0.75,0.8+0.8*0.75) {$2^{+\frac{1}{2}}$};
\vertex [particle] (i3) at (0.9+0.7*0.75,0.8-0.8*0.75) {$3^{-1}$};
\vertex (v1) at (0.9,0.8);
\vertex [ringdot,red] (v3) at (0.9+0.7*0.2,0.8-0.8*0.2) {};
\graph{(i1)--[plain,cyan,very thick] (v1)};
\graph{(i2)--[plain,cyan,very thick] (v1)};
\graph{(i3)--[plain,red] (v3)--[plain,red] (v1)};
\draw plot[mark=x,mark size=3.5,mark options={rotate=0}] coordinates {(0.75,0.8)};
\end{feynhand} \end{tikzpicture}$=\bar{x}^{-1}[\eta_1 2]$ \\
\end{tabular}
\end{equation}
Here we also list the 3-pt amplitudes with a negative helicity photon. 

\section{Factorized Massive Helicity-Chirality Amplitudes}\label{sec:facAmp}

As discussed in the previous section, the 3-pt amplitudes are determined by the $U(2)$ LG covariance, which reflects the symmetry of this EFT. In order to describe the higher-point amplitudes, one can suitably glue the 3-pt amplitudes together, which is based on the following properties of the singularity structure of scattering amplitudes with the total $\delta^{(4)}$-momentum conservation stripped out:
\bit 
\item Analyticity: the singularities are restricted to only having poles and branch points;
\item Locality: the singularities are further restricted to having only simple poles and branch cuts;
\item Unitarity: when approaching the singularities, one or more particles go on-shell and the amplitude factorizes into lower-point amplitudes.
\eit
Therefore, the gluing technique of constructing higher-point tree-level amplitudes from lower ones is based on
\begin{equation} \label{eq:FactorizeMassless}
\lim_{P^2\rightarrow 0} \lim_{\substack{2P\cdot\eta \rightarrow \mathbf{m}^2\\ \eta^2\rightarrow 0}} P^2 \mathcal{M}
=\sum_{h_P} M^L\times M^R,
\end{equation}
where $P$, $\eta$, $\mathbf{m}$ and $h_P$ are the momenta, mass and helicity of the on-shell propagator, $M^L$ and $M^R$ are sub-amplitudes.

\subsection{Gluing massive amplitudes}

We focus on amplitudes that involve only helicity mass insertions, as chirality mass insertions can be easily represented by multiplying $m$ and $\tilde{m}$ for internal and external particles. Consider the massive $F\bar{F}VS$ amplitude with helicity category $(+\frac{1}{2}, -\frac{1}{2}, +1, 0)$. For renormalizable interactions, the transversality can be $(\pm\frac{1}{2}, \pm\frac{1}{2}, 0, 0)$ or $(\pm\frac{1}{2}, \mp\frac{1}{2}, 0, 0)$. We first select the transversality $(+\frac{1}{2}, -\frac{1}{2}, 0, 0)$ and the $s_{14}$ channel, in which the internal line corresponds to a fermion propagator.
The relevant 3-pt sub-amplitudes are
\begin{equation}\label{eq:subampffvs} \begin{aligned}
\begin{tikzpicture}[baseline=0.7cm] \begin{feynhand}
\vertex [particle] (i1) at (-0.2,0.8) {$2^{-\frac{1}{2}}$};
\vertex [particle] (i2) at (1.6,1.6) {$P^{+\frac{1}{2}}$};
\vertex [particle] (i3) at (1.6,0) {$3^{+1}$};
\vertex (v1) at (0.9,0.8);
\graph{(i1)--[plain,cyan,very thick] (v1)};
\graph{(i2)--[plain,cyan,very thick] (v1)};
\graph{(i3)--[plain,brown,very thick] (v1)};
\draw plot[mark=x,mark size=2.7,mark options={rotate=45}] coordinates {(0.9+0.7*0.31,0.8-0.8*0.31)};
\end{feynhand} \end{tikzpicture}
&=-\langle 2 \eta_3\rangle [3P_{14}],&  
\begin{tikzpicture}[baseline=0.7cm] \begin{feynhand}
\vertex [particle] (i1) at (-0.2,0.8) {$2^{-\frac{1}{2}}$};
\vertex [particle] (i2) at (1.6,1.6) {$P^{+\frac{1}{2}}$};
\vertex [particle] (i3) at (1.6,0) {$3^{+1}$};
\vertex (v1) at (0.9,0.8);
\graph{(i1)--[plain,cyan,very thick] (v1)};
\graph{(i2)--[plain,cyan,very thick] (v1)};
\graph{(i3)--[plain,brown,very thick] (v1)};
\draw plot[mark=x,mark size=2.7,mark options={rotate=45}] coordinates {(0.9+0.7*0.31,0.8-0.8*0.31)};
\draw plot[mark=x,mark size=2.7,mark options={rotate=45}] coordinates {(0.9+0.7*0.31,0.8+0.8*0.31)};
\end{feynhand} \end{tikzpicture}
&=-\langle 2 \eta_3\rangle [3\eta_{14}], \\
\begin{tikzpicture}[baseline=0.7cm] \begin{feynhand}
\vertex [particle] (i1) at (-0.2,0.8) {$1^{+\frac{1}{2}}$};
\vertex [particle] (i2) at (1.6,1.6) {$P^{-\frac{1}{2}}$};
\vertex [particle] (i3) at (1.6,0) {$4^{0}$};
\vertex (v1) at (0.9,0.8);
\graph{(i1)--[plain,cyan,very thick] (v1)};
\graph{(i2)--[plain,red,very thick] (v1)};
\graph{(i3)--[plain,brown,very thick] (v1)};
\end{feynhand} \end{tikzpicture} &=[1P_{14}], &
\begin{tikzpicture}[baseline=0.7cm] \begin{feynhand}
\vertex [particle] (i1) at (-0.2,0.8) {$1^{+\frac{1}{2}}$};
\vertex [particle] (i2) at (1.6,1.6) {$P^{-\frac{1}{2}}$};
\vertex [particle] (i3) at (1.6,0) {$4^{0}$};
\vertex (v1) at (0.9,0.8);
\graph{(i1)--[plain,cyan,very thick] (v1)};
\graph{(i2)--[plain,red,very thick] (v1)};
\graph{(i3)--[plain,brown,very thick] (v1)};
\end{feynhand} 
\draw plot[mark=x,mark size=2.7,mark options={rotate=45}] coordinates {(0.9+0.7*0.31,0.8+0.8*0.31)};
\end{tikzpicture}&=[1\eta_{14}],
\end{aligned} \end{equation}
where $P_{14}\equiv p_1+p_4$, $\eta_{14}\equiv \eta_1+\eta_4$, and particle $P$ is taken to be the internal fermion.
Taking the factorized limit, we find that the transversality of the internal fermion should be $(+\frac{1}{2}, +\frac{1}{2})$, but we need to sum over the helicities of the internal fermion
\begin{equation} \begin{aligned}
\begin{tikzpicture}[baseline=0.7cm] \begin{feynhand}
\vertex [dot] (v1) at (0,0.8) {};
\vertex [dot] (v2) at (2,0.8) {};
\vertex (v3) at (1,0.8);
\node at (0.7,1) {\tiny $-\frac{1}{2}$};
\node at (1.35,1) {\tiny $+\frac{1}{2}$};
\graph{(v1)--[plain,cyan,very thick](v3)--[plain,red,slash={[style=black]0},very thick] (v2)};
\draw plot[mark=x,mark size=2.7] coordinates {(0.3,0.8)};
\end{feynhand} \end{tikzpicture} \quad + \quad
\begin{tikzpicture}[baseline=0.7cm] \begin{feynhand}
\vertex [dot] (v1) at (0,0.8) {};
\vertex [dot] (v2) at (2,0.8) {};
\vertex (v3) at (1,0.8);
\node at (0.7,1) {\tiny $-\frac{1}{2}$};
\node at (1.35,1) {\tiny $+\frac{1}{2}$};
\graph{(v1)--[plain,cyan,very thick](v3)--[plain,red,slash={[style=black]0},very thick] (v2)};
\draw plot[mark=x,mark size=2.7] coordinates {(1.7,0.8)};
\end{feynhand} \end{tikzpicture}, 
\end{aligned} \end{equation}
where the $\pm \frac12$ next to the slash denotes the helicity of the internal fermion at the factorization limit. 

Gluing the 3-pt amplitudes given by eq.~\eqref{eq:subampffvs} gives
\begin{equation}
\begin{tikzpicture}[baseline=0.8cm] \begin{feynhand}
\vertex [particle] (i1) at (0,1.8) {$1^{+\frac{1}{2}}$};
\vertex [particle] (i2) at (0,0) {$2^{-\frac{1}{2}}$};
\vertex [particle] (i3) at (1.8,0) {$3^{+1}$};
\vertex [particle] (i4) at (1.8,1.8) {$4^0$};
\vertex (v1) at (0.9,0.6);
\vertex (v2) at (0.9,1.2);
\vertex (v3) at (0.9,0.9);
\graph{(i1)--[plain,cyan,very thick] (v2)--[plain,brown,very thick] (i4)};
\graph{(i2)--[plain,cyan,very thick] (v1)--[plain,brown,very thick] (i3)};
\graph{(v1)--[plain,cyan,very thick] (v3)--[plain,red,slash={[style=black]0},very thick] (v2)};
\draw plot[mark=x,mark size=2.7] coordinates {(0.9,1.05)};
\draw plot[mark=x,mark size=2.7,mark options={rotate=45}] coordinates {(0.9+0.9*0.3,0.6-0.6*0.3)};
\end{feynhand} \end{tikzpicture}+
\begin{tikzpicture}[baseline=0.8cm] \begin{feynhand}
\vertex [particle] (i1) at (0,1.8) {$1^{+\frac{1}{2}}$};
\vertex [particle] (i2) at (0,0) {$2^{-\frac{1}{2}}$};
\vertex [particle] (i3) at (1.8,0) {$3^{+1}$};
\vertex [particle] (i4) at (1.8,1.8) {$4^0$};
\vertex (v1) at (0.9,0.6);
\vertex (v2) at (0.9,1.2);
\vertex (v3) at (0.9,0.9);
\graph{(i1)--[plain,cyan,very thick] (v2)--[plain,brown,very thick] (i4)};
\graph{(i2)--[plain,cyan,very thick] (v1)--[plain,brown,very thick] (i3)};
\graph{(v1)--[plain,cyan,very thick] (v3)--[plain,red,slash={[style=black]0},very thick] (v2)};
\draw plot[mark=x,mark size=2.7] coordinates {(0.9,0.75)};
\draw plot[mark=x,mark size=2.7,mark options={rotate=45}] coordinates {(0.9+0.9*0.3,0.6-0.6*0.3)};
\end{feynhand} \end{tikzpicture} \Rightarrow
\begin{tikzpicture}[baseline=0.8cm] \begin{feynhand}
\vertex [particle] (i1) at (0,1.8) {$1^{+\frac{1}{2}}$};
\vertex [particle] (i2) at (0,0) {$2^{-\frac{1}{2}}$};
\vertex [particle] (i3) at (1.8,0) {$3^{+1}$};
\vertex [particle] (i4) at (1.8,1.8) {$4^0$};
\vertex (v1) at (0.9,0.6);
\vertex (v2) at (0.9,1.2);
\vertex (v3) at (0.9,0.9);
\graph{(i1)--[plain,cyan,very thick] (v2)--[plain,brown,very thick] (i4)};
\graph{(i2)--[plain,cyan,very thick] (v1)--[plain,brown,very thick] (i3)};
\graph{(v1)--[plain,cyan,very thick] (v3)--[plain,red,very thick] (v2)};
\draw[very thick] plot[mark=x,mark size=2.7] coordinates {(0.9,0.9)};
\draw plot[mark=x,mark size=2.7,mark options={rotate=45}] coordinates {(0.9+0.9*0.3,0.6-0.6*0.3)};
\end{feynhand} \end{tikzpicture}.
\end{equation}
After the gluing process, the helicity mass insertion of the particles in each 3-pt sub-amplitude, labeled by $P_{14}$, transforms into a chirality mass insertion for the internal particle in the 4-pt factorized amplitude.
The numerator can be read out
\begin{equation}
[1\eta_{14}]\times[P_{14}3]\langle\eta_3 2\rangle-[1P_{14}]\times[\eta_{14} 3]\langle\eta_3 2\rangle=-\tilde{m}_{14}[13]\langle\eta_3 2\rangle,
\end{equation}
where $\tilde{m}_{14}$ is the spurion mass of the internal fermion. Similarly, the $s_{13}$ channel gives $m_{13} [13]\langle\eta_3 2\rangle$. Adding back the denominators, the result takes the form
\begin{equation} 
\begin{aligned} \label{eq:FFVS-s13-s14}
\varepsilon_\eta^2 \left(c_1\frac{m_{13}[13]\langle\eta_3 2\rangle}{s_{13}}+c_2\frac{\tilde{m}_{14}[13]\langle\eta_3 2\rangle}{s_{14}}\right),
\end{aligned} 
\end{equation}
where $\varepsilon_\eta^2$ denotes that there are two mass insertions in the diagrams. The coefficients $c_1$ and $c_2$ are equal to a product of the ones from two 3-pt sub-amplitudes.

Next, we consider the $s_{12}$ channel, where the internal particle is a spin-1 boson. Given renormalizable interactions, the transversality of the internal particle is $(0,0)$. This can lead to contributions in three different orders, corresponding to zero, two, and four mass insertions:
\bea
&& \begin{tikzpicture}[baseline=0.7cm] \begin{feynhand}
\vertex [dot] (v1) at (0,0.8) {};
\vertex [dot] (v2) at (2,0.8) {};
\node at (0.7,1) {\tiny $0$};
\node at (1.35,1) {\tiny $0$};
\graph{(v2)--[plain,slash={[style=black]0.5},brown,very thick] (v1)};
\end{feynhand} \end{tikzpicture} 
\sim \frac{1}{2} \tilde{\lambda}_{\dot{\alpha}} \lambda_{\alpha} \tilde{\lambda}_{\dot{\beta}} \lambda_{\dot{\beta}} = \frac{1}{2} p_{\alpha \dot{\alpha}} p_{\beta\dot{\beta}} ,\label{eq:InternalVector0} \\
&& \begin{tikzpicture}[baseline=0.7cm] \begin{feynhand}
\vertex [dot] (v1) at (0,0.8) {};
\vertex [dot] (v2) at (2,0.8) {};
\node at (0.7,1) {\tiny $-1$};
\node at (1.35,1) {\tiny $+1$};
\graph{(v2)--[plain,slash={[style=black]0.5},brown,very thick] (v1)};
\draw plot[mark=x,mark size=2.7] coordinates {(0.3,0.8)};
\draw plot[mark=x,mark size=2.7] coordinates {(1.7,0.8)};
\end{feynhand} \end{tikzpicture}+
\begin{tikzpicture}[baseline=0.7cm] \begin{feynhand}
\vertex [dot] (v1) at (0,0.8) {};
\vertex [dot] (v2) at (2,0.8) {};
\node at (0.7,1) {\tiny $+1$};
\node at (1.35,1) {\tiny $-1$};
\graph{(v2)--[plain,slash={[style=black]0.5},brown,very thick] (v1)};
\draw plot[mark=x,mark size=2.7] coordinates {(0.3,0.8)};
\draw plot[mark=x,mark size=2.7] coordinates {(1.7,0.8)};
\end{feynhand} \end{tikzpicture}+
\begin{tikzpicture}[baseline=0.7cm] \begin{feynhand}
\vertex [dot] (v1) at (0,0.8) {};
\vertex [dot] (v2) at (2,0.8) {};
\node at (0.8,1) {\tiny $0$};
\node at (1.3,1) {\tiny $0$};
\graph{(v2)--[plain,slash={[style=black]0.5},brown,very thick] (v1)};
\draw plot[mark=x,mark size=2.7] coordinates {(0.3,0.8)};
\draw plot[mark=x,mark size=2.7] coordinates {(0.6,0.8)};
\end{feynhand} \end{tikzpicture}+
\begin{tikzpicture}[baseline=0.7cm] \begin{feynhand}
\vertex [dot] (v1) at (0,0.8) {};
\vertex [dot] (v2) at (2,0.8) {};
\node at (0.8,1) {\tiny $0$};
\node at (1.3,1) {\tiny $0$};
\graph{(v2)--[plain,slash={[style=black]0.5},brown,very thick] (v1)};
\draw plot[mark=x,mark size=2.7] coordinates {(1.7,0.8)};
\draw plot[mark=x,mark size=2.7] coordinates {(1.4,0.8)};
\end{feynhand} \end{tikzpicture} \nonumber \\
&\sim &  
-\tilde{\eta}_{\dot{\alpha}} \lambda_{\alpha} \tilde{\lambda}_{\dot{\beta}} \eta_{\beta}
-\tilde{\lambda}_{\dot{\alpha}} \eta_{\alpha} \tilde{\eta}_{\dot{\beta}} \lambda_{\beta}  
+\frac{1}{2} \tilde{\eta}_{\dot{\alpha}} \eta_{\alpha} \tilde{\lambda}_{\dot{\beta}} \lambda_{\beta}
+\frac{1}{2} \tilde{\lambda}_{\dot{\alpha}} \lambda_{\alpha} \tilde{\eta}_{\dot{\beta}} \eta_{\beta} \label{eq:InternalVector2} \\
&=& \mathbf{m}^2 \epsilon_{\alpha\beta} \epsilon_{\dot{\alpha}\dot{\beta}} 
-\frac{1}{2} p_{\alpha\dot{\alpha}} \eta_{\beta\dot{\beta}} 
-\frac{1}{2} \eta_{\alpha\dot{\alpha}} p_{\beta\dot{\beta}}, \nonumber \\
&& \begin{tikzpicture}[baseline=0.7cm] \begin{feynhand}
\vertex [dot] (v1) at (0,0.8) {};
\vertex [dot] (v2) at (2,0.8) {};
\node at (0.8,1) {\tiny $0$};
\node at (1.3,1) {\tiny $0$};
\graph{(v2)--[plain,slash={[style=black]0.5},brown,very thick] (v1)};
\draw plot[mark=x,mark size=2.7] coordinates {(0.3,0.8)};
\draw plot[mark=x,mark size=2.7] coordinates {(0.6,0.8)};
\draw plot[mark=x,mark size=2.7] coordinates {(1.7,0.8)};
\draw plot[mark=x,mark size=2.7] coordinates {(1.4,0.8)};
\end{feynhand} \end{tikzpicture}
\sim \frac{1}{2} \tilde{\eta}_{\dot{\alpha}} \eta_{\alpha} \tilde{\eta}_{\dot{\beta}} \eta_{\dot{\beta}} = \frac{1}{2} \eta_{\alpha \dot{\alpha}} \eta_{\beta\dot{\beta}}.\label{eq:InternalVector4}
\eea

Note that the contribution to $F\bar{F}V$ without mass insertion, given by eq.~\eqref{eq:InternalVector0}, vanishes due to the 3-particle kinematics. The diagram of the internal particle with two mass insertions given by eq.~\eqref{eq:InternalVector2} is
\begin{equation}
\begin{tikzpicture}[baseline=0.8cm] \begin{feynhand}
\vertex [particle] (i1) at (0,1.8) {$1^{+\frac{1}{2}}$};
\vertex [particle] (i2) at (0,0) {$2^{-\frac{1}{2}}$};
\vertex [particle] (i3) at (1.8,0) {$3^{+1}$};
\vertex [particle] (i4) at (1.8,1.8) {$4^0$};
\vertex (v1) at (0.6,0.9);
\vertex (v2) at (1.2,0.9);
\graph{(i1)--[plain,cyan,very thick] (v1)--[plain,cyan,very thick] (i2)};
\graph{(i4)--[plain,brown,very thick] (v2)--[plain,brown,very thick] (i3)};
\graph{(v1)--[plain,brown,slash={[style=black]0.5},very thick] (v2)};
\draw plot[mark=x,mark size=2.7] coordinates {(0.75,0.9)};
\draw plot[mark=x,mark size=2.7] coordinates {(1.05,0.9)};
\draw plot[mark=x,mark size=2.7,mark options={rotate=45}] coordinates {(1.2+0.6*0.35,0.9-0.9*0.35)};
\end{feynhand} \end{tikzpicture}+
\begin{tikzpicture}[baseline=0.8cm] \begin{feynhand}
\vertex [particle] (i1) at (0,1.8) {$1^{+\frac{1}{2}}$};
\vertex [particle] (i2) at (0,0) {$2^{-\frac{1}{2}}$};
\vertex [particle] (i3) at (1.8,0) {$3^{+1}$};
\vertex [particle] (i4) at (1.8,1.8) {$4^0$};
\vertex (v1) at (0.6,0.9);
\vertex (v2) at (1.2,0.9);
\graph{(i1)--[plain,cyan,very thick] (v1)--[plain,cyan,very thick] (i2)};
\graph{(i4)--[plain,brown,very thick] (v2)--[plain,brown,very thick] (i3)};
\graph{(v1)--[plain,brown,slash={[style=black]0.8},very thick] (v2)};
\draw plot[mark=x,mark size=2.7] coordinates {(0.7,0.9)};
\draw plot[mark=x,mark size=2.7] coordinates {(0.9,0.9)};
\draw plot[mark=x,mark size=2.7,mark options={rotate=45}] coordinates {(1.2+0.6*0.35,0.9-0.9*0.35)};
\end{feynhand} \end{tikzpicture} \Rightarrow
\begin{tikzpicture}[baseline=0.8cm] \begin{feynhand}
\vertex [particle] (i1) at (0,1.8) {$1^{+\frac{1}{2}}$};
\vertex [particle] (i2) at (0,0) {$2^{-\frac{1}{2}}$};
\vertex [particle] (i3) at (1.8,0) {$3^{+1}$};
\vertex [particle] (i4) at (1.8,1.8) {$4^0$};
\vertex (v1) at (0.6,0.9);
\vertex (v2) at (1.2,0.9);
\graph{(i1)--[plain,cyan,very thick] (v1)--[plain,cyan,very thick] (i2)};
\graph{(i4)--[plain,brown,very thick] (v2)--[plain,brown,very thick] (i3)};
\graph{(v1)--[plain,brown,very thick] (v2)};
\draw plot[mark=x,mark size=2.7] coordinates {(0.75,0.9)};
\draw plot[mark=x,mark size=2.7] coordinates {(1.05,0.9)};
\draw plot[mark=x,mark size=2.7,mark options={rotate=45}] coordinates {(1.2+0.6*0.35,0.9-0.9*0.35)};
\end{feynhand} \end{tikzpicture}.
\end{equation}
Thus the numerator can be read out
\begin{equation} \begin{aligned}
&-2\mathbf{m}^2_{12}[13]\langle2\eta_3\rangle+[1|\eta_{12}|2\rangle[3|P_{12}|\eta_3\rangle,
\end{aligned} \end{equation}
where $\mathbf{m}_{12}$ is the physical mass of the internal vector.

The diagram with four mass insertions in the internal particle, eq.~\eqref{eq:InternalVector4}, is
\begin{equation}
\begin{tikzpicture}[baseline=0.8cm] \begin{feynhand}
\vertex [particle] (i1) at (0,1.8) {$1^{+\frac{1}{2}}$};
\vertex [particle] (i2) at (0,0) {$2^{-\frac{1}{2}}$};
\vertex [particle] (i3) at (2.2,0) {$3^{+1}$};
\vertex [particle] (i4) at (2.2,1.8) {$4^0$};
\vertex (v1) at (0.6,0.9);
\vertex (v2) at (1.6,0.9);
\graph{(i1)--[plain,cyan,very thick] (v1)--[plain,cyan,very thick] (i2)};
\graph{(i4)--[plain,brown,very thick] (v2)--[plain,brown,very thick] (i3)};
\graph{(v1)--[plain,brown,slash={[style=black]0.5},very thick] (v2)};
\draw plot[mark=x,mark size=2.7] coordinates {(0.75,0.9)};
\draw plot[mark=x,mark size=2.7] coordinates {(0.95,0.9)};
\draw plot[mark=x,mark size=2.7] coordinates {(1.45,0.9)};
\draw plot[mark=x,mark size=2.7] coordinates {(1.25,0.9)};
\draw plot[mark=x,mark size=2.7,mark options={rotate=45}] coordinates {(1.6+0.6*0.35,0.9-0.9*0.35)};
\end{feynhand} \end{tikzpicture} \Rightarrow
\begin{tikzpicture}[baseline=0.8cm] \begin{feynhand}
\vertex [particle] (i1) at (0,1.8) {$1^{+\frac{1}{2}}$};
\vertex [particle] (i2) at (0,0) {$2^{-\frac{1}{2}}$};
\vertex [particle] (i3) at (2.2,0) {$3^{+1}$};
\vertex [particle] (i4) at (2.2,1.8) {$4^0$};
\vertex (v1) at (0.6,0.9);
\vertex (v2) at (1.6,0.9);
\graph{(i1)--[plain,cyan,very thick] (v1)--[plain,cyan,very thick] (i2)};
\graph{(i4)--[plain,brown,very thick] (v2)--[plain,brown,very thick] (i3)};
\graph{(v1)--[plain,brown,very thick] (v2)};
\draw plot[mark=x,mark size=2.7] coordinates {(0.75,0.9)};
\draw plot[mark=x,mark size=2.7] coordinates {(0.98,0.9)};
\draw plot[mark=x,mark size=2.7] coordinates {(1.45,0.9)};
\draw plot[mark=x,mark size=2.7] coordinates {(1.22,0.9)};
\draw plot[mark=x,mark size=2.7,mark options={rotate=45}] coordinates {(1.6+0.6*0.35,0.9-0.9*0.35)};
\end{feynhand} \end{tikzpicture}.
\end{equation}
The numerator takes the form
\begin{equation}
-[1|\eta_{12}|2\rangle[3|\eta_{12}|\eta_3\rangle.
\end{equation}
Adding back the denominator $s_{12}$ and summing over two kinds of contributions in this channel, we obtain
\begin{equation} \begin{aligned} \label{eq:FFVS-s12}
\varepsilon_\eta^3 c_3\frac{2\mathbf{m}^2_{12}[13]\langle2\eta_3\rangle-[1|\eta_{12}|2\rangle[3|P_{12}|\eta_3\rangle}{s_{12}}+\varepsilon_\eta^5 c_3 \frac{-[1|\eta_{12}|2\rangle[3|\eta_{12}|\eta_3\rangle}{s_{12}}.
\end{aligned} \end{equation}
In total, we obtain the helicity amplitudes in this transversality $(+\frac{1}{2},-\frac{1}{2},0,0)$:
\begin{equation} \begin{aligned} \label{eq:FFVS1}
\mathcal{M}_{(+\frac{1}{2},-\frac{1}{2},0,0)}
=&\varepsilon_\eta^2 \left(c_1\frac{m_{13}[13]\langle\eta_3 2\rangle}{s_{13}}+c_2\frac{\tilde{m}_{14}[13]\langle\eta_3 2\rangle}{s_{14}}\right)\\
&+\varepsilon_\eta^3 c_3\frac{2\mathbf{m}^2_{12}[13]\langle2\eta_3\rangle-[1|\eta_{12}|2\rangle[3|P_{12}|\eta_3\rangle}{s_{12}}\\
&+\varepsilon_\eta^5 c_3\frac{-[1|\eta_{12}|2\rangle[3|\eta_{12}|\eta_3\rangle}{s_{12}}. 
\end{aligned} \end{equation}

Similarly, we can calculate the amplitude with other transversality. In the factorized limit of the $s_{14}$ channel, we find that the transversality of the renormalizable amplitude should be $(+\frac{1}{2}, -\frac{1}{2},0,0)$. Therefore, we have: 
\begin{equation}
\begin{tikzpicture}[baseline=0.8cm] \begin{feynhand}
\vertex [particle] (i1) at (0,1.8) {$1^{+\frac{1}{2}}$};
\vertex [particle] (i2) at (0,0) {$2^{-\frac{1}{2}}$};
\vertex [particle] (i3) at (1.8,0) {$3^{+1}$};
\vertex [particle] (i4) at (1.8,1.8) {$4^0$};
\vertex (v1) at (0.9,0.6);
\vertex (v2) at (0.9,1.2);
\vertex (v3) at (0.9,0.9);
\graph{(i1)--[plain,cyan,very thick] (v2)--[plain,brown,very thick] (i4)};
\graph{(i2)--[plain,red,very thick] (v1)--[plain,brown,very thick] (i3)};
\graph{(v1)--[plain,red,very thick] (v3)--[plain,red,slash={[style=black]0},very thick] (v2)};
\draw plot[mark=x,mark size=2.7,mark options={rotate=45}] coordinates {(0.9+0.9*0.3,0.6-0.6*0.3)};
\draw plot[mark=x,mark size=2.7,mark options={rotate=45}] coordinates {(0.9-0.9*0.3,0.6-0.6*0.3)};
\end{feynhand} \end{tikzpicture} \Rightarrow
\begin{tikzpicture}[baseline=0.8cm] \begin{feynhand}
\vertex [particle] (i1) at (0,1.8) {$1^{+\frac{1}{2}}$};
\vertex [particle] (i2) at (0,0) {$2^{-\frac{1}{2}}$};
\vertex [particle] (i3) at (1.8,0) {$3^{+1}$};
\vertex [particle] (i4) at (1.8,1.8) {$4^0$};
\vertex (v1) at (0.9,0.6);
\vertex (v2) at (0.9,1.2);
\vertex (v3) at (0.9,0.9);
\graph{(i1)--[plain,cyan,very thick] (v2)--[plain,brown,very thick] (i4)};
\graph{(i2)--[plain,red,very thick] (v1)--[plain,brown,very thick] (i3)};
\graph{(v1)--[plain,red,very thick] (v3)--[plain,red,very thick] (v2)};
\draw plot[mark=x,mark size=2.7,mark options={rotate=45}] coordinates {(0.9+0.9*0.3,0.6-0.6*0.3)};
\draw plot[mark=x,mark size=2.7,mark options={rotate=45}] coordinates {(0.9-0.9*0.3,0.6-0.6*0.3)};
\end{feynhand} \end{tikzpicture}. 
\end{equation}
The numerator can be read out
\begin{equation}
[1P_{14}]\times\langle P_{14}\eta_3\rangle[3\eta_2]=[1|P_{14}|\eta_3\rangle[3\eta_2]. 
\end{equation}

There are higher-order mass insertions in this $s_{14}$ channel, which can be calculated similarly. The $s_{13}$ channel is also calculated in a similar manner, while the $s_{12}$ channel is forbidden by the transversality of particles 1 and 2. Therefore, the amplitude is simpler than eq.~\eqref{eq:FFVS1}.
We have
\begin{equation} \begin{aligned}
\mathcal{M}_{(+\frac{1}{2},+\frac{1}{2},0,0)}
=&\varepsilon_\eta^2 \left(c_4\frac{[\eta_2|P_{13}|\eta_3\rangle[31]}{s_{13}}+c_5\frac{[1|P_{14}|\eta_3\rangle[3\eta_2]}{s_{14}}\right)\\
&+\varepsilon_\eta^4 \left(c_4\frac{[\eta_2|\eta_{13}|\eta_3\rangle[31]}{s_{13}}+c_5\frac{[1|\eta_{14}|\eta_3\rangle[3\eta_2]}{s_{14}}\right). 
\end{aligned} \end{equation}

Similarly, we can derive the amplitudes in the other two transversality $(-\frac12,\pm\frac12,0,0)$. Summing over all transversality, we obtain the final result for this helicity category:
\begin{equation} \begin{aligned}
&\mathcal{M}^{\mathcal{H}=(+\frac{1}{2},-\frac{1}{2},+1,0)}_{F\bar{F}VS}\\
=&\mathcal{M}^\mathcal{H}_{(+\frac12,+\frac12,0,0)}+\mathcal{M}^\mathcal{H}_{(+\frac12,-\frac12,0,0)}+\mathcal{M}^\mathcal{H}_{(-\frac12,+\frac12,0,0)}+\mathcal{M}^\mathcal{H}_{(-\frac12,-\frac12,0,0)}\\
=&\varepsilon_\eta^2\left(c_1\frac{m_{13}[13]\langle\eta_32\rangle}{s_{13}}+c_2\frac{\tilde{m}_{14}[13]\langle\eta_32\rangle}{s_{14}}+ \bar{c}_1\frac{\tilde{m}_{13}\langle1\eta_3\rangle[32]}{s_{13}}+\bar{c}_2\frac{m_{14}\langle1\eta_3\rangle[32]}{s_{14}}\right.\\
&+\left.c_4\frac{[\eta_2|P_{13}|\eta_3\rangle[31]}{s_{13}}+c_5\frac{[1|P_{13}|\eta_3\rangle[3\eta_2]}{s_{14}}+\bar{c}_4\frac{\langle2|P_{13}|3]\langle\eta_3\eta_1\rangle}{s_{13}}+\bar{c}_5\frac{\langle\eta_1|P_{14}|3]\langle\eta_32\rangle}{s_{14}}\right)\\
&+\varepsilon_\eta^3 \left(c_3\frac{2\mathbf{m}^2_{12}[13]\langle2\eta_3\rangle-[1|\eta_{12}|2\rangle[3|p_4|\eta_3\rangle}{s_{12}}+\bar{c}_3\frac{-\langle\eta_1|P_{12}|\eta_2]\langle\eta_3|p_{4}|3]}{s_{12}}\right)\\
&+\varepsilon_\eta^4\left(c_4\frac{[\eta_2|\eta_{13}|\eta_3\rangle[31]}{s_{13}}+c_5\frac{[1|\eta_{13}|\eta_3\rangle[3\eta_2]}{s_{14}}+\bar{c}_4\frac{\langle2|P_{13}|3]\langle\eta_3\eta_1\rangle}{s_{13}}+\bar{c}_5\frac{\langle\eta_1|P_{14}|3]\langle\eta_32\rangle}{s_{14}}\right)\\
&+\varepsilon_\eta^5 \left(c_3\frac{-[1|\eta_{12}|2\rangle[3|\eta_{12}|\eta_3\rangle}{s_{12}}+\bar{c}_3\frac{2\mathbf{m}^2_{12}\langle\eta_1\eta_3\rangle[\eta_23]-\langle\eta_1|P_{12}|\eta_2]\langle\eta_3|\eta_{12}|3]}{s_{12}}\right),\\
\end{aligned} \end{equation}
where $\bar{c}_i$ are the coefficients from transversality $(-\frac12,\pm\frac12,0,0)$.

\subsection{Mass insertion for internal on-shell particles}\label{sec:massInsert}

For all-massive amplitudes, the kinematic structure of internal particles is determined by the extended Poincar\'e symmetry $ISO(2) \times ISO(3,1)$. In this sense, the internal off-shell particles can be effectively viewed as on-shell without additional transformation, because the on-shell and off-shell massive one-pt states share the same form, although they have different degrees of freedom. In this subsection, we want to summarize the calculation for internal particles, and give a systematical study based on the point view of group theory. Group theory suggests that the spin-$s$ propagator of a scattering amplitude serves as a projector of $SO(2) \times SO(3,1)$, and is LG invariant, 
\eq{
& \sum_{h=-s}^{s} |\mathbf{P},m,\tilde{m},s,h\rangle \langle \mathbf{P},m,\tilde{m},s,-h| \\
\sim& \tilde{\lambda}_{\dot{\alpha}_1}^{(I_1} \cdots \tilde{\lambda}_{\dot{\alpha}_{s+t}}^{I_{s+t}} \lambda_{\alpha_1}^{I_{s+t+1}}\cdots \lambda_{\alpha_{s-t}}^{I_{2s})}
\tilde{\lambda}_{\dot{\beta}_1 I_1} \cdots \tilde{\lambda}_{\dot{\beta}_{s+t'} I_{s+t'}} \lambda_{\beta_1 I_{s+t'+1}} \cdots \lambda_{\beta_{s-t'} I_{2s}},
}
where the contraction of LG indices indicates a sum over helicities of the spin-$s$ internal particle. The gluing technique of the propagator proposed in eq.~\eqref{eq:FactorizeMassless} is equivalent to summing over the inner product of two helicity-conjugated particle states.

\subsubsection{Internal on-shell Fermion}\label{sec:FerPropag}

When we glue two sub-amplitudes, two of their external particles will become an internal one, in which the helicity of these two external particles should be opposite, a property called helicity conjugation. For convenience, let us consider the representations without chirality flip. For spin-$\frac{1}{2}$ particle, we have four ways to glue the fermion and anti-fermion as
\begin{equation}\label{eq:gluefer}
\sum_{h_P=\pm\frac{1}{2}} M^L\times M^R\Rightarrow
\begin{pmatrix}
\begin{tikzpicture}[baseline=0.7cm] \begin{feynhand}
\vertex [dot] (v1) at (0,0.8) {};
\vertex [dot] (v2) at (2,0.8) {};
\vertex (v3) at (1,0.8);
\node at (0.7,1) {\tiny $-\frac{1}{2}$};
\node at (1.35,1) {\tiny $+\frac{1}{2}$};
\graph{(v1)--[plain,red,very thick](v3)--[plain,cyan,slash={[style=black]0},very thick] (v2)};
\draw plot[mark=x,mark size=2.7] coordinates {(1.7,0.8)};
\end{feynhand} \end{tikzpicture}+
\begin{tikzpicture}[baseline=0.7cm] \begin{feynhand}
\vertex [dot] (v1) at (0,0.8) {};
\vertex [dot] (v2) at (2,0.8) {};
\vertex (v3) at (1,0.8);
\node at (0.7,1) {\tiny $+\frac{1}{2}$};
\node at (1.35,1) {\tiny $-\frac{1}{2}$};
\graph{(v1)--[plain,red,very thick](v3)--[plain,cyan,slash={[style=black]0},very thick] (v2)};
\draw plot[mark=x,mark size=2.7] coordinates {(0.3,0.8)};
\draw plot coordinates {(2.5,0.8)};
\end{feynhand} \end{tikzpicture} &
\begin{tikzpicture}[baseline=0.7cm] \begin{feynhand}
\vertex [dot] (v1) at (0,0.8) {};
\vertex [dot] (v2) at (2,0.8) {};
\vertex (v3) at (1,0.8);
\node at (0.7,1) {\tiny $+\frac{1}{2}$};
\node at (1.35,1) {\tiny $-\frac{1}{2}$};
\graph{(v2)--[plain,cyan,very thick] (v3)--[plain,slash={[style=black]0},cyan,very thick] (v1)};
\end{feynhand} \end{tikzpicture}+
\begin{tikzpicture}[baseline=0.7cm] \begin{feynhand}
\vertex [dot] (v1) at (0,0.8) {};
\vertex [dot] (v2) at (2,0.8) {};
\vertex (v3) at (1,0.8);
\node at (0.7,1) {\tiny $-\frac{1}{2}$};
\node at (1.35,1) {\tiny $+\frac{1}{2}$};
\graph{(v2)--[plain,cyan,very thick] (v3)--[plain,slash={[style=black]0},cyan,very thick] (v1)};
\draw plot[mark=x,mark size=2.7] coordinates {(0.3,0.8)};
\draw plot[mark=x,mark size=2.7] coordinates {(1.7,0.8)};
\end{feynhand} \end{tikzpicture} \\
\begin{tikzpicture}[baseline=0.7cm] \begin{feynhand}
\vertex [dot] (v1) at (0,0.8) {};
\vertex [dot] (v2) at (2,0.8) {};
\vertex (v3) at (1,0.8);
\node at (0.7,1) {\tiny $-\frac{1}{2}$};
\node at (1.35,1) {\tiny $+\frac{1}{2}$};
\graph{(v2)--[plain,red,very thick] (v3)--[plain,slash={[style=black]0},red,very thick] (v1)};
\end{feynhand} \end{tikzpicture}+
\begin{tikzpicture}[baseline=0.7cm] \begin{feynhand}
\vertex [dot] (v1) at (0,0.8) {};
\vertex [dot] (v2) at (2,0.8) {};
\vertex (v3) at (1,0.8);
\node at (0.7,1) {\tiny $+\frac{1}{2}$};
\node at (1.35,1) {\tiny $-\frac{1}{2}$};
\graph{(v2)--[plain,red,very thick] (v3)--[plain,slash={[style=black]0},red,very thick] (v1)};
\draw plot[mark=x,mark size=2.7] coordinates {(0.3,0.8)};
\draw plot[mark=x,mark size=2.7] coordinates {(1.7,0.8)};
\draw plot coordinates {(2.5,0.8)};
\end{feynhand} \end{tikzpicture} &
\begin{tikzpicture}[baseline=0.7cm] \begin{feynhand}
\vertex [dot] (v1) at (0,0.8) {};
\vertex [dot] (v2) at (2,0.8) {};
\vertex (v3) at (1,0.8);
\node at (0.7,1) {\tiny $+\frac{1}{2}$};
\node at (1.35,1) {\tiny $-\frac{1}{2}$};
\graph{(v1)--[plain,cyan,very thick](v3)--[plain,red,slash={[style=black]0},very thick] (v2)};
\draw plot[mark=x,mark size=2.7] coordinates {(1.7,0.8)};
\end{feynhand} \end{tikzpicture}+
\begin{tikzpicture}[baseline=0.7cm] \begin{feynhand}
\vertex [dot] (v1) at (0,0.8) {};
\vertex [dot] (v2) at (2,0.8) {};
\vertex (v3) at (1,0.8);
\node at (0.7,1) {\tiny $-\frac{1}{2}$};
\node at (1.35,1) {\tiny $+\frac{1}{2}$};
\graph{(v1)--[plain,cyan,very thick](v3)--[plain,red,slash={[style=black]0},very thick] (v2)};
\draw plot[mark=x,mark size=2.7] coordinates {(0.3,0.8)};
\end{feynhand} \end{tikzpicture} \\
\end{pmatrix}.
\end{equation}
In each diagram, the left side corresponds to a fermion while the right side corresponds to an anti-fermion.

After some calculation, we find that the structures of the internal fermion are
\begin{equation}
\begin{pmatrix}
m\delta_\alpha^\beta & p_{\dot{\alpha}}^\beta+\eta_{\dot{\alpha}}^\beta \\
p_{\alpha}^{\dot{\beta}}+\eta_{\alpha}^{\dot{\beta}} & \tilde{m}\delta_{\dot{\alpha}}^{\dot{\beta}} \\
\end{pmatrix},
\end{equation}
where the lower and upper indices correspond to the fermion and anti-fermion, respectively. This is the on-shell expression, but we can generalize it to the off-shell case by requiring $p_{\alpha \dot{\alpha}} \eta^{\dot{\alpha} \alpha} \neq m \tilde{m}$. Now, it becomes the completeness relation for the massive fermion.

This gluing result can be viewed as a single-particle state with $(\Delta,J_1,J_2)=(1,0,0)$, so we can use the representation defined in subsection~\ref{sec:littlegroup} to directly construct it. Note that the internal particle is a spin-$0$ state, so the helicity flip caused by $J^\pm$ annihilates the state. In this case, the highest weight construction of $J^{\pm}$ does not work. 

To construct all the possible internal fermions, we start from the wave function with maximal transversality. Since the two sides of the internal particle correspond to particle and anti-particle, we label the spinor indices explicitly to distinguish them. So we have
\bea
(t=+1):\quad
\begin{tikzpicture}[baseline=0.7cm] \begin{feynhand}
\vertex [dot] (i1) at (1.6,0.8) {};
\vertex [dot] (v1) at (0,0.8) {};
\vertex (v2) at (0.8,0.8);
\graph{(v1)--[plain,cyan,very thick] (v2)--[plain,red,very thick] (i1)};
\draw[very thick] plot[mark=x,mark size=2.7] coordinates {(0.8,0.8)};
\end{feynhand} \end{tikzpicture}
=\tilde{m} \delta_{\dot{\alpha}}^{\dot{\beta}}. \label{eq:ferpropag+1}
\eea
Then we perform transversality flip $T^-$ to obtain other components
\begin{equation}\label{eq:tflip}
\includegraphics[width=0.3\linewidth,valign=c]{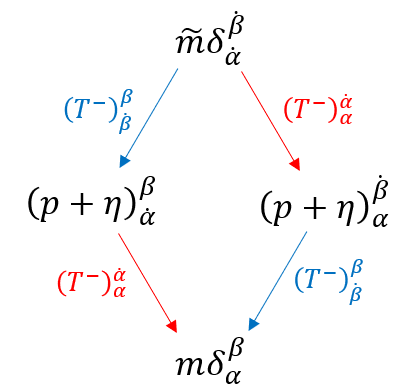}
\end{equation}

Now we have the internal line with all transversality in $\Delta=1$ case. Diagrammatically, they are
\bea
(t=+1) && \begin{tikzpicture}[baseline=0.7cm] \begin{feynhand}
\vertex [dot] (i1) at (1.6,0.8) {};
\vertex [dot] (v1) at (0,0.8) {};
\vertex (v2) at (0.8,0.8);
\graph{(v1)--[plain,cyan,very thick] (v2)--[plain,red,very thick] (i1)};
\draw[very thick] plot[mark=x,mark size=2.7] coordinates {(0.8,0.8)};
\end{feynhand} \end{tikzpicture} \sim \tilde{m} \delta_{\dot{\alpha}}^{\dot{\beta}} ,\label{eq:ferpropag01} \\
(t=0) && \begin{tikzpicture}[baseline=0.7cm] \begin{feynhand}
\vertex [dot] (v1) at (0,0.8) {};
\vertex [dot] (v2) at (1.6,0.8) {};
\graph{(v2)--[plain,red,very thick](v1)};
\end{feynhand} \end{tikzpicture} \sim -p_{\alpha}^{\dot{\beta}},\quad
\begin{tikzpicture}[baseline=0.7cm] \begin{feynhand}
\vertex [dot] (v1) at (0,0.8) {};
\vertex [dot] (v2) at (1.6,0.8) {};
\graph{(v2)--[plain,red,very thick](v1)};
\draw plot[mark=x,mark size=2.7] coordinates {(0.48,0.8)};
\draw plot[mark=x,mark size=2.7] coordinates {(1.12,0.8)};
\end{feynhand} \end{tikzpicture} \sim -\eta_{\alpha}^{\dot{\beta}},\label{eq:ferpropag02}\\
(t=-1) && \begin{tikzpicture}[baseline=0.7cm] \begin{feynhand}
\vertex [dot] (i1) at (1.6,0.8) {};
\vertex [dot] (v1) at (0,0.8) {};
\vertex (v2) at (0.8,0.8);
\graph{(v1)--[plain,red,very thick] (v2)--[plain,cyan,very thick] (i1)};
\draw[very thick] plot[mark=x,mark size=2.7] coordinates {(0.8,0.8)};
\end{feynhand} \end{tikzpicture} \sim m \delta_{\alpha}^\beta.
\label{eq:ferpropag-1}
\eea

Equivalently, the fermion propagator of $\Delta=1$ can also be interpreted by the Young diagram formulation proposed in appendix~\ref{app:Young}. Take $t = +1$ as an example, whose result consists of two terms of helicity-conjugated states,
\eq{
\begin{tikzpicture}[baseline=0.7cm] \begin{feynhand}
\vertex [dot] (i1) at (1.6,0.8) {};
\vertex [dot] (v1) at (0,0.8) {};
\vertex (v2) at (0.8,0.8);
\graph{(v1)--[plain,cyan,very thick] (v2)--[plain,red,very thick] (i1)};
\draw[very thick] plot[mark=x,mark size=2.7] coordinates {(0.8,0.8)};
\end{feynhand} \end{tikzpicture}
\sim \ {\color{orange} \yng(1,1)} = {\color{orange} \young(+)} \odot {\color{orange} \young(-)} +{\color{orange} \young(-)} \odot {\color{orange} \young(+)} \ ,
}
where $\odot$ denotes the inner product of the two states.
Each term corresponds to a gluing diagram,
\eq{
\begin{tikzpicture}[baseline=0.7cm] \begin{feynhand}
\vertex [dot] (v1) at (0,0.8) {};
\vertex [dot] (v2) at (2,0.8) {};
\vertex (v3) at (1,0.8);
\node at (0.7,1) {\tiny $-\frac{1}{2}$};
\node at (1.35,1) {\tiny $+\frac{1}{2}$};
\graph{(v1)--[plain,cyan,very thick](v3)--[plain,red,slash={[style=black]0},very thick] (v2)};
\draw plot[mark=x,mark size=2.7] coordinates {(0.3,0.8)};
\end{feynhand} \end{tikzpicture}
\sim& \ {\color{orange} \young(-)} \odot {\color{orange} \young(+)} \sim 
\epsilon_{IJ} \tilde{\eta}_{\dot{\alpha}} \zeta^{-I} \tilde{\lambda}_{\dot{\beta}} \zeta^{+J} = -\tilde{\eta}_{\dot{\alpha}} \tilde{\lambda}_{\dot{\beta}}, \\
\begin{tikzpicture}[baseline=0.7cm] \begin{feynhand}
\vertex [dot] (v1) at (0,0.8) {};
\vertex [dot] (v2) at (2,0.8) {};
\vertex (v3) at (1,0.8);
\node at (0.7,1) {\tiny $+\frac{1}{2}$};
\node at (1.35,1) {\tiny $-\frac{1}{2}$};
\graph{(v1)--[plain,cyan,very thick](v3)--[plain,red,slash={[style=black]0},very thick] (v2)};
\draw plot[mark=x,mark size=2.7] coordinates {(1.7,0.8)};
\end{feynhand} \end{tikzpicture}
\sim& \ {\color{orange} \young(+)} \odot {\color{orange} \young(-)} \sim \epsilon_{IJ} \tilde{\lambda}_{\dot{\alpha}} \zeta^{+I} \tilde{\eta}_{\dot{\beta}} \zeta^{-J}
= \tilde{\lambda}_{\dot{\alpha}} \tilde{\eta}_{\dot{\beta}}, 
}
where the inner product is about the contraction of $\zeta^{\pm I}$.

Similar to eqs.~(\ref{eq:ferpropag+1}-\ref{eq:ferpropag-1}), the Young diagram also shows that there are three types of fermion propagators, categorized according to their respective transversality, 
\bea
(t=+1) && {\color{orange} \yng(1,1)} = {\color{orange} \yng(1)} \odot {\color{orange} \yng(1)} \sim \tilde{\lambda}_{\dot{\alpha}}^{I} \tilde{\lambda}_{\dot{\beta} I} = -\tilde{m} \epsilon_{\dot{\alpha} \dot{\beta}}, \\
(t=0) && 
\begin{cases}
\arraycolsep=0pt\def\arraystretch{0} \begin{array}{c}
{\color{orange} \yng(1)} \\
\yng(1)
\end{array} = {\color{orange} \yng(1)} \odot {\yng(1)} 
\sim -\mathbf{p}_{\beta\dot{\alpha}}, \\ \\
\arraycolsep=0pt\def\arraystretch{0} \begin{array}{c}
\yng(1) \\ 
{\color{orange} \yng(1)}
\end{array} = {\color{orange} \yng(1)} \odot {\yng(1)} 
\sim \mathbf{p}_{\alpha\dot{\beta}},
\end{cases} \\
(t=-1) && \yng(1,1) = m \epsilon_{\alpha\beta} .
\eea

We point out that the power counting of the internal fermion is
\begin{equation}
\begin{tabular}{c|ccc}
Transversality & $1$ & $\varepsilon_\eta$ & $\varepsilon_\eta^2$ \\
\hline
$+1$ & - & $\tilde{m}\delta_{\dot{\alpha}}^{\dot{\beta}}$ & - \\
\hline
\multirow{2}{*}{$0$} & $p_{\dot{\alpha}}^{\beta}$ & - & $\eta_{\dot{\alpha}}^{\beta}$ \\
& $p_{\alpha}^{\dot{\beta}}$ & - & $\eta_{\alpha}^{\dot{\beta}}$ \\
\hline
$-1$ & - & $m\delta_{\alpha}^{\beta}$ & - \\
\end{tabular}\label{eq:FerPropag}
\end{equation}

\subsubsection{Internal on-shell vector}

For spin-1 particle, we can consider the single-particle state with $(\Delta,J_1,J_2)=(2,0,0)$. Similarly, we begin with the component with maximal transversality,
\bea
\begin{tikzpicture}[baseline=0.7cm] \begin{feynhand}
\vertex [dot] (i1) at (1.6,0.8) {};
\vertex [dot] (v1) at (0,0.8) {};
\vertex (v2) at (0.53,0.8);
\vertex (v3) at (1.07,0.8);
\graph{(v1)--[plain,red,very thick] (v2)--[plain,brown,very thick] (v3)--[plain,cyan,very thick] (i1)};
\draw[very thick] plot[mark=x,mark size=2.7] coordinates {(v2)};
\draw[very thick] plot[mark=x,mark size=2.7] coordinates {(v3)};
\end{feynhand} \end{tikzpicture}
=& \tilde{m}^2 \epsilon_{(\dot{\alpha}_1}^{\dot{\beta}_1} \delta_{\dot{\alpha}_2)}^{\dot{\beta}_2},
\eea
where $(\cdots)$ still stands for the symmetrization of spinor indices.

Then we perform the transversality flip $T^-$ to derive other components
\begin{equation}
\includegraphics[width=0.8\linewidth,valign=c]{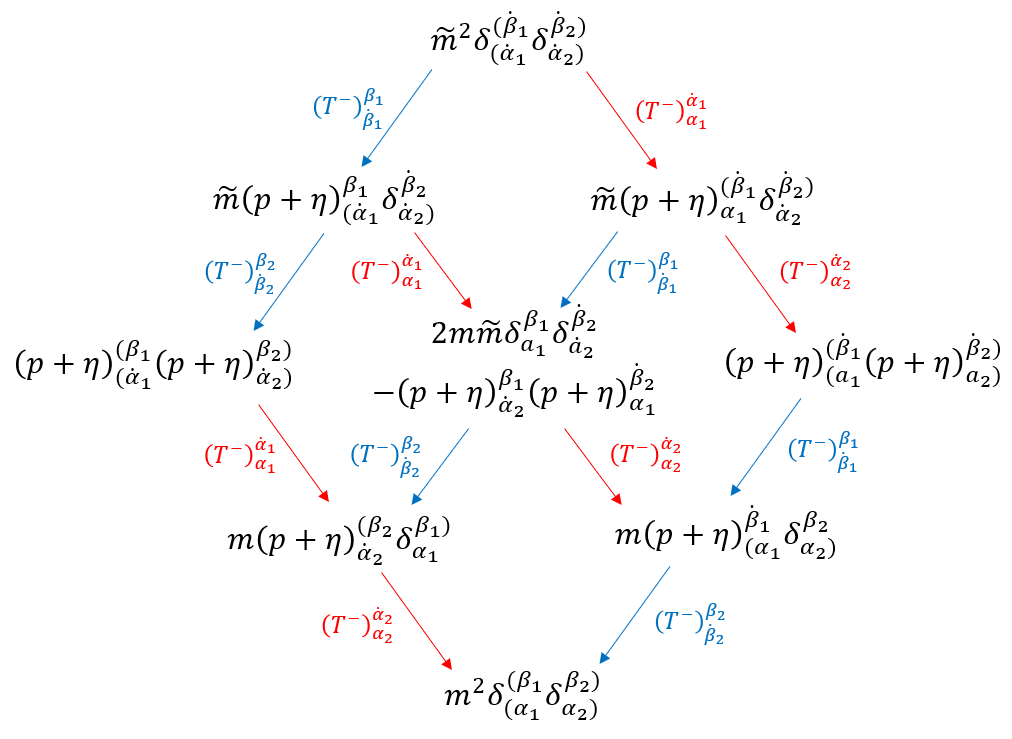}\label{eq:vecpropag}
\end{equation}
The $(2,0,0)$ representation can also be interpreted by the Young diagram formulation, with 
\eq{
\includegraphics[width=0.7\linewidth]{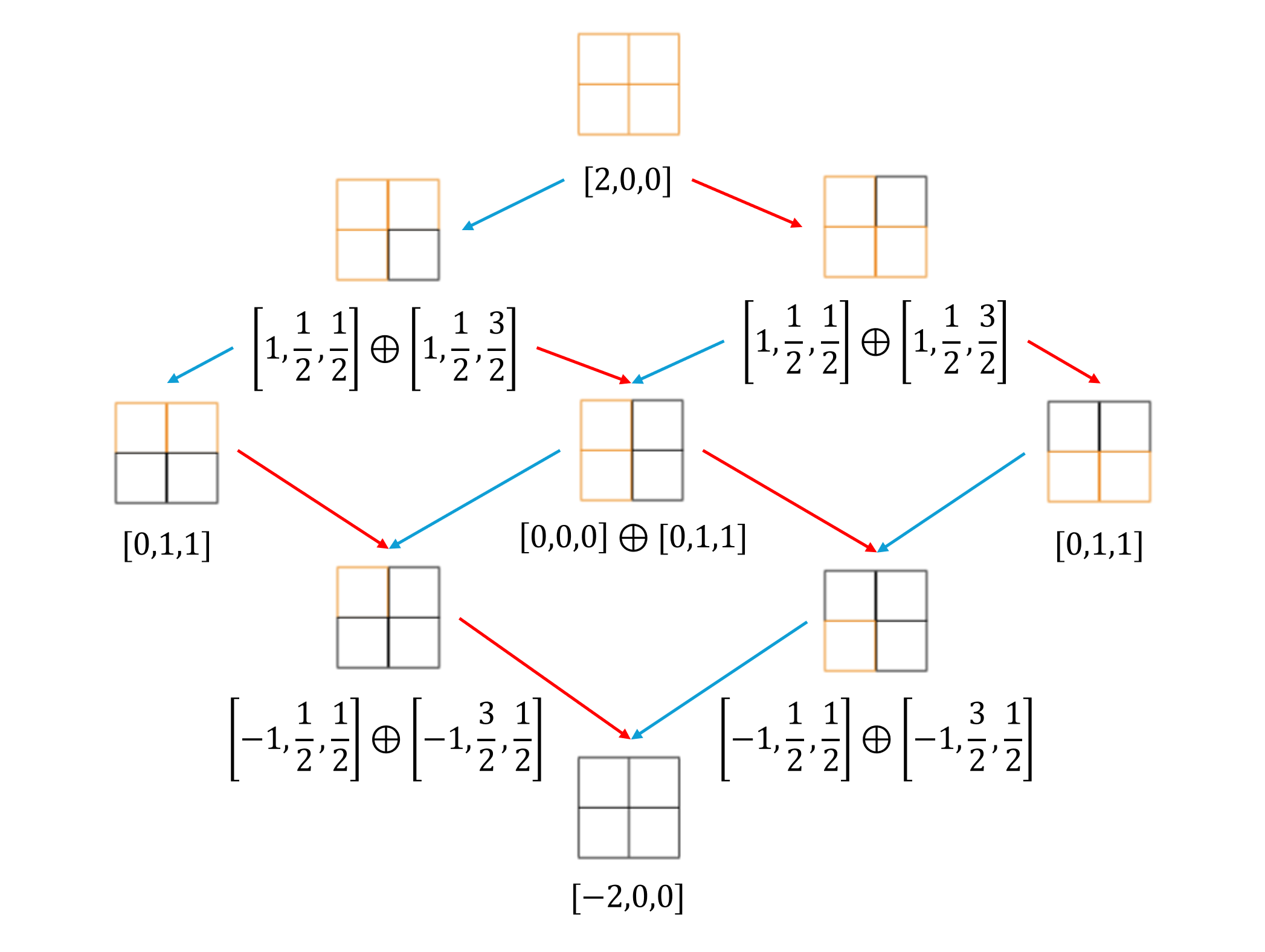}\label{eq:vecpropagYD}
}
where the red line represents the $T^{-}$ operator of the first row, the blue line represents the $T^{-}$ operator of the second row, and the $SO(2) \times SO(3,1)$ representations are listed under every Young diagram.
eq.~\eqref{eq:vecpropag} and eq.~\eqref{eq:vecpropagYD} also have a one-to-one correspondence. 

The complete result of the spin-1 particle in the form of the gluing diagram is
\begin{equation} \begin{aligned} \tiny
\begin{pmatrix}
\begin{tikzpicture}[baseline=0.7cm] \begin{feynhand}
\vertex [dot] (i1) at (1,0.8) {};
\vertex [dot] (v1) at (0,0.8) {};
\vertex (v2) at (0.333,0.8);
\vertex (v3) at (0.667,0.8);
\graph{(v1)--[plain,red,very thick] (v2)--[plain,brown,very thick] (v3)--[plain,cyan,very thick] (i1)};
\draw[very thick] plot[mark=x,mark size=2.7] coordinates {(v2)};
\draw[very thick] plot[mark=x,mark size=2.7] coordinates {(v3)};
\end{feynhand} \end{tikzpicture} &
\begin{tikzpicture}[baseline=0.7cm] \begin{feynhand}
\vertex [dot] (v1) at (0,0.8) {};
\vertex [dot] (v2) at (1,0.8) {};
\vertex (v3) at (0.5,0.8);
\graph{(v2)--[plain,cyan,very thick] (v3)--[plain,brown,very thick] (v1)};
\draw[very thick] plot[mark=x,mark size=2.7] coordinates {(v3)};
\end{feynhand} \end{tikzpicture}+ 
\begin{tikzpicture}[baseline=0.7cm] \begin{feynhand}
\vertex [dot] (v1) at (0,0.8) {};
\vertex [dot] (v2) at (1,0.8) {};
\vertex (v3) at (0.5,0.8);
\graph{(v2)--[plain,cyan,very thick] (v3)--[plain,brown,very thick] (v1)};
\draw[very thick] plot[mark=x,mark size=2.7] coordinates {(v3)};
\draw plot[mark=x,mark size=2.7] coordinates {(0.25,0.8)};
\draw plot[mark=x,mark size=2.7] coordinates {(0.75,0.8)};
\end{feynhand} \end{tikzpicture} &
\begin{tikzpicture}[baseline=0.7cm] \begin{feynhand}
\vertex [dot] (v1) at (0,0.8) {};
\vertex [dot] (v2) at (1,0.8) {};
\vertex (v3) at (0.5,0.8);
\graph{(v2)--[plain,cyan,very thick] (v3)--[plain,cyan,very thick] (v1)};
\end{feynhand} \end{tikzpicture}+
\begin{tikzpicture}[baseline=0.7cm] \begin{feynhand}
\vertex [dot] (v1) at (0,0.8) {};
\vertex [dot] (v2) at (1,0.8) {};
\vertex (v3) at (0.5,0.8);
\graph{(v2)--[plain,cyan,very thick] (v3)--[plain,cyan,very thick] (v1)};
\draw plot[mark=x,mark size=2.7] coordinates {(0.3,0.8)};
\draw plot[mark=x,mark size=2.7] coordinates {(0.7,0.8)};
\end{feynhand} \end{tikzpicture}+
\begin{tikzpicture}[baseline=0.7cm] \begin{feynhand}
\vertex [dot] (v1) at (0,0.8) {};
\vertex [dot] (v2) at (1,0.8) {};
\vertex (v3) at (0.5,0.8);
\graph{(v2)--[plain,cyan,very thick] (v3)--[plain,cyan,very thick] (v1)};
\draw plot[mark=x,mark size=2.7] coordinates {(0.2,0.8)};
\draw plot[mark=x,mark size=2.7] coordinates {(0.4,0.8)};
\draw plot[mark=x,mark size=2.7] coordinates {(0.6,0.8)};
\draw plot[mark=x,mark size=2.7] coordinates {(0.8,0.8)};
\end{feynhand} \end{tikzpicture} \\ && \\
\begin{tikzpicture}[baseline=0.7cm] \begin{feynhand}
\vertex [dot] (v1) at (0,0.8) {};
\vertex [dot] (v2) at (1,0.8) {};
\vertex (v3) at (0.5,0.8);
\graph{(v2)--[plain,brown,very thick] (v3)--[plain,red,very thick] (v1)};
\draw[very thick] plot[mark=x,mark size=2.7] coordinates {(v3)};
\end{feynhand} \end{tikzpicture}+ 
\begin{tikzpicture}[baseline=0.7cm] \begin{feynhand}
\vertex [dot] (v1) at (0,0.8) {};
\vertex [dot] (v2) at (1,0.8) {};
\vertex (v3) at (0.5,0.8);
\graph{(v2)--[plain,brown,very thick] (v3)--[plain,red,very thick] (v1)};
\draw[very thick] plot[mark=x,mark size=2.7] coordinates {(v3)};
\draw plot[mark=x,mark size=2.7] coordinates {(0.25,0.8)};
\draw plot[mark=x,mark size=2.7] coordinates {(0.75,0.8)};
\end{feynhand} \end{tikzpicture} & 
\begin{tikzpicture}[baseline=0.7cm] \begin{feynhand}
\vertex [dot] (v1) at (0,0.8) {};
\vertex [dot] (v2) at (1,0.8) {};
\graph{(v2)--[plain,brown,very thick] (v1)};
\end{feynhand} \end{tikzpicture}+
\begin{tikzpicture}[baseline=0.7cm] \begin{feynhand}
\vertex [dot] (v1) at (0,0.8) {};
\vertex [dot] (v2) at (1,0.8) {};
\graph{(v2)--[plain,brown,very thick] (v1)};
\draw plot[mark=x,mark size=2.7] coordinates {(0.333,0.8)};
\draw plot[mark=x,mark size=2.7] coordinates {(0.667,0.8)};
\end{feynhand} \end{tikzpicture}+
\begin{tikzpicture}[baseline=0.7cm] \begin{feynhand}
\vertex [dot] (v1) at (0,0.8) {};
\vertex [dot] (v2) at (1,0.8) {};
\graph{(v2)--[plain,brown,very thick] (v1)};
\draw plot[mark=x,mark size=2.7] coordinates {(0.2,0.8)};
\draw plot[mark=x,mark size=2.7] coordinates {(0.4,0.8)};
\draw plot[mark=x,mark size=2.7] coordinates {(0.6,0.8)};
\draw plot[mark=x,mark size=2.7] coordinates {(0.8,0.8)};
\end{feynhand} \end{tikzpicture} & 
\begin{tikzpicture}[baseline=0.7cm] \begin{feynhand}
\vertex [dot] (v1) at (0,0.8) {};
\vertex [dot] (v2) at (1,0.8) {};
\vertex (v3) at (0.5,0.8);
\graph{(v2)--[plain,brown,very thick] (v3)--[plain,cyan,very thick] (v1)};
\draw[very thick] plot[mark=x,mark size=2.7] coordinates {(v3)};
\end{feynhand} \end{tikzpicture}+ 
\begin{tikzpicture}[baseline=0.7cm] \begin{feynhand}
\vertex [dot] (v1) at (0,0.8) {};
\vertex [dot] (v2) at (1,0.8) {};
\vertex (v3) at (0.5,0.8);
\graph{(v2)--[plain,brown,very thick] (v3)--[plain,cyan,very thick] (v1)};
\draw[very thick] plot[mark=x,mark size=2.7] coordinates {(v3)};
\draw plot[mark=x,mark size=2.7] coordinates {(0.25,0.8)};
\draw plot[mark=x,mark size=2.7] coordinates {(0.75,0.8)};
\end{feynhand} \end{tikzpicture} \\ && \\
\begin{tikzpicture}[baseline=0.7cm] \begin{feynhand}
\vertex [dot] (v1) at (0,0.8) {};
\vertex [dot] (v2) at (1,0.8) {};
\vertex (v3) at (0.5,0.8);s
\graph{(v2)--[plain,red,very thick] (v3)--[plain,red,very thick] (v1)};
\end{feynhand} \end{tikzpicture}+
\begin{tikzpicture}[baseline=0.7cm] \begin{feynhand}
\vertex [dot] (v1) at (0,0.8) {};
\vertex [dot] (v2) at (1,0.8) {};
\vertex (v3) at (0.5,0.8);
\graph{(v2)--[plain,red,very thick] (v3)--[plain,red,very thick] (v1)};
\draw plot[mark=x,mark size=2.7] coordinates {(0.333,0.8)};
\draw plot[mark=x,mark size=2.7] coordinates {(0.667,0.8)};
\end{feynhand} \end{tikzpicture}+
\begin{tikzpicture}[baseline=0.7cm] \begin{feynhand}
\vertex [dot] (v1) at (0,0.8) {};
\vertex [dot] (v2) at (1,0.8) {};
\vertex (v3) at (0.5,0.8);
\graph{(v2)--[plain,red,very thick] (v3)--[plain,red,very thick] (v1)};
\draw plot[mark=x,mark size=2.7] coordinates {(0.2,0.8)};
\draw plot[mark=x,mark size=2.7] coordinates {(0.4,0.8)};
\draw plot[mark=x,mark size=2.7] coordinates {(0.6,0.8)};
\draw plot[mark=x,mark size=2.7] coordinates {(0.8,0.8)};
\end{feynhand} \end{tikzpicture} & 
\begin{tikzpicture}[baseline=0.7cm] \begin{feynhand}
\vertex [dot] (v1) at (0,0.8) {};
\vertex [dot] (v2) at (1,0.8) {};
\vertex (v3) at (0.5,0.8);
\graph{(v2)--[plain,red,very thick] (v3)--[plain,brown,very thick] (v1)};
\draw[very thick] plot[mark=x,mark size=2.7] coordinates {(v3)};
\end{feynhand} \end{tikzpicture}+ 
\begin{tikzpicture}[baseline=0.7cm] \begin{feynhand}
\vertex [dot] (v1) at (0,0.8) {};
\vertex [dot] (v2) at (1,0.8) {};
\vertex (v3) at (0.5,0.8);
\graph{(v2)--[plain,red,very thick] (v3)--[plain,brown,very thick] (v1)};
\draw[very thick] plot[mark=x,mark size=2.7] coordinates {(v3)};
\draw plot[mark=x,mark size=2.7] coordinates {(0.25,0.8)};
\draw plot[mark=x,mark size=2.7] coordinates {(0.75,0.8)};
\end{feynhand} \end{tikzpicture} & 
\begin{tikzpicture}[baseline=0.7cm] \begin{feynhand}
\vertex [dot] (i1) at (1,0.8) {};
\vertex [dot] (v1) at (0,0.8) {};
\vertex (v2) at (0.333,0.8);
\vertex (v3) at (0.667,0.8);
\graph{(v1)--[plain,cyan,very thick] (v2)--[plain,brown,very thick] (v3)--[plain,red,very thick] (i1)};
\draw[very thick] plot[mark=x,mark size=2.7] coordinates {(v2)};
\draw[very thick] plot[mark=x,mark size=2.7] coordinates {(v3)};
\end{feynhand} \end{tikzpicture} \\
\end{pmatrix},
\end{aligned} \end{equation}
with the corresponding expressions:
\begin{equation} \begin{aligned}
\begin{pmatrix}
m^2\delta_{(\alpha_1}^{(\beta_1}\delta_{\alpha_2)}^{\beta_2)} & m(p+\eta)_{\dot{\alpha}_2}^{(\beta_2}\delta_{\alpha_1}^{\beta_1)} & (p+\eta)_{(\dot{\alpha}_1}^{(\beta_1}(p+\eta)_{\dot{\alpha}_2)}^{\beta_2)} \\
m(p+\eta)_{(\alpha_1}^{\dot{\beta}_1}\delta_{\alpha_2)}^{\beta_2} & 2\mathbf{m}^2\delta_{\alpha_1}^{\beta_1}\delta_{\dot{\alpha}_2}^{\dot{\beta}_2}-(p+\eta)_{\dot{\alpha}_2}^{\beta_1}(p+\eta)_{\alpha_1}^{\dot{\beta}_2} & \tilde{m}(p+\eta)_{(\dot{\alpha}_1}^{\beta_1}\delta_{\dot{\alpha}_2)}^{\dot{\beta}_2} \\
(p+\eta)_{(\alpha_1}^{(\dot{\beta}_1}(p+\eta)_{\alpha_2)}^{\dot{\beta}_2)} & \tilde{m}(p+\eta)_{\alpha_1}^{(\dot{\beta}_1}\delta_{\dot{\alpha}_2}^{\dot{\beta}_2)} & \tilde{m}^2\delta_{(\dot{\alpha}_1}^{(\dot{\beta}_2}\delta_{\dot{\alpha}_2)}^{\dot{\beta}_2)} \\
\end{pmatrix}.
\end{aligned} \end{equation}

Finally, the power counting of the internal vector boson is 
\begin{equation}
\begin{tabular}{c|ccccc}
$t$ & $1$ & $\varepsilon_\eta$ & $\varepsilon_\eta^2$ & $\varepsilon_\eta^3$ & $\varepsilon_\eta^4$ \\
\hline
$+2$ & - & - & $\tilde{m}^2\delta_{(\dot{\alpha}_1}^{(\dot{\beta}_2}\delta_{\dot{\alpha}_2)}^{\dot{\beta}_2)}$ & - & - \\
\hline
\multirow{2}{*}{$+1$} & - & $\tilde{m}p_{(\dot{\alpha}_1}^{\beta_1}\delta_{\dot{\alpha}_2)}^{\dot{\beta}_2}$ & - & $\tilde{m}p_{(\dot{\alpha}_1}^{\beta_1}\delta_{\dot{\alpha}_2)}^{\dot{\beta}_2}$ & - \\
& - & $\tilde{m}\eta_{\alpha_1}^{(\dot{\beta}_1}\delta_{\dot{\alpha}_2}^{\dot{\beta}_2)}$ & - & $\tilde{m}\eta_{\alpha_1}^{(\dot{\beta}_1}\delta_{\dot{\alpha}_2}^{\dot{\beta}_2)}$ & - \\
\hline
\multirow{3}{*}{$0$}  & $p_{(\dot{\alpha}_1}^{(\beta_1}p_{\dot{\alpha}_2)}^{\beta_2)}$ & - & $2p_{(\dot{\alpha}_1}^{(\beta_1}\eta_{\dot{\alpha}_2)}^{\beta_2)}$ & - & $\eta_{(\dot{\alpha}_1}^{(\beta_1}\eta_{\dot{\alpha}_2)}^{\beta_2)}$ \\
& $p_{(\alpha_1}^{(\dot{\beta}_1}p_{\alpha_2)}^{\dot{\beta}_2)}$ & - & $2p_{(\alpha_1}^{(\dot{\beta}_1}\eta_{\alpha_2)}^{\dot{\beta}_2)}$ & - & $\eta_{(\alpha_1}^{(\dot{\beta}_1}\eta_{\alpha_2)}^{\dot{\beta}_2)}$ \\
\cline{2-6}
& $-p_{\dot{\alpha}_2}^{\beta_1}p_{\alpha_1}^{\dot{\beta}_2}$ & - & \makecell{$2\mathbf{m}^2\delta_{\alpha_1}^{\beta_1}\delta_{\dot{\alpha}_2}^{\dot{\beta}_2}$\\$-p_{\dot{\alpha}_2}^{\beta_1}\eta_{\alpha_1}^{\dot{\beta}_2}-\eta_{\dot{\alpha}_2}^{\beta_1}p_{\alpha_1}^{\dot{\beta}_2}$} & - & $-\eta_{\dot{\alpha}_2}^{\beta_1}\eta_{\alpha_1}^{\dot{\beta}_2}$ \\
\hline
\multirow{2}{*}{$-1$} & - & $\tilde{m}p_{(\dot{\alpha}_1}^{\beta_1}\delta_{\dot{\alpha}_2)}^{\dot{\beta}_2}$ & - & $\tilde{m}\eta_{(\dot{\alpha}_1}^{\beta_1}\delta_{\dot{\alpha}_2)}^{\dot{\beta}_2}$ & - \\
& - & $m p_{(\alpha_1}^{\dot{\beta}_1}\delta_{\alpha_2)}^{\beta_2}$ & - & $m\eta_{(\alpha_1}^{\dot{\beta}_1}\delta_{\alpha_2)}^{\beta_2}$ & - \\
\hline
$-2$ & - & - & $m^2\delta_{(\alpha_1}^{(\beta_1}\delta_{\alpha_2)}^{\beta_2)}$ & - & - \\
\end{tabular}
\end{equation}

\subsection{4-pt factorized amplitudes}


Given the rules on the internal particle in subsection above, let us reconsider the $F\bar{F}VS$ amplitude with helicity category $(+\frac{1}{2},-\frac{1}{2},+1,0)$ and transversality $(-\frac{1}{2},\pm\frac{1}{2},0,0)$. 

The massive amplitude can be written as a direct product of the single-particle states and internal structures, which include internal lines and vertices. For example, the $F\bar{F}V$ amplitude can be decomposed into
\begin{equation}
\begin{tikzpicture}[baseline=0.7cm] \begin{feynhand}
\label{eq:decompose3pt}
\setlength{\feynhandblobsize}{3.5mm}
\vertex [particle] (i1) at (-0.2,0.8) {$1^{-\frac{1}{2}}$};
\vertex [particle] (i2) at (1.6,1.6) {$2^{+\frac{1}{2}}$};
\vertex [particle] (i3) at (1.6,0) {$3^{+1}$};
\vertex [ringblob] (v1) at (0.9,0.8) {$\epsilon$};
\graph{(i1)--[plain,red,very thick] (v1)};
\graph{(i2)--[plain,red,very thick] (v1)};
\graph{(i3)--[plain,brown,very thick] (v1)};
\draw plot[mark=x,mark size=2.7,mark options={rotate=45}] coordinates {(0.9+0.7*0.38,0.8+0.8*0.38)};
\draw plot[mark=x,mark size=2.7,mark options={rotate=45}] coordinates {(0.9+0.7*0.38,0.8-0.8*0.38)};
\end{feynhand} \end{tikzpicture}
=|1\rangle^{\alpha} |\eta_2]^{\dot{\beta}} |3]^{\dot{\gamma}}|\eta_3\rangle^{\gamma}\times \epsilon_{\alpha\gamma}\epsilon_{\dot{\beta}\dot{\gamma}}. 
\end{equation}
Similarly, we can do the decomposition for 4-pt amplitude. Consider the amplitude with transversality $(-\frac{1}{2},+\frac{1}{2},0,0)$. It has the form
\begin{equation}
\begin{tikzpicture}[baseline=0.8cm] \begin{feynhand}
\setlength{\feynhandblobsize}{6mm}
\vertex [particle] (i1) at (0,1.8) {$1^{+\frac{1}{2}}$};
\vertex [particle] (i2) at (0,0) {$2^{-\frac{1}{2}}$};
\vertex [particle] (i3) at (1.8,0) {$3^{+1}$};
\vertex [particle] (i4) at (1.8,1.8) {$4^0$};
\vertex [ringblob] (v1) at (0.9,0.9) {$\mathcal{I}$};
\graph{(i1)--[plain,red,very thick] (v1)--[plain,brown,very thick] (i4)};
\graph{(i2)--[plain,red,very thick] (v1)--[plain,brown,very thick] (i3)};
\draw plot[mark=x,mark size=2.7,mark options={rotate=45}] coordinates {(1.35,0.45)};
\draw plot[mark=x,mark size=2.7,mark options={rotate=45}] coordinates {(0.45,1.35)};
\draw plot[mark=x,mark size=2.7,mark options={rotate=45}] coordinates {(0.45,0.45)};
\end{feynhand} \end{tikzpicture}
=|\eta_1\rangle^{\alpha} |\eta_2]^{\dot{\beta}} |3]^{\dot{\gamma}}|\eta_3\rangle^{\gamma}\times \mathcal{I}_{\alpha\dot{\beta}\dot{\gamma}\gamma},
\end{equation}
where $\mathcal{I}$ stands for the internal strucutre. 

Diagrammatically, the leading contributions in the $s_{12}$ and $s_{14}$ channels are
\begin{equation} \begin{aligned}
\begin{tikzpicture}[baseline=0.8cm] \begin{feynhand}
\vertex [particle] (i1) at (0,1.8) {$1^{+\frac{1}{2}}$};
\vertex [particle] (i2) at (0,0) {$2^{-\frac{1}{2}}$};
\vertex [particle] (i3) at (1.8,0) {$3^{+1}$};
\vertex [particle] (i4) at (1.8,1.8) {$4^0$};
\vertex (v1) at (0.9,0.6);
\vertex (v2) at (0.9,1.2);
\vertex (v3) at (0.9,0.9);
\graph{(i1)--[plain,red,very thick] (v2)--[plain,brown,very thick] (i4)};
\graph{(i2)--[plain,red,very thick] (v1)--[plain,brown,very thick] (i3)};
\graph{(v1)--[plain,red,very thick] (v3)--[plain,cyan,very thick] (v2)};
\draw[very thick] plot[mark=x,mark size=2.7] coordinates {(0.9,0.9)};
\draw plot[mark=x,mark size=2.7,mark options={rotate=45}] coordinates {(0.9+0.9*0.3,0.6-0.6*0.3)};
\draw plot[mark=x,mark size=2.7,mark options={rotate=45}] coordinates {(0.9-0.9*0.3,0.6-0.6*0.3)};
\draw plot[mark=x,mark size=2.7,mark options={rotate=45}] coordinates {(0.9-0.9*0.3,1.2+0.6*0.3)};
\end{feynhand} \end{tikzpicture}+
\begin{tikzpicture}[baseline=0.8cm] \begin{feynhand}
\vertex [particle] (i1) at (0,1.8) {$1^{+\frac{1}{2}}$};
\vertex [particle] (i2) at (0,0) {$2^{-\frac{1}{2}}$};
\vertex [particle] (i3) at (1.8,0) {$3^{+1}$};
\vertex [particle] (i4) at (1.8,1.8) {$4^0$};
\vertex (v1) at (0.6,0.9);
\vertex (v2) at (1.2,0.9);
\graph{(i1)--[plain,red,very thick] (v1)--[plain,red,very thick] (i2)};
\graph{(i4)--[plain,brown,very thick] (v2)--[plain,brown,very thick] (i3)};
\graph{(v1)--[plain,brown,very thick] (v2)};
\draw plot[mark=x,mark size=2.7] coordinates {(0.75,0.9)};
\draw plot[mark=x,mark size=2.7] coordinates {(1.05,0.9)};
\draw plot[mark=x,mark size=2.7,mark options={rotate=45}] coordinates {(1.2+0.6*0.35,0.9-0.9*0.35)};
\draw plot[mark=x,mark size=2.7,mark options={rotate=45}] coordinates {(0.6-0.6*0.35,0.9-0.9*0.35)};
\draw plot[mark=x,mark size=2.7,mark options={rotate=45}] coordinates {(0.6-0.6*0.35,0.9+0.9*0.35)};
\end{feynhand} \end{tikzpicture}.
\end{aligned} \end{equation}
The internal structure reads out to be
\bea
\mathcal{I}_{\alpha\dot{\beta}\dot{\gamma}\gamma}\supset\frac{m_{14}\epsilon_{\alpha\gamma}}{s_{14}}\times \epsilon_{\dot{\beta}\dot{\gamma}}+\frac{2\mathbf{m}^2_{12} \epsilon_{\alpha\gamma} \epsilon_{\dot{\beta}\dot{\gamma}} 
- P_{12,\alpha\dot{\beta}} \eta_{12,\gamma\dot{\gamma}} 
- \eta_{12,\alpha\dot{\beta}} P_{12,\gamma\dot{\gamma}}}{s_{12}}, 
\eea
where $\epsilon_{\dot{\beta}\dot{\gamma}}$ comes from the $F\bar{F}V$ vertex in eq.~\eqref{eq:decompose3pt}, and the other structures with pole come from the internal lines. The calculation for the $s_{13}$ channel is similar to the $s_{14}$ channel. 

The sub-leading contributions can also be calculated by this method. The final result of this transversality is 
\begin{equation} \begin{aligned}
\mathcal{M}_{(-\frac{1}{2},+\frac{1}{2},0,0)}
=&\varepsilon_\eta^3 \bar{c}_3\frac{-\langle\eta_1|P_{12}|\eta_2]\langle\eta_3|P_{12}|3]}{s_{12}}\\
&+\varepsilon_\eta^4 \left(\bar{c}_1\frac{\tilde{m}_{13}\langle\eta_1\eta_3\rangle[3 \eta_2]}{s_{13}}+\bar{c}_2\frac{m_{14}\langle\eta_1\eta_3\rangle[3 \eta_2]}{s_{14}}\right)\\
&+\varepsilon_\eta^5 \bar{c}_3\frac{2\mathbf{m}^2_{12}\langle\eta_1\eta_3\rangle[\eta_2 3]-\langle\eta_1|P_{12}|\eta_2]\langle\eta_3|\eta_{12}|3]}{s_{12}}.
\end{aligned} \end{equation}

The last transversality is $(-\frac{1}{2},-\frac{1}{2},0,0)$. The leading contribution of the $s_{14}$ channel is
\begin{equation}
\begin{tikzpicture}[baseline=0.8cm] \begin{feynhand}
\vertex [particle] (i1) at (0,1.8) {$1^{+\frac{1}{2}}$};
\vertex [particle] (i2) at (0,0) {$2^{-\frac{1}{2}}$};
\vertex [particle] (i3) at (1.8,0) {$3^{+1}$};
\vertex [particle] (i4) at (1.8,1.8) {$4^0$};
\vertex (v1) at (0.9,0.6);
\vertex (v2) at (0.9,1.2);
\vertex (v3) at (0.9,0.9);
\graph{(i1)--[plain,red,very thick] (v2)--[plain,brown,very thick] (i4)};
\graph{(i2)--[plain,cyan,very thick] (v1)--[plain,brown,very thick] (i3)};
\graph{(v1)--[plain,cyan,very thick] (v3)--[plain,cyan,very thick] (v2)};
\draw plot[mark=x,mark size=2.7,mark options={rotate=45}] coordinates {(0.9+0.9*0.3,0.6-0.6*0.3)};
\draw plot[mark=x,mark size=2.7,mark options={rotate=45}] coordinates {(0.9-0.9*0.3,1.2+0.6*0.3)};
\end{feynhand} \end{tikzpicture}
=|\eta_1\rangle_{\alpha} |2\rangle_{\beta} |3]_{\dot{\gamma}}|\eta_3\rangle_{\gamma}\times
\frac{P_{14,\alpha\dot{\gamma}}}{s_{14}}\times \epsilon_{\beta\gamma}.
\end{equation}
Similarly, we can derive the other contributions in this transversality. The final result is
\begin{equation} \begin{aligned}
\mathcal{M}_{(-\frac{1}{2},-\frac{1}{2},0,0)}
=&\varepsilon_\eta^2 \left(\bar{c}_4\frac{\langle2|P_{13}|3]\langle\eta_3\eta_1\rangle}{s_{13}}+\bar{c}_5\frac{\langle\eta_1|P_{14}|3]\langle\eta_3 2\rangle}{s_{14}}\right)\\
&+\varepsilon_\eta^4 \left(\bar{c}_4\frac{\langle2|\eta_{13}|3]\langle\eta_3\eta_1\rangle}{s_{13}}+\bar{c}_5\frac{\langle\eta_1|\eta_{14}|3]\langle\eta_3 2\rangle}{s_{14}}\right). 
\end{aligned} \end{equation}

Among the four transversality, we have calculated two, and the other two with opposite transversality can be derived from the reversal $t\rightarrow -t$. 
Since we do not impose the CP conservation in the amplitude, the coefficients will change (i.e. $c_i\leftrightarrow \bar{c}_i$). 

Summing over all amplitudes with different transversality, we obtain the final results for this helicity category: 
\begin{equation}
\mathcal{M}^{\mathcal{H}=(+\frac{1}{2},-\frac{1}{2},+1,0)}=\sum_{t_1,t_2=\pm\frac{1}{2}}\mathcal{M}^\mathcal{H}_{(t_1,t_2,0,0)}.
\end{equation}
Similar to 3-pt amplitudes, given the amplitudes for all the transversality in one helicity category, we can utilize the LG covariance to obtain the massive amplitude with bolded notation. 

Notice that the complete higher-point amplitudes would include the non-locality for internal particles, because we identify $P^2$ instead of $\mathbf{P}^2-\mathbf{m}^2$ as the inverse of the propagator. Actually, the factorized limit in eq.~\eqref{eq:FactorizeMassless} is equivalent to $\mathbf{P}^2\rightarrow\mathbf{m}^2$. Consider another factorized limit with relaxed constraint, we find the multi-pole structures as
\begin{equation}
\lim_{P^2\rightarrow 0} P^2 \mathcal{M}
=\sum_{h_P} \left(M^L\times M^R+M^L\times\frac{2P\cdot\eta-\mathbf{m}^2}{P^2}\times M^R+M^L\times\frac{\eta^2}{P^2}\times M^R+\cdots\right), 
\end{equation}
where $2P\cdot\eta-\mathbf{m}^2$ and $\eta^2$ can be viewed as 2-pt interaction. The terms with extra poles $\frac{1}{P^2}$ are non-local. 

Finally, the massive amplitudes at the IR can be obtained by recovering the LG covariance. Resumming the multi-pole structure for $P^2$, we will find the single-pole structure for $\mathbf{P}^2-\mathbf{m}^2$ as
\begin{equation}
\mathcal{M}\sim \frac{\sum_{h_P} M^L\times M^R}{(P+\eta)^2-\mathbf{m}^2}. 
\end{equation}
Then we recover the LG covariance of the numerator by performing $\lambda,\eta\rightarrow\lambda^I$ and $\tilde{\lambda},\tilde{\eta}\rightarrow\tilde{\lambda}^I$. The LG covariant $F\bar{F}VS$ amplitude is
\begin{equation} \begin{aligned}
\mathcal{M}
=&c_3\frac{-2\mathbf{m}^2_{12}[\mathbf{13}]\langle\mathbf{23}\rangle+[\mathbf{1}|\mathbf{P}_{12}|\mathbf{2}\rangle[\mathbf{3}|\mathbf{P}_{12}|\mathbf{3}\rangle}{\mathbf{s}_{12}-\mathbf{m}_{12}^2}\\
&+\bar{c}_3\frac{2\mathbf{m}^2_{12}\langle\mathbf{13}\rangle[\mathbf{23}]-\langle\mathbf{1}|\mathbf{P}_{12}|\mathbf{2}]\langle\mathbf{3}|\mathbf{P}_{12}|\mathbf{3}]}{\mathbf{s}_{12}-\mathbf{m}_{12}^2}\\
&-c_1\frac{\mathbf{m}_{13}[\mathbf{13}]\langle\mathbf{32}\rangle}{\mathbf{s}_{13}-\mathbf{m}_{13}^2}-c_2\frac{\mathbf{m}_{14}[\mathbf{13}]\langle\mathbf{32}\rangle}{\mathbf{s}_{14}-\mathbf{m}_{14}^2}
+\bar{c}_1\frac{\mathbf{m}_{13}\langle\mathbf{13}\rangle[\mathbf{32}]}{\mathbf{s}_{13}-\mathbf{m}_{13}^2}+\bar{c}_2\frac{\mathbf{m}_{14}\langle\mathbf{13}\rangle[\mathbf{32}]}{\mathbf{s}_{14}-\mathbf{m}_{14}^2}\\
&+c_4\frac{[\mathbf{2}|\mathbf{p}_{13}|\mathbf{3}\rangle[\mathbf{31}]}{\mathbf{s}_{13}-\mathbf{m}_{13}^2}+c_5\frac{[\mathbf{1}|\mathbf{p}_{14}|\mathbf{3}\rangle[\mathbf{32}]}{\mathbf{s}_{14}-\mathbf{m}_{14}^2}
-\bar{c}_4\frac{\langle\mathbf{2}|\mathbf{p}_{13}|\mathbf{3}]\langle\mathbf{31}\rangle}{\mathbf{s}_{13}-\mathbf{m}_{13}^2}-\bar{c}_5\frac{\langle\mathbf{2}|\mathbf{p}_{14}|\mathbf{3}]\langle\mathbf{31}\rangle}{\mathbf{s}_{14}-\mathbf{m}_{14}^2}. \\
\end{aligned} \end{equation}

\subsection{Massive QED amplitudes}

Now we consider the construction for higher-point massive amplitude with massless gauge bosons. In this case, we should glue the 1-massless-2-massive amplitude with equal mass, which includes spurious poles from $x$-factor. When the gauge boson is an internal particle, it cannot be viewed as a spin-$0$ momentum structure because of the spurious poles. Therefore, higher-point amplitude with gauge bosons cannot be read from the diagram. Since the 3-pt amplitudes do not have manifest locality, we can reduce them to the form in subsection~\ref{sec:generic3pt}. Based on the locality and power counting for higher-point amplitudes, we can deform the product of lower-point amplitudes to give the correct result.

\noindent \textbf{Scalar QED}

As a warm-up, let us consider a scattering process in the scalar QED, e.g. $\phi \phi^*\rightarrow\Phi \Phi^*$ with transversality $(0,0,0,0)$. This process has only one channel $s_{12}$, and the internal line is a photon. 

Since the building blocks in the massive QED are the 1-massless-2-massive amplitudes with equal mass, the constraint from 3-particle kinematics is stronger than the all-massive case. When we go to the factorized limit, the 4-particle kinematics will degenerate into two 3-particle kinematics, and some Lorentz structures may vanish in such kinematics. Basically, we need to calculate the amplitude with different kinematics to give the complete 
amplitudes. 

For 4-pt amplitudes, there are $2\times 2$ possible kinematics in the factorized limit, but two of them are enough to give the 4-fermion amplitude. We first choose
\begin{equation} \begin{aligned} \label{eq:kinematicQED1}
|1\rangle\propto|2\rangle\propto&|P_{12}\rangle\propto|\eta_3\rangle\propto|\eta_4\rangle, \\
|3]\propto|4]\propto&|P_{12}]\propto|\eta_1]\propto|\eta_2]. 
\end{aligned} \end{equation} 
This kinematics involves two contributions:
\beq
\begin{tikzpicture}[baseline=0.8cm] \begin{feynhand}
\setlength{\feynhandblobsize}{2mm}
\vertex [particle] (i1) at (0,1.8) {$1^{0}$};
\vertex [particle] (i2) at (0,0) {$2^{0}$};
\vertex [particle] (i3) at (2.2,0) {$3^{0}$};
\vertex [particle] (i4) at (2.2,1.8) {$4^{0}$};
\vertex (v1) at (0.6,0.9);
\vertex [ringblob,color=cyan,fill=white] (v2) at (0.9,0.9) {};
\vertex (v5) at (1.1,0.9);
\vertex [ringblob,color=red,fill=white] (v3) at (1.3,0.9) {};
\vertex (v4) at (1.6,0.9);
\graph{(i1)--[plain,brown,very thick](v1)--[plain,brown,very thick] (i2)};
\graph{(i4)--[plain,brown,very thick](v4)--[plain,brown,very thick](i3)};
\graph{(v1)--[plain,cyan](v2)--[plain,cyan](v5)--[plain,red,slash={[style=black]0}](v3)--[plain,red](v4)};
\end{feynhand} \end{tikzpicture}+
\begin{tikzpicture}[baseline=0.8cm] \begin{feynhand}
\setlength{\feynhanddotsize}{1mm}
\vertex [particle] (i1) at (0,1.8) {$1^{0}$};
\vertex [particle] (i2) at (0,0) {$2^{0}$};
\vertex [particle] (i3) at (2.2,0) {$3^{0}$};
\vertex [particle] (i4) at (2.2,1.8) {$4^{0}$};
\vertex (v1) at (0.6,0.9);
\vertex [crossdot,color=red,fill=white] (v2) at (0.9,0.9) {};
\vertex (v5) at (1.1,0.9);
\vertex [crossdot,color=cyan,fill=white] (v3) at (1.3,0.9) {};
\vertex (v4) at (1.6,0.9);
\graph{(i1)--[plain,brown,very thick](v1)--[plain,brown,very thick] (i2)};
\graph{(i4)--[plain,brown,very thick](v4)--[plain,brown,very thick](i3)};
\graph{(v1)--[plain,red](v2)--[plain,red](v5)--[plain,cyan,slash={[style=black]0}](v3)--[plain,cyan](v4)};
\end{feynhand} \end{tikzpicture}.
\eeq
The numerator of this 4-pt amplitude is just the product of two $x$-factors with opposite helicity. Here we choose eq.~\eqref{eq:t0x} to define the $x$-factor. The product can be reduced to
\begin{equation} \begin{aligned}
&x_{12}\times \bar{x}_{34}^{-1}+x_{12}^{-1}\times \bar{x}_{34} \\
=&\frac{1}{\mathbf{m}_e}\frac{[1P_{12}][2P_{12}]}{[12]}\times\frac{1}{\mathbf{m}_\mu}\frac{\langle P_{12}3\rangle\langle P_{12}4\rangle}{\langle 34\rangle}+\mathbf{m}_e\frac{[12]}{[1P_{12}][2P_{12}]}\times\mathbf{m}_\mu\frac{\langle 34\rangle}{\langle P_{12}3\rangle\langle P_{12}4\rangle}\\
=&\frac{1}{\mathbf{m}_e \mathbf{m}_\mu}[13]\langle31\rangle+\frac{\mathbf{m}_e \mathbf{m}_\mu}{[13]\langle31\rangle}=\frac{1}{\mathbf{m}_e \mathbf{m}_\mu}([13]\langle31\rangle+[\eta_1 \eta_3]\langle\eta_3 \eta_1\rangle).
\end{aligned} \end{equation} 
In the last line, we reduce the gluing result to a local expression.

Then we consider another kinematics
\begin{equation} \begin{aligned} \label{eq:kinematicQED2}
|1\rangle\propto|2\rangle\propto&|P_{12}\rangle\propto|3\rangle\propto|4\rangle,\\
|\eta_1]\propto|\eta_2]\propto&|P_{12}]\propto|\eta_3]\propto|\eta_4].
\end{aligned} \end{equation} 
It involves the other two contributions
\beq
\begin{tikzpicture}[baseline=0.8cm] \begin{feynhand}
\setlength{\feynhandblobsize}{2mm}
\setlength{\feynhanddotsize}{1mm}
\vertex [particle] (i1) at (0,1.8) {$1^{0}$};
\vertex [particle] (i2) at (0,0) {$2^{0}$};
\vertex [particle] (i3) at (2.2,0) {$3^{0}$};
\vertex [particle] (i4) at (2.2,1.8) {$4^{0}$};
\vertex (v1) at (0.6,0.9);
\vertex [ringblob,color=cyan,fill=white] (v2) at (0.9,0.9) {};
\vertex (v5) at (1.1,0.9);
\vertex [crossdot,color=red,fill=white] (v3) at (1.3,0.9) {};
\vertex (v4) at (1.6,0.9);
\graph{(i1)--[plain,brown,very thick](v1)--[plain,brown,very thick] (i2)};
\graph{(i4)--[plain,brown,very thick](v4)--[plain,brown,very thick](i3)};
\graph{(v1)--[plain,cyan](v2)--[plain,cyan](v5)--[plain,red,slash={[style=black]0}](v3)--[plain,red](v4)};
\end{feynhand} \end{tikzpicture}+
\begin{tikzpicture}[baseline=0.8cm] \begin{feynhand}
\setlength{\feynhandblobsize}{2mm}
\setlength{\feynhanddotsize}{1mm}
\vertex [particle] (i1) at (0,1.8) {$1^{0}$};
\vertex [particle] (i2) at (0,0) {$2^{0}$};
\vertex [particle] (i3) at (2.2,0) {$3^{0}$};
\vertex [particle] (i4) at (2.2,1.8) {$4^{0}$};
\vertex (v1) at (0.6,0.9);
\vertex [crossdot,color=red,fill=white] (v2) at (0.9,0.9) {};
\vertex (v5) at (1.1,0.9);
\vertex [ringblob,color=cyan,fill=white] (v3) at (1.3,0.9) {};
\vertex (v4) at (1.6,0.9);
\graph{(i1)--[plain,brown,very thick](v1)--[plain,brown,very thick] (i2)};
\graph{(i4)--[plain,brown,very thick](v4)--[plain,brown,very thick](i3)};
\graph{(v1)--[plain,red](v2)--[plain,red](v5)--[plain,cyan,slash={[style=black]0}](v3)--[plain,cyan](v4)};
\end{feynhand} \end{tikzpicture}.
\eeq
Taking the product of two 3-pt amplitudes we obtain
\begin{equation} \begin{aligned}
&x_{12}\times x_{34}^{-1}+x_{12}^{-1}\times x_{34} \\
=&\frac{1}{\mathbf{m}_e}\frac{[1P_{12}][2P_{12}]}{[12]}\times\mathbf{m}_\mu\frac{[34]}{[3P_{12}][4P_{12}]}+\mathbf{m}_e\frac{[12]}{[1P_{12}][2P_{12}]}\times\frac{1}{\mathbf{m}_\mu}\frac{[3P_{12}][4P_{12}]}{[34]}\\
=&\frac{\mathbf{m}_\mu^2}{\mathbf{m}_e \mathbf{m}_\mu}\frac{[1P_{12}]}{[3P_{12}]}\frac{[2P_{12}][34]}{[4P_{12}][12]}+\frac{\mathbf{m}_\mu^2}{\mathbf{m}_e \mathbf{m}_\mu}\frac{[3P_{12}]}{[1P_{12}]}\frac{[4P_{12}][12]}{[2P_{12}][34]}=\frac{1}{\mathbf{m}_e \mathbf{m}_\mu}([1\eta_3]\langle\eta_3 1\rangle+[\eta_1 3]\langle\eta_1 3\rangle). 
\end{aligned} \end{equation} 
Recovering the LG covariance and adding the pole structure $\mathbf{s}_{12}=(\mathbf{p}_1+\mathbf{p}_2)^2$, we have
\begin{equation} \begin{aligned}
\mathcal{M}(\phi \phi^*\rightarrow\Phi \Phi^*)
=\frac{1}{\mathbf{m}_e \mathbf{m}_\mu}\frac{(\mathbf{p}_1\cdot\mathbf{p}_3-\mathbf{p}_1\cdot\mathbf{p}_4)}{\mathbf{s}_{12}}. 
\end{aligned} \end{equation} 

\noindent \textbf{Spinor QED}

Then we consider the scattering process $e\bar{e}\rightarrow\mu\bar{\mu}$ with helicity category $(+\frac{1}{2},-\frac{1}{2},+\frac{1}{2},-\frac{1}{2})$. It also has one channel $s_{12}$. In the spinor QED, the fermion and anti-fermion in the 3-pt amplitudes have the same chirality. Therefore, the transversality can be $(\pm\frac{1}{2},\mp\frac{1}{2},\pm\frac{1}{2},\mp\frac{1}{2})$ and $(\pm\frac{1}{2},\mp\frac{1}{2},\mp\frac{1}{2},\pm\frac{1}{2})$. Note that there are two 3-particle kinematics choices as shown in eq.~\eqref{eq:kinematicQED1} and eq.~\eqref{eq:kinematicQED2}, although the 4-particle kinematics are the same for these two choices. In each kinematics, we can obtain results for all four transversality. Since the 4-particle kinematics are the same, we only need to consider two different transversality in each kinematics. In the following for the kinematics in eq.~\eqref{eq:kinematicQED1}, two transversality $(\pm\frac{1}{2},\mp\frac{1}{2},\pm\frac{1}{2},\mp\frac{1}{2})$ are chosen, while for the kinematics in eq.~\eqref{eq:kinematicQED2}, the other two transversality  $(\pm\frac{1}{2},\mp\frac{1}{2},\pm\frac{1}{2},\mp\frac{1}{2})$ are chosen.


Let us consider the first kinematics in eq.~\eqref{eq:kinematicQED1}. Note that the definition of $x$-factor is not unique as discussed in the above section. Our first choice would be using the $x$-factor with $t_1=t_2=0$, as shown in eq.~\eqref{eq:t0x}, we can obtain the following two diagrams
\bea
\begin{tikzpicture}[baseline=0.8cm] \begin{feynhand}
\setlength{\feynhandblobsize}{2mm}
\vertex [particle] (i1) at (0,1.8) {$1^{+\frac{1}{2}}$};
\vertex [particle] (i2) at (0,0) {$2^{-\frac{1}{2}}$};
\vertex [particle] (i3) at (2.2,0) {$3^{-\frac{1}{2}}$};
\vertex [particle] (i4) at (2.2,1.8) {$4^{+\frac{1}{2}}$};
\vertex (a1) at (0.6-0.6*0.4,0.9+0.9*0.4);
\vertex (a2) at (1.6+0.6*0.4,0.9+0.9*0.4);
\vertex (a3) at (1.1,0.9);
\vertex (v1) at (0.6,0.9);
\vertex [ringblob,color=cyan,fill=white] (v2) at (0.9,0.9) {};
\vertex [ringblob,color=red,fill=white] (v3) at (1.3,0.9) {};
\vertex (v4) at (1.6,0.9);
\graph{(v1)--[plain,cyan] (v2)--[plain,cyan] (a3)--[plain,red,slash={[style=black]0}] (v3)--[plain,red] (v4)};
\graph{(i1)--[plain,cyan,very thick] (a1)--[plain,red,very thick] (v1)--[plain,cyan,very thick] (i2)};
\graph{(i4)--[plain,cyan,very thick] (a2)--[plain,red,very thick] (v4)--[plain,cyan,very thick] (i3)};
\draw plot[mark=x,mark size=2.7,mark options={rotate=30}] coordinates {(0.6-0.6*0.18,0.9+0.9*0.18)};
\draw plot[mark=x,mark size=2.7,mark options={rotate=60}] coordinates {(1.6+0.6*0.18,0.9+0.9*0.18)};
\draw[very thick] plot[mark=x,mark size=2.7,mark options={rotate=30}] coordinates {(a1)};
\draw[very thick] plot[mark=x,mark size=2.7,mark options={rotate=60}] coordinates {(a2)};
\end{feynhand} \end{tikzpicture}+
\begin{tikzpicture}[baseline=0.8cm] \begin{feynhand}
\setlength{\feynhanddotsize}{1mm}
\vertex [particle] (i1) at (0,1.8) {$1^{+\frac{1}{2}}$};
\vertex [particle] (i2) at (0,0) {$2^{-\frac{1}{2}}$};
\vertex [particle] (i3) at (2.2,0) {$3^{-\frac{1}{2}}$};
\vertex [particle] (i4) at (2.2,1.8) {$4^{+\frac{1}{2}}$};
\vertex (a1) at (0.6-0.6*0.33,0.9+0.9*0.33);
\vertex (a2) at (1.6+0.6*0.33,0.9+0.9*0.33);
\vertex (a3) at (1.1,0.9);
\vertex (v1) at (0.6,0.9);
\vertex [crossdot,color=red,fill=white] (v2) at (0.9,0.9) {};
\vertex [crossdot,color=cyan,fill=white] (v3) at (1.3,0.9) {};
\vertex (v4) at (1.6,0.9);
\graph{(v1)--[plain,red] (v2)--[plain,red] (a3)--[plain,cyan,slash={[style=black]0}] (v3)--[plain,cyan] (v4)};
\graph{(i1)--[plain,red,very thick] (a1)--[plain,cyan,very thick] (v1)--[plain,red,very thick] (i2)};
\graph{(i4)--[plain,red,very thick] (a2)--[plain,cyan,very thick] (v4)--[plain,red,very thick] (i3)};
\draw plot[mark=x,mark size=2.7,mark options={rotate=60}] coordinates {(0.6-0.6*0.33,0.9-0.9*0.33)};
\draw plot[mark=x,mark size=2.7,mark options={rotate=30}] coordinates {(1.6+0.6*0.33,0.9-0.9*0.3)};
\draw[very thick] plot[mark=x,mark size=2.7,mark options={rotate=30}] coordinates {(a1)};
\draw[very thick] plot[mark=x,mark size=2.7,mark options={rotate=60}] coordinates {(a2)};
\end{feynhand} \end{tikzpicture}.
\eea
The first diagram carries the transversality $(+\frac{1}{2},-\frac{1}{2},+\frac{1}{2},-\frac{1}{2})$, which satifies the condition $h_i=t_i$. From the helicity-chirality unification, it corresponds to the 4-pt UV massless amplitude of the leading order $\varepsilon_\eta^0$. However, from the diagram, we note that there are several mass insertions, with $\circ\sim\varepsilon_\eta^{-1}$ and $ \times \sim\varepsilon_\eta$, and thus the naive power counting for this diagram is $\varepsilon_\eta^{2}$. The mismatch in the power counting needs to be compensated by $\varepsilon_\eta^{-2}$ in the coefficients of the 3-pt amplitudes.

To have a simple power counting rule on the diagrams, we can consider another definition for $x$-factor, as shown in eq.~\eqref{eq:neq0x}. With this new definition, the leading power counting is order $1$, agreed with the helicity-chirality unification. For two 3-pt amplitudes, the $x$-factor takes the form
\begin{equation} \begin{aligned}
x_{12}&=\frac{[2P_{12}][P_{12}1]}{m_1 [12]},&
\bar{x}_{12}&=\frac{m_2\langle12\rangle}{\langle2P_{12}\rangle\langle P_{12}1\rangle},&\\
x_{34}&=\frac{[3P_{12}][P_{12}4]}{m_3 [34]},&
\bar{x}_{34}&=\frac{m_4\langle34\rangle}{\langle3P_{12}\rangle\langle P_{12}4\rangle}.&
\end{aligned} \end{equation}
Therefore, the relevant 3-pt sub-amplitudes should take the result in eq.~\eqref{eq:equalFFV1}. Gluing these 3-pt amplitudes, we have the following two diagrams with two transversality:
\bea
\begin{tikzpicture}[baseline=0.8cm] \begin{feynhand}
\setlength{\feynhandblobsize}{2mm}
\vertex [particle] (i1) at (0,1.8) {$1^{+\frac{1}{2}}$};
\vertex [particle] (i2) at (0,0) {$2^{-\frac{1}{2}}$};
\vertex [particle] (i3) at (2.2,0) {$3^{+\frac{1}{2}}$};
\vertex [particle] (i4) at (2.2,1.8) {$4^{-\frac{1}{2}}$};
\vertex (a1) at (0.6-0.6*0.33,0.9+0.9*0.33);
\vertex (a2) at (1.6+0.6*0.33,0.9+0.9*0.33);
\vertex (v1) at (0.6,0.9);
\vertex [ringblob,color=cyan,fill=white] (v2) at (0.9,0.9) {};
\vertex (v5) at (1.1,0.9);
\vertex [ringblob,color=red,fill=white] (v3) at (1.3,0.9) {};
\vertex (v4) at (1.6,0.9);
\graph{(i1)--[plain,cyan,very thick](v1)--[plain,cyan,very thick] (i2)};
\graph{(i4)--[plain,cyan,very thick](v4)--[plain,cyan,very thick](i3)};
\graph{(v1)--[plain,cyan](v2)--[plain,cyan](v5)--[plain,red,slash={[style=black]0}](v3)--[plain,red](v4)};
\draw plot[mark=x,mark size=2.7,mark options={rotate=30}] coordinates {(a1)};
\draw plot[mark=x,mark size=2.7,mark options={rotate=60}] coordinates {(a2)};
\end{feynhand} \end{tikzpicture}+
\begin{tikzpicture}[baseline=0.8cm] \begin{feynhand}
\setlength{\feynhanddotsize}{1mm}
\vertex [particle] (i1) at (0,1.8) {$1^{+\frac{1}{2}}$};
\vertex [particle] (i2) at (0,0) {$2^{-\frac{1}{2}}$};
\vertex [particle] (i3) at (2.2,0) {$3^{+\frac{1}{2}}$};
\vertex [particle] (i4) at (2.2,1.8) {$4^{-\frac{1}{2}}$};
\vertex (a1) at (0.6-0.6*0.33,0.9-0.9*0.33);
\vertex (a2) at (1.6+0.6*0.33,0.9-0.9*0.33);
\vertex (v1) at (0.6,0.9);
\vertex [crossdot,color=red,fill=white] (v2) at (0.9,0.9) {};
\vertex (v5) at (1.1,0.9);
\vertex [crossdot,color=cyan,fill=white] (v3) at (1.3,0.9) {};
\vertex (v4) at (1.6,0.9);
\graph{(i1)--[plain,red,very thick](v1)--[plain,red,very thick] (i2)};
\graph{(i4)--[plain,red,very thick](v4)--[plain,red,very thick](i3)};
\graph{(v1)--[plain,red](v2)--[plain,red](v5)--[plain,cyan,slash={[style=black]0}](v3)--[plain,cyan](v4)};
\draw plot[mark=x,mark size=2.7,mark options={rotate=60}] coordinates {(a1)};
\draw plot[mark=x,mark size=2.7,mark options={rotate=30}] coordinates {(a2)};
\end{feynhand} \end{tikzpicture}. 
\eea
Notice that the colors of fermion and anti-fermion lines are different from eq.~\eqref{eq:equalFFV}, because we choose the $x$-factor which has non-zero transversality. Since the $x$-factor contains the spurious pole, we can use the independent Lorentz structures (namely only $\lambda$ or $\tilde{\lambda}$) to express the 3-pt sub-amplitudes. The direct product of the sub-amplitudes is
\begin{equation} \begin{aligned} \label{eq:num1}
&-x_{12}\langle\eta_1 2\rangle\times \bar{x}_{34}^{-1}[3 \eta_4]-x_{12}^{-1}[1\eta_2]\times \bar{x}_{34} \langle\eta_3 4\rangle\\
=&-\frac{[1P_{12}]^2}{[12]}\times\frac{\langle P_{12}4\rangle^2}{\langle 34\rangle}-m_1 \tilde{m}_2\frac{[12]}{[2P_{12}]^2}\times m_3\tilde{m}_4\frac{\langle 34\rangle}{\langle P_{12} 3\rangle^2}\\
=&[13]\langle24\rangle+\frac{m_1 \tilde{m}_2 m_3\tilde{m}_4}{[24]\langle13\rangle}=[13]\langle24\rangle+[\eta_2\eta_4]\langle\eta_1\eta_3\rangle.
\end{aligned} \end{equation}  
This result shows the numerator of the 4-pt amplitudes in order $1$ and $\varepsilon_{\eta}^4$.

Take another 3-particle kinematics in \eqref{eq:kinematicQED2}, and analyze the other two transversality $(\pm\frac{1}{2},\mp\frac{1}{2},\mp\frac{1}{2},\pm\frac{1}{2})$. Diagrammatically we have two contributions
\bea
\begin{tikzpicture}[baseline=0.8cm] \begin{feynhand}
\setlength{\feynhandblobsize}{2mm}
\setlength{\feynhanddotsize}{1mm}
\vertex [particle] (i1) at (0,1.8) {$1^{+\frac{1}{2}}$};
\vertex [particle] (i2) at (0,0) {$2^{-\frac{1}{2}}$};
\vertex [particle] (i3) at (2.2,0) {$3^{+\frac{1}{2}}$};
\vertex [particle] (i4) at (2.2,1.8) {$4^{-\frac{1}{2}}$};
\vertex (a1) at (0.6-0.6*0.33,0.9+0.9*0.33);
\vertex (a2) at (1.6+0.6*0.33,0.9+0.9*0.33);
\vertex (v1) at (0.6,0.9);
\vertex [ringblob,color=cyan,fill=white] (v2) at (0.9,0.9) {};
\vertex (v5) at (1.1,0.9);
\vertex [crossdot,color=red,fill=white] (v3) at (1.3,0.9) {};
\vertex (v4) at (1.6,0.9);
\graph{(i1)--[plain,cyan,very thick](v1)--[plain,cyan,very thick] (i2)};
\graph{(i4)--[plain,red,very thick](v4)--[plain,red,very thick](i3)};
\graph{(v1)--[plain,cyan](v2)--[plain,cyan](v5)--[plain,red,slash={[style=black]0}](v3)--[plain,red](v4)};
\draw plot[mark=x,mark size=2.7,mark options={rotate=30}] coordinates {(a1)};
\draw plot[mark=x,mark size=2.7,mark options={rotate=60}] coordinates {(a2)};
\end{feynhand} \end{tikzpicture}+
\begin{tikzpicture}[baseline=0.8cm] \begin{feynhand}
\setlength{\feynhandblobsize}{2mm}
\setlength{\feynhanddotsize}{1mm}
\vertex [particle] (i1) at (0,1.8) {$1^{+\frac{1}{2}}$};
\vertex [particle] (i2) at (0,0) {$2^{-\frac{1}{2}}$};
\vertex [particle] (i3) at (2.2,0) {$3^{+\frac{1}{2}}$};
\vertex [particle] (i4) at (2.2,1.8) {$4^{-\frac{1}{2}}$};
\vertex (a1) at (0.6-0.6*0.33,0.9-0.9*0.33);
\vertex (a2) at (1.6+0.6*0.33,0.9-0.9*0.33);
\vertex (v1) at (0.6,0.9);
\vertex [crossdot,color=red,fill=white] (v2) at (0.9,0.9) {};
\vertex (v5) at (1.1,0.9);
\vertex [ringblob,color=cyan,fill=white] (v3) at (1.3,0.9) {};
\vertex (v4) at (1.6,0.9);
\graph{(i1)--[plain,red,very thick](v1)--[plain,red,very thick] (i2)};
\graph{(i4)--[plain,cyan,very thick](v4)--[plain,cyan,very thick](i3)};
\graph{(v1)--[plain,red](v2)--[plain,red](v5)--[plain,cyan,slash={[style=black]0}](v3)--[plain,cyan](v4)};
\draw plot[mark=x,mark size=2.7,mark options={rotate=60}] coordinates {(a1)};
\draw plot[mark=x,mark size=2.7,mark options={rotate=30}] coordinates {(a2)};
\end{feynhand} \end{tikzpicture}.
\eea
Similarly, we obtain the numerator of the 4-pt amplitudes in order $\varepsilon_{\eta}^2$
\begin{equation} \begin{aligned} \label{eq:num2}
&-x_{12}\langle\eta_1 2\rangle\times x_{34}^{-1}[3 \eta_4]-x_{12}^{-1}[1\eta_2]\times x_{34} \langle\eta_3 4\rangle\\
=&\frac{[1P_{12}]^2}{[12]}\times m_3 \tilde{m}_4\frac{[34]}{[P_{12}4]^2}+m_1 \tilde{m}_2\frac{[12]}{[2P_{12}]^2}\times \frac{[P_{12}3]^2}{[34]}\\
=&-[1\eta_4]\langle2\eta_3\rangle-[\eta_2 3]\langle\eta_1 4\rangle. 
\end{aligned} \end{equation} 

After obtaining the numerators, we can return to the 4-particle kinematics. Summing over all transversality and adding the pole structure, we obtain the amplitude
\begin{equation} \begin{aligned}
\mathcal{M}({e\bar{e}\to\mu\bar{\mu}}) &=
\mathcal{M}_{(+\frac12,-\frac12,+\frac12,-\frac12)}+\mathcal{M}_{(-\frac12,+\frac12,-\frac12,+\frac12)}+\mathcal{M}_{(+\frac12,-\frac12,-\frac12,+\frac12)}+\mathcal{M}_{(-\frac12,+\frac12,+\frac12,-\frac12)} \\
&=\frac{[13]\langle24\rangle+[\eta_2\eta_4]\langle\eta_1\eta_3\rangle}{s_{12}} + \frac{-[\eta_2 3]\langle\eta_1 4\rangle-[1\eta_4]\langle2\eta_3\rangle}{s_{12}}.
\end{aligned} \end{equation} 
Finally, recovering the LG covariance, we get the correct result agreeing with the standard QED textbook
\begin{equation} \begin{aligned}
\mathcal{M}({e\bar{e}\to\mu\bar{\mu}})=\frac{[\mathbf{13}]\langle\mathbf{24}\rangle+[\mathbf{24}]\langle\mathbf{13}\rangle+[\mathbf{14}]\langle\mathbf{23}\rangle+[\mathbf{23}]\langle\mathbf{14}\rangle}{\mathbf{s}_{12}}. 
\end{aligned} \end{equation}


\section{Summary}
\label{sec:sum}

In this work, we have introduced a new massive spinor-helicity formalism with a new quantum number, transversality, closely related to chirality. The massive AHH spin-spinor is extended to the ST spinor, which can be decomposed into large and small components $\lambda$ and $\eta$ carrying the transversality quantum number, denoted as the helicity-transversality spinor. An extended Poincar\'e symmetry is recognized to relate different transversality in one multiplet, similar to the Poincar\'e symmetry which relates different helicity. This symmetry also entirely determines the 3-pt massive amplitudes through the primary and descendant representations with the ladder operators, including all-massive, 2-massive-1-massless, 2-massive-1-massless and all massless cases.

The massless helicity-transversality spinors $\lambda$ and $\eta$ are the large and small components of a massive ST spinors following the $\lambda \sim \sqrt{E}, \eta \sim \mathbf{m}/\sqrt{E}$ expansion order by order. This formulates the building blocks, symmetries and power counting rules of an effective theory, denoted as the large energy effective theory. In this framework, the helicity-transversality spinor can be diagrammatically expressed by the on-shell mass insertion of the massless spinors, denoted as the helicity flip and chirality flip. Therefore, the helicity and chirality flip counts the order of amplitudes in the large energy effective theory.

Although helicity and chirality are two seperated quantum numbers for a massive spinor, helicity and chirality should be unified for the massless spinor. Therefore, the chirality and helicity unification tells that a massive amplitude has a one-to-one correspondence to the massless amplitude at the UV, with/without additional Higgs insertion. This would benefit the massive amplitudes for a given UV. For example, the charged pion decay and top quark decay have specific electroweak couplings and thus a mass enhancement exists. In this case, the selected UV gives the mass enhancement. The massive $F\bar{F}\gamma$ amplitude not only describes the QED but also other UVs. Selecting the correct QED amplitudes would give rise to the correct 3-pt QED amplitudes. This builds a one-to-one massless massive correspondence.

The massless-massive correspondence also tells us that we can utilize the unitarity techniques developed in the massless on-shell method to construct the higher-point massive amplitudes. For all massive amplitudes, we find that it is convenient to build amplitudes from listing all helicity-transversality spinors for both external and internal on-shell particles, similar to the Feynman rules. With the selected $F\bar{F}\gamma$ 3-pt amplitudes, we can construct the $e^+e^- \to \mu^+\mu^-$ amplitudes with correct results.

This helicity-chirality unification could be systematically applied to the electroweak standard model and effective field theories. Because of the massless-massive correspondence, the massive electroweak amplitudes can have a one-to-one correspondence to the massless amplitudes in the unbroken phase of the standard model, leading to further understanding of the Higgs mechanism, and Goldstone equivalence theorem, unitarity, etc. All of these deserve further study in the future. 

\acknowledgments

We thank Bo Feng, Song He, Xiao-Gang He, Mingxing Luo, Yue-Liang Wu and Gang Yang for their valuable discussions. This work is supported by the National Science Foundation of China under Grants No. 12347105, No. 12375099 and No. 12047503, and the National Key Research and Development Program of China Grant No. 2020YFC2201501, No. 2021YFA0718304. 

\appendix

\section{The spinor formalism of the $ISO(5,1)$ generators}
\label{app:6D}

Let us first review the generators of the Poincar\'e group, $ISO(3,1)$, which include $L_{\mu\nu} \sigma^\mu_{\alpha\dot{\alpha}} \sigma^\nu_{\beta\dot{\beta}} = \epsilon_{\alpha\beta} \widetilde{L}_{\dot{\alpha}\dot{\beta}} + {\tilde{\epsilon}}_{\dot{\alpha}\dot{\beta}} L_{\alpha\beta}$ and $P_{\alpha\dot{\alpha}}$. For massive particles, they are
\bea
P_{\alpha\dot{\alpha}} &=& p_{\alpha\dot{\alpha}} + \eta_{\alpha\dot{\alpha}} = \lambda_{\alpha} \tilde{\lambda}_{\dot{\alpha}} + \eta_{\alpha} \tilde{\eta}_{\dot{\alpha}} , \\
L_{\alpha \beta} &=& -i\left(\lambda_{\alpha}^{I}\frac{\partial}{\partial \lambda^{\beta I}} + \lambda_{\beta}^{I}\frac{\partial}{\partial \lambda^{\alpha I}}\right) = -2i \left(\lambda_{(\alpha} \frac{\partial}{\partial \lambda^{\beta)}} + \eta_{(\alpha} \frac{\partial}{\partial \eta^{\beta)}} \right), \\ 
{\widetilde{L}}_{\dot{\alpha} \dot{\beta}} &=& -i\left(\tilde{\lambda}_{\dot{\alpha}}^{I}\frac{\partial}{\partial \tilde{\lambda}^{\dot{\beta} I}} + \tilde{\lambda}_{\dot{\beta}}^{I}\frac{\partial}{\partial \tilde{\lambda}^{\dot{\alpha} I}}\right) = -2i \left(\tilde{\lambda}_{(\dot{\alpha}} \frac{\partial}{\partial \tilde{\lambda}^{\dot{\beta})}} + \tilde{\eta}_{(\dot{\alpha}} \frac{\partial}{\partial \tilde{\eta}^{\dot{\beta})}}\right). 
\eea
In the spinor-helicity formalism, the commutators are
\bea
[L_{\alpha\beta}, L_{\gamma\delta}] &=& i (\epsilon_{\gamma\beta} L_{\beta\delta} + \epsilon_{\delta\beta} L_{\alpha\gamma} + \epsilon_{\gamma\alpha} L_{\beta\delta} + \epsilon_{\delta\alpha} L_{\beta \gamma}), \\
{[{\widetilde{L}}_{\dot{\alpha}\dot{\beta}}, {\widetilde{L}}_{\dot{\gamma}\dot{\delta}}]} &=& -i({\tilde{\epsilon}}_{\dot{\gamma}\dot{\beta}} {\widetilde{L}}_{\dot{\beta}\dot{\delta}} + {\tilde{\epsilon}}_{\dot{\delta}\dot{\beta}} {\widetilde{L}}_{\dot{\alpha}\dot{\gamma}} + {\tilde{\epsilon}}_{\dot{\gamma}\dot{\alpha}} {\widetilde{L}}_{\dot{\beta}\dot{\delta}} + {\tilde{\epsilon}}_{\dot{\delta}\dot{\alpha}} {\tilde{M}}_{\dot{\beta}\dot{\gamma}}), \\
{[P_{\alpha\dot{\alpha}}, L_{\beta\gamma}]} &=& i(\epsilon_{\alpha\gamma} P_{\beta\dot{\alpha}} +\epsilon_{\alpha\beta} P_{\gamma\dot{\alpha}}), \\
{[P_{\alpha\dot{\alpha}}, \widetilde{L}_{\dot{\beta}\dot{\gamma}}]} &=& -i(\tilde{\epsilon}_{\dot{\alpha}\dot{\gamma}} P_{\alpha\dot{\beta}} + \tilde{\epsilon}_{\dot{\alpha}\dot{\beta}} P_{\alpha\dot{\gamma}}), 
\eea
and the two Casimir operators are
\bea
P^2 &=& \tilde{m} m = \mathbf{m}^2, \\
W^2 &=&\frac{1}{8} {P}^2(\mathrm{Tr}[L^2] + \mathrm{Tr}[\widetilde{L}^2]) - \frac14 \mathrm{Tr}[{P}^\intercal L P \widetilde{L}]. 
\eea

The extended $SO(2)\times ISO(3,1)$ still contains the two Casimir operators above, as well as an additional Casimir operator $D_-$.
These Casimir operators acting on an arbitrary particle state of the representation $(t,s)$ gives
\bea
P^2 |\mathbf{p}, t, s, h\rangle &=& \mathbf{p}^2 |\mathbf{p}, t, s, h\rangle, \\
W^2 |\mathbf{p}, t, s, h\rangle &=& -s(s+1) \mathbf{p}^2 |\mathbf{p}, t, s, h\rangle, \\
D_- |\mathbf{p}, t, s, h\rangle &=& t |\mathbf{p}, t, s, h\rangle. 
\eea
The Casimir operators, along with their eigenvalues, uniquely label a representation of $SO(2) \times ISO(3,1)$. We can thus obtain the corresponding representations by solving the eigenfunctions. The helicity-chirality states introduced in subsection~\ref{sec:induce} span a linear space for each pair $(t,h)$, allowing us to use them as an eigenbasis for the $W^2$ operator.
For example, eq.~\eqref{eq:h0t0state} contains two $(t,h)=(0,0)$ states, spanning a dim-2 linear space. $W^2$ acting on eq.~\eqref{eq:h0t0state} gives
\bea
W^2 \begin{pmatrix}
\eta_{\alpha_1}\tilde{\eta}_{\dot{\alpha}_2} -\lambda_{\alpha_1}\tilde{\lambda}_{\dot{\alpha}_2} \\ \eta_{\alpha_1}\tilde{\eta}_{\dot{\alpha}_2}+\lambda_{\alpha_1}\tilde{\lambda}_{\dot{\alpha}_2}
\end{pmatrix} 
= \begin{pmatrix}
-2 (2 \lambda \cdot \eta) & 0 \\
0 & 0
\end{pmatrix}
\begin{pmatrix}
\eta_{\alpha_1}\tilde{\eta}_{\dot{\alpha}_2}-\lambda_{\alpha_1}\tilde{\lambda}_{\dot{\alpha}_2} \\ \eta_{\alpha_1}\tilde{\eta}_{\dot{\alpha}_2}+\lambda_{\alpha_1}\tilde{\lambda}_{\dot{\alpha}_2}
\end{pmatrix}, 
\eea
separating the spin-1 and spin-0 states.

The Lorentz algebra spanned by the generators 
$L_{\mu\nu}$
can be extended to $SO(5,1)$, which is isomorphic to the 4D conformal algebra, by introducing the transversality generators $\{ T^{\pm}, D_- \}$. The key feature of the conformal algebra lies in the commutators, which are 
\bea
[(T^+)_{\alpha\dot{\alpha}}, (T^-)_{\beta\dot{\beta}} ] &=& -\frac{i}{2}( \epsilon_{\alpha\beta} \tilde{L}_{\dot{\alpha}\dot{\beta}} +\tilde{\epsilon}_{\dot{\alpha}\dot{\beta}} L_{\alpha\beta} ) -\frac{1}{2}\epsilon_{\alpha\beta}\tilde{\epsilon}_{\dot{\alpha}\dot{\beta}} D_-, \\
{[D_-, {(T^{\pm})}_{\alpha\dot{\alpha}} ]} &=& \pm{(T^{\pm})}_{\alpha\dot{\alpha}}, \\
{[L_{\beta\gamma},(T^{\pm})_{\alpha\dot{\alpha}}}] &=& -i\epsilon_{\beta\alpha}(T^{\pm})_{\gamma\dot{\alpha}}-i\epsilon_{\gamma\alpha}(T^{\pm})_{\beta\dot{\alpha}}.
\eea
After a suitable redefinition, all the generators can be written into a 6D antisymmetric matrix
\bea
\label{eq:matrix-form-of-generators}
L_{MN} = 
\left(\begin{array}{c|cc}
L_{\mu\nu} & L_{\mu 4} & L_{\mu 5}\\
\hline
L_{4\nu}    & 0 & L_{45}\\
L_{5\nu} & L_{54} & 0
\end{array}
\right), 
\eea
where $M,N = 0, 1, 2, 3, 4, 5$. The components of eq.~\eqref{eq:matrix-form-of-generators} are
\eq{
L_{54}&\equiv D_{-}  = \frac{1}{2} \left(\tilde{\lambda}_{\dot{\alpha}} \frac{\partial}{\partial \tilde{\lambda}_{\dot{\alpha}}} + \tilde{\eta}_{\dot{\alpha}} \frac{\partial}{\partial \tilde{\eta}_{\dot{\alpha}}} \right) -\frac{1}{2} \left( \lambda_{\alpha} \frac{\partial}{\partial \lambda_{\alpha}} + \eta_{\alpha} \frac{\partial}{\partial \eta_{\alpha}} \right) , \\
L_{5\mu}&\equiv -i\left(T^{+}_{\mu} -T^{-}_{\mu}\right),\\
L_{4\mu}&\equiv T^{+}_{\mu} + T^{-}_{\mu}. 
}
The algebra defined by the commutators of $\{T^{\pm}, D_{-}, L\}$ can be reorganized into the form of the regular $SO(5, 1)$ algebra with generators $ \{L_{MN} \} $:
\begin{equation}
            [L_{QR},L_{MN}]=i\left(\eta_{QM}L_{NR}-\eta_{QN}L_{MR}-\eta_{RM}L_{NQ}+\eta_{RN}L_{MQ}\right), 
\end{equation}
where $\eta_{MN}$ is the metric in the 6D spacetime, chosen to be $\mathrm{diag}(1, -1, -1, -1, -1, -1)$.

Then, we introduce the translation generators $\{P_{\alpha \dot{\alpha}},m,\tilde{m}\}$ to the $SO(5,1)$ algebra, extending it to the $ISO(5,1)$ algebra defined by the following commutators: 
\bea
{[P_{\alpha\dot{\alpha}}, {T^+}^{\beta}_{\dot{\beta}}]} &=& -\tilde{m} \delta_{\alpha}^{\beta} \tilde{\epsilon}_{\dot{\alpha}\dot{\beta}}, \\
{[P_{\alpha\dot{\alpha}}, {T^-}^{\dot{\beta}}_{\beta}]} &=& -m \tilde{\delta}^{\dot{\beta}}_{\dot{\alpha}} \epsilon_{\alpha\beta}, \\
{[m, {T^+}^{\alpha}_{\dot{\alpha}}]} &=& -P^{\alpha}_{\dot{\alpha}}, \\
{[\tilde{m}, {T^-}^{\dot{\alpha}}_{\alpha}]} &=& P^{\dot{\alpha}}_{\alpha},\\
{[m, D_{-}]} &=&  m, \\
{[\tilde{m}, D_{-}]} &=& - \tilde{m}.
\eea
The 4th and 5th components of a 6D momentum $p_{M}$ are defined as
\eq{
p_4 = \frac{1}{2} (\tilde{m}+m), \ 
p_5 = \frac{i}{2} (\tilde{m}-m). 
}
Therefore, the EOM of the 6D momentum is $p_M p^M = 0$, indicating that the 4D massive particle becomes a massless particle in the extended 6D spacetime. This suggests that the massive 4D amplitudes carry a hidden 6D symmetry.

\section{Massive Dirac/Majorana fermion and massive vector}
\label{app:particle}

Usually, people use both $(\frac12, 0)$ and $(0, \frac12)$ representations of the $SL(2,\mathbbm{C})$ group to describe a massive fermion. This description has four degrees of freedom and is called the Dirac fermion. The helicity states are given by four-component spinors as 
\begin{equation} \begin{aligned}
u_{h=-\frac{1}{2}}^D&=\begin{pmatrix}
    \lambda  \\ \tilde{\eta}
\end{pmatrix},&
u_{h=+\frac{1}{2}}^D&=\begin{pmatrix}
\eta  \\ \tilde{\lambda}
\end{pmatrix},\\
v_{h=-\frac{1}{2}}^D&=\begin{pmatrix}
-\eta  \\ \tilde{\lambda}
\end{pmatrix},&
v_{h=+\frac{1}{2}}^D&=\begin{pmatrix}
\lambda  \\ -\tilde{\eta}
\end{pmatrix}.
\end{aligned} \end{equation}
where the superscript $D$ represents the Dirac fermion. The massive momentum can be written as
\bea
(\mathbf{p}^D)_{\alpha\dot{\alpha}}=(p^D)_{\alpha\dot{\alpha}}+(\eta^D)_{\alpha\dot{\alpha}}
=\lambda_{\alpha}\tilde{\lambda}_{\dot{\alpha}}+\eta_{\alpha}\tilde{\eta}_{\dot{\alpha}}.
\eea

The Dirac fermion should satisfy an EOM called the Dirac equation:
\bea
\begin{cases}
    (p^D)_{\alpha\dot{\alpha}} \tilde{\eta}^{\dot{\alpha}}=-\tilde{m}^D \lambda_{\alpha}\\
    (\eta^D)_{\alpha\dot{\alpha}} \tilde{\lambda}^{\dot{\alpha}}=\tilde{m}^D \eta_{\alpha}\\
\end{cases},\quad 
\begin{cases}
    (p^D)_{\alpha\dot{\alpha}} \eta^{\alpha}=-m^D \tilde{\lambda}_{\dot{\alpha}}\\
    (\eta^D)_{\alpha\dot{\alpha}} \lambda^{\alpha}=m^D \tilde{\eta}_{\dot{\alpha}}\\
\end{cases},\label{eq:EOMD}
\eea
where the spurion masses are
\bea
m^D=\langle p\eta\rangle,\quad \tilde{m}^D=[\eta p].
\eea

For a spin-$1/2$ fermion with momentum $p^\mu=(E,P\sin\theta\cos\varphi,P\sin\theta\sin\varphi,P\cos\theta)$, the helicity spinors are chosen to be the eigenstates of the helicity operator $h = \frac{\vec{p} \cdot \vec{\sigma}}{|\vec{p}|}$,
\bea
\chi_{h= -} = \binom{-e^{-i \varphi} \sin \frac{\theta}{2}}{\cos \frac{\theta}{2}}, \quad
\chi_{h = +} = \binom{\cos \frac{\theta}{2}}{e^{i \varphi} \sin \frac{\theta}{2}}.
\eea
At the same time, each chirality can be written as
\begin{equation} \begin{aligned} \label{eq:helicity-spinor-rep}
&h = -\frac12:& \lambda_\alpha (t = -\frac12) &= \sqrt{E+P} \chi_{h= -},& 
\tilde{\eta}_\alpha (t = \frac12) &=  \frac{m}{\sqrt{E+P}} \chi_{h= -}, \\
&h = +\frac12:&  \tilde{\lambda}_\alpha (t = \frac12) &=\sqrt{E+P} \chi_{h= +},&
\eta_\alpha(t = -\frac12) &= \frac{m}{\sqrt{E+P}} \chi_{h= +}.
\end{aligned} \end{equation}
We obtain the helicity states for massive spin-$1/2$ fermion: 
\bea
u_{h=-1/2} = \lambda + \tilde{\eta} \to 
\begin{pmatrix}
\lambda \\ \tilde{\eta}
\end{pmatrix}, \quad
u_{h=+1/2} = \tilde{\lambda} + \eta \to \begin{pmatrix}
\eta  \\ \tilde{\lambda}
\end{pmatrix},  \\
v_{h=-1/2} =  \tilde{\lambda} - \eta \to \begin{pmatrix}
-\eta  \\ \tilde{\lambda}
\end{pmatrix}, \quad
v_{h=+1/2} = \lambda - \tilde{\eta} \to 
\begin{pmatrix}
\lambda \\ -\tilde{\eta}
\end{pmatrix}.
\eea
These are the building blocks for the helicity amplitude calculations.

However, there is another type of massive fermion called the Majorana fermion, which only contains $(\frac12, 0)$ or $(0, \frac12)$ representation. For example, we choose $(0, \frac12)$ to represent the massive vector, i.e. $\tilde{\lambda}$ and $\tilde{\eta}$. Since the complex conjugate representations $\tilde{\lambda}^*$ and $\tilde{\eta}^*$ transform as the $(\frac12, 0)$ representation under the $SL(2,\mathbbm{C})$ Lorentz group, we can construct the four-component spinor: 
\begin{equation} \begin{aligned} 
u_{h=-\frac{1}{2}}^M &=\begin{pmatrix}
\tilde{\lambda}^*  \\ \tilde{\eta}
\end{pmatrix},&
u_{h=+\frac{1}{2}}^M &=\begin{pmatrix}
\tilde{\eta}^*  \\ \tilde{\lambda}
\end{pmatrix}, \\
v_{h=-\frac{1}{2}}^M &=\begin{pmatrix}
-\tilde{\eta}^*  \\ \tilde{\lambda}
\end{pmatrix},&
v_{h=+\frac{1}{2}}^M &=\begin{pmatrix}
\tilde{\lambda}^*  \\ -\tilde{\eta}
\end{pmatrix},
\end{aligned} \end{equation}
where the superscript $M$ represents the Majorana fermion. In this case, the massive momentum can be written as
\bea
(\mathbf{p}^M)_{\dot{\alpha}}^{\dot{\beta}}=(p^M)_{\dot{\alpha}}^{\dot{\beta}}+(\eta^M)_{\dot{\alpha}}^{\dot{\beta}}
=\tilde{\lambda}_{\dot{\alpha}}^*\tilde{\lambda}^{\dot{\beta}}+\tilde{\eta}_{\dot{\alpha}}^*\tilde{\eta}^{\dot{\beta}}.
\eea
Therefore, the momentum of Majorana fermion is always real. We cannot generalize its LG from $SU(2)$ or $U(2)$ into $SL(2,\mathbbm{C})$ or $GL(2,\mathbbm{C})$. 

The Majorana fermion should satisfy the EOM other than eq.~\eqref{eq:EOMD},
\bea
\begin{cases}
    (p^M)_{\dot{\alpha}}^{\dot{\beta}} \tilde{\eta}_{\dot{\beta}}=-\tilde{m}^M \tilde{\lambda}^{*\dot{\alpha}}\\
    (\eta^M)_{\dot{\alpha}}^{\dot{\beta}} \tilde{\lambda}_{\dot{\beta}}=\tilde{m}^M \tilde{\eta}^{*\dot{\alpha}}\\
\end{cases},\quad 
\begin{cases}
    (p^M)_{\dot{\alpha}}^{\dot{\beta}} \tilde{\eta}^{*\dot{\alpha}}=-m^M \tilde{\lambda}^{\dot{\beta}}\\
    (\eta^M)_{\dot{\alpha}}^{\dot{\beta}} \tilde{\lambda}^{*\dot{\alpha}}=m^M \tilde{\eta}^{\dot{\beta}}\\
\end{cases}.
\eea
where the spurion masses have a different definition
\bea
m^M=[p\eta]^*,\quad \tilde{m}^M=[p\eta].
\eea
It means that the spurion mass of a Majorana fermion has a phase rotation, different from the uncharged physical mass $\mathbf{m}$.

The massive vector has three representations of the $SL(2,\mathbbm{C})$ Lorentz group: $(1,0)$, $(\frac{1}{2},\frac{1}{2})$ and $(0,1)$. In analogy to the Dirac fermion, we can use them to define a series of six-component helicity states for massive vectors as 
\bea \label{eq:DiracVector}
u_{h=-1}^D=\begin{pmatrix}
\lambda_{\alpha_1}\lambda_{\alpha_2} \\ -\lambda_{\alpha_1}\tilde{\eta}_{\dot{\alpha}_2} \\ \tilde{\eta}_{\dot{\alpha}_1} \tilde{\eta}_{\dot{\alpha}_2}
\end{pmatrix},
u_{h=0}^D=\begin{pmatrix}
-\lambda_{\alpha_1}\eta_{\alpha_2} -\eta_{\alpha_1}\lambda_{\alpha_2}  \\ \eta_{\alpha_1}\tilde{\eta}_{\dot{\alpha}_2}-\lambda_{\alpha_1}\tilde{\lambda}_{\dot{\alpha}_2}  \\ -\tilde{\lambda}_{\dot{\alpha}_1}\tilde{\eta}_{\dot{\alpha}_2}-\tilde{\eta}_{\dot{\alpha}_1}\tilde{\lambda}_{\dot{\alpha}_2}
\end{pmatrix},
u_{h=+1}^D=\begin{pmatrix}
\eta_{\alpha_1}\eta_{\alpha_2}  \\ \eta_{\alpha_1}\tilde{\lambda}_{\dot{\alpha}_2} \\ \tilde{\lambda}_{\dot{\alpha}_1} \tilde{\lambda}_{\dot{\alpha}_2}
\end{pmatrix}. 
\eea
For $h=0$, we eliminate the $s=0$ state
\bea
\begin{pmatrix}
m\epsilon_{\alpha_1\alpha_2}  \\ \eta_{\alpha_1}\tilde{\eta}_{\dot{\alpha}_2}+\lambda_{\alpha_1}\tilde{\lambda}_{\dot{\alpha}_2}  \\ \tilde{m}\epsilon_{\dot{\alpha}_1\dot{\alpha}_2}
\end{pmatrix}. 
\eea
Only the state with transversality $0$ generates the Goldstone boson. The states with $t=\pm 1$ become subleading at high energy. 

For helicity $+1$, the EOM is
\bea
\begin{pmatrix}
    0 & \eta_{\alpha_2\dot{\alpha}_2} & 0 \\
    p_{\alpha_2\dot{\alpha}_2} & 0 & \eta_{\alpha_1\dot{\alpha}_1} \\
    0 & p_{\alpha_1\dot{\alpha}_1} & 0 \\
\end{pmatrix}
\begin{pmatrix}
\eta_{\alpha_1}\eta_{\alpha_2}  \\
\eta_{\alpha_1}\tilde{\lambda}_{\dot{\alpha}_2} \\
\tilde{\lambda}_{\alpha_1} \tilde{\lambda}_{\alpha_2}
\end{pmatrix}=
\begin{pmatrix}
    \tilde{m} & 0 & 0 \\
    0 & \tilde{m}+m & 0 \\
    0 & 0 & m \\
\end{pmatrix}
\begin{pmatrix}
\eta_{\alpha_1}\eta_{\alpha_2}  \\
\eta_{\alpha_1}\tilde{\lambda}_{\dot{\alpha}_2} \\
\tilde{\lambda}_{\alpha_1} \tilde{\lambda}_{\alpha_2}
\end{pmatrix}. 
\eea

For helicity $0$, the EOM is
\begin{equation} \begin{aligned}
&\begin{pmatrix}
    0 & (p+\eta)_{\alpha_2}^{\dot{\alpha}_2} & 0 \\
    (p+\eta)_{\dot{\alpha}_2}^{\alpha_2} & 0 & (p+\eta)_{\dot{\alpha}_1}^{\alpha_1} \\
    0 & (p+\eta)_{\dot{\alpha}_1}^{\alpha_1} & 0 \\
\end{pmatrix}
\begin{pmatrix}
-\lambda_{\alpha_1}\eta_{\alpha_2} -\eta_{\alpha_1}\lambda_{\alpha_2}  \\ \eta_{\alpha_1}\tilde{\eta}_{\dot{\alpha}_2}-\lambda_{\alpha_1}\tilde{\lambda}_{\dot{\alpha}_2}  \\ -\tilde{\lambda}_{\dot{\alpha}_1}\tilde{\eta}_{\dot{\alpha}_2}-\tilde{\eta}_{\dot{\alpha}_1}\tilde{\lambda}_{\dot{\alpha}_2}
\end{pmatrix}\\
=&\begin{pmatrix}
    \tilde{m} & 0 & 0 \\
    0 & \tilde{m}+m & 0 \\
    0 & 0 & m \\
\end{pmatrix}
\begin{pmatrix}
-\lambda_{\alpha_1}\eta_{\alpha_2} -\eta_{\alpha_1}\lambda_{\alpha_2}  \\ \eta_{\alpha_1}\tilde{\eta}_{\dot{\alpha}_2}-\lambda_{\alpha_1}\tilde{\lambda}_{\dot{\alpha}_2}  \\ -\tilde{\lambda}_{\dot{\alpha}_1}\tilde{\eta}_{\dot{\alpha}_2}-\tilde{\eta}_{\dot{\alpha}_1}\tilde{\lambda}_{\dot{\alpha}_2}
\end{pmatrix}. 
\end{aligned} \end{equation}

If we only consider the leading components in eq.~\eqref{eq:DiracVector}, we can derive the common description for massive vector at IR. After normalization, we find the polarization vector as
\bea
\varepsilon_{h=-1}=\lambda_{\alpha_1}\lambda_{\alpha_2},\quad
\varepsilon_{h=0}=\eta_{\alpha_1}\tilde{\eta}_{\dot{\alpha}_2}-\lambda_{\alpha_1}\tilde{\lambda}_{\dot{\alpha}_2},\quad
\varepsilon_{h=+1}=\tilde{\lambda}_{\dot{\alpha}_1} \tilde{\lambda}_{\dot{\alpha}_2}.
\eea

In the SM, only the $t=0$ massive vectors couple to other particles, so $t=\pm 1$ states should be generated by chirality flip. For example, we act momentum on the state with $h=t=+1$ and get
\bea
\mathbf{p} \circ (\lambda_{\alpha_1} \lambda_{\alpha_2}) = -m \lambda_{\alpha} \tilde{\eta}_{\dot{\alpha}}. 
\eea
The transversality is flipped to $0$.

On the other hand, we can only use two-component spinors and their complex conjugation to describe the massive vector. If we begin with $|t|=1$, the construction is similar to the Majorana fermion. For example, the $t=-1$ massive vector can give
\bea
u_{h=-1}^M=\begin{pmatrix}
\lambda_{\alpha_1}\lambda_{\alpha_2} \\ -\lambda_{\alpha_1}\eta^*_{\alpha_2} \\ \eta^*_{\alpha_1} \eta^*_{\alpha_2}
\end{pmatrix},\quad
u_{h=0}^M=\begin{pmatrix}
-\lambda_{\alpha_1}\eta_{\alpha_2} -\eta_{\alpha_1}\lambda_{\alpha_2}  \\ \eta_{\alpha_1}\eta^*_{\alpha_2}-\lambda_{\alpha_1}\lambda^*_{\alpha_2}  \\ -\lambda^*_{\alpha_1}\eta^*_{\alpha_2}-\eta^*_{\alpha_1}\lambda^*_{\alpha_2}
\end{pmatrix},\quad
u_{h=+1}^M=\begin{pmatrix}
\eta_{\alpha_1}\eta_{\alpha_2}  \\ \eta_{\alpha_1}\lambda^*_{\alpha_2} \\ \lambda^*_{\alpha_1} \lambda^*_{\alpha_2}
\end{pmatrix}. 
\eea
In this case, a massive vector can be represented by only the $(0,\frac{1}{2})$ spinors.

\section{3-pt building blocks and diagrammatic formulation}
\label{app:3pt}

For completeness, here we list the 3-pt amplitudes for $F\bar{F}V$ and $VVS$, which will be used in section 4. 
We apply the highest-weight construction, and flip helicity to derive other Lorentz structures.
For the $F\bar{F}V$ amplitude, we have
\begin{equation}  \label{eq:FFV-hflip}
\begin{tabular}{c|cccc}
\diagbox[width=2.4cm,innerleftsep=.4cm,innerrightsep=.4cm]{$h_i$}{$t_i$} & $(+\frac{1}{2},+\frac{1}{2},+1)$ & $(-\frac{1}{2},+\frac{1}{2},0)$ & $(+\frac{1}{2},-\frac{1}{2},0)$ & $(-\frac{1}{2},-\frac{1}{2},-1)$ \\
\hline
$(+\frac{1}{2},+\frac{1}{2},+1)$ & 
\makecell{\Amp{+\frac{1}{2}}{+\frac{1}{2}}{+1}{cyan}{red}{cyan}{}\\$[23][31]$} & 
\makecell{\Amp{+\frac{1}{2}}{+\frac{1}{2}}{+1}{red}{red}{brown}{\one \three}\\$[23]\langle\eta_3 \eta_1\rangle$} & 
\makecell{\Amp{+\frac{1}{2}}{+\frac{1}{2}}{+1}{cyan}{cyan}{brown}{\two \three}\\$\langle\eta_2 \eta_3\rangle[31]$} & 
\makecell{\Amp{+\frac{1}{2}}{+\frac{1}{2}}{+1}{red}{cyan}{red}{\one \two \threethree}\\$\langle\eta_2 \eta_3\rangle\langle\eta_3 \eta_1\rangle$} \\
\hline
$(-\frac{1}{2},+\frac{1}{2},+1)$ & 
\makecell{\Amp{-\frac{1}{2}}{+\frac{1}{2}}{+1}{cyan}{red}{cyan}{\one}\\$[23][3\eta_1]$} & 
\makecell{\Amp{-\frac{1}{2}}{+\frac{1}{2}}{+1}{red}{red}{brown}{\three}\\$-[23]\langle\eta_31\rangle$}  & 
\makecell{\Amp{-\frac{1}{2}}{+\frac{1}{2}}{+1}{cyan}{cyan}{brown}{\one \two \three}\\$\langle\eta_2 \eta_3\rangle[3\eta_1]$} & 
\makecell{\Amp{-\frac{1}{2}}{+\frac{1}{2}}{+1}{red}{red}{cyan}{\two \threethree}\\$-\langle\eta_2 \eta_3\rangle\langle\eta_3 1\rangle$}  \\
\hline
$(+\frac{1}{2},-\frac{1}{2},+1)$ & 
\makecell{\Amp{+\frac{1}{2}}{-\frac{1}{2}}{+1}{cyan}{red}{cyan}{\two}\\$[\eta_23][31]$} & 
\makecell{\Amp{+\frac{1}{2}}{-\frac{1}{2}}{+1}{red}{red}{brown}{\one \two \three}\\$[\eta_23]\langle\eta_3\eta_1\rangle$} & 
\makecell{\Amp{+\frac{1}{2}}{-\frac{1}{2}}{+1}{cyan}{cyan}{brown}{\three}\\$-\langle2\eta_3\rangle[31]$} & 
\makecell{\Amp{+\frac{1}{2}}{-\frac{1}{2}}{+1}{red}{red}{cyan}{\one \threethree}\\$-\langle2\eta_3\rangle\langle\eta_3 \eta_1\rangle$} \\
\hline
$(-\frac{1}{2},-\frac{1}{2},+1)$ & 
\makecell{\Amp{-\frac{1}{2}}{-\frac{1}{2}}{+1}{cyan}{red}{cyan}{\one \two}\\$[\eta_23][3\eta_1]$} & 
\makecell{\Amp{-\frac{1}{2}}{-\frac{1}{2}}{+1}{red}{red}{brown}{\two \three}\\$-[\eta_23]\langle\eta_31\rangle$} & 
\makecell{\Amp{-\frac{1}{2}}{-\frac{1}{2}}{+1}{cyan}{cyan}{brown}{\one \three}\\$-\langle2\eta_3\rangle[3\eta_1]$} & 
\makecell{\Amp{-\frac{1}{2}}{-\frac{1}{2}}{+1}{red}{red}{cyan}{\threethree}\\$\langle2\eta_3\rangle\langle\eta_31\rangle$} \\
\hline
$(+\frac{1}{2},+\frac{1}{2},0)$ & 
\makecell{\Amp{+\frac{1}{2}}{+\frac{1}{2}}{0}{cyan}{red}{cyan}{\three}\\$[2\eta_3][31]$\\$+[23][\eta_31]$} & 
\makecell{\Amp{+\frac{1}{2}}{+\frac{1}{2}}{0}{red}{red}{brown}{\one}\\$-[23]\langle3\eta_1\rangle$\\ \Amp{+\frac{1}{2}}{+\frac{1}{2}}{0}{red}{red}{brown}{\one \threethree}\\$[2\eta_3]\langle\eta_3\eta_1\rangle$} & 
\makecell{\Amp{+\frac{1}{2}}{+\frac{1}{2}}{0}{cyan}{cyan}{brown}{\two}\\$-\langle\eta_23\rangle[31]$\\ \Amp{+\frac{1}{2}}{+\frac{1}{2}}{0}{cyan}{cyan}{brown}{\two \threethree}\\$\langle\eta_2\eta_3\rangle[\eta_31]$} & 
\makecell{\Amp{+\frac{1}{2}}{+\frac{1}{2}}{0}{red}{red}{cyan}{\one \two \three}\\$\langle\eta_2\eta_3\rangle\langle3\eta_1\rangle$\\$+\langle\eta_23\rangle\langle\eta_3\eta_1\rangle$} \\
\hline
$(-\frac{1}{2},+\frac{1}{2},0)$ & 
\makecell{\Amp{-\frac{1}{2}}{+\frac{1}{2}}{0}{cyan}{red}{cyan}{\one \three}\\$[2\eta_3][3\eta_1]$\\$+[23][\eta_3\eta_1]$} & 
\makecell{\Amp{-\frac{1}{2}}{+\frac{1}{2}}{0}{red}{red}{brown}{}\\$[23]\langle31\rangle$\\ \Amp{-\frac{1}{2}}{+\frac{1}{2}}{0}{red}{red}{brown}{\threethree}\\$-[2\eta_3]\langle\eta_31\rangle$} & 
\makecell{\Amp{-\frac{1}{2}}{+\frac{1}{2}}{0}{cyan}{cyan}{brown}{\one \two}\\$-\langle\eta_23\rangle[3\eta_1]$\\ \Amp{-\frac{1}{2}}{+\frac{1}{2}}{0}{cyan}{cyan}{brown}{\one \two \threethree}\\$\langle\eta_2\eta_3\rangle[\eta_3\eta_1]$} & 
\makecell{\Amp{-\frac{1}{2}}{+\frac{1}{2}}{0}{red}{red}{cyan}{\two \three}\\$\langle\eta_2\eta_3\rangle\langle31\rangle$\\$+\langle\eta_23\rangle\langle\eta_3 1\rangle$} \\
\end{tabular}
\end{equation}
where we show half of the helicity categories of the $F\bar{F}V$ amplitude, and the other half can be easily obtained by the Hermitian conjugation of the amplitudes in the table above.

For the $VVS$ amplitude, we have
\begin{equation}
\begin{tabular}{c|ccc}
\diagbox[width=2.4cm,innerleftsep=.4cm,innerrightsep=.4cm]{$h_i$}{$t_i$} & $(+1,+1,0)$ & $(0,0,0)$ & $(-1,-1,0)$ \\
\hline
$(+1,+1,0)$ & 
\makecell{\Amp{+1}{+1}{0}{cyan}{red}{brown}{}\\$[12]^2$} & 
\makecell{\Amp{+1}{+1}{0}{brown}{brown}{brown}{\one \two}\\$[12]\langle\eta_1\eta_2\rangle$} & 
\makecell{\Amp{+1}{+1}{0}{red}{cyan}{brown}{\oneone \twotwo}\\$\langle\eta_1\eta_2\rangle^2$} \\
\hline
$(0,+1,0)$ & 
\makecell{\Amp{0}{+1}{0}{cyan}{red}{brown}{\one}\\$[12][\eta_12]$} & 
\makecell{\Amp{0}{+1}{0}{brown}{brown}{brown}{\two}\\$-[12]\langle1\eta_2\rangle$\\ \Amp{0}{+1}{0}{brown}{brown}{brown}{\oneone \two}\\$[\eta_12]\langle\eta_1\eta_2\rangle$} & 
\makecell{\Amp{0}{+1}{0}{red}{cyan}{brown}{\one \twotwo}\\$-\langle1\eta_2\rangle\langle\eta_1\eta_2\rangle$} \\
\hline
$(-1,+1,0)$ & 
\makecell{\Amp{-1}{+1}{0}{cyan}{red}{brown}{\oneone}\\$[\eta_1 2]^2$} & 
\makecell{\Amp{-1}{+1}{0}{brown}{brown}{brown}{\one \two}\\$-[\eta_12]\langle1\eta_2\rangle$} & 
\makecell{\Amp{-1}{+1}{0}{red}{cyan}{brown}{\twotwo}\\$\langle1\eta_2\rangle^2$} \\
\hline
$(+1,0,0)$ & 
\makecell{\Amp{+1}{0}{0}{cyan}{red}{brown}{\two}\\$[12][1\eta_2]$} & 
\makecell{\Amp{+1}{0}{0}{brown}{brown}{brown}{\one}\\$-[12]\langle\eta_12\rangle$\\ \Amp{+1}{0}{0}{brown}{brown}{brown}{\one \twotwo}\\$[1\eta_2]\langle\eta_1\eta_2\rangle$} & 
\makecell{\Amp{+1}{0}{0}{red}{cyan}{brown}{\oneone \two}\\$-\langle1\eta_2\rangle\langle\eta_1\eta_2\rangle$} \\
\hline
$(0,0,0)$ & 
\makecell{\Amp{0}{0}{0}{cyan}{red}{brown}{\one \two}\\$[12][\eta_1\eta_2]$\\$+[\eta_12][1\eta_2]$} & 
\makecell{\Amp{0}{0}{0}{brown}{brown}{brown}{}\\$[12]\langle12\rangle$\\ \Amp{0}{0}{0}{brown}{brown}{brown}{\oneone}\\$-[\eta_12]\langle\eta_12\rangle$}
\makecell{\Amp{0}{0}{0}{brown}{brown}{brown}{\twotwo}\\$-[1\eta_2]\langle1\eta_2\rangle$\\ \Amp{0}{0}{0}{brown}{brown}{brown}{\oneone \twotwo}\\$[\eta_1\eta_2]\langle\eta_1\eta_2\rangle$} & 
\makecell{\Amp{0}{0}{0}{red}{cyan}{brown}{\one \two}\\$\langle12\rangle\langle\eta_1\eta_2\rangle$\\$+\langle\eta_12\rangle\langle1\eta_2\rangle$} \\
\end{tabular}
\end{equation}


We then introduce a new notation to illustrate the chirality of massive particles, which has the potential to describe arbitrary spin. We use the solid and dashed lines to represent the right-handed and left-handed chirality respectively, instead of colored lines. 
For massive fermion, we have
\begin{equation} \begin{aligned}
\begin{tikzpicture}[baseline=0.7cm] \begin{feynhand}
\vertex [particle] (i1) at (1.5,0.8) {$i^{+\frac{1}{2}}$};
\vertex [dot] (v1) at (0,0.8) {};
\graph{(v1)--[plain] (i1)};
\end{feynhand} \end{tikzpicture}&=|i],&
\begin{tikzpicture}[baseline=0.7cm] \begin{feynhand}
\vertex [particle] (i1) at (1.5,0.8) {$i^{+\frac{1}{2}}$};
\vertex [dot] (v1) at (0,0.8) {};
\graph{(v1)--[sca] (i1)};
\end{feynhand} \end{tikzpicture}&=-|\eta_i\rangle,\\
\begin{tikzpicture}[baseline=0.7cm] \begin{feynhand}
\vertex [particle] (i1) at (1.5,0.8) {$i^{-\frac{1}{2}}$};
\vertex [dot] (v1) at (0,0.8) {};
\graph{(v1)--[plain] (i1)};
\end{feynhand} \end{tikzpicture}&=|\eta_i],&
\begin{tikzpicture}[baseline=0.7cm] \begin{feynhand}
\vertex [particle] (i1) at (1.5,0.8) {$i^{-\frac{1}{2}}$};
\vertex [dot] (v1) at (0,0.8) {};
\graph{(v1)--[sca] (i1)};
\end{feynhand} \end{tikzpicture}&=|i\rangle.
\end{aligned} \end{equation}
Therefore, the chirality mass insertion can be represented by 
\begin{equation} \begin{aligned}
\begin{tikzpicture}[baseline=0.7cm] \begin{feynhand}
\vertex [particle] (i1) at (1.5,0.8) {$i^{+\frac{1}{2}}$};
\vertex [dot] (v1) at (0,0.8) {};
\vertex (v2) at (0.6,0.8);
\graph{(v1)--[sca,insertion=1] (v2)--[plain] (i1)};
\end{feynhand} \end{tikzpicture}
&=-|\eta_i\rangle \tilde{m}_i,&
\begin{tikzpicture}[baseline=0.7cm] \begin{feynhand}
\vertex [particle] (i1) at (1.5,0.8) {$i^{+\frac{1}{2}}$};
\vertex [dot] (v1) at (0,0.8) {};
\vertex (v2) at (0.6,0.8);
\graph{(v1)--[plain,insertion=1] (v2)--[sca] (i1)};
\end{feynhand} \end{tikzpicture}
&=|i] m_i,\\
\begin{tikzpicture}[baseline=0.7cm] \begin{feynhand}
\vertex [particle] (i1) at (1.5,0.8) {$i^{-\frac{1}{2}}$};
\vertex [dot] (v1) at (0,0.8) {};
\vertex (v2) at (0.6,0.8);
\graph{(v1)--[sca,insertion=1] (v2)--[plain] (i1)};
\end{feynhand} \end{tikzpicture}
&=|i\rangle \tilde{m}_i,&
\begin{tikzpicture}[baseline=0.7cm] \begin{feynhand}
\vertex [particle] (i1) at (1.5,0.8) {$i^{-\frac{1}{2}}$};
\vertex [dot] (v1) at (0,0.8) {};
\vertex (v2) at (0.6,0.8);
\graph{(v1)--[plain,insertion=1] (v2)--[sca] (i1)};
\end{feynhand} \end{tikzpicture}
&=|\eta_i] m_i.\\
\end{aligned} \end{equation}

For massive vector bosons, we use a double line to represent the single-particle state. The massive vector with helicity $+1$ is represented by
\begin{equation} \begin{aligned}
\begin{tikzpicture}[baseline=-0.1cm] \begin{feynhand}
\vertex (a1) at (0,0.05);
\vertex (a3) at (1.15,0.05);
\vertex (b1) at (0,-0.05);
\vertex (b3) at (1.15,-0.05);
\graph{(a1)--[plain] (a3)};
\graph{(b1)--[plain] (b3)};
\vertex [particle] (i1) at (1.5,0) {$i^{+1}$};
\vertex [dot] (v1) at (0,0) {};
\end{feynhand} \end{tikzpicture}=|i]^2,\quad
\begin{tikzpicture}[baseline=-0.1cm] \begin{feynhand}
\vertex (a1) at (0,0.05);
\vertex (a3) at (1.15,0.05);
\vertex (b1) at (0,-0.05);
\vertex (b3) at (1.15,-0.05);
\graph{(a1)--[plain] (a3)};
\graph{(b1)--[sca] (b3)};
\vertex [particle] (i1) at (1.5,0) {$i^{+1}$};
\vertex [dot] (v1) at (0,0) {};
\end{feynhand} \end{tikzpicture}=-|i]|\eta_i\rangle,\quad
\begin{tikzpicture}[baseline=-0.1cm] \begin{feynhand}
\vertex (a1) at (0,0.05);
\vertex (a3) at (1.15,0.05);
\vertex (b1) at (0,-0.05);
\vertex (b3) at (1.15,-0.05);
\graph{(a1)--[sca] (a3)};
\graph{(b1)--[sca] (b3)};
\vertex [particle] (i1) at (1.5,0) {$i^{+1}$};
\vertex [dot] (v1) at (0,0) {};
\end{feynhand} \end{tikzpicture}=-|\eta_i\rangle^2.
\end{aligned} \end{equation}
There are four kinds of chirality mass insertions:
\begin{equation} \begin{aligned}
\begin{tikzpicture}[baseline=-0.1cm] \begin{feynhand}
\vertex (a1) at (0,0.05);
\vertex (a2) at (0.6,0.05);
\vertex (a3) at (1.15,0.05);
\vertex (b1) at (0,-0.05);
\vertex (b3) at (1.15,-0.05);
\graph{(a1)--[sca,insertion=1] (a2)--[plain] (a3)};
\graph{(b1)--[sca] (b3)};
\vertex [particle] (i1) at (1.5,0) {$i^{+1}$};
\vertex [dot] (v1) at (0,0) {};
\end{feynhand} \end{tikzpicture}
&=|\eta_i\rangle^2 \tilde{m}_i,\\
\begin{tikzpicture}[baseline=-0.1cm] \begin{feynhand}
\vertex (a1) at (0,0.05);
\vertex (a2) at (0.6,0.05);
\vertex (a3) at (1.15,0.05);
\vertex (b1) at (0,-0.05);
\vertex (b2) at (0.7,-0.05);
\vertex (b3) at (1.15,-0.05);
\graph{(a1)--[plain,insertion=1] (a2)--[sca] (a3)};
\graph{(b1)--[plain] (b3)};
\vertex [particle] (i1) at (1.5,0) {$i^{+1}$};
\vertex [dot] (v1) at (0,0) {};
\end{feynhand} \end{tikzpicture}
&=|i]^2 m_i,\\
\begin{tikzpicture}[baseline=-0.1cm] \begin{feynhand}
\vertex (a1) at (0,0.05);
\vertex (a2) at (0.6,0.05);
\vertex (a3) at (1.15,0.05);
\vertex (b1) at (0,-0.05);
\vertex (b3) at (1.15,-0.05);
\graph{(a1)--[sca,insertion=1] (a2)--[plain] (a3)};
\graph{(b1)--[plain] (b3)};
\vertex [particle] (i1) at (1.5,0) {$i^{+1}$};
\vertex [dot] (v1) at (0,0) {};
\end{feynhand} \end{tikzpicture}
&=|i]|\eta_i\rangle \tilde{m}_i,\\
\begin{tikzpicture}[baseline=-0.1cm] \begin{feynhand}
\vertex (a1) at (0,0.05);
\vertex (a2) at (0.6,0.05);
\vertex (a3) at (1.15,0.05);
\vertex (b1) at (0,-0.05);
\vertex (b3) at (1.15,-0.05);
\graph{(a1)--[plain,insertion=1] (a2)--[sca] (a3)};
\graph{(b1)--[sca] (b3)};
\vertex [particle] (i1) at (1.5,0) {$i^{+1}$};
\vertex [dot] (v1) at (0,0) {};
\end{feynhand} \end{tikzpicture}
&=|i]|\eta_i\rangle m_i.\\
\end{aligned} \end{equation}
In the last two lines, the $t=\pm 1$ states are generated by acting with spurion mass on the $t=0$ state. The chirality mass insertion can act on the different spinors of the single-particle states, so we can act $\times$ on both lines 
\begin{equation} \begin{aligned}
\begin{tikzpicture}[baseline=-0.1cm] \begin{feynhand}
\vertex (a1) at (0,0.05);
\vertex (a2) at (0.5,0.05);
\vertex (a3) at (1.15,0.05);
\vertex (b1) at (0,-0.05);
\vertex (b2) at (0.8,-0.05);
\vertex (b3) at (1.15,-0.05);
\graph{(a1)--[sca,insertion=1] (a2)--[plain] (a3)};
\graph{(b1)--[sca,insertion=1] (b2)--[plain] (b3)};
\vertex [particle] (i1) at (1.5,0) {$i^{+1}$};
\vertex [dot] (v1) at (0,0) {};
\end{feynhand} \end{tikzpicture}
&=|\eta_i\rangle^2 \tilde{m}_i^2,\\
\begin{tikzpicture}[baseline=-0.1cm] \begin{feynhand}
\vertex (a1) at (0,0.05);
\vertex (a2) at (0.5,0.05);
\vertex (a3) at (1.15,0.05);
\vertex (b1) at (0,-0.05);
\vertex (b2) at (0.7,-0.05);
\vertex (b3) at (1.15,-0.05);
\graph{(a3)--[sca] (a2)--[plain,insertion=0] (a1)};
\graph{(b3)--[sca] (b2)--[plain,insertion=0] (b1)};
\vertex [particle] (i1) at (1.5,0) {$i^{+1}$};
\vertex [dot] (v1) at (0,0) {};
\end{feynhand} \end{tikzpicture}
&=|i]^2 m_i^2.\\
\end{aligned} \end{equation}
There is no $t=0$ state, because double mass insertions $m\tilde{m}$ do not change the chirality.

For the $F\bar{F}V$ amplitude, the highest weight representations are
\begin{equation} \begin{aligned}
\begin{tikzpicture}[baseline=-0.1cm] \begin{feynhand}
\vertex [particle] (a1) at (-1.1,0) {$1^{+\frac{1}{2}}$};
\vertex [particle] (b1) at (0.05*1.732+0.5,0.05+0.5*1.732) {$2^{+\frac{1}{2}}$};
\vertex [particle] (i3) at (0.55,-0.8) {$3^{+1}$};
\vertex (a2) at (0,0);
\vertex (a3) at (0.35,-0.35*1.732);
\vertex (b2) at (0.05*1.732,0.05);
\vertex (b3) at (0.35+0.05*1.732,-0.35*1.732+0.05);
\graph{(a1)--[plain] (a2)--[plain] (a3)};
\graph{(b1)--[plain] (b2)--[plain] (b3)};
\end{feynhand} \end{tikzpicture}&=[23][31],&
\begin{tikzpicture}[baseline=-0.1cm] \begin{feynhand}
\vertex [particle] (a1) at (-1.1,0) {$1^{-\frac{1}{2}}$};
\vertex [particle] (b1) at (0.05*1.732+0.5,0.05+0.5*1.732) {$2^{+\frac{1}{2}}$};
\vertex [particle] (i3) at (0.55,-0.8) {$3^{0}$};
\vertex (a2) at (0,0);
\vertex (a3) at (0.35,-0.35*1.732);
\vertex (b2) at (0.05*1.732,0.05);
\vertex (b3) at (0.35+0.05*1.732,-0.35*1.732+0.05);
\graph{(a1)--[sca] (a2)--[sca] (a3)};
\graph{(b1)--[plain] (b2)--[plain] (b3)};
\end{feynhand} \end{tikzpicture}&=[23]\langle31\rangle,&\\
\begin{tikzpicture}[baseline=-0.1cm] \begin{feynhand}
\vertex [particle] (a1) at (-1.1,0) {$1^{+\frac{1}{2}}$};
\vertex [particle] (b1) at (0.05*1.732+0.5,0.05+0.5*1.732) {$2^{-\frac{1}{2}}$};
\vertex [particle] (i3) at (0.55,-0.8) {$3^{0}$};
\vertex (a2) at (0,0);
\vertex (a3) at (0.35,-0.35*1.732);
\vertex (b2) at (0.05*1.732,0.05);
\vertex (b3) at (0.35+0.05*1.732,-0.35*1.732+0.05);
\graph{(a1)--[plain] (a2)--[plain] (a3)};
\graph{(b1)--[sca] (b2)--[sca] (b3)};
\end{feynhand} \end{tikzpicture}&=\langle23\rangle[31],&
\begin{tikzpicture}[baseline=-0.1cm] \begin{feynhand}
\vertex [particle] (a1) at (-1.1,0) {$1^{-\frac{1}{2}}$};
\vertex [particle] (b1) at (0.05*1.732+0.5,0.05+0.5*1.732) {$2^{-\frac{1}{2}}$};
\vertex [particle] (i3) at (0.55,-0.8) {$3^{-1}$};
\vertex (a2) at (0,0);
\vertex (a3) at (0.35,-0.35*1.732);
\vertex (b2) at (0.05*1.732,0.05);
\vertex (b3) at (0.35+0.05*1.732,-0.35*1.732+0.05);
\graph{(a1)--[sca] (a2)--[sca] (a3)};
\graph{(b1)--[sca] (b2)--[sca] (b3)};
\end{feynhand} \end{tikzpicture}&=\langle23\rangle\langle31\rangle.&\\
\end{aligned} \end{equation}

\section{The Young diagram formulation of single-particle states}
\label{app:Young}

In this section, we introduce a new formulation of single-particle states and wave functions, realized by the semi-standard Young diagrams. The Young diagrams are used to label both the $SU(2)$ LG representations and the $SO(2) \times SO(3,1)$ reps, with the building blocks given by
\eq{
{\color{orange} \yng(1)} \sim \tilde{\lambda}_{\dot{\alpha}}^I, \quad \yng(1) \sim \lambda_{\alpha}^I,
}
where each box represents an $SU(2)$ index $I$ and a spinor index $\alpha$ or $\dot{\alpha}$, and the color distinguishes the transversality. 

The generic Young diagram consists of at most two rows, and the orange boxes are always placed ahead of the black ones,\footnote{The requirement is for each row, so there is a possibility that $\tilde{n}_1< \tilde{n}_2$, where some black boxes are placed on the top of some orange ones. } \newline
\eq{\label{eq:YoungD}
&{\arraycolsep=0pt\def\arraystretch{0}
\begin{array}{ccccccccccc}
\overmat{\tilde{n}_1}{ {\color{orange}\yng(1)} & \cdots & {\color{orange}\yng(1)} & {\color{orange}\yng(1)} & \cdots & {\color{orange}\yng(1)} } & \overmat{n_1}{ \yng(1) & \cdots & \yng(1) & \cdots & \yng(1)} \\ 
\undermat{\tilde{n}_2}{ {\color{orange}\yng(1)} & \cdots & {\color{orange}\yng(1)}} & \undermat{n_2}{ \yng(1) & \cdots & \yng(1) & \yng(1) & \cdots & \yng(1)} &&
\end{array}}\\ \\
\sim& \tilde{\lambda}_{\dot{\alpha}_1}^{(I_1} \cdots \tilde{\lambda}_{\dot{\alpha}_{\tilde{n}_1}}^{I_{\tilde{n}_1}} \lambda_{\alpha_1}^{I_{\tilde{n}_1+1}} \cdots \lambda_{\alpha_{n_1}}^{I_{\tilde{n}_1+n_1})} \tilde{\lambda}_{\dot{\beta}_1 I_1} \cdots \tilde{\lambda}_{\dot{\beta}_{\tilde{n}_2} I_{\tilde{n}_2}} \lambda_{\beta_1 I_{\tilde{n}_2+1}} \cdots \lambda_{\beta_{n_2} I_{\tilde{n}_2+n_2}}, \\
&\mbox{with } t=\frac{\tilde{n}_1 + \tilde{n}_2 - n_1 - n_2}{2}, s = \frac{\tilde{n}_1 + n_1 - \tilde{n}_2 - n_2}{2}. \quad
}
It can be read from the structure that there are $(\tilde{n}_1 + n_1 - \tilde{n}_2 - n_2)$ permuted uncontracted $SU(2)$ indices, corresponding to spin-$\left(\frac{\tilde{n}_1 + n_1 - \tilde{n}_2 - n_2}{2}\right)$ structures. The transversality is obtained by subtracting the number of black boxes $(n_1 + n_2)$ from the number of orange boxes $(\tilde{n}_1 + \tilde{n}_2)$, resulting in $\left(\frac{\tilde{n}_1 + \tilde{n}_2 - n_1 - n_2}{2}\right)$ transversality. 

The permutation of the $SU(2)$ indices would be transferred to the spinor indices, such that the Young diagram represents both the representation of $SU(2)$ and the one of $SO(2) \times SO(3,1)$. Therefore, a Young diagram denotes an induced representation of $ISO(2) \times ISO(3,1)$ from the $SU(2)$ LG.

After building the correspondence between the Young diagram and spinor structures, now we discuss the $SO(2) \times SO(3,1)$ representation with the one-row Young diagram
\eq{\label{eq:YD1row}
[t,j_1,j_2] = \left[\frac{\tilde{n}-n}{2}, \frac{n}{2}, \frac{\tilde{n}}{2}\right] \quad
\arraycolsep=0pt\def\arraystretch{1.2}
\begin{array}{cccccc}
  \undermat{\tilde{n}}{ {\color{orange} \yng(1)}\ &\ldots{}&\ {\color{orange}\yng(1)}}  & \undermat{n}{\yng(1)&\ \dots{}\  &\yng(1)} \\
  &&&&&
 \end{array} \sim \tilde{\lambda}_{\dot{\alpha}_1}^{(I_1} \cdots \tilde{\lambda}_{\dot{\alpha}_{\tilde{n}}}^{I_{\tilde{n}}} \lambda_{\alpha_1}^{I_{\tilde{n}+1}} \cdots \lambda_{\alpha_{n}}^{I_{\tilde{n}+n})}, 
}
which is also a spin-$\left(\frac{n+\tilde{n}}{2}\right)$ wave function. The relationship between the wave function and its field is
\eq{
\Psi_{(\dot{\alpha}_1 \cdots \dot{\alpha}_{\tilde{n}}) (\alpha_1 \cdots \alpha_{n})} = \int \frac{d^4 \mathbf{p}}{(2\pi)^4} \delta(\mathbf{p}^2-\mathbf{m}^2) \Big( & \tilde{\lambda}_{\dot{\alpha}_1}^{(I_1} \cdots \tilde{\lambda}_{\dot{\alpha}_{\tilde{n}}}^{I_{\tilde{n}}} \lambda_{\alpha_1}^{I_{\tilde{n}+1}} \cdots \lambda_{\alpha_{n}}^{I_{\tilde{n}+n})} a_{I_1 \cdots I_{\tilde{n}+n}} e^{-i \mathbf{p} \cdot x} \\ & +\tilde{\lambda}_{\dot{\alpha}_1 (I_1} \cdots \tilde{\lambda}_{\dot{\alpha}_{\tilde{n}} I_{\tilde{n}}} \lambda_{\alpha_1 I_{\tilde{n}+1}} \cdots \lambda_{\alpha_{n} I_{\tilde{n}+n})} (a^{\dagger})^{I_1 \cdots I_{\tilde{n}+n}} e^{i \mathbf{p} \cdot x} \Big), 
}
where $a$ and $a^{\dagger}$ are the annihilating and creating operators of a particle.

The two-row structures like eq.~\eqref{eq:YoungD} involve contraction of the $SU(2)$ indices, leading to factors of the 6D momentum 
\eq{
p_{AB} = \begin{pmatrix}
m \epsilon_{\alpha \beta} & -\mathbf{p}_{\beta \dot{\alpha}} \\
\mathbf{p}_{\alpha \dot{\beta}} & -\tilde{m} \tilde{\epsilon}_{\dot{\alpha} \dot{\beta}}
\end{pmatrix},
}
so they are single-particle states instead of wave functions. 
Specifically, the one-column-two-row structures are 
\begin{equation} \begin{aligned}
 {\color{orange} \yng(1,1) }\, &\sim \tilde{\lambda}_{\dot{\alpha}}^{I_1} \tilde{\lambda}_{\dot{\beta} I_1} = -\tilde{m} \tilde{\epsilon}_{\dot{\alpha} \dot{\beta}},& 
 \yng(1,1)\, &\sim \lambda_{\alpha}^{I_1} \lambda_{\beta I_1} = m \epsilon_{\alpha\beta}, \\ 
\renewcommand{\arraystretch}{0}\begin{array}{c}
{\color{orange} \yng(1)} \\
 \yng(1)
\end{array} &\sim \tilde{\lambda}_{\dot{\alpha}}^{I_1} \lambda_{\beta I_1} = -\mathbf{p}_{\beta \dot{\alpha}},& 
\renewcommand{\arraystretch}{0}\begin{array}{c}
 \yng(1) \\ 
{\color{orange} \yng(1)} 
\end{array} &\sim \lambda_{\alpha}^{I_1} \tilde{\lambda}_{\dot{\beta} I_1} = \mathbf{p}_{\alpha \dot{\beta}}. \label{eq:LGsinglet}
\end{aligned} \end{equation}

The two-row structures like eq.~\eqref{eq:YoungD} usually do not belong to a single irreducible representation of $SO(2) \times SO(3,1)$ when $n_1 \geq n_2$ or $\tilde{n}_1 \geq \tilde{n}_2$, but instead a direct sum of irreducible representations,\newline
\eq{\label{eq:YD-so4_sum}
& \arraycolsep=0pt\def\arraystretch{0}
\begin{array}{ccccccccccc}
\overmat{\tilde{n}_1}{ {\color{orange}\yng(1)} & \cdots & {\color{orange}\yng(1)} & {\color{orange}\yng(1)} & \cdots & {\color{orange}\yng(1)} } & \overmat{n_1}{ \yng(1) & \cdots & \yng(1) & \cdots & \yng(1)} \\
\undermat{\tilde{n}_2}{ {\color{orange}\yng(1)} & \cdots & {\color{orange}\yng(1)}} & \undermat{n_2}{ \yng(1) & \cdots & \yng(1) & \yng(1) & \cdots & \yng(1)} &&
\end{array} \\ & \\
\sim& \arraycolsep=1pt\def\arraystretch{1}
\begin{cases}
\mathop{\bigoplus}\limits_{0 \leq k \leq n_1-n_2} 
\begin{cases}
\left[ \frac{\tilde{n}_1 + \tilde{n}_2 - n_1 - n_2}{2}, 
\frac{n_1-n_2}{2}, 
\frac{\tilde{n}_1 -\tilde{n}_2 +2(k-n_2)}{2}
\right] & k \geq n_2 \\
\left[ \frac{\tilde{n}_1 + \tilde{n}_2 - n_1 - n_2}{2}, 
\frac{n_1+n_2-2k}{2}, 
\frac{\tilde{n}_1-\tilde{n}_2}{2}
\right] & k < n_2
\end{cases} & n_1 \geq n_2 \\
\mathop{\bigoplus}\limits_{0 \leq k \leq \tilde{n}_1-\tilde{n}_2}
\begin{cases}
\left[ \frac{\tilde{n}_1 + \tilde{n}_2 - n_1 - n_2}{2}, 
\frac{n_1-n_2}{2}, 
\frac{\tilde{n}_1+\tilde{n}_2-2k}{2}
\right] & k < \tilde{n}_2 \\
\left[ \frac{\tilde{n}_1 + \tilde{n}_2 - n_1 - n_2}{2}, 
\frac{n_1 -n_2 +2(k-\tilde{n}_2)}{2}, 
\frac{\tilde{n}_1-\tilde{n}_2}{2}
\right] & k \geq \tilde{n}_2 
\end{cases}
& \tilde{n}_1 \geq \tilde{n}_2
\end{cases}
}
which is inferred from the correspondence between the Young diagram and its spinor structure given by eq.~\eqref{eq:YoungD}. For example, 
\eq{
\arraycolsep=0pt\def\arraystretch{0}
\begin{array}{ccc}
{\color{orange} \yng(1)} & \yng(1) & \yng(1) \\
{\color{orange} \yng(1)} & 
\end{array}
\sim \tilde{\lambda}_{\dot{\alpha}}^{(I_1} \lambda_{\alpha_1}^{I_2} \lambda_{\alpha_2}^{I_3)} \tilde{\lambda}_{\dot{\beta} I_1}
= -\frac{1}{3} \tilde{m} \tilde{\epsilon}_{\dot{\alpha} \dot{\beta}} \lambda_{\alpha_1}^{(I_2} \lambda_{\alpha_2}^{I_3)}
+\frac{2}{3} \tilde{\lambda}_{\dot{\alpha}}^{(I_2} \lambda_{(\alpha_1}^{I_3)} \mathbf{p}_{\alpha_2) \dot{\beta}}
\sim [0,1,0] \oplus [0,1,1]. }
The one-to-one correspondence between the Young diagram and irreducible representation only happens when $n_1=0$, $\tilde{n}_1 =0$ or $n_2=\tilde{n}_2=0$, \newline\newline
\eq{
[t,j_1,j_2] 
\sim & \begin{cases}
\arraycolsep=0pt\def\arraystretch{0}
\begin{array}{cccccccc}
\overmat{j_1 +j_2 -t}{\yng(1) & \cdots & \yng(1) & \yng(1) & \cdots & \yng(1) & \cdots & \yng(1)} \\
\undermat{2 j_2}{ {\color{orange} \yng(1)} & \cdots & {\color{orange} \yng(1)}} & \undermat{\tiny j_2 -j_1 -t}{\yng(1) & \cdots & \yng(1)} &&
\end{array} & \tilde{n}_1=0, \\
& \\ & \\
\arraycolsep=0pt\def\arraystretch{0}
\begin{array}{cccccccc}
\overmat{j_1 +j_2 +t}{ {\color{orange}\yng(1)} & \cdots & {\color{orange}\yng(1)} & {\color{orange}\yng(1)} & \cdots & {\color{orange}\yng(1)} & \cdots & {\color{orange}\yng(1)} } \\
\undermat{\tiny j_1 -j_2 +t}{ {\color{orange}\yng(1)} & \cdots & {\color{orange}\yng(1)}} & \undermat{2 j_1}{ \yng(1) & \cdots & \yng(1)} &&
\end{array} & n_1=0, \\
& \\ & \\
\arraycolsep=0pt\def\arraystretch{0}
\begin{array}{cccccc}
\overmat{j_2}{{\color{orange} \yng(1)} & \cdots & {\color{orange} \yng(1)} } & \overmat{j_1}{\yng(1) & \cdots & \yng(1)}
\end{array} & n_2=\tilde{n}_2=0.
\end{cases} 
}

The single-particle state $|t,s,h\rangle$ can be obtained by filling the Young tableau of the irreducible representation $[t,j_1,j_2]$ with appropriate values. There are only two possible values, $+$ and $-$, corresponding to the $SU(2)$ states $\zeta^{+I}$ and $\zeta^{-I}$ respectively: 
\eq{
{\color{orange} \young(+)} \sim \tilde{\lambda}_{\dot{\alpha}} \zeta^{+I}, \ 
{\color{orange} \young(-)} \sim \tilde{\eta}_{\dot{\alpha}} \zeta^{-I}, \\ 
\young(-) \sim -\lambda_{\alpha} \zeta^{-I}, \ 
\young(+) \sim \eta_{\alpha} \zeta^{+I}. 
}
The rule for filling the Young tableau, subject to the constraints of Young's orthogonality relations, requires that $+$ always appears to the left of $-$ in each row, and no column can contain repeated values.

We illustrate the Young tableau using examples for spin-1, showing how the states are arranged according to Young's orthogonality relations. The $[t,j_1,j_2]=\left[-1,1,0\right]$ representation with all possible helicity states is given by
\eq{
\begin{tabular}{ccl}
$h=+1$ & $\young(++)$ & $\sim \lambda_{\alpha_1}^{(+} \lambda_{\alpha_2}^{+)} = \eta_{(\alpha_1} \eta_{\alpha_2)} \zeta^{+(I_1} \zeta^{+I_2)}$, \\
$h=0$ & $\young(+-)$ & $\sim \lambda_{\alpha_1}^{(+} \lambda_{\alpha_2}^{-)} = -\lambda_{(\alpha_1} \eta_{\alpha_2)} \zeta^{+(I_1} \zeta^{-I_2)}$, \\
$h=-1$ & $\young(--)$ & $\sim \lambda_{\alpha_1}^{(-} \lambda_{\alpha_2}^{-)} = \lambda_{(\alpha_1} \lambda_{\alpha_2)} \zeta^{-(I_1} \zeta^{-I_2)}$.
\end{tabular}
}
where $\begin{cases}
\lambda_{\alpha}^{+} \equiv \eta_{\alpha} \zeta^{+I} \\
\lambda_{\alpha}^{-} \equiv -\lambda_{\alpha} \zeta^{-I}
\end{cases} $ labels the components of the little group index $I$.
Similarly, the helicity states of $\left[0,\frac{1}{2},\frac{1}{2}\right]$ are given by
\eq{
\begin{tabular}{ccl}
$h=+1$ & $\arraycolsep=0pt\def\arraystretch{0}\begin{array}{cc}
{\color{orange} \young(+)} & \young(+)
\end{array}$ & $\sim \tilde{\lambda}_{\dot{\alpha}}^{(+} \lambda_{\alpha}^{+)} = \tilde{\lambda}_{\dot{\alpha}} \eta_{\alpha} \zeta^{+(I_1} \zeta^{+I_2)} $, \\
$h=0$ & $\arraycolsep=0pt\def\arraystretch{0}\begin{array}{cc}
{\color{orange} \young(+)} & \young(-)
\end{array}$ & $\sim \tilde{\lambda}_{\dot{\alpha}}^{(+} \lambda_{\alpha}^{-)} = (\tilde{\eta}_{\dot{\alpha}} \eta_{\alpha} -\tilde{\lambda}_{\dot{\alpha}} \lambda_{\alpha}) \zeta^{+(I_1} \zeta^{-I_2)} $, \\
$h=-1$ & $\arraycolsep=0pt\def\arraystretch{0}\begin{array}{cc}
{\color{orange} \young(-)} & \young(-)
\end{array}$ & $\sim \tilde{\lambda}_{\dot{\alpha}}^{(-} \lambda_{\alpha}^{-)} = -\tilde{\eta}_{\dot{\alpha}} \lambda_{\alpha} \zeta^{-(I_1} \zeta^{-I_2)} $.
\end{tabular}
}
where $\begin{cases}
\tilde{\lambda}_{\dot{\alpha}}^{+} \equiv \tilde{\lambda}_{\dot{\alpha}} \zeta^{+I} \\
\tilde{\lambda}_{\dot{\alpha}}^{-} \equiv \tilde{\eta}_{\dot{\alpha}} \zeta^{-I}
\end{cases}$. 


\bibliographystyle{JHEP}
\bibliography{ref}

\end{document}